\documentclass[11pt]{article}

\RequirePackage{amssymb}
\RequirePackage{amsmath}
\RequirePackage{latexsym}
\RequirePackage{mathrsfs}

\DeclareMathOperator*{\argmax}{argmax}
\DeclareMathOperator*{\supp}{supp}

\newcommand{\A}{\ensuremath{\mathbf{A}}}
\newcommand{\Z}{\ensuremath{\mathbf{Z}}}

\def\avg{\frac{1}{N} \sum_{i=1}^N}
\renewcommand{\P}{\ensuremath{\mathbb{P}}}

\newcommand{\R}{\ensuremath{\mathbb{R}}}
\newcommand{\Exp}{\ensuremath{\mathbb{E}}}
\newcommand{\Prob}{\ensuremath{\mathbb{P}}}

\usepackage{bbm}
\newcommand{\indicator}{\ensuremath{\mathbbm{1}}}
\newcommand{\ind}{\ensuremath{\mathbbm{1}}}

\newcommand\independent{\protect\mathpalette{\protect\independenT}{\perp}}
\def\independenT#1#2{\mathrel{\rlap{$#1#2$}\mkern2mu{#1#2}}}
\newcommand\indep{\independent}

\usepackage{color}

\usepackage{wrapfig}
\usepackage{tikz-cd}
\usepackage{tikz}
\usepackage{pgfplots}

\RequirePackage{amsthm}

\theoremstyle{definition}

\newtheorem{proposition}{Proposition}
\newtheorem{lemma}{Lemma}

\newtheorem{definition}{Definition}
\newtheorem{theorem}{Theorem}
\newtheorem{corollary}{Corollary}

\newcounter{partialIndepSection}
\newcommand{\aAssump}{A\arabic{partialIndepSection}}

\newtheoremstyle{theoremSuppressedNumber}{}{}{}{}{\bfseries}{.}{ }{\thmname{#1}\thmnote{ (\mdseries #3)}}
\theoremstyle{theoremSuppressedNumber}
\newtheorem{partialIndepAssump}{Assumption \aAssump \addtocounter{partialIndepSection}{1}}

\usepackage{setspace}
\usepackage[longnamesfirst]{natbib}
\usepackage{ulem}
\usepackage{subfig}
\usepackage{wrapfig}
\usepackage{float}

\usepackage{epstopdf}

\pdfoutput=1
\usepackage{hyperref}
\hypersetup{
    colorlinks=true,
    linkcolor=black,
    citecolor=black,
    filecolor=black,
    urlcolor=black,
}

\usepackage{xr-hyper}
\title{\textbf{Inference on Breakdown Frontiers}\footnote{This paper was presented at the 2017 IAAE Conference, the 2017 ``Inference in Nonstandard Problems'' CEME Conference, the 2017 University of Chicago Interactions Workshop, the 2017 Midwest Econometrics Group, the 2017 Southern Economic Association Meetings, the 2018 Winter Meeting of the Econometric Society, Duke, UC Irvine, the University of Pennsylvania, UC Berkeley, Yale, the University of Georgia, Auburn University, the University of Colorado Boulder, Georgetown University, UNC-Chapel Hill, Rice University,  Boston University, University College London, University of Cambridge, and the LSE. We thank audiences at those conferences and seminars, as well as Federico Bugni, Ivan Canay, Joachim Freyberger, Guido Imbens, Ying-Ying Lee, Chuck Manski, Anna Mikusheva, Jim Powell, Pedro Sant'Anna, and Andres Santos for helpful conversations and comments. We thank Margaux Luflade for excellent research assistance. This paper uses data from phase 1 of SWAY, the Survey of War Affected Youth in northern Uganda. We thank the SWAY principal researchers Jeannie Annan and Chris Blattman for making their data publicly available and for answering our questions.}}

\author{Matthew A. Masten\footnote{Department of Economics, Duke University,
        \texttt{matt.masten@duke.edu}} \qquad Alexandre Poirier\thanks{
    Department of Economics, Georgetown University,
    \texttt{alexandre.poirier@georgetown.edu}}
}

\begin{document}
\maketitle
\begin{abstract}
Given a set of baseline assumptions, a breakdown frontier is the boundary between the set of assumptions which lead to a specific conclusion and those which do not. In a potential outcomes model with a binary treatment, we consider two conclusions: First, that ATE is at least a specific value (e.g., nonnegative) and second that the proportion of units who benefit from treatment is at least a specific value (e.g., at least 50\%). For these conclusions, we derive the breakdown frontier for two kinds of assumptions: one which indexes relaxations of the baseline random assignment of treatment assumption, and one which indexes relaxations of the baseline rank invariance assumption. These classes of assumptions nest both the point identifying assumptions of random assignment and rank invariance and the opposite end of no constraints on treatment selection or the dependence structure between potential outcomes. This frontier provides a quantitative measure of robustness of conclusions to relaxations of the baseline point identifying assumptions. We derive $\sqrt{N}$-consistent sample analog estimators for these frontiers. We then provide two asymptotically valid bootstrap procedures for constructing lower uniform confidence bands for the breakdown frontier. As a measure of robustness, estimated breakdown frontiers and their corresponding confidence bands can be presented alongside traditional point estimates and confidence intervals obtained under point identifying assumptions. We illustrate this approach in an empirical application to the effect of child soldiering on wages. We find that sufficiently weak conclusions are robust to simultaneous failures of rank invariance and random assignment, while some stronger conclusions are fairly robust to failures of rank invariance but not necessarily to relaxations of random assignment.
\end{abstract}



\onehalfspacing
\normalsize

\newpage
\onehalfspacing

\section{Introduction}\label{sec:intro}

Traditional empirical analysis combines the observed data with a set of assumptions to draw conclusions about a parameter of interest. \emph{Breakdown frontier} analysis reverses this ordering. It begins with a fixed conclusion and a set of baseline assumptions and asks, `What are the weakest assumptions needed to draw that conclusion, given the observed data?' For example, consider the impact of a binary treatment on some outcome variable. The traditional approach might assume random assignment, point identify the average treatment effect (ATE), and then report the obtained value. The breakdown frontier approach instead begins with a conclusion about ATE, like `ATE is positive', and reports the weakest assumption---relative to random assignment---on the relationship between treatment assignment and potential outcomes needed to obtain this conclusion, when such an assumption exists. When more than one kind of assumption is considered, this approach leads to a curve, representing the weakest combinations of assumptions which lead to the desired conclusion. This curve is the breakdown frontier.

At the population level, the difference between the traditional approach and the breakdown frontier approach is a matter of perspective: an answer to one question is an answer to the other. This relationship has long been present in the literature initiated by Manski on partial identification (for example, see \citealt{Manski2007} or section 3 of \citealt{Manski2013}). In finite samples, however, which approach one chooses has important implications for how one does statistical inference. Specifically, the traditional approach estimates the parameter or its identified set. Here we instead estimate the breakdown frontier. The traditional approach then performs inference on the parameter or its identified set. Here we instead perform inference on the breakdown frontier. Thus the breakdown frontier approach puts the weakest assumptions necessary to draw a conclusion at the center of attention. Consequently, by construction, this approach avoids the non-tight bounds critique of partial identification methods (for example, see section 7.2 of \citealt{HoRosen2016}). One distinction is that the traditional approach may require inference on a partially identified parameter. The breakdown frontier approach, however, only requires inference on a point identified object.

The breakdown frontier we study generalizes the concept of an ``identification breakdown point'' introduced by \cite{HorowitzManski1995}, a one dimensional breakdown frontier.\footnote{The \emph{identification} breakdown point is distinct from the breakdown point introduced earlier by \cite{Hampel1968,Hampel1971} in the robust statistics literature that began with \cite{Huber1964}; also see \cite{DonohoHuber1983}. \cite{HorowitzManski1995} give a detailed comparison of the two concepts. Throughout this paper we use the term ``breakdown'' in the same sense as Horowitz and Manski's identification breakdown point.} Their breakdown point was further studied and generalized by \cite{Stoye2005,Stoye2010}. Our emphasis on inference on the breakdown frontier follows \cite{KlineSantos2013}, who proposed doing inference on a breakdown point. Finally, our focus on multi-dimensional frontiers builds on the graphical sensitivity analysis of \cite{Imbens2003} and the multi-dimensional sensitivity analysis of \cite{ManskiPepper2018}. We discuss these papers and others in detail in appendix \ref{sec:relatedLit}.

\subsection*{The breakdown frontier approach}

The breakdown frontier approach requires six main steps: (a) specify a parameter of interest, (b) specify a set of baseline assumptions, (c) define a class of assumptions indexed by a sensitivity parameter which deliver a nested sequence of identified sets, with the baseline assumptions obtained at one extreme and the no assumptions bounds obtained at the other, (d) characterize identified sets for the parameter of interest as a function of the sensitivity parameter, (e) use those identified sets to define the breakdown frontier for a conclusion of interest, and (f) develop estimation and inference procedures for that frontier based on its characterization.

In principle this analysis can be done for a general class of models, for example, by using the general identification analysis in \cite{ChesherRosen2015} or \cite{Torgovitsky2015} for step (d), and then applying general tools for nonparametric estimation and inference for step (f). While such a general analysis is an important next step for future work, in this paper we focus on just one important and widely used model: the potential outcomes model with a binary treatment. By focusing on a single concrete model we can clearly illustrate how to do the six main steps (a)--(f) required for a breakdown frontier analysis in any model. While the mathematical analysis will differ from model to model, the general ideas and approach do not.

In the rest of this section and this paper, we use the binary treatment potential outcomes model to explain and illustrate the breakdown frontier approach. Our main parameter of interest is the proportion of units who benefit from treatment. Under random assignment of treatment and rank invariance, this parameter is point identified. One may be concerned, however, that these two assumptions are too strong. We relax rank invariance by supposing that there are two types of units in the population: one type for which rank invariance holds and another type for which it may not. The proportion $t$ of the second type measures the relaxation of rank invariance. We relax random assignment using a propensity score distance $c \geq 0$ as in our previous work, \cite{MastenPoirier2017}. We give more details on both of these relaxations in section \ref{sec:Model}. We derive the identified set for $\Prob(Y_1 > Y_0)$ as a function of $(c,t)$. For a specific conclusion, such as $\Prob(Y_1 > Y_0) \geq 0.5$, this identification result defines a breakdown frontier.

\begin{figure}[t]
\centering
\includegraphics[width=65mm]{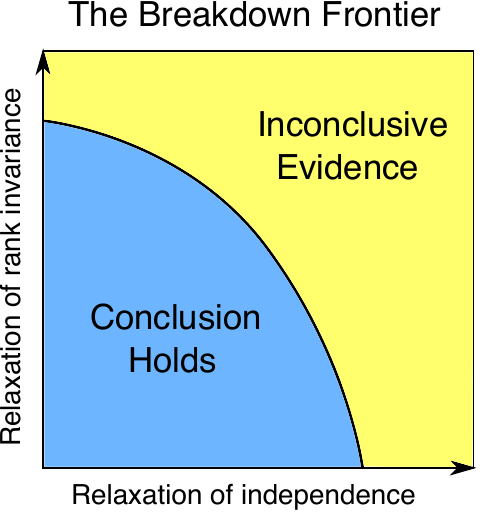}
\caption{An example breakdown frontier, partitioning the space of assumptions into the set for which our conclusion of interest holds (the robust region) and the set for which our evidence is inconclusive.}
\label{BF_illustration}
\end{figure}

Figure \ref{BF_illustration} illustrates this breakdown frontier. The horizontal axis measures $c$, the relaxation of the random assignment assumption. The vertical axis measures $t$, the relaxation of rank invariance. The origin represents the baseline point identifying assumptions of random assignment and rank invariance. Points along the vertical axis represent random assignment paired with various relaxations of rank invariance. Points along the horizontal axis represent rank invariance paired with various relaxations of random assignment. Points in the interior of the box represent relaxations of both assumptions. The points in the lower left region are pairs of assumptions $(c,t)$ such that the data allow us to draw our desired conclusion: $\Prob(Y_1 > Y_0) \geq 0.5$. We call this set the \emph{robust region}. Specifically, no matter what value of $(c,t)$ we choose in this region, the identified set for $\Prob(Y_1 > Y_0)$ always lies completely above 0.5. The points in the upper right region are pairs of assumptions that do not allow us to draw this conclusion. For these pairs $(c,t)$ the identified set for $\Prob(Y_1 > Y_0)$ contains elements smaller than 0.5. The boundary between these two regions is precisely the breakdown frontier. The area under the breakdown frontier---the robust region---is a quantitative measure of robustness.

\begin{figure}[t]
  \centering
  \caption{Example breakdown frontiers}
  \subfloat[For the claim $\Prob(Y_1 > Y_0) \geq \underline{p}$, for five different values of $\underline{p}$.]{
  \label{BF_illustration2}
\includegraphics[width=30mm]{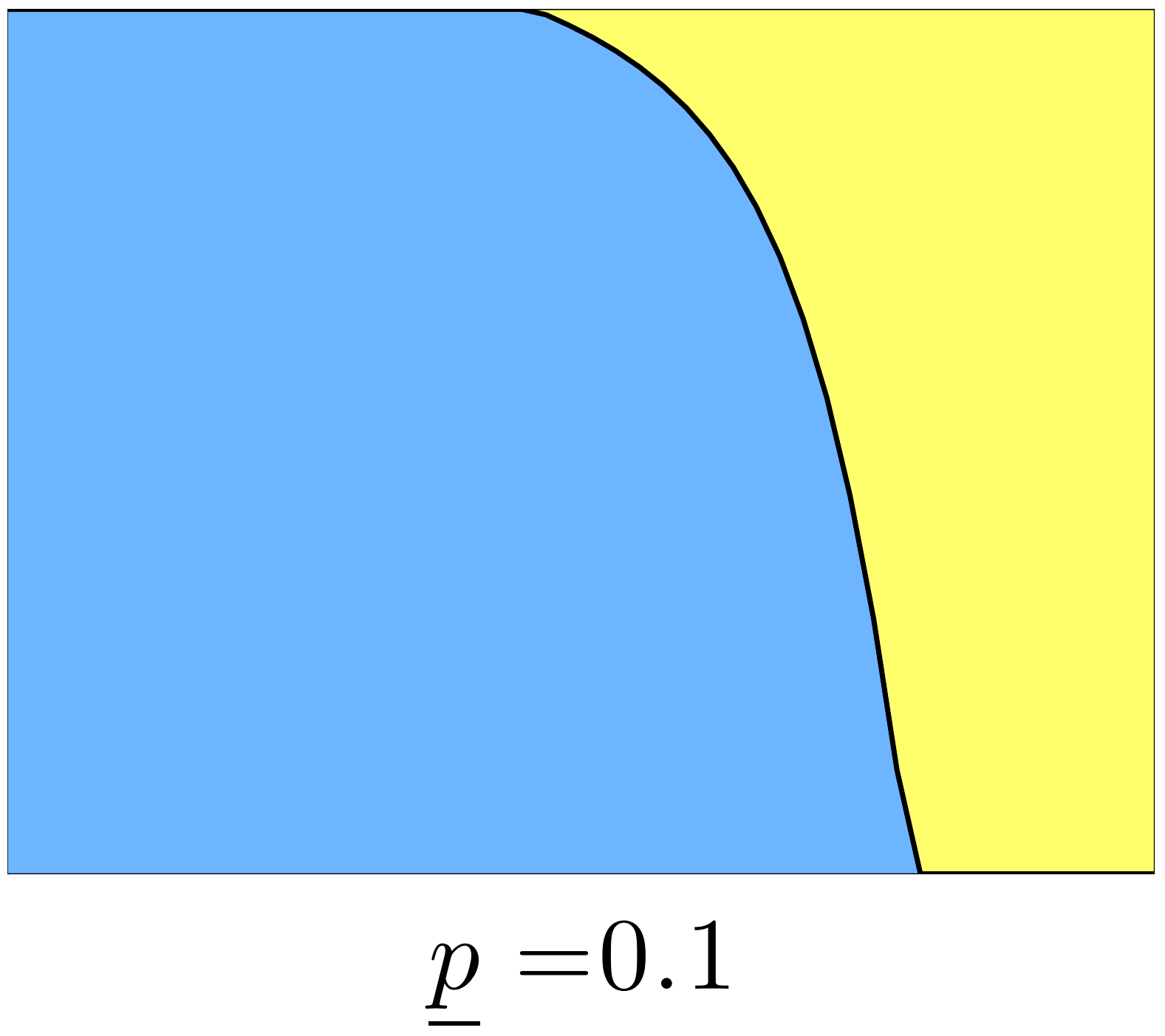}
\includegraphics[width=30mm]{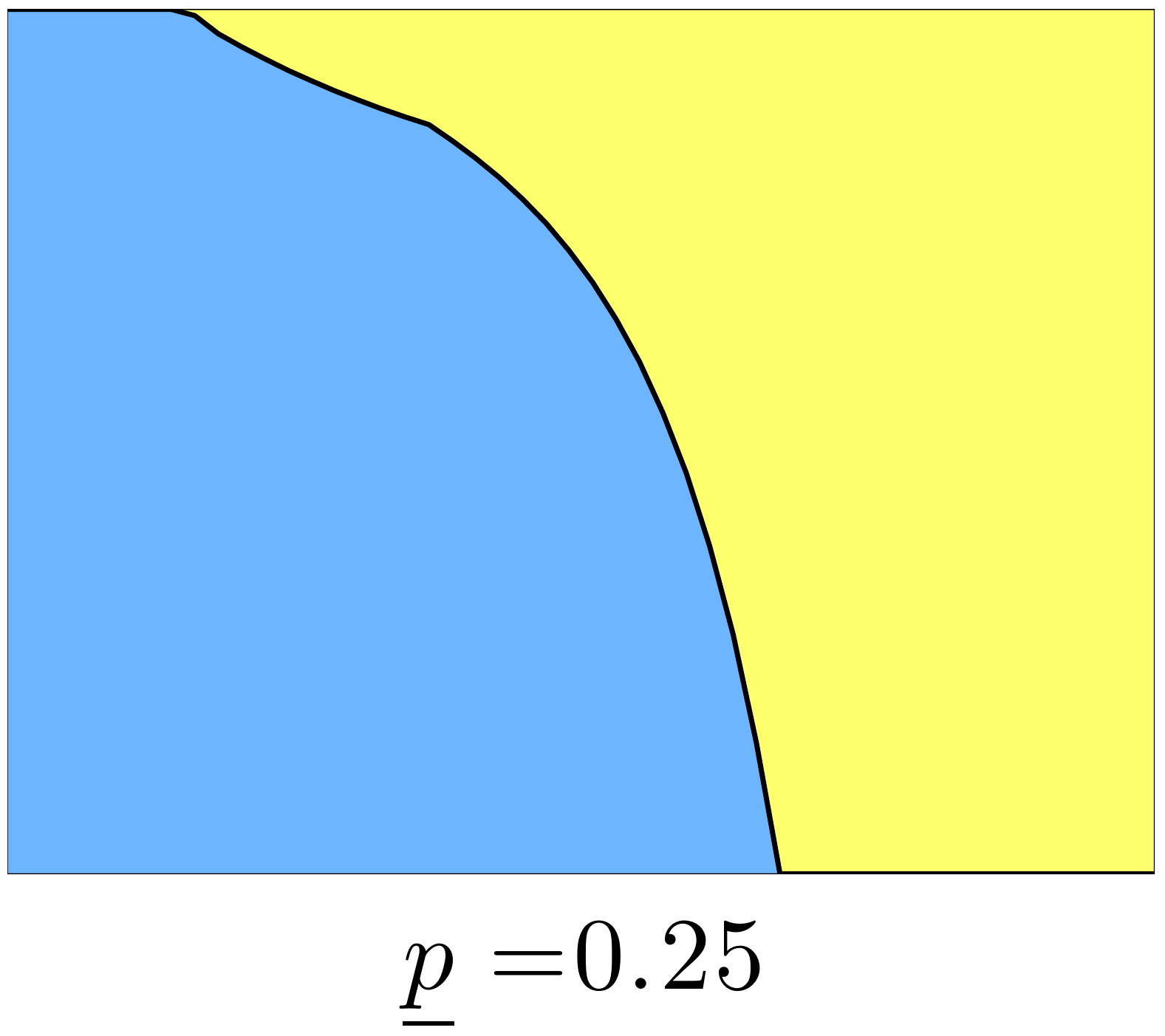}
\includegraphics[width=30mm]{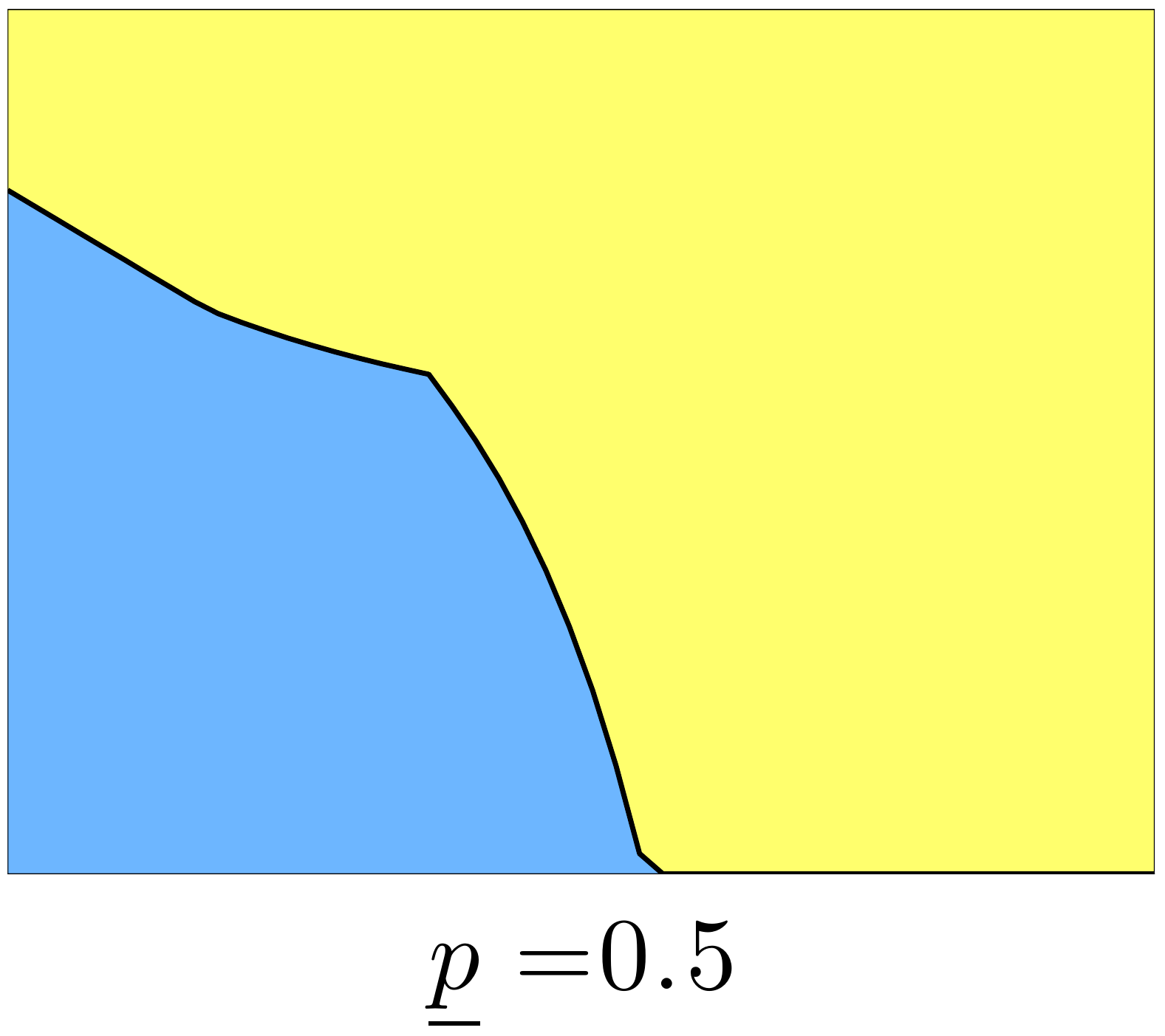}
\includegraphics[width=30mm]{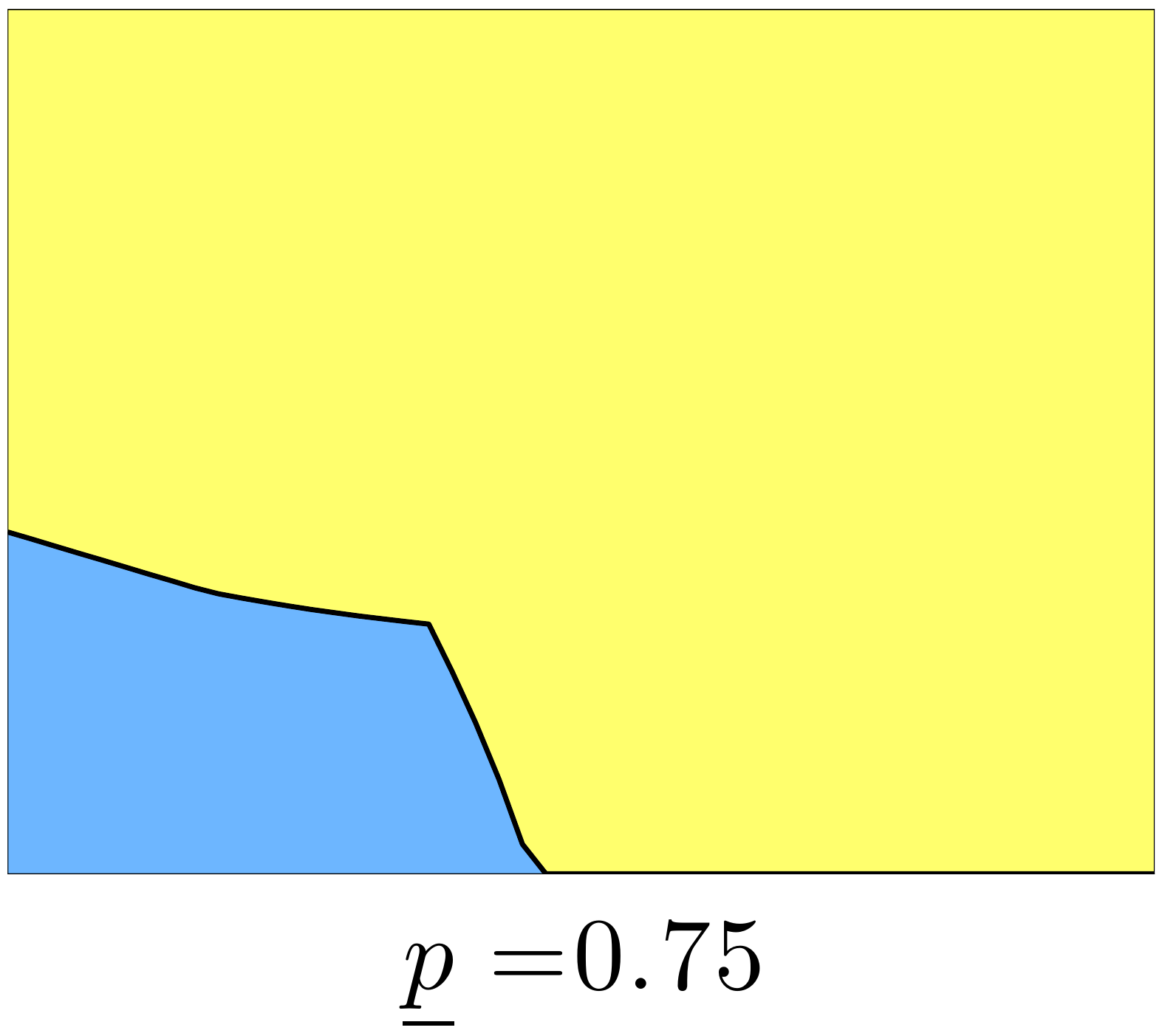}
\includegraphics[width=30mm]{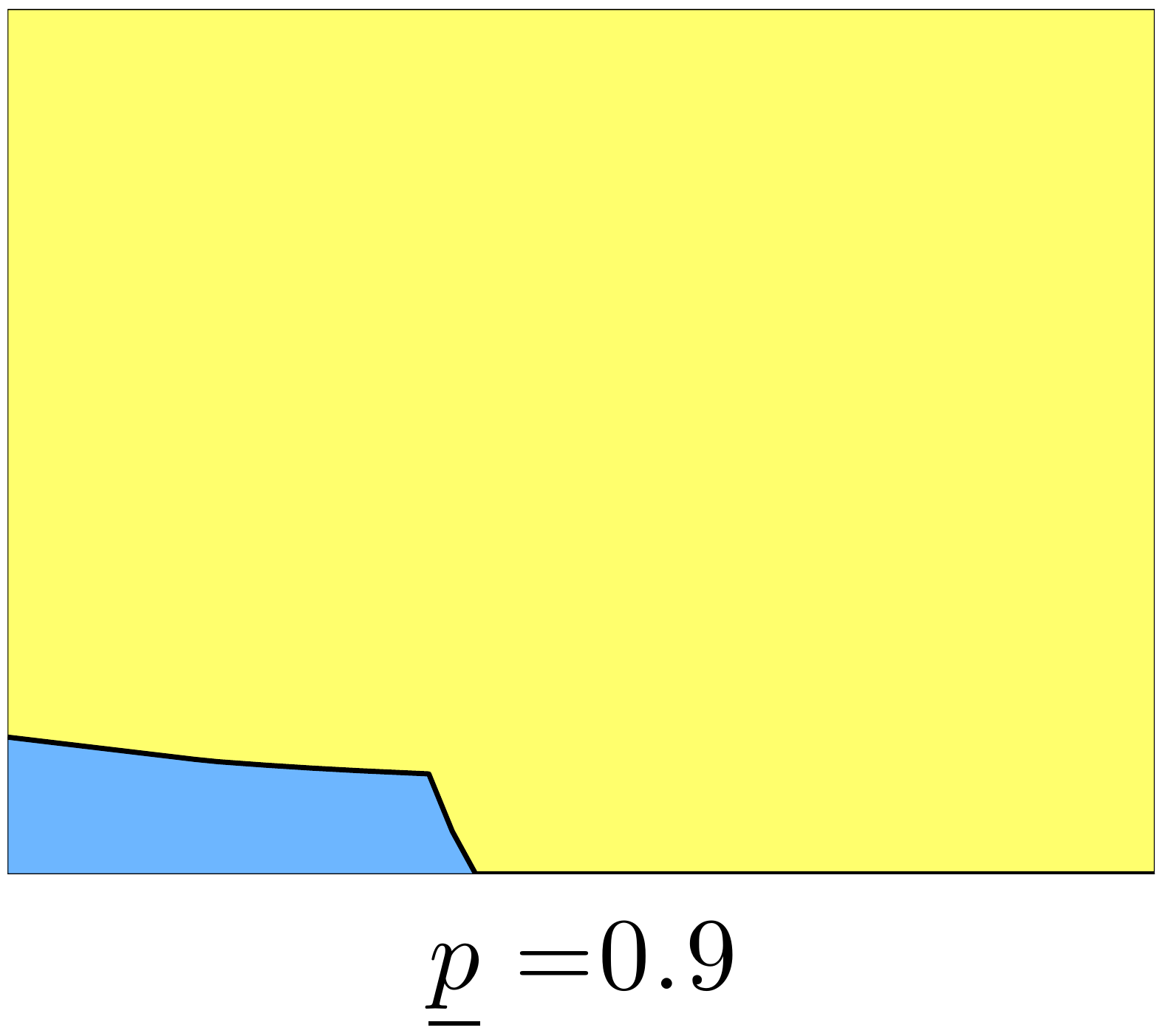}
  }
  \\
  \subfloat[For the claim $\text{ATE} \geq \mu$, for five different values of $\mu$.]{
  \label{BF_illustration3}
\includegraphics[width=30mm]{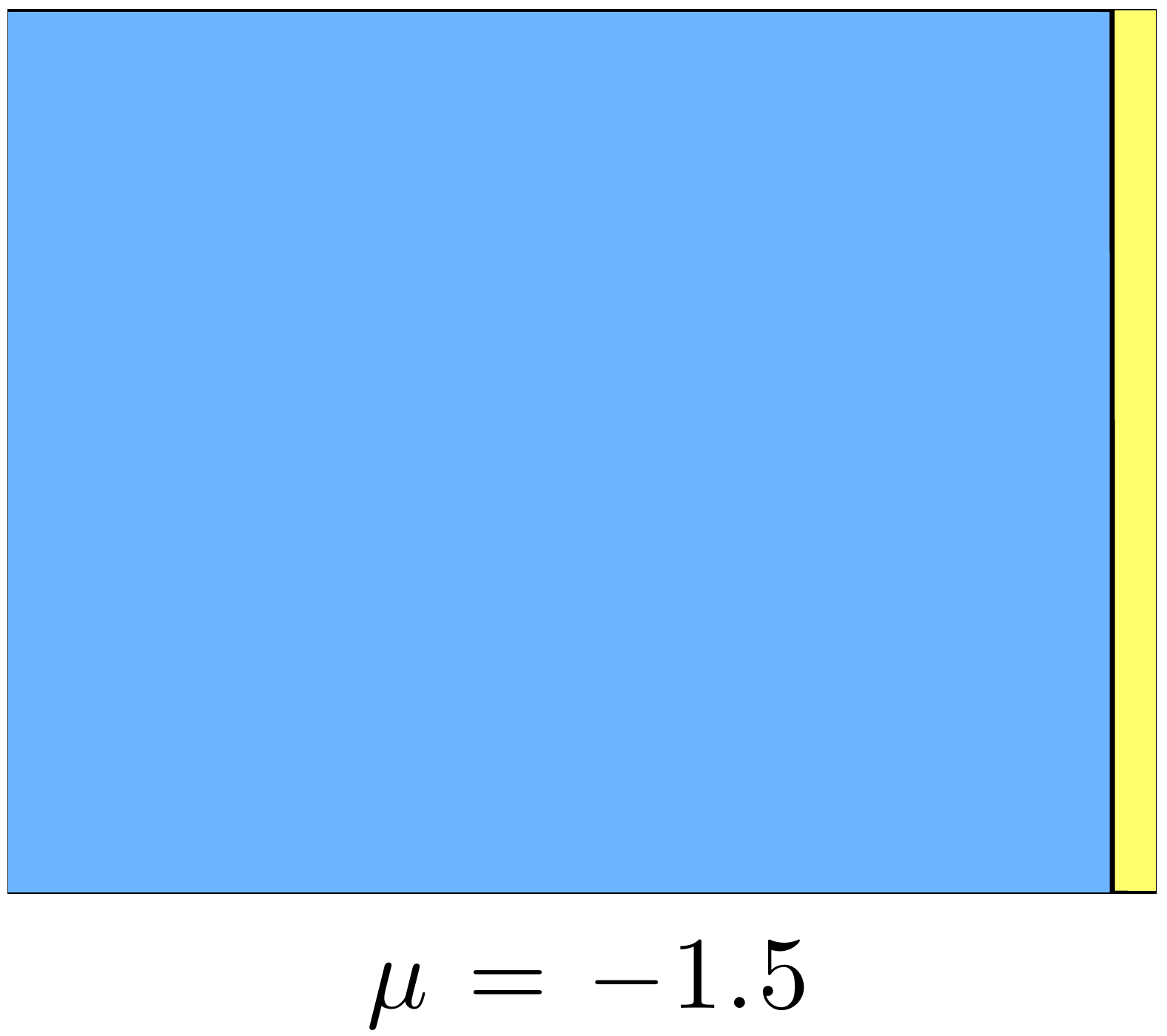}
\includegraphics[width=30mm]{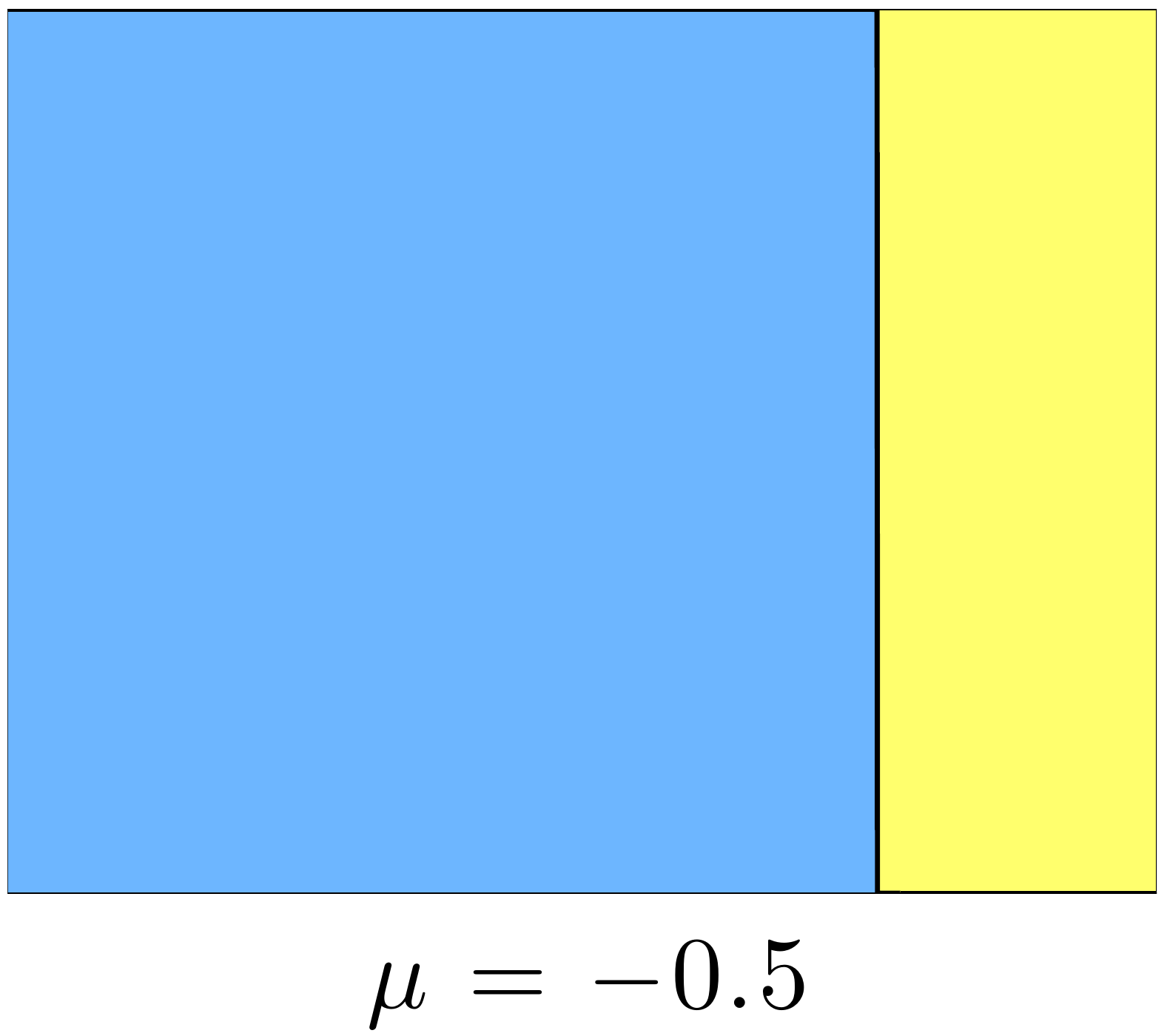}
\includegraphics[width=30mm]{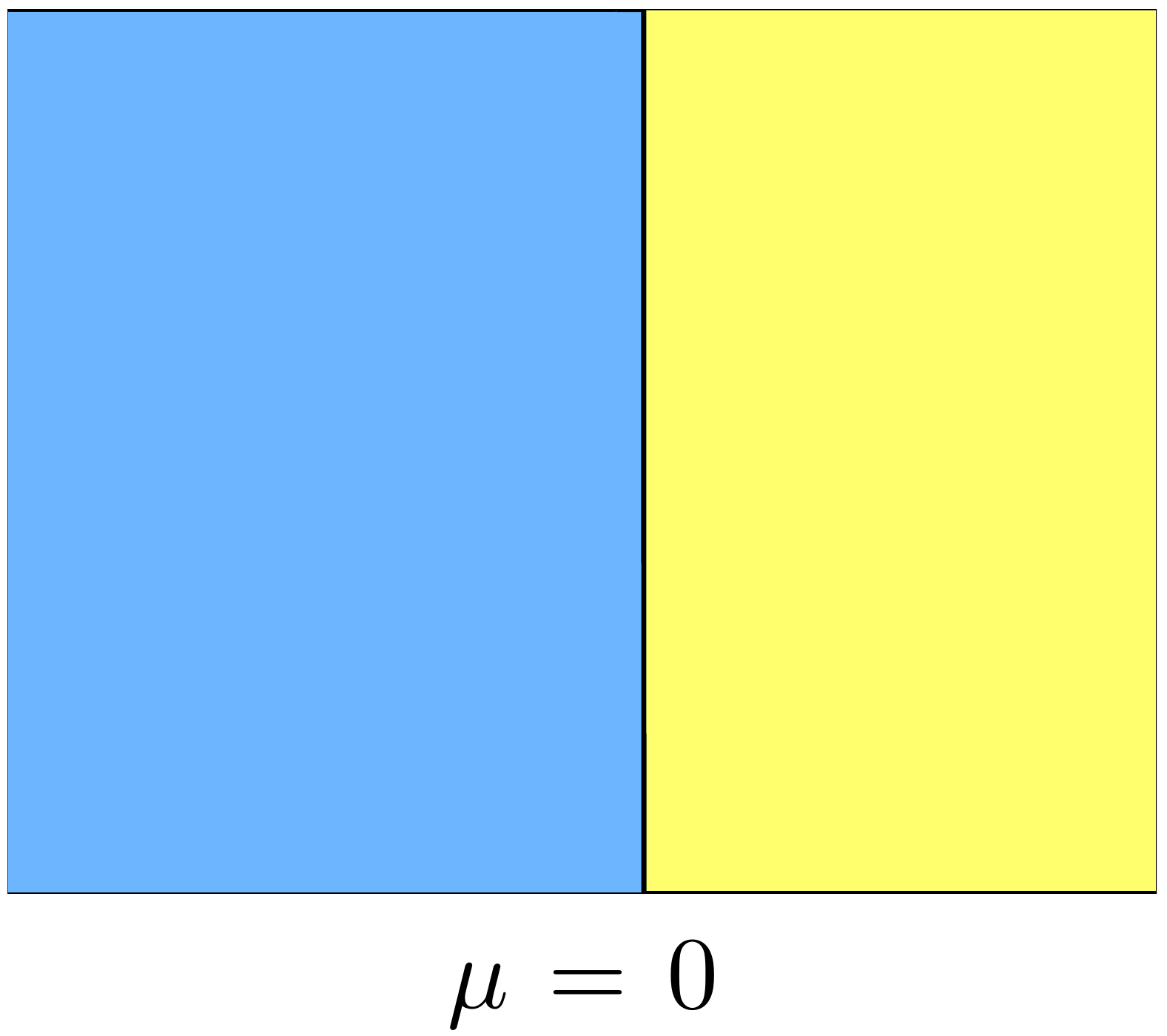}
\includegraphics[width=30mm]{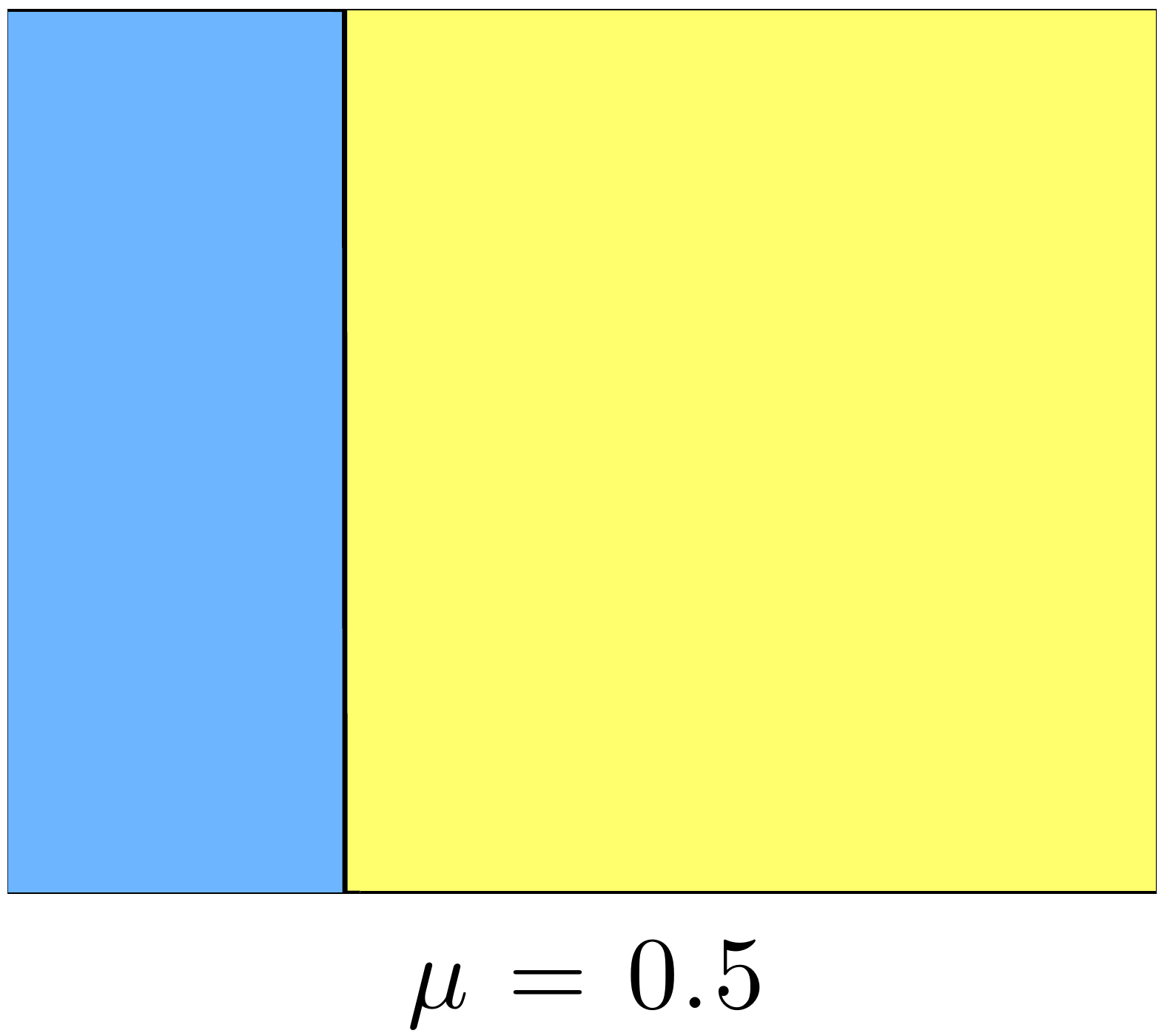}
\includegraphics[width=30mm]{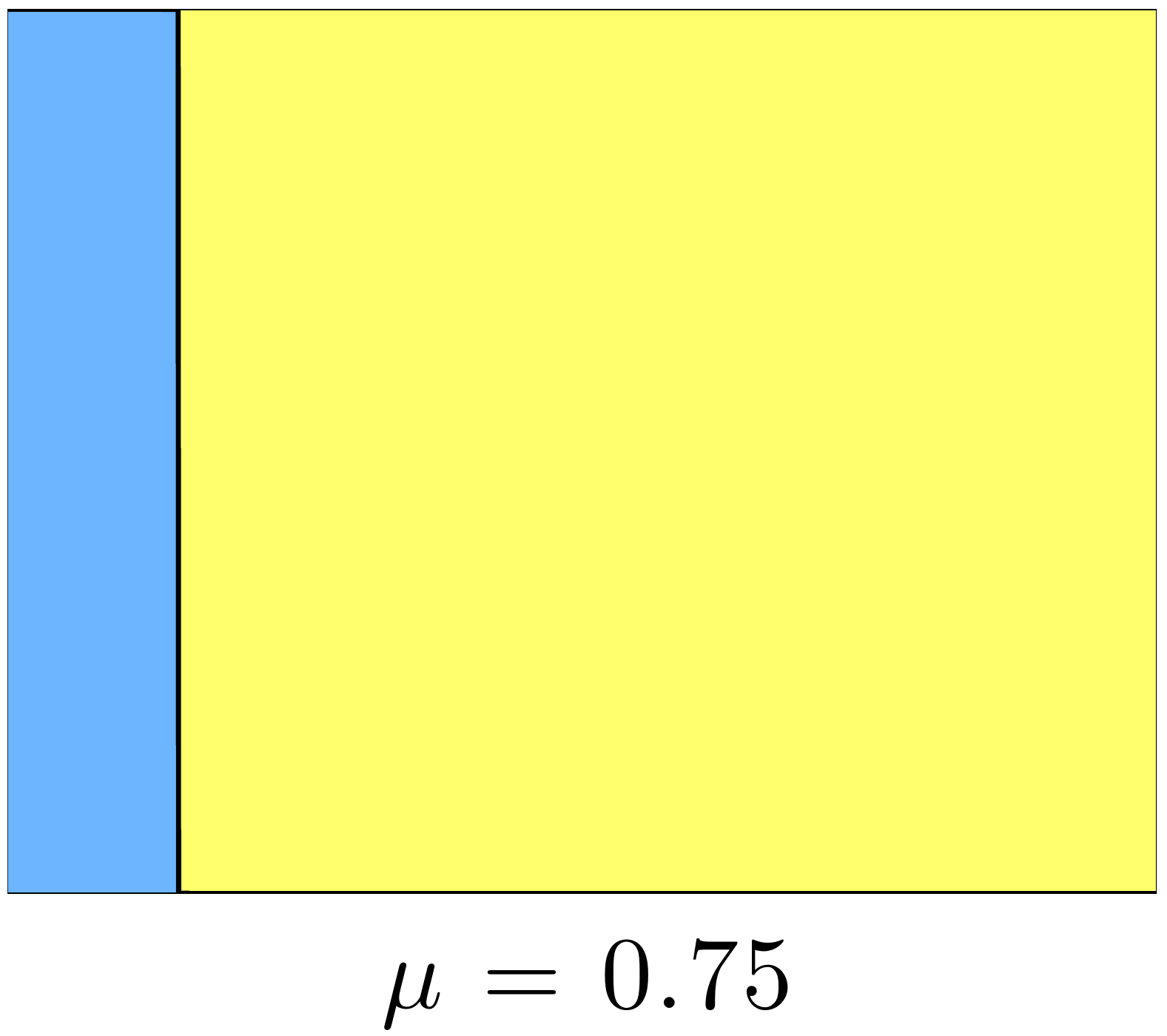}
  }
  \\
  \subfloat[For the joint claim $\Prob(Y_1 > Y_0) \geq \underline{p}$ and $\text{ATE} \geq 0$, for five different values of $\underline{p}$.]{
  \label{BF_illustration4}
\includegraphics[width=30mm]{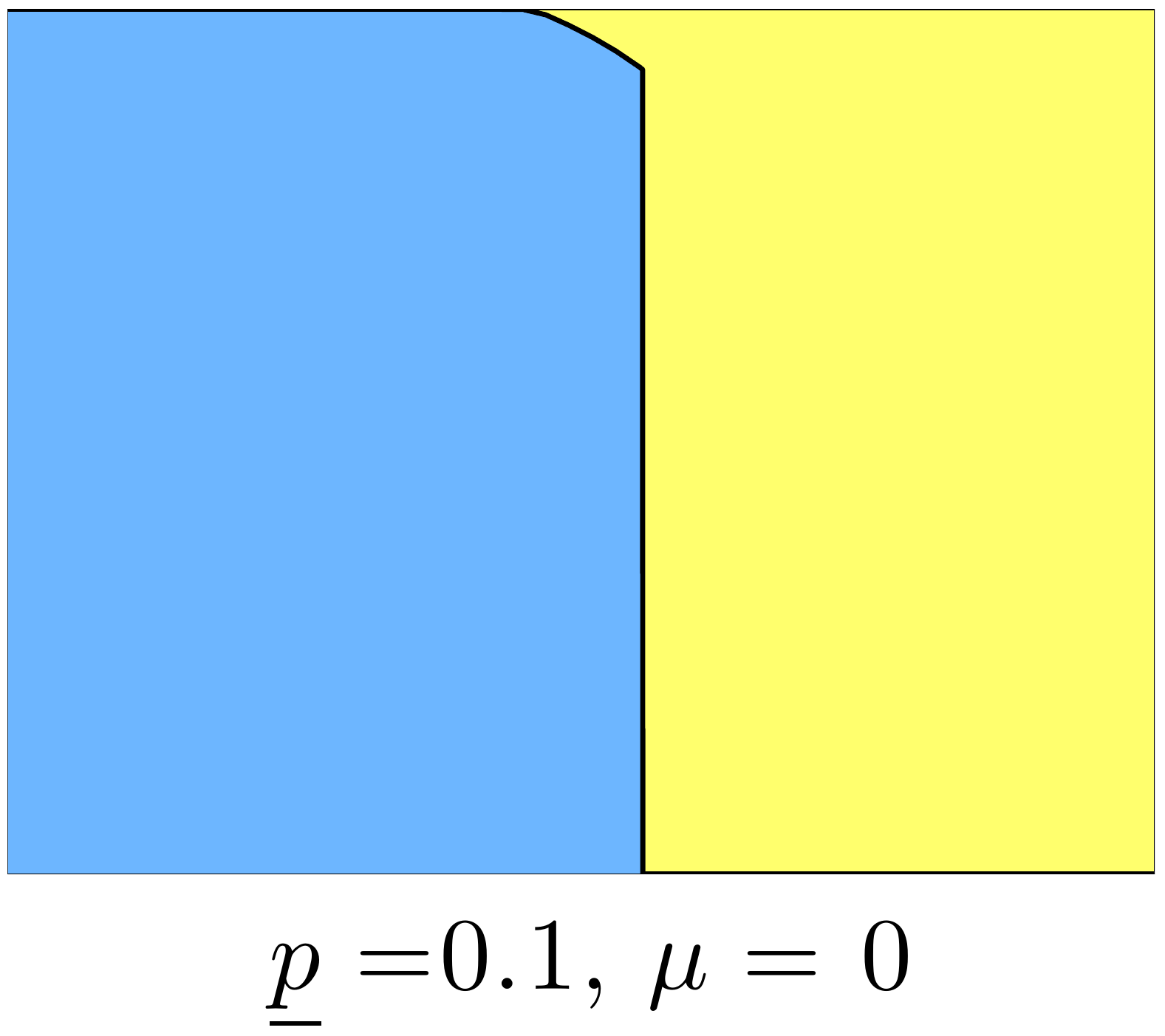}
\includegraphics[width=30mm]{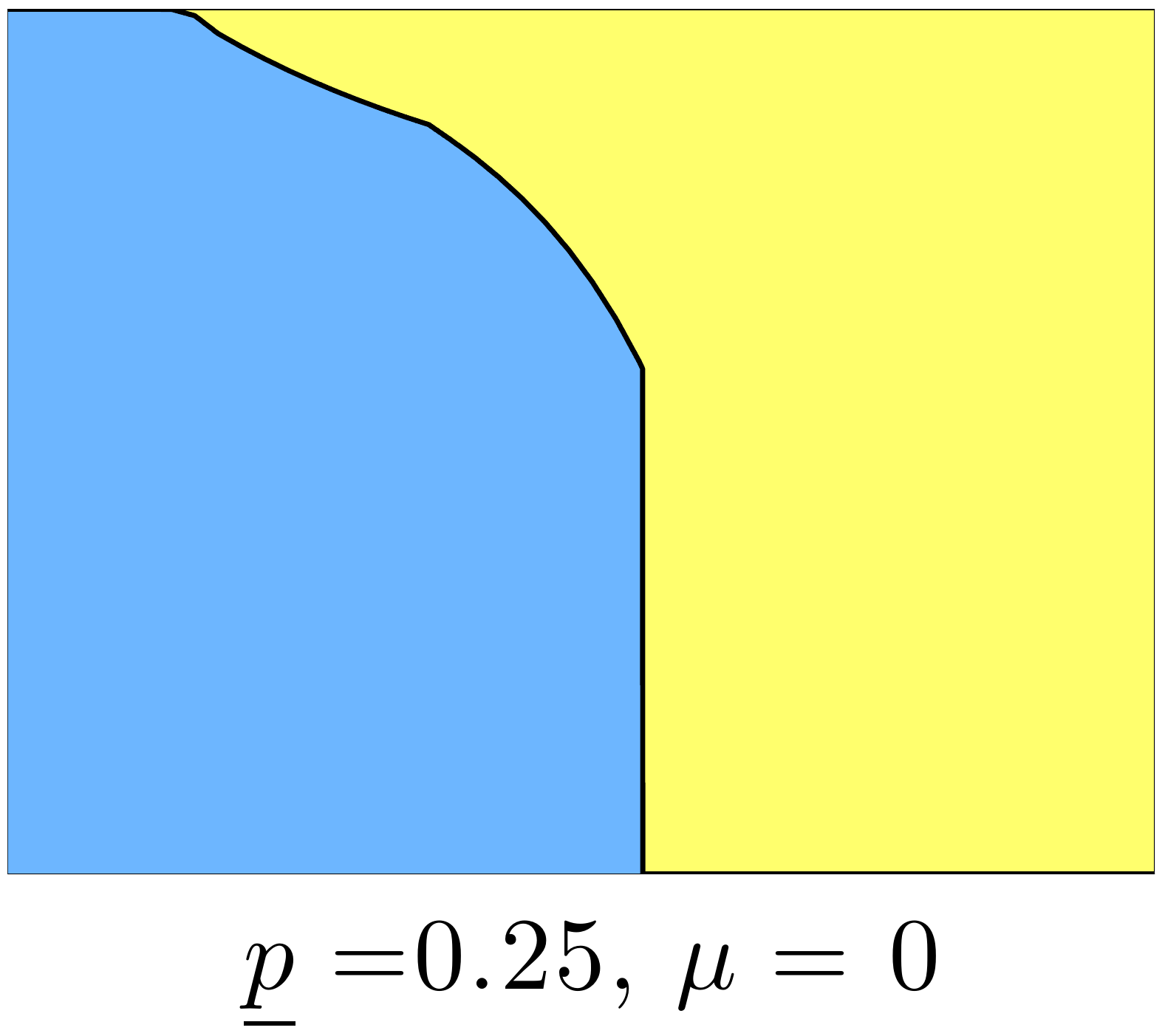}
\includegraphics[width=30mm]{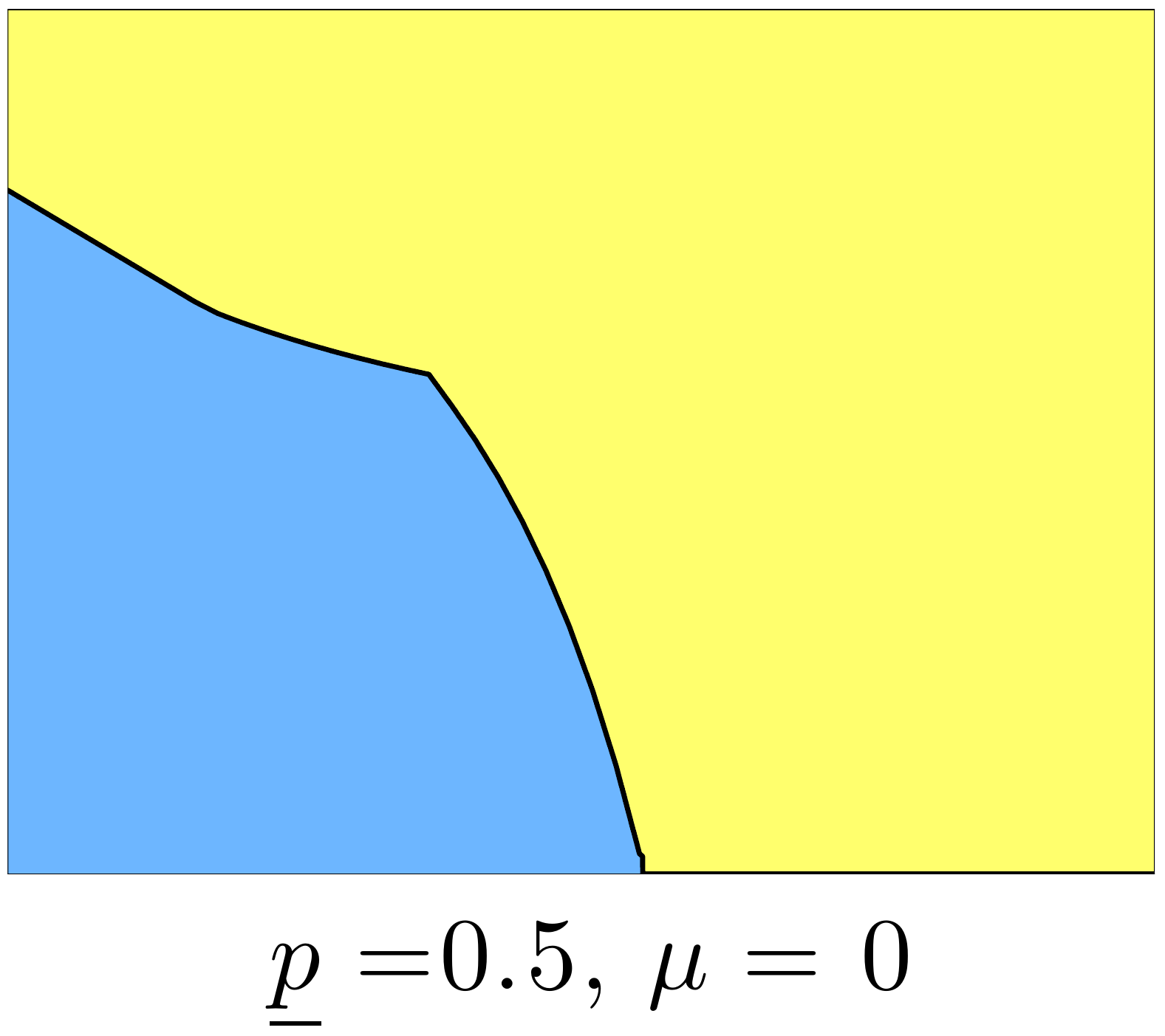}
\includegraphics[width=30mm]{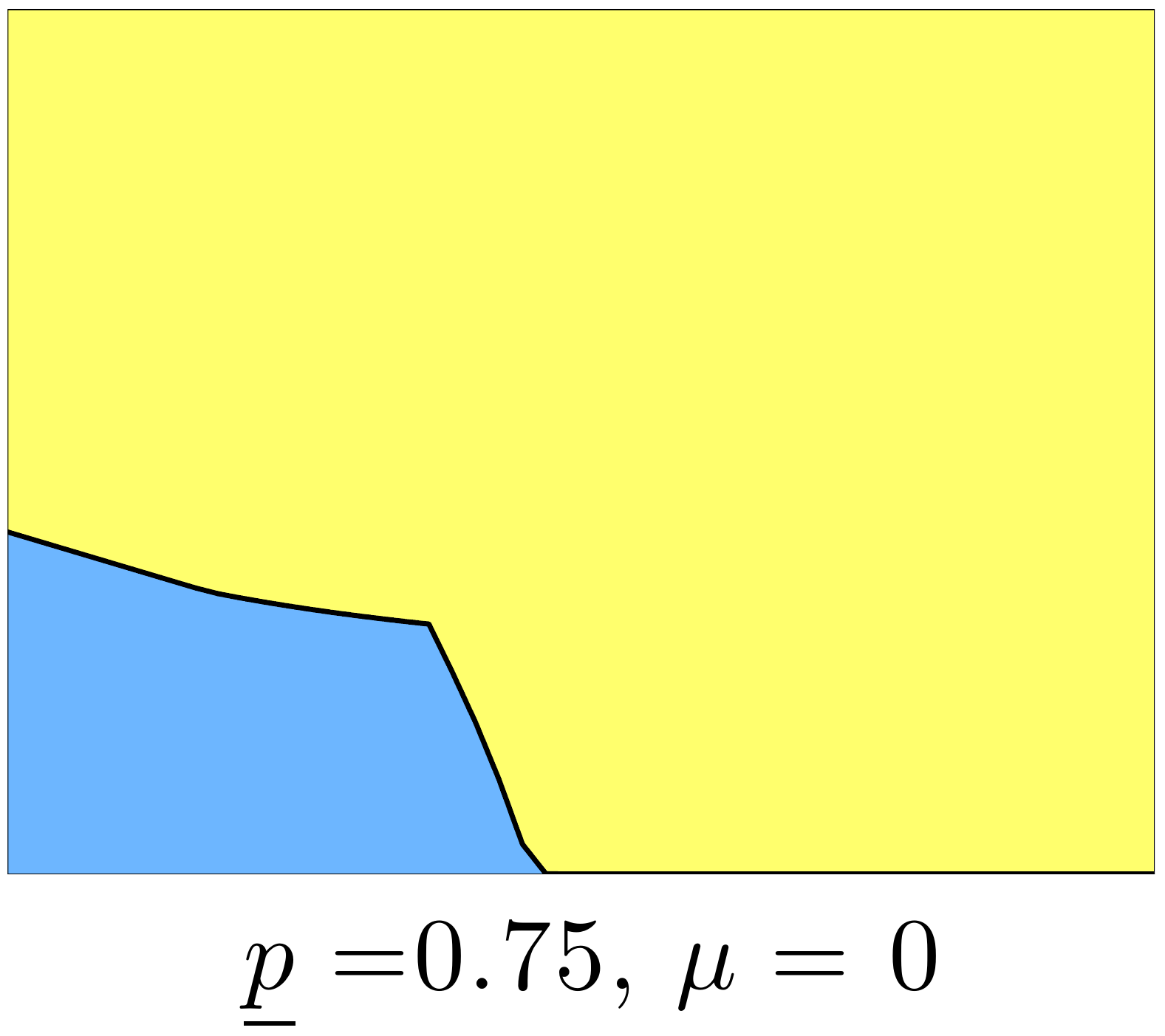}
\includegraphics[width=30mm]{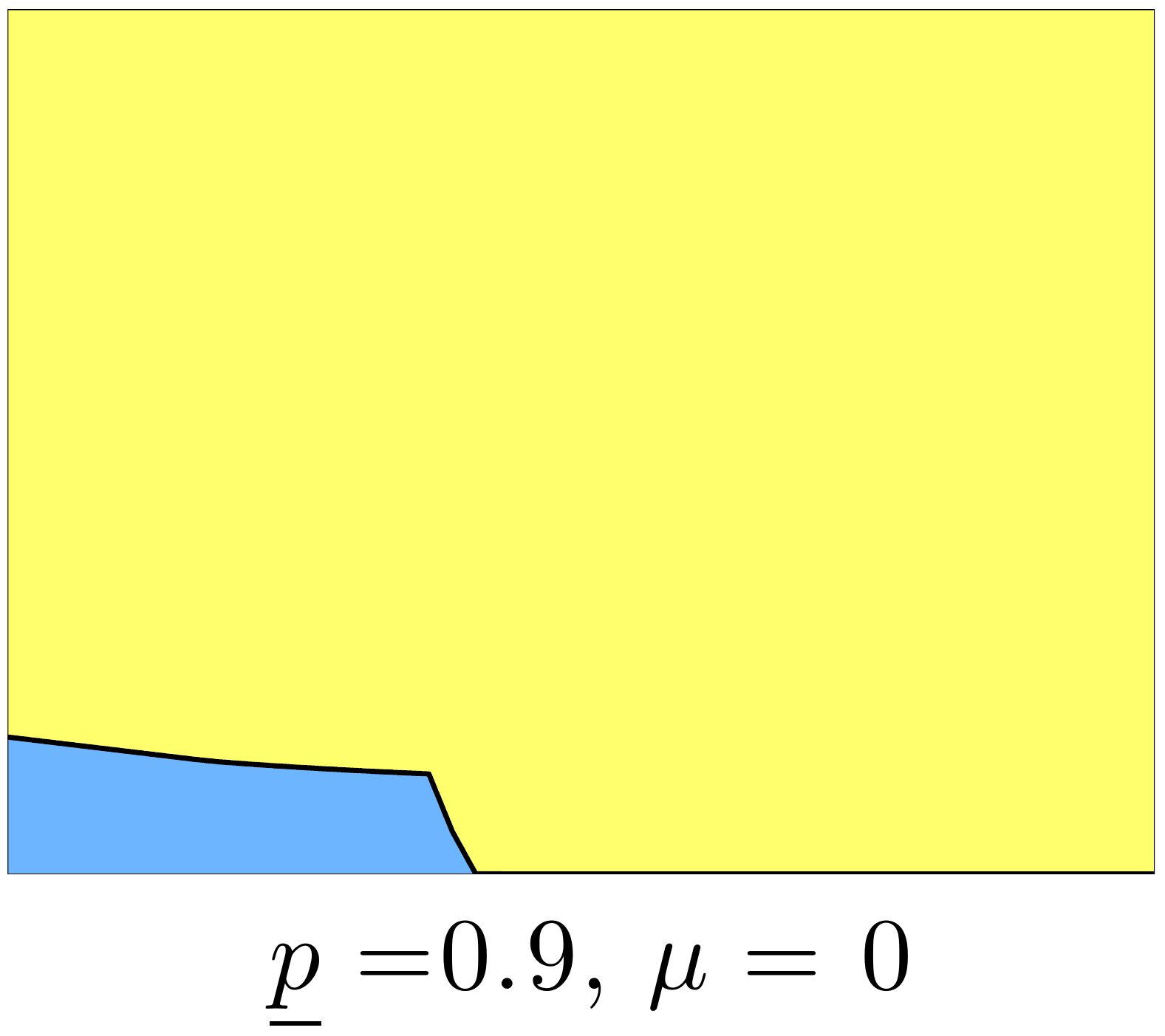}
  }
\end{figure}

Figure \ref{BF_illustration2} illustrates how the breakdown frontier changes as our conclusion of interest changes. Specifically, consider the conclusion that
\[
	\Prob(Y_1 > Y_0) \geq \underline{p}
\]
for five different values for $\underline{p}$. The figure shows the corresponding breakdown frontiers. As $\underline{p}$ increases towards one, we are making a stronger claim about the true parameter and hence the set of assumptions for which the conclusion holds shrinks. For strong enough claims, the claim may be refuted even with the strongest assumptions possible. Conversely, as $\underline{p}$ decreases towards zero, we are making progressively weaker claims about the true parameter, and hence the set of assumptions for which the conclusion holds grows larger.

Under the strongest assumptions of $(c,t) = (0,0)$, the parameter $\P(Y_1 > Y_0)$ is point identified. Let $p_{0,0}$ denote its value. The value $p_{0,0}$ is often strictly less than 1. In this case, any $\underline{p} \in (p_{0,0},1]$ yields a degenerate breakdown frontier: This conclusion is refuted under the point identifying assumptions. Even if $p_{0,0}<\underline{p}$, the conclusion $\Prob(Y_1 > Y_0) \geq \underline{p}$ may still be correct. This follows since, for strictly positive values of $c$ and $t$, the identified sets for $\Prob(Y_1 > Y_0)$ do contain values larger than $p_{0,0}$. But they also contain values smaller than $p_{0,0}$. Hence there do not exist any assumptions for which we can draw the desired conclusion.

The breakdown frontier is similar to the classical Pareto frontier from microeconomic theory: It shows the trade-offs between different assumptions in drawing a specific conclusion. For example, consider figure \ref{BF_illustration}. If we are at the top left point, where the breakdown frontier intersects the vertical axis, then any relaxation of random assignment requires strengthening the rank invariance assumption in order to still be able to draw our desired conclusion. The breakdown frontier specifies the precise marginal rate of substitution between the two assumptions. If we can interpret the units of each relaxation separately, we can interpret trade-offs between them. We discuss these individual interpretations in section \ref{sec:Model}. It can sometimes be helpful to measure the two different relaxations in common units. This is not necessary, however. For example, in labor supply models agents trade off leisure hours for consumption, although time and quantities of goods are measured in fundamentally different units. Many common rates outside of economics, like kilometers per hour or beats per minute, also do not have common units.

We also study breakdown frontiers and derive asymptotic distributional results for the average treatment effect $\text{ATE} = \Exp(Y_1 - Y_0)$ and quantile treatment effects $\text{QTE}(\tau) = Q_{Y_1}(\tau) - Q_{Y_0}(\tau)$ for $\tau \in (0,1)$. Because the breakdown frontier analysis for these two parameters is nearly identical, we focus on ATE for brevity. Since ATE does not rely on the dependence structure between potential outcomes, assumptions regarding rank invariance do not affect the identified set. Hence, for a specific conclusion like $\text{ATE} \geq 0$, the breakdown frontiers are vertical lines. This was not the case for the breakdown frontier for claims about $\Prob(Y_1 > Y_0)$, where the assumptions on independence of treatment assignment and on rank invariance interact.

Figure \ref{BF_illustration3} illustrates how the breakdown frontier for ATE changes as our conclusion of interest changes. Specifically, consider the conclusion that
\[
	\text{ATE} \geq \mu
\]
for five different values for $\mu$. The figure shows the corresponding breakdown frontiers. As $\mu$ increases, we are making a stronger claim about the true parameter and hence the set of assumptions for which the conclusion holds shrinks. For strong enough claims, the claim may be refuted even with the strongest assumptions possible. This happens when $\Exp(Y \mid X=1) - \Exp(Y \mid X=0) < \mu$. Conversely, as $\mu$ decreases, we are making progressively weaker claims about the true parameter, and hence the set of assumptions for which the conclusion holds grows larger.

Breakdown frontiers can be defined for many simultaneous claims. For example, figure \ref{BF_illustration4} illustrates breakdown frontiers for the joint claim that $\Prob(Y_1 > Y_0) \geq \underline{p}$ and $\text{ATE} \geq 0$, for five different values of $\underline{p}$. Compare these plots to those of figures \ref{BF_illustration2} and \ref{BF_illustration3}. For small $\underline{p}$, many pairs $(c,t)$ lead to the conclusion $\Prob(Y_1 > Y_0) \geq \underline{p}$. But many of these pairs do not also lead to the conclusion $\Exp(Y_1 - Y_0) \geq 0$, and hence those pairs are removed. As $\underline{p}$ increases, however, the constraint that we also want to conclude $\text{ATE} \geq 0$ eventually does not bind. Similarly, if we look at the breakdown frontier solely for our claim about ATE, many points $(c,t)$ are included which are ruled out when we also wish to make our claim about $\Prob(Y_1 > Y_0)$. This is especially true as $\underline{p}$ gets larger.

Although we focus on one-sided claims like $\Prob(Y_1 > Y_0) \geq \underline{p}$, one may also be interested in the simultaneous claim that $\Prob(Y_1 > Y_0) \geq \underline{p}$ and $\Prob(Y_1 > Y_0) \leq \overline{p}$, where $\underline{p}<\overline{p}$. Similarly to the joint one-sided claims about $\Prob(Y_1 > Y_0)$ and ATE discussed above, the breakdown frontier for this two-sided claim on $\Prob(Y_1 > Y_0)$ can be obtained by taking the minimum of the two breakdown frontier functions for each one-sided claim separately. Equivalently, the area under the breakdown frontier for the two-sided claim is just the intersection of the areas under the breakdown frontiers for the one-sided claims.

\subsection*{Inference}

As mentioned above, although identified sets are used in its definition, the breakdown frontier is always point identified. Hence for estimation and inference we use mostly standard nonparametric techniques. We first propose a consistent sample-analog estimator of the breakdown frontier. We then do inference via two asymptotically valid bootstrap procedures. Our main approach, discussed in section \ref{sec:EstimationAndInference}, relies on important work by \cite{Dumbgen1993}, \cite{FangSantos2014}, and \cite{HongLi2015}. We discuss an alternative approach in supplemental appendix D. In both approaches we construct one-sided lower uniform confidence bands for the breakdown frontier. A one-sided band is appropriate because our inferential goal is to determine the set of assumptions for which we can still draw our conclusion. If $\alpha$ denotes the nominal size, then approximately $100(1-\alpha)$\% of the time, all elements $(c,t)$ which are below the confidence band lead to population level identified sets for the parameters of interest which allow us to conclude that our conclusions of interest hold; we also discuss a testing interpretation in supplemental appendix A. We examine the finite sample performance of our main approach in several Monte Carlo simulations.

\subsection*{An empirical illustration of the breakdown frontier approach}

Finally, we use our results to examine the role of assumptions in determining the effects of child soldiering on wages, as studied in \cite{BlattmanAnnan2010}. We illustrate how our results can be used as a sensitivity analysis within a larger empirical study. Specifically, we begin by first estimating and doing inference on parameters under point identifying assumptions. Next, we estimate breakdown frontiers for several claims about these parameters. We present our one-sided confidence bands as well. We then use these breakdown frontier point estimates and confidence bands to judge the sensitivity of our conclusion to the point identifying assumptions: A large inner confidence set for the robust region which includes many different points $(c,t)$ suggests robust results while a small inner confidence set close to the origin suggests non-robust results.

\subsection*{Motivation for the parameter $\Prob(Y_1 > Y_0)$}

Our empirical illustration uses data from \cite{BlattmanAnnan2010}, which is part of a large literature on the impact of compulsory military service on wages. As \citet[pages 57--58]{CardCardoso2012} note,
\begin{quote}
``\cite{Angrist1990} showed that Vietnam-era draftees had lower earnings than non-draftees, a finding he attributed to the low value of military experience in the civilian labor market. Subsequent research in the United States and other countries, however, has uncovered a surprisingly mixed pattern of impacts.''
\end{quote}
On one hand, enlistees learn basic skills and receive occupational training in the military. On the other hand, they forgo civilian schooling and work experience, and may experience debilitating psychological trauma. Which of these two explanations is correct? Most likely, both. By using models where the treatment effects $Y_1 - Y_0$ are heterogeneous, we allow some people to have overall positive effects, and thus satisfy the first explanation, and some people to have overall negative effects, and thus satisfy the second explanation. The parameter $\Prob(Y_1 > Y_0)$ tells us precisely which proportion of the population primarily satisfies the first versus the second explanation. Thus it gives researchers a more nuanced understanding of treatment effects, by quantitatively measuring the prevalence of two opposing explanations of the impact of treatment on outcomes. We suspect this parameter can be particularly helpful in literatures which find a ``mixed pattern of impacts,'' like the work on compulsory military service and wages discussed above.

That said, one may be concerned that the rank invariance assumption used to point identify this parameter is too strong. As in \cite{HeckmanSmithClements1997}, this concern motivates our sensitivity analysis. For further motivation of this parameter in various settings, see \cite{BedoyaEtAl2017} and \cite{Mullahy2018}.

\section{Model and identification}\label{sec:Model}

To illustrate the breakdown frontier approach, we study the standard potential outcomes model with a binary treatment. We focus on two key assumptions: random assignment and rank invariance. We discuss how to relax these two assumptions and derive identified sets for various parameters under these relaxations. While there are many different ways to relax these assumptions, our goal is to illustrate the breakdown frontier methodology and hence we focus on just one kind of relaxation for each assumption.

\subsection*{Setup}

Let $Y_1$ and $Y_0$ denote the unobserved potential outcomes. Let $X \in \{ 0,1\}$ be an observed binary treatment. Let $W\in \supp(W)$ be a vector of observed covariates. This vector may contain discrete covariates, continuous covariates, or both. We observe the scalar outcome variable
\begin{equation}\label{eq:potential outcomes}
	Y = X Y_1 + (1-X) Y_0.
\end{equation}
Let $p_{x|w} = \Prob(X=x \mid W=w)$ denote the observed propensity score. We maintain the following assumption on the joint distribution of $(Y_1,Y_0,X,W)$ throughout.

\setcounter{partialIndepSection}{1}
\begin{partialIndepAssump}\label{assn:continuity}
For each $x,x' \in \{0,1\}$ and $w \in \supp(W)$:
\begin{enumerate}
\item \label{A1_1} $Y_x \mid X=x', W=w$ has a strictly increasing and  continuous distribution function on its support, $\supp(Y_x \mid X=x', W=w)$.

\item \label{A1_3} $\supp(Y_x \mid X=x',W=w) = \supp(Y_x \mid W=w) = [\underline{y}_x(w),\overline{y}_x(w)]$ where $-\infty \leq \underline{y}_x(w) < \overline{y}_x(w) \leq \infty$.

\item \label{A1_4} $p_{x|w} > 0$.
\end{enumerate}
\end{partialIndepAssump}

Via A\ref{assn:continuity}.\ref{A1_1}, we restrict attention to continuously distributed potential outcomes. A\ref{assn:continuity}.\ref{A1_3} states that the support of $Y_x \mid X=x',W=w$ does not depend on $x'$, and is a possibly infinite closed interval. This assumption implies that the endpoints $\underline{y}_x(w)$ and $\overline{y}_x(w)$ are point identified. We maintain A\ref{assn:continuity}.\ref{A1_3} for simplicity, but it can be relaxed using similar derivations as in \cite{MastenPoirier2016}. A\ref{assn:continuity}.\ref{A1_4} is an overlap assumption.

Define the \emph{conditional rank} random variables $U_0 = F_{Y_0 \mid W}(Y_0 \mid W)$ and $U_1 = F_{Y_1 \mid W}(Y_1 \mid W)$. Since $F_{Y_1 \mid W}(\cdot \mid w)$ and $F_{Y_0 \mid W}(\cdot \mid w)$ are strictly increasing (by A\ref{assn:continuity}.\ref{A1_1}), $U_0 \mid W$ and $U_1 \mid W$ are uniformly distributed on $[0,1]$. The value of unit $i$'s conditional rank random variable $U_x$ tells us where unit $i$ lies in the distribution of $Y_x \mid W$.

\subsection*{Identifying assumptions}

It is well known that the joint conditional distribution of potential outcomes $(Y_1,Y_0) \mid W$ is point identified under two assumptions:
\begin{enumerate}
\item Conditional random assignment of treatment: $X \indep Y_1 \mid W$ and $X \indep Y_0 \mid W$.

\item Conditional rank invariance: Conditional on $W$, $U_1 = U_0$ almost surely.
\end{enumerate}
Note that the joint conditional independence assumption $X \indep (Y_1,Y_0) \mid W$ provides no additional identifying power beyond the marginal conditional independence assumption stated above. Any functional of $F_{Y_1,Y_0 \mid W}$ is point identified under these random assignment and rank invariance assumptions. The goal of our identification analysis is to study what can be said about such functionals when one or both of these point-identifying assumptions fails. To do this we define two classes of assumptions: one which indexes the relaxation of random assignment of treatment, and one which indexes the relaxation of rank invariance. These classes of assumptions nest both the point identifying assumptions of random assignment and rank invariance and the opposite end of no constraints on treatment selection or the dependence structure between potential outcomes. 

A key feature of the relaxations we use is that they are `orthogonal' in the sense that we can relax each of the two assumptions separately: The amount by which we relax one assumption does not constrain the amount by which we can relax the other assumption. This feature is important since a key goal of our analysis is to quantify the trade-off between relaxations of these two assumptions.

We begin with our measure of distance from conditional independence.

\begin{definition}\label{def:c-indep}
Let $c$ be a scalar between 0 and 1. Say $X$ is \emph{conditionally $c$-dependent} with $Y_x$ given $W$ if
\begin{equation}\label{eq:c-indep1}
	\sup_{y \in \supp(Y_x \mid W=w)} | \P(X=x \mid Y_x = y, W=w) - \P(X=x\mid W=w) | \leq c
\end{equation}
for $x \in \{ 0,1\}$ and $w\in \supp(W)$.
\end{definition}

For $c=0$, conditional $c$-dependence implies $X \indep Y_1 \mid W$ and $X \indep Y_0 \mid W$. For $c > 0$, however, it allows for some deviations from conditional independence. Specifically, it allows the unobserved treatment assignment probability $\P(X=1 \mid Y_x = y, W=w)$ to be at most $c$ probability units away from the observed propensity score $p_{1|w}$. We discuss one way to interpret the magnitude of $c$ on page \pageref{paragraph:leaveOutVariableK}. We give further discussion in our previous paper \cite{MastenPoirier2017}. 

Our second class of assumptions constrains the dependence structure between $Y_1$ and $Y_0$, conditional on $W$. By Sklar's Theorem (\citealt{Sklar1959}), write
\[
	F_{Y_1,Y_0 \mid W}(y_1,y_0 \mid w) = C(F_{Y_1 \mid W}(y_1 \mid w),F_{Y_0 \mid W}(y_0 \mid w) \mid w)
\]
where $C(\cdot,\cdot \mid w)$ is a unique conditional copula function. See \cite{Nelsen2006} for an overview of copulas and \cite{FanPatton2014} for a survey of their use in econometrics. Restrictions on $C$ constrain the dependence between potential outcomes. For example, if
\[
	C(u_1,u_0 \mid w) = \min\{u_1,u_0\} \equiv M(u_1,u_0),
\]
then $U_1 = U_0$ almost surely, conditional on $W$. Thus conditional rank invariance holds. In this case the potential outcomes $Y_1$ and $Y_0$ are sometimes called \emph{conditionally comonotonic} and $M$ is called the \emph{comonotonicity} copula. At the opposite extreme, when $C$ is an arbitrary copula, the dependence between $Y_1 \mid W$ and $Y_0 \mid W$ is constrained only by the Fr\'echet-Hoeffding bounds, which state that
\[
	\max\{u_1 + u_0 -1, 0\} \leq C(u_1,u_0 \mid w) \leq M(u_1,u_0).
\]
We next define a class of assumptions which includes both conditional rank invariance and no assumptions on the dependence structure as special cases. 

\begin{definition}\label{def:t-rankinvariance}
The potential outcomes $(Y_1,Y_0)$ satisfy \emph{(1$-t)$-percent conditional rank invariance} given $W$ if for all $w \in \supp(W)$ their conditional copula $C$ satisfies
\begin{equation}\label{eq:t-rankinvariance}
	C(u_1,u_0 \mid w) = (1-t) M(u_1,u_0) + t H(u_1,u_0 \mid w)
\end{equation}
where $M(u_1,u_0) = \min\{ u_1,u_0 \}$ and $H$ is some conditional copula.
\end{definition}

This assumption says that within each covariate cell the population is a mixture of two parts: In one part, rank invariance holds. This part contains $100 \cdot t \,$\% of the overall population in that cell. In the second part, rank invariance fails in an arbitrary, unknown way. Hence, for this part, the dependence structure is unconstrained beyond the Fr\'echet-Hoeffding bounds. This part contains $100(1-t)$\% of the overall population in that cell. Thus for $t=0$ the usual conditional rank invariance assumption holds, while for $t=1$ no assumptions are made about the dependence structure. For $t \in (0,1)$ we obtain a kind of partial conditional rank invariance. Note that by exercise 2.3 on page 14 of \cite{Nelsen2006}, a mixture of copulas like that in equation \eqref{eq:t-rankinvariance} is also a copula.

To see this mixture interpretation formally, let $T$ follow a Bernoulli distribution with $\P(T=1 \mid W=w) = t$, where $T\indep Y_1 \mid W$ and $T \indep Y_0 \mid W$, but $T$ is not independent of $(Y_1,Y_0) \mid W$ jointly. This implies that $T$ has an effect on the dependence structure of $(Y_1,Y_0) \mid W$ but not on their conditional marginal distributions. Suppose that individuals for whom $T_i = 1$ have an arbitrary dependence structure, while those with $T_i=0$ have conditionally rank invariant potential outcomes. Then by the law of total probability,
\begin{align*}
	&F_{Y_1,Y_0 \mid W}(y_1,y_0 \mid w) \\
		&= (1-t)F_{Y_1,Y_0 \mid W,T}(y_1,y_0 \mid w,0) + t F_{Y_1,Y_0 \mid W,T}(y_1,y_0 \mid w,1)\\
		&= (1-t)C \big( F_{Y_1 \mid W,T}(y_1 \mid w,0),F_{Y_0 \mid W,T}(y_0 \mid w,0) \mid w,0 \big)\\
		&\qquad + t \, C \big(F_{Y_1 \mid W,T}(y_1 \mid w,1),F_{Y_0 \mid W,T}(y_0 \mid w,1) \mid w,1 \big)\\
		&= (1-t)M(F_{Y_1 \mid W}(y_1 \mid w),F_{Y_0 \mid W}(y_0 \mid w)) + t H(F_{Y_1 \mid W}(y_1 \mid w),F_{Y_0 \mid W}(y_0 \mid w) \mid w).
\end{align*}

Our approach to relaxing rank invariance is an example of a more general approach. In this approach we take a weak assumption and a stronger assumption and use them to define a continuous class of assumptions by considering the population as a mixture of two subpopulations. The weak assumption holds in one subpopulation while the stronger assumption holds in the other subpopulation. The mixing proportion $t$ continuously spans the two distinct assumptions we began with. This approach was used earlier by \cite{HorowitzManski1995} in their analysis of the contaminated sampling model. While this general approach may not always be the most natural way to relax an assumption, it is always available and hence can be used to facilitate breakdown frontier analyses.

Throughout the rest of this section we impose both conditional $c$-dependence and (1$-t$)-percent conditional rank invariance given $W$.

\begin{partialIndepAssump}\label{assn:independenceAndRankInvariance}
\hfill
\begin{enumerate}
\item \label{A2_1} $X$ is conditionally $c$-dependent with the potential outcomes $Y_x$ given $W$, where for all $w \in \supp(W)$ we have $c < \min\{ p_{1|w}, p_{0|w} \}$.

\item \label{A2_2} $(Y_1,Y_0)$ satisfies (1$-t$)-percent conditional rank invariance given $W$, where $t \in [0,1]$.
\end{enumerate}
\end{partialIndepAssump}

\medskip

\noindent For brevity we focus on the case $c < \min \{ p_{1|w}, p_{0|w} \}$ throughout this paper. This allows us to explain the key ideas while keeping the notation and derivations relatively simple. All of our results, however, can be relaxed to the general case where $c \in [0,1]$.

\subsection*{Partial identification under relaxations of independence and rank invariance}

We next study identification under the deviations from full independence and rank invariance defined above. We begin by briefly recalling results from \cite{MastenPoirier2017} on identification of the conditional quantile treatment effect $\text{CQTE}(\tau \mid w) =Q_{Y_1 \mid W}(\tau \mid w) - Q_{Y_0}(\tau \mid w)$, the conditional average treatment effect $\text{CATE}(w) = \Exp(Y_1 - Y_0 \mid W=w)$, and the conditional marginal cdfs of potential outcomes under $c$-dependence. We then derive new identification results for the distribution of treatment effects (DTE), $F_{Y_1 - Y_0}(z)$, and its related parameter $\Prob(Y_1 > Y_0)$.

In \cite{MastenPoirier2017}, we showed that A\ref{assn:continuity} and A\ref{assn:independenceAndRankInvariance}.\ref{A2_1} imply that the identified set for $\text{CQTE}(\tau \mid w)$ is\footnote{In that paper we also extended the bounds in equations \eqref{eq:QTE bounds}--\eqref{eq:cdf lowerbound} to the $c \geq \min \{ p_{1|w}, p_{0|w} \}$ case. As noted earlier, here we focus on the case $c < \min \{ p_{1|w}, p_{0|w} \}$ case for brevity.}
\begin{align}
	&\Big[ \underline{\text{CQTE}}(\tau, c \mid w), \overline{\text{CQTE}}(\tau, c \mid w) \Big] \notag\\
	&\qquad \equiv \left[\underline{Q}_{Y_1 \mid W}^c (\tau \mid w) - \overline{Q}_{Y_0 \mid W}^c (\tau \mid w), \,
	\overline{Q}_{Y_1 \mid W}^c (\tau \mid w) - \underline{Q}_{Y_0 \mid W}^c (\tau \mid w)\right] \label{eq:QTE bounds}
\end{align}
where
\begin{align}\label{eq:quantilebounds_smallc}
	\overline{Q}_{Y_x \mid W}^c(\tau \mid w)
		&= Q_{Y \mid X,W}\left(\tau + \frac{c}{p_{x|w}}\min\left\{\tau,1-\tau\right\} \mid x,w \right) \\
    	\underline{Q}_{Y_x \mid W}^c(\tau \mid w) 
		&= Q_{Y \mid X,W}\left(\tau - \frac{c}{p_{x|w}}\min\left\{\tau,1-\tau\right\} \mid x,w\right).\notag
\end{align}
We further showed that, assuming $\Exp(| Y | \mid X=x, W=w) < \infty$ for $x \in \{0,1\}$ and all $w \in \supp(W)$, the identified set for $\text{CATE}(w)$ is
\begin{equation}\label{eq:ATE bounds}
	[ \underline{\text{CATE}}(c \mid w), \overline{\text{CATE}}(c \mid w)]
	\equiv
	\left[
	\int_0^1 \underline{\text{CQTE}}(\tau, c \mid w) \; d\tau, \,
  	\int_0^1 \overline{\text{CQTE}}(\tau, c \mid w) \; d\tau
	\right]
\end{equation}
and that
\begin{equation}\label{eq:cdf upperbound}
	\overline{F}^c_{Y_x \mid W}(y \mid w)
	= \min \left\{ \frac{F_{Y \mid X,W}(y \mid x,w)p_{x|w}}{p_{x|w}-c}, \; \frac{F_{Y \mid X,W}(y \mid x,w)p_{x|w}+c}{p_{x|w}+c}\right\} 
\end{equation}
and
\begin{equation}\label{eq:cdf lowerbound}
	\underline{F}^c_{Y_x \mid W}(y \mid w)
	=
	\max \left\{ \frac{F_{Y \mid X,W}(y \mid x,w) p_{x|w}}{p_{x|w} + c}, \; \frac{F_{Y \mid X,w}(y \mid x,w) p_{x|w} - c}{p_{x|w} - c} \right\}
\end{equation}
are functionally sharp bounds on the cdf $F_{Y_x \mid W}$. We use these conditional cdf bounds in our DTE bounds. Bounds on the corresponding unconditional parameters, like $\text{ATE} = \Exp[\text{CATE}(W)]$, can be obtained by integrating the conditional bounds over the marginal distribution of $W$. These results are unchanged if we further impose A\ref{assn:independenceAndRankInvariance}.\ref{A2_2}. That is, assumptions on rank invariance have no identifying power for functionals of the marginal distributions of potential outcomes.

We next derive the identified set for the distribution of treatment effects (DTE), the cdf
\[
	\text{DTE}(z) 
	= \Prob(Y_1 - Y_0 \leq z).
\]
To do this, we first derive the identified set for the conditional distribution of treatment effects (CDTE), the cdf
\[
	\text{CDTE}(z \mid w) 
	= \Prob(Y_1 - Y_0 \leq z \mid W=w).
\]
By the law of iterated expectations,
\[
	\text{DTE}(z) = \Exp[ \text{CDTE}(z \mid W)].
\]
Thus we will obtain the identified set for the DTE by averaging bounds for the CDTE. While the ATE only depends on the conditional marginal distributions of potential outcomes, the CDTE depends on the joint distribution of $(Y_1,Y_0) \mid W$. Consequently, as we'll see below, the identified set for the CDTE depends on the value of $t$.

For any $z \in \R$ define $\mathcal{Y}_z(w) = [\underline{y}_1(w), \overline{y}_1(w)] \cap [\underline{y}_0(w) + z, \overline{y}_0(w) + z]$. Note that $\supp(Y_1 - Y_0 \mid W=w) \subseteq [\underline{y}_1(w) - \overline{y}_0(w), \overline{y}_1(w) - \underline{y}_0(w)]$. Let $z$ be an element of $[\underline{y}_1(w) - \overline{y}_0(w), \overline{y}_1(w) - \underline{y}_0(w)]$ such that $\mathcal{Y}_z(w)$ is nonempty. If $z$ is such that $\mathcal{Y}_z(w)$ is empty, then the CDTE is either 0 or 1 depending solely on the relative location of the two supports, which is point identified by A\ref{assn:continuity}.\ref{A1_3}. In this case, define $\overline{\text{CDTE}}(z,c,t \mid w)$ and $\underline{\text{CDTE}}(z,c,t \mid w)$ to equal this point identified value. If $z > \overline{y}_1(w) - \underline{y}_0(w)$, define these CDTE bounds to equal 1. If $z < \underline{y}_1(w) - \overline{y}_0(w)$, define these CDTE bounds to equal 0.

If $\mathcal{Y}_z(w)$ is nonempty, define
\begin{multline*}
	\overline{\text{CDTE}}(z,c,t \mid w) 
	= (1-t) \P(\underline{Q}^c_{Y_1 \mid W}(U \mid w) - \overline{Q}^c_{Y_0 \mid W}(U \mid w) \leq z) \\
		+ t \left(	1+ \min \left \{	\inf_{y \in \mathcal{Y}_z(w)}	(\overline{F}^c_{Y_1 \mid W}(y \mid w) - \underline{F}^c_{Y_0 \mid W}(y-z \mid w)), 0
	\right\}
	\right)
\end{multline*}
\begin{multline*}
	\underline{\text{CDTE}}(z,c,t \mid w)
	= (1-t) \P(\overline{Q}^c_{Y_1 \mid W}(U \mid w) - \underline{Q}^c_{Y_0 \mid W}(U \mid w) \leq z) \\
		+ t \max \left \{
		\sup_{y \in \mathcal{Y}_z(w)} (\underline{F}^c_{Y_1 \mid W}(y \mid w) 
		- \overline{F}^c_{Y_0 \mid W}(y-z \mid w) ), 0
	\right\}
\end{multline*}
where $U \sim \text{Unif}[0,1]$. The following result shows that (a) these are sharp bounds on the CDTE, and (b) the integral of these bounds over the marginal distribution of $W$ yields sharp bounds on the DTE, defined as $\Prob(Y_1 - Y_0 \leq z)$.

\begin{theorem}[DTE bounds]\label{thm:dte_bounds}
Suppose the joint distribution of $(Y,X,W)$ is known. Suppose A\ref{assn:continuity} and A\ref{assn:independenceAndRankInvariance} hold. Let $z \in \R$. Then the identified set for $\P(Y_1 - Y_0 \leq z \mid W=w)$ is
\[
	[ \underline{\text{CDTE}}(z,c,t \mid w), \overline{\text{CDTE}}(z,c,t \mid w) ].
\]
Moreover, the identified set for $\P(Y_1 - Y_0 \leq z)$ is
\begin{multline*}
	[ \underline{\text{DTE}}(z,c,t), \overline{\text{DTE}}(z,c,t) ] \\
	= \left[ \int_{\supp(W)} \underline{\text{CDTE}}(z,c,t \mid w) \; dF_W(w), \int_{\supp(W)}\overline{\text{CDTE}}(z,c,t \mid w) \; dF_W(w) \right].
\end{multline*}
\end{theorem}

The bound functions $\underline{\text{DTE}}(z,\cdot,\cdot)$ and $\overline{\text{DTE}}(z,\cdot,\cdot)$ are continuous and monotonic in both arguments. When both conditional random assignment ($c=0$) and conditional rank invariance ($t=0$) hold, these bounds collapse to a single point and we obtain point identification. If we impose conditional random assignment ($c=0$) but allow arbitrary dependence between $Y_1$ and $Y_0$ ($t=1$) then we obtain a conditional version of the well known Makarov \citeyearpar{Makarov1982} bounds. For example, see equation (2) of \cite{FanPark2010}. DTE bounds have been studied extensively by Fan and coauthors; see the introduction of \cite{FanGuerreZhu2017} for a recent and comprehensive discussion of this literature.

Theorem \ref{thm:dte_bounds} immediately implies that the identified set for $\P(Y_1 - Y_0 > z)$ is
\[
	\P(Y_1 - Y_0 > z) 
	\in [1 - \overline{\text{DTE}}(z,c,t), 1-\underline{\text{DTE}}(z,c,t)].
\]
In particular, setting $z=0$ yields the proportion who benefit from treatment, $\P(Y_1 > Y_0)$. Thus theorem \ref{thm:dte_bounds} allows us to study the sensitivity of this parameter to relaxations of full conditional independence and conditional rank invariance.

Finally, notice that all of the bounds and identified sets discussed in this section are analytically tractable and depend on just three functions identified from the population---the conditional cdf $F_{Y \mid X,W}$, the propensity scores $p_{x|w}$, and the marginal distribution of covariates $F_W$. This suggests a plug-in estimation approach which we study in section \ref{sec:EstimationAndInference}.

\subsection*{Breakdown frontiers}

We now formally define the \emph{breakdown frontier}, which generalizes the scalar breakdown point to multiple assumptions or dimensions. We also define the \emph{robust region}, the area below the breakdown frontier. These objects can be defined for different conclusions about different parameters in various models. For concreteness, however, we focus on just a few conclusions about $\Prob(Y_1 - Y_0 > z)$ and ATE in the potential outcomes model discussed above.

We begin with the conclusion that $\Prob(Y_1 - Y_0 > z) \geq \underline{p}$ for a fixed $\underline{p} \in [0,1]$ and $z \in \R$. For example, if $z = 0$ and $\underline{p} = 0.5$, then this conclusion states that at least 50\% of people have higher outcomes with treatment than without. If we impose conditional random assignment and conditional rank invariance, then $\Prob(Y_1 - Y_0 > z)$ is point identified and hence we can directly check whether this conclusion holds. But the breakdown frontier approach asks: Relative to these baseline assumptions, what are the \emph{weakest} assumptions that allow us to draw this conclusion, given the observed distribution of $(Y,X,W)$? Specifically, since larger values of $c$ and $t$ correspond to weaker assumptions, what are the \emph{largest} values of $c$ and $t$ such that we can still definitively conclude that $\Prob(Y_1 - Y_0 > z) \geq \underline{p}$?

We answer this question in two steps. First we gather \emph{all} values of $c$ and $t$ such that the conclusion holds. We call this set the \emph{robust region}. Since the lower bound of the identified set for $\Prob(Y_1 - Y_0 > z)$ is $1 - \overline{\text{DTE}}(z,c,t)$ (by theorem \ref{thm:dte_bounds}), the robust region for the conclusion that $\Prob(Y_1 - Y_0 > z) \geq \underline{p}$ is
\begin{align*}
	\text{RR}(z,\underline{p})
	&= \{ (c,t) \in [0,1]^2: 1-\overline{\text{DTE}}(z,c,t) \geq \underline{p} \} \\
	&= \{ (c,t) \in [0,1]^2: \overline{\text{DTE}}(z,c,t) \leq 1-\underline{p} \}.
\end{align*}
The robust region is simply the set of all $(c,t)$ which deliver an identified set for $\Prob(Y_1 - Y_0 > z)$ which lies on or above $\underline{p}$. See pages 60--61 of \cite{Stoye2005} for similar definitions in the scalar assumption case in a different model. Since $\overline{\text{DTE}}(z,c,t)$ is increasing in $c$ and $t$, the robust region will be empty if $\overline{\text{DTE}}(z,0,0) > 1-\underline{p}$, and non-empty if $\overline{\text{DTE}}(z,0,0) \leq 1-\underline{p}$. That is, if the conclusion of interest does not hold under the point identifying assumptions, it certainly will not hold under weaker assumptions. From here on we only consider the first case, where the conclusion of interest holds under the point identifying assumptions. That is, we suppose $\overline{\text{DTE}}(z,0,0) \leq 1-\underline{p}$ so that $\text{RR}(z,\underline{p}) \neq \emptyset$. 

The \emph{breakdown frontier} is the set of points $(c,t)$ on the boundary of the robust region. Specifically, for the conclusion that $\P(Y_1 > Y_0) \geq \underline{p}$, this frontier is the set
\[
	\text{BF}(\underline{p}) 
	= \{ (c,t) \in[0,1]^2 : \overline{\text{DTE}}(0,c,t) = 1-\underline{p} \}.
\]
Solving for $t$ in the equation $\overline{\text{DTE}}(0,c,t) = 1-\underline{p}$ yields
\begin{equation}\label{eq:popbf}
	\text{bf}(c,\underline{p})
	= 
	\frac{\text{num}}{\text{denom}}
\end{equation}
where
\[
	\text{num} =
		1 - \underline{p} 
		- \int_{\supp(W)}\P(\underline{Q}^c_{Y_1 \mid W}(U \mid w) - \overline{Q}^c_{Y_0 \mid W}(U \mid w) \leq 0) \; dF_W(w)
\]
and
\begin{multline*}
	\text{denom} =
	1 + \int_{\supp(W)} \bigg[\min \left\{ 
	\inf_{y \in \mathcal{Y}_0(w)} 
	(\overline{F}^c_{Y_1 \mid W}(y \mid w) - \underline{F}^c_{Y_0 \mid W}(y-0 \mid w)), 0 
	\right\} \\
	- \P(\underline{Q}^c_{Y_1 \mid W}(U \mid w) - \overline{Q}^c_{Y_0 \mid W}(U \mid w) \leq 0)
	\bigg] \, dF_W(w).
\end{multline*}
Thus we obtain the following analytical expression for the  breakdown frontier as a function of $c$: 
\[
	\text{BF}(c,\underline{p})
	= 
	\min\{ \max \{ \text{bf}(c,\underline{p}), 0 \}, 1 \}.
\]
This frontier provides the largest relaxations $c$ and $t$ which still allow us to conclude that $\Prob(Y_1 > Y_0) \geq \underline{p}$. It thus provides a quantitative measure of robustness of this conclusion to relaxations of the baseline point identifying assumptions of conditional random assignment and conditional rank invariance. Moreover, the shape of this frontier allows us to understand the trade-off between these two types of relaxations in drawing our conclusion. We illustrate this trade-off between assumptions in our empirical illustration of section \ref{sec:empiricalApplication}.

We next consider breakdown frontiers for ATE. Consider the conclusion that $\text{ATE} \geq \mu$ for some $\mu \in \R$. Analogously to above, the robust region for this conclusion is
\[
	\text{RR}_\textsc{ate}(\mu) = \{ (c,t) \in [0,1]^2 : \underline{\text{ATE}}(c) \geq \mu \}
\]
and the breakdown frontier is
\[
	\text{BF}_\textsc{ate}(\mu) = \{ (c,t) \in [0,1]^2 : \underline{\text{ATE}}(c) = \mu \}.
\]
These sets are nonempty if $\underline{\text{ATE}}(0) \geq \mu$; that is, if our conclusion holds under the point identifying assumptions. As we mentioned earlier, rank invariance has no identifying power for ATE, and hence the breakdown frontier is a vertical line at the point 
\[
	c^* = \inf\{ c \in [0,1] : \underline{\text{ATE}}(c) \leq \mu \}.
\]
This point $c^*$ is a breakdown point for the conclusion that $\text{ATE} \geq \mu$. Note that continuity of $\underline{\text{ATE}}(\cdot)$ implies $\underline{\text{ATE}}(c^*) = \mu$. Thus we've seen two kinds of breakdown frontiers so far: The first had nontrivial curvature, which indicates a trade-off between the two assumptions. The second was vertical in one direction, indicating a lack of identifying power of that assumption.

We can also derive robust regions and breakdown frontiers for more complicated joint conclusions. For example, suppose we are interested in concluding that both $\Prob(Y_1 > Y_0) \geq \underline{p}$ and $\text{ATE} \geq \mu$ hold. Then the robust region for this joint conclusion is just the intersection of the two individual robust regions:
\[
	\text{RR}(0,\underline{p}) \cap \text{RR}_\textsc{ate}(\mu).
\]
The breakdown frontier for the joint conclusion is just the boundary of this intersected region. Viewing these frontiers as functions mapping $c$ to $t$, the breakdown frontier for this joint conclusion can be computed as the minimum of the two individual frontier functions. For example, see figure \ref{BF_illustration4} on page \pageref{BF_illustration4}.

Above we focused on one-sided conclusions about the parameters of interest. Another natural joint conclusion is the two-sided conclusion that $\Prob(Y_1 - Y_0 > z) \geq \underline{p}$ and $\Prob(Y_1 - Y_0 > z) \leq \overline{p}$, for $0 \leq \underline{p} < \overline{p} \leq 1$. No new issues arise here: the robust region for this joint conclusion is still the intersection of the two separate robust regions. Keep in mind, though, that whether we look at a one-sided or a two-sided conclusion is unrelated to the fact that we use \emph{lower} confidence bands in section \ref{sec:EstimationAndInference}.

Finally, the bootstrap procedures we propose in section \ref{sec:EstimationAndInference} can also be used to do inference on these joint breakdown frontiers. For simplicity, though, in that section we focus on the case where we are only interested in the conclusion $\Prob(Y_1 - Y_0 > z) \geq \underline{p}$.

\section{Estimation and inference}\label{sec:EstimationAndInference}

In this section we study estimation and inference on the breakdown frontier defined above. The breakdown frontier is a known functional of the conditional cdf of outcomes given treatment and covariates, the probability of treatment given covariates, and the marginal distribution of the covariates. Hence we propose nonparametric sample analog plug-in estimators of the breakdown frontier. We derive $\sqrt{N}$-consistency and asymptotic distributional results using a delta method for directionally differentiable functionals. We then use a bootstrap procedure to construct asymptotically valid lower confidence bands for the breakdown frontier. We conclude by discussing selection of the tuning parameter for this bootstrap procedure.

Although we focus on inference on the breakdown frontier, one might also be interested in doing inference directly on the parameters of interest. If we fix $c$ and $t$ a priori then we obtain identified sets for ATE, QTE, and the DTE from section \ref{sec:Model}. Our asymptotic results below may be used as inputs to traditional inference on partially identified parameters. See \cite{CanayShaikh2017} for a survey of this literature.

To establish our main asymptotic results, we present a sequence of results. Each result focuses on a different component of the overall breakdown frontier: (1) the bounds on the marginal distributions of potential outcomes conditional on $W$, (2) the CQTE bounds, (3) the CATE and ATE bounds, (4) breakdown points for ATE, (5) the CDTE under conditional rank invariance but without full conditional independence, (6) the DTE without either conditional rank invariance or full conditional independence, and finally (7) the breakdown frontier itself. 

We first suppose we observe a random sample of data.

\begin{partialIndepAssump}\label{assn:iid}
The random variables $\{ (Y_i,X_i,W_i) \}_{i=1}^N$ are independently and identically distributed according to the distribution of $(Y,X,W)$.
\end{partialIndepAssump}

We assume the support of $W$ is discrete. We sketch an approach to handling continuous covariates in appendix \ref{sec:contcovariates}. Note that $W$ may still be a vector.

\begin{partialIndepAssump}\label{assn:discreteW} 
The support of $W$ is discrete and finite. Let $\supp(W) = \{w_1,\ldots, w_K\}$.
\end{partialIndepAssump}

All parameters of interest are defined as functionals of the underlying parameters $F_{Y \mid X,W}(y \mid x,w)$, $p_{x|w} = \Prob(X=x \mid W=w)$, and $q_w = \P(W=w)$. Let
\[
	\widehat{F}_{Y \mid X,W}(y \mid x,w) = \frac{\frac{1}{N} \sum_{i=1}^N \ind(Y_i \leq y) \ind(X_i = x ,W_i = w) }{ \frac{1}{N} \sum_{i=1}^N \ind(X_i = x, W_i = w)},
\]
\[
	\widehat{p}_{x|w} = \frac{\frac{1}{N} \sum_{i=1}^N \ind(X_i = x, W_i = w)}{\frac{1}{N} \sum_{i=1}^N \ind(W_i = w)},
\]
and
\[
	\widehat{q}_w = \frac{1}{N} \sum_{i=1}^N \ind(W_i = w)
\]
denote the sample analog estimators of these three quantities, which converge uniformly to a Gaussian process at a $\sqrt{N}$-rate; see lemma \ref{lemma:cdf_conv} in appendix \ref{sec:proofsForEstimation}.

Next consider the bounds \eqref{eq:cdf upperbound} and \eqref{eq:cdf lowerbound} on the marginal distributions of potential outcomes. These population bounds are a functional $\phi_1$ evaluated at $(F_{Y \mid X,W}(\cdot \mid \cdot, \cdot),p_{(\cdot|\cdot)},q_{(\cdot)})$ where $p_{(\cdot|\cdot)}$ denotes the probability $p_{x|w}$ as a function of $(x,w)\in\{0,1\}\times \supp(W)$, and $q_{(\cdot)}$ denotes $q_w$ as a function of $w \in \supp(W)$. We estimate these bounds by a plug-in estimator $\phi_1(\widehat{F}_{Y \mid X,W}(\cdot \mid \cdot, \cdot),\widehat{p}_{(\cdot|\cdot)}, \widehat{q}_{(\cdot)})$. If this functional is differentiable in an appropriate sense, $\sqrt{N}$-convergence in distribution of its arguments will carry over to the functional by the delta method. The type of differentiability we require is \emph{Hadamard directional differentiability}, first defined by \cite{Shapiro1990} and \cite{Dumbgen1993}, and further studied in \cite{FangSantos2014}.

\begin{definition}\label{def:Hadamard_direc_diff}
Let $\mathbb{D}$ and $\mathbb{E}$ be Banach spaces with norms $\|\cdot\|_{\mathbb{D}}$ and $\|\cdot\|_{\mathbb{E}}$. Let $\mathbb{D}_\phi \subseteq \mathbb{D}$ and $\mathbb{D}_0 \subseteq \mathbb{D}$. The map $\phi:\mathbb{D}_\phi \rightarrow\mathbb{E}$ is \emph{Hadamard directionally differentiable} at $\theta\in \mathbb{D}_\phi$ tangentially to $\mathbb{D}_0$ if there is a continuous map $\phi'_\theta:\mathbb{D}_0 \rightarrow \mathbb{E}$ such that
\begin{align*}
  \lim_{n\rightarrow\infty} \left\| \frac{\phi(\theta + t_nh_n) - \phi(\theta)}{t_n} - \phi'_\theta(h)\right\|_{\mathbb{E}} &= 0
\end{align*}
\noindent for all sequences $\{h_n\}\subset \mathbb{D}$ and $\{t_n\} \in \R_+$ such that $t_n \searrow 0$, $\|h_n - h\|_{\mathbb{D}}\rightarrow 0$, $h\in\mathbb{D}_0$ as $n\rightarrow\infty$, and $\theta + t_nh_n \in \mathbb{D}_\phi$ for all $n$.
\end{definition}

If we further have that $\phi_\theta'$ is linear, then we say $\phi$ is \emph{Hadamard differentiable} (proposition 2.1 of \citealt{FangSantos2014}). Not every Hadamard directional derivative $\phi_\theta'$ must be linear, however.

We use the functional delta method for Hadamard directionally differentiable mappings (e.g., theorem 2.1 in \citealt{FangSantos2014}) to show convergence in distribution of our estimators. Such convergence is usually to a non-Gaussian limiting process. We do not use this distribution to do inference since obtaining analytical asymptotic confidence bands would be challenging. Instead, we use a bootstrap procedure to obtain asymptotically valid uniform confidence bands for our breakdown frontier and associated estimators.

Returning to our population bounds \eqref{eq:cdf upperbound} and \eqref{eq:cdf lowerbound}, we estimate these by
\begin{align}\label{eq:cdfbounds_estimates}
	\widehat{\overline{F}}^c_{Y_x \mid W}(y \mid w)
	&= \min\left\{ \frac{\widehat{F}_{Y \mid X,W}(y \mid x,w)\widehat{p}_{x|w}}{\widehat{p}_{x|w}-c}, \,
	\frac{\widehat{F}_{Y \mid X,W}(y \mid x,w)\widehat{p}_{x|w}+c}{\widehat{p}_{x|w}+c} \right\}\\
	\widehat{\underline{F}}^c_{Y_x \mid W}(y \mid w)
	&= \max\left\{ \frac{\widehat{F}_{Y \mid X,W}(y \mid x,w)\widehat{p}_{x|w}}{\widehat{p}_{x|w}+c}, \,
	\frac{\widehat{F}_{Y \mid X}(y \mid x,w)\widehat{p}_{x|w}-c}{\widehat{p}_{x|w}-c} \right\}.\notag
\end{align}
Note that these estimators may not perform well when $c$ is close to $p_{x|w}$. In our analysis we assume $c$ is bounded away from $p_{x|w}$. 

In addition to assumption A\ref{assn:continuity}, we make the following regularity assumptions.

\begin{partialIndepAssump}\label{assn:support_c_y}
\hfill
\begin{enumerate}
\item For each $x \in \{0,1 \}$ and $w\in\supp(W)$, $-\infty < \underline{y}_x(w) < \overline{y}_x(w) < +\infty$. \label{A4_1}

\item For each $x\in\{0,1\}$ and $w\in\supp(W)$, $F_{Y \mid X,W}(y \mid x,w)$ is continuously differentiable everywhere with density $f_{Y \mid X,W}(y \mid x,w)$ uniformly continuous in $y$, uniformly bounded from above, and uniformly bounded away from zero on $\supp(Y \mid X=x,W=w)$. \label{A4_2}

\end{enumerate}
\end{partialIndepAssump}

A\ref{assn:support_c_y}.\ref{A4_1} combined with our earlier assumption A\ref{assn:continuity}.\ref{A1_3} constrain the potential outcomes to have compact support. This compact support assumption is not used to analyze our cdf bounds estimators \eqref{eq:cdfbounds_estimates}, but we use it later to obtain estimates of the corresponding quantile function bounds uniformly over their arguments $u \in (0,1)$, which we then use to estimate the bounds on $\Prob(Q_{Y_1 \mid W}(U \mid w) - Q_{Y_0 \mid W}(U \mid w) \leq z)$. This is a well known issue when estimating quantile processes; for example, see \cite{Vaart2000} lemma 21.4(ii). A\ref{assn:support_c_y}.\ref{A4_2} requires the density of $Y \mid X,W$ to be bounded away from zero uniformly. This ensures that conditional quantiles of $Y \mid X,W$ are uniquely defined. It also implies that the limiting distribution of the estimated quantile bounds will be well-behaved. Uniform continuity of the density implies that the derivatives of the conditional quantile function with respect to $\tau$ are uniformly continuous.

For some of our main results in this section (lemmas \ref{lemma:conv_cdf_PI}, \ref{lemma:conv_intpiece}, \ref{lemma:conv_DTEbounds} and theorem \ref{thm:BF convergence}) we establish convergence uniformly over $c \in \mathcal{C}$ for some finite grid $\mathcal{C} =\{c_1,c_2,\ldots,c_J\} \subset [0, \min \{p_{1|w}, p_{0|w}\} )$. We discuss the choice of these grid points on page \pageref{sec:choosingGridPoints}. We constrain this grid to be below $\min \{p_{1|w}, p_{0|w}\}$ solely for simplicity, as all our results can be extended to grids $\mathcal{C}\subset [0,1]$ by combining our present bound estimates with estimates based on the $c \geq \min \{ p_{1|w}, p_{0|w} \}$ case given in \cite{MastenPoirier2017}. Weak convergence does not hold uniformly over an interval of $c$ since some of the functionals below are not Hadamard directionally differentiable when their codomain is a set of functions on that interval. To resolve this issue, we propose two ways of conducting inference on the breakdown frontier uniformly over intervals of $c$. The first is to use the fixed grid and monotonicity of the breakdown frontier to construct a uniform band. The second is to smooth the population breakdown frontier such that it is Hadamard differentiable when viewed as a function of $c$. We use the first approach in this section and the second approach in supplemental appendix D.

The next result establishes convergence in distribution of the cdf bound estimators. Here and below we use the following notation: For an arbitrary set $\mathcal{A}$ and a Banach space $\mathcal{B}$, $\ell^\infty(\mathcal{A},\mathcal{B})$ denotes the set of all maps $z: \mathcal{A} \rightarrow \mathcal{B}$ with finite sup-norm $\| z \| = \sup_{a \in \mathcal{A}} \| z(a) \|_{\mathcal{B}}$, equipped with this norm. For example, see \citet[page 381]{VaartWellner1996}.

\begin{lemma}\label{lemma:conv_cdf_PI}
Suppose A\ref{assn:continuity}, A\ref{assn:iid}, and A\ref{assn:discreteW} hold. Let $\mathcal{Y} \subset \R$ be a finite grid of points. Then
\begin{align*}
  \sqrt{N}
  \begin{pmatrix}
              \widehat{\overline{F}}^c_{Y_x \mid W}(y \mid w)  - \overline{F}^c_{Y_x \mid W}(y \mid w) \\
              \widehat{\underline{F}}^c_{Y_x \mid W}(y \mid w) - \underline{F}^c_{Y_x \mid W}(y \mid w)
	\end{pmatrix}
	&\rightsquigarrow \mathbf{Z}_2(y,x,w,c),
          \end{align*}
  \noindent a tight random element of $\ell^\infty(\mathcal{Y} \times \{0,1\} \times \supp(W)\times \mathcal{C},\R^2)$.
\end{lemma}

$\mathbf{Z}_2$ is not Gaussian itself, but it is a continuous transformation of Gaussian processes. For given $(x,c,w)$, the limit will be Gaussian at all values of $y$ except for
\[
	y \in\left\{ Q_{Y \mid X,W}\left(\frac{p_{x|w} - c}{2p_{x|w}} \mid x,w\right), \ Q_{Y \mid X,W}\left(\frac{p_{x|w} + c}{2p_{x|w}} \mid x,w\right)\right\}.
\]
A characterization of this limiting process is given in the proof in appendix \ref{sec:proofsForEstimation}.

Next consider the conditional quantile bounds \eqref{eq:quantilebounds_smallc}, which we estimate by
\begin{align*}
    \widehat{\overline{Q}}_{Y_x \mid W}^c(\tau)  &= \widehat{Q}_{Y \mid X,W}\left(\tau + \frac{c}{\widehat{p}_{x|w}}\min\{\tau,1-\tau\}  \mid x,w\right) \\
    \widehat{\underline{Q}}_{Y_x \mid W}^c(\tau) &= \widehat{Q}_{Y \mid X,W}\left(\tau - \frac{c}{\widehat{p}_{x|w}}\min\{\tau,1-\tau\}  \mid x,w\right).
\end{align*}
The next result establishes uniform convergence in distribution of these quantile bounds estimators. For the following results, let $\overline{C} \in (0,\min\{p_{1|w},p_{0|w}\})$ for all $w \in \supp(W)$.

\begin{lemma}\label{lemma:conv_quantiles_PI}
Suppose A\ref{assn:continuity}, A\ref{assn:iid}, A\ref{assn:discreteW}, and A\ref{assn:support_c_y} hold. Then
\begin{align*}
  \sqrt{N}
  \begin{pmatrix}
              \widehat{\overline{Q}}^c_{Y_x \mid W}(\tau \mid w)  - \overline{Q}^c_{Y_x \mid W}(\tau \mid w) \\
              \widehat{\underline{Q}}^c_{Y_x \mid W}(\tau \mid w) - \underline{Q}^c_{Y_x \mid W}(\tau \mid w)
	\end{pmatrix}
&\rightsquigarrow \mathbf{Z}_3(\tau,x,w,c),
\end{align*}
a mean-zero Gaussian process in $\ell^\infty((0,1) \times \{0,1\} \times \supp(W) \times [0,\overline{C}],\R^2)$ with continuous paths.
\end{lemma}

This result is uniform in $c$ on an interval, in $x\in\{0,1\}$, $w\in\supp(W)$, and in $\tau\in(0,1)$. This result directly implies convergence over $c\in\mathcal{C}$ as well. Unlike the distribution of the cdf bounds estimators, this process is Gaussian. This follows by Hadamard differentiability of the mapping between $\theta_0 \equiv (F_{Y \mid X,W}(\cdot \mid \cdot,\cdot), p_{(\cdot|\cdot)}, q_{(\cdot)})$ and the conditional quantile bounds. By applying the functional delta method, we can show asymptotic normality of smooth functionals of these bounds. A first set of functionals are the CQTE bounds of equation \eqref{eq:QTE bounds}, which are a linear combination of the quantile bounds. Let
\[
	\widehat{\underline{\text{CQTE}}}(\tau,c \mid w) = \widehat{\underline{Q}}^c_{Y_1 \mid W}(\tau \mid w) - \widehat{\overline{Q}}^c_{Y_0 \mid W}(\tau \mid w)
\]
and
\[
	\widehat{\overline{\text{CQTE}}}(\tau,c \mid w) = \widehat{\overline{Q}}^c_{Y_1 \mid W}(\tau \mid w) - \widehat{\underline{Q}}^c_{Y_0 \mid W}(\tau \mid w).
\]
Then,
\[
	 \sqrt{N}
	 \begin{pmatrix}
               \widehat{\overline{\text{CQTE}}}(\tau,c \mid w) - \overline{\text{CQTE}}(\tau,c \mid w) \\
              \widehat{\underline{\text{CQTE}}}(\tau,c \mid w) - \underline{\text{CQTE}}(\tau,c \mid w)
	\end{pmatrix}
          \rightsquigarrow
          \begin{pmatrix}
          	\mathbf{Z}_3^{(1)}(\tau,1,w,c) - \mathbf{Z}_3^{(2)}(\tau,0,w,c) \\
          	\mathbf{Z}_3^{(2)}(\tau,1,w,c) - \mathbf{Z}_3^{(1)}(\tau,0,w,c)
          \end{pmatrix},
\]
where the superscript $\mathbf{Z}^{(j)}$ denotes the $j$th component of the vector $\mathbf{Z}$.

A second set of functionals are the CATE bounds from equation \eqref{eq:ATE bounds}. These bounds are smooth linear functionals of the CQTE bounds. Therefore the joint asymptotic distribution of these bounds can be established by the continuous mapping theorem. Let
\[
	\widehat{\underline{\text{CATE}}}(c \mid w) = \int_0^1 \widehat{\underline{\text{CQTE}}}(u,c \mid w) \; du
	\qquad \text{and} \qquad
	\widehat{\overline{\text{CATE}}}(c \mid w) = \int_0^1 \widehat{\overline{\text{CQTE}}}(u,c \mid w) \; du.
\]
Then, by the linearity of the integral operator, these estimated CATE bounds converge to their population counterpart at a $\sqrt{N}$-rate and therefore
\begin{align*}
             \sqrt{N}
             \begin{pmatrix}
              \widehat{\overline{\text{CATE}}}(c \mid w) - \overline{\text{CATE}}(c \mid w) \\
              \widehat{\underline{\text{CATE}}}(c \mid w) - \underline{\text{CATE}}(c \mid w)
              \end{pmatrix}
          &\rightsquigarrow
          \begin{pmatrix}
                                   \int_0^1 (\mathbf{Z}_3^{(1)}(u,1,w,c) - \mathbf{Z}_3^{(2)}(u,0,w,c)) \; du \\
                                   \int_0^1 (\mathbf{Z}_3^{(2)}(u,1,w,c) - \mathbf{Z}_3^{(1)}(u,0,w,c)) \; du
	\end{pmatrix},
\end{align*}
a mean-zero Gaussian process in $\ell^\infty(\supp(W) \times [0,\overline{C}],\R^2)$ with continuous paths. 

We estimate the unconditional $\text{ATE}$ bounds by integrating over the empirical distribution of the covariates $W$: Let
\[
	\widehat{\underline{\text{ATE}}}(c) = \avg \widehat{\underline{\text{CATE}}}(c \mid W_i)
	\qquad \text{and} \qquad
	\widehat{\overline{\text{ATE}}}(c) = \avg \widehat{\overline{\text{CATE}}}(c \mid W_i).
\]
The following decomposition implies that the estimated ATE upper bound converges weakly to a Gaussian element:
\begin{align*}
	&\sqrt{N}\left(\widehat{\overline{\text{ATE}}}(c) - \overline{\text{ATE}}(c)\right) \\
	&= \sqrt{N}\sum_{k=1}^K \widehat{q}_{w_k}\left( \widehat{\overline{\text{CATE}}}(c \mid w_k) - \overline{\text{CATE}}(c \mid w_k) \right) + \sum_{k=1}^K \overline{\text{CATE}}(c \mid w_k) \sqrt{N}(\widehat{q}_{w_k} - q_{w_k})\\
	&\rightsquigarrow \sum_{k=1}^K q_{w_k} \int_0^1 (\mathbf{Z}_3^{(1)}(u,1,w_k,c) - \mathbf{Z}_3^{(2)}(u,0,w_k,c)) \; du + \sum_{k=1}^K \overline{\text{CATE}}(c \mid w_k) \mathbf{Z}_1^{(3)}(0,0,w_k).
\end{align*}
A similar result holds for the estimated ATE lower bound.

Next consider estimation of the breakdown point for the claim that $\text{ATE}\geq \mu$ where $\mu \in \R$. To focus on the nondegenerate case, suppose the population value of ATE obtained under full independence is greater than $\mu$, $\underline{\text{ATE}}(0) > \mu$ which implies $c^* > 0$. Let
\[
	\widehat{c}^* = \inf\{c\in[0,1]: \widehat{\underline{\text{ATE}}}(c) \leq \mu\}
\]
be the estimated breakdown point. This is the estimated smallest relaxation of independence such that we cannot conclude that the ATE is strictly greater than $\mu$. By the properties of the quantile bounds as a function of $c$, the function $\underline{\text{ATE}}(c)$ is non-increasing and differentiable in $c$. We now present a result about the asymptotic distribution of $\widehat{c}^*$. 

\begin{proposition}\label{prop:conv_breakdownpoint}
Suppose A\ref{assn:continuity}, A\ref{assn:iid}, A\ref{assn:discreteW}, and A\ref{assn:support_c_y} hold. Assume $c^*\in(0,\overline{C}]$. Then $\sqrt{N}(\widehat{c}^* - c^*) \rightsquigarrow \mathbf{Z}_{4}$, a Gaussian random variable.
\end{proposition}

The assumption that $c^*\in(0,\overline{C}]$ can again be relaxed to the general case where $c^*\in(0,1]$ but we maintain the stronger assumption for brevity. 

Under conditional rank invariance, we can also establish asymptotic normality of bounds for $\P(Q_{Y_1 \mid W}(U \mid w) - Q_{Y_0 \mid W}(U \mid w) \leq z)$ for a fixed $z \in \R$. These bounds are given by
\[
	(\underline{P}(c \mid w),\overline{P}(c \mid w)) \equiv (\underline{\text{CDTE}}(z,c,0 \mid w),\overline{\text{CDTE}}(z,c,0 \mid w)).
\]
We keep $z$ implicit in the notation for these bounds. Estimates for these quantities are provided by
\begin{align*}
	\widehat{\underline{P}}(c \mid w) 
	&= \int_0^1 \ind(\widehat{\overline{Q}}^c_{Y_1 \mid W}(u \mid w) - \widehat{\underline{Q}}^c_{Y_0 \mid W}(u \mid w) \leq z) \; du \\
	\widehat{\overline{P}}(c \mid w) 
	&= \int_0^1 \ind(\widehat{\underline{Q}}^c_{Y_1 \mid W}(u \mid w) - \widehat{\overline{Q}}^c_{Y_0 \mid W}(u \mid w) \leq z) \; du.
\end{align*}
Asymptotic normality can be established using the Hadamard directional differentiability of the mapping from the differences in quantile bounds to the bounds $(\underline{P}(c \mid w),\overline{P}(c \mid w))$. This mapping is called the \emph{pre-rearrangement operator}. \cite{ChernozhukovFernandez-ValGalichon2010} showed that this operator was Hadamard differentiable when the quantile functions are continuously differentiable for all $u \in (0,1)$. In our case, the underlying quantile functions are continuously differentiable on $(0,1/2)\cup(1/2,1)$, and continuous but not differentiable at $u=1/2$. At this value, the left and right derivatives exist and are finite, but are generally different from one another. We extend the result of \cite{ChernozhukovFernandez-ValGalichon2010} to the case where the quantile function has a point of non-differentiability by showing Hadamard directional differentiability of this mapping.

To do so, we make additional assumptions on the behavior of these quantile functions. 
\begin{partialIndepAssump}\label{assn:smoothness_intpiece}
For each $c \in \mathcal{C}$ and $w\in\supp(W)$,
\begin{enumerate}
\item \label{A5_1} The number of elements in each of the sets
\begin{align*}
	\mathcal{U}^*_1(c \mid w) = \{u\in(0,1) &: \partial_u^- (\overline{Q}^c_{Y_1 \mid W}(u \mid w) - \underline{Q}^c_{Y_0 \mid W}(u \mid w)) = 0 \\
	&\quad \text{ or } \partial_u^+ (\overline{Q}^c_{Y_1 \mid W}(u \mid w) - \underline{Q}^c_{Y_0 \mid W}(u \mid w)) = 0\}
\\[0.5em]
\mathcal{U}^*_2(c \mid w) = \{u\in(0,1) &: \partial_u^- (\underline{Q}^c_{Y_1 \mid W}(u \mid w) - \overline{Q}^c_{Y_0 \mid W}(u \mid w)) = 0 \\
&\quad \text{ or } \partial_u^+ (\underline{Q}^c_{Y_1 \mid W}(u \mid w) - \overline{Q}^c_{Y_0 \mid W}(u \mid w)) = 0\}
\end{align*}
is finite.

\item \label{A5_2} The following hold.
\begin{enumerate}
\item For any $u\in\mathcal{U}^*_1(c \mid w)$, $\overline{Q}^c_{Y_1 \mid W}(u \mid w) - \underline{Q}^c_{Y_0 \mid W}(u \mid w) \neq z$. 

\item For any $u\in\mathcal{U}^*_2(c \mid w)$, $\underline{Q}^c_{Y_1 \mid W}(u \mid w) - \overline{Q}^c_{Y_0 \mid W}(u \mid w) \neq z$.
\end{enumerate}
\end{enumerate}
\end{partialIndepAssump}

These assumptions imply that the respective function's derivatives change signs a finite number of times, and therefore they cross the horizontal line at $z$ a finite number of times. These functions are continuously differentiable in $u$ everywhere on $(0,1/2)\cup(1/2,1)$, and are directionally differentiable at $1/2$. The second assumption rules out the functions being flat when exactly valued at $z$. Failure of the second condition in this assumption implies that convergence will hold uniformly over any compact subset that excludes these values, which typically form a measure-zero set. Therefore this assumption can be satisfied by considering convergence for values of $c$ which exclude those where the second part of assumption A\ref{assn:smoothness_intpiece} fails. Without knowing a priori at which values this assumption may fail, selecting grid points randomly from a continuous distribution ensures that these values are selected with probability zero.\label{paragraph:randomChoiceOfGridPoints}

An alternative approach to inference if the second condition fails for some values of $c$ is to smooth the population function using methods described in supplemental appendix D. Like in \citet[corollary 4]{ChernozhukovFernandez-ValGalichon2010}, we require a tuning parameters to control the level of smoothing. We show that $\sqrt{N}$-convergence holds for all parameter values when introducing any amount of fixed smoothing. 

Finally, note that A\ref{assn:smoothness_intpiece} is refutable, since it is expressed as a function of identified quantities, namely the CQTE bounds for all $u \in (0,1)$.  
 
With this additional assumption we can show $\sqrt{N}$-convergence of the bounds uniformly in $\supp(W)\times \mathcal{C}$.

\begin{lemma}\label{lemma:conv_intpiece}
Suppose A\ref{assn:continuity}, A\ref{assn:iid}, A\ref{assn:discreteW}, A\ref{assn:support_c_y}, and A\ref{assn:smoothness_intpiece} hold. Then
\[
	\sqrt{N}
	\begin{pmatrix}
              \widehat{\overline{P}}(c \mid w) - \overline{P}(c \mid w) \\
              \widehat{\underline{P}}(c \mid w) - \underline{P}(c \mid w)
	\end{pmatrix}
	\rightsquigarrow \mathbf{Z}_{5}(w,c),
\]
a tight random element in $\ell^\infty(\supp(W)\times\mathcal{C},\R^2)$.
\end{lemma}

If conditional random assignment holds ($c=0$) in addition to conditional rank invariance ($t=0$), then the CDTE is point identified and lemma \ref{lemma:conv_intpiece} gives the asymptotic distribution of the sample analog CDTE estimator (in this case the upper and lower bound functions are equal). This can be considered an estimator of the CDTE in one of the models of \cite{Matzkin2003}.

We now establish the limiting distribution of the CDTE bounds uniformly in $(c,t,w)$. Let
\begin{align*}
	\widehat{\underline{\text{CDTE}}}(z,c,t \mid w)
	&= (1-t)\widehat{\underline{P}}(c \mid w) 
	+ t \max\left\{\sup_{y\in\mathcal{Y}_z(w)} (\widehat{\underline{F}}^c_{Y_1 \mid W}(a \mid w) - \widehat{\overline{F}}^c_{Y_0 \mid W}(y-z \mid w)),0\right\} \\
	\widehat{\overline{\text{CDTE}}}(z,c,t \mid w) 
	&= (1-t)\widehat{\overline{P}}(c \mid w)
	+ t \left(1+ \min\left\{\inf_{y\in\mathcal{Y}_z(w)}(\widehat{\overline{F}}^c_{Y_1 \mid W}(y \mid w) - \widehat{\underline{F}}^c_{Y_0 \mid W}(y-z \mid w)),0\right\}\right).
\end{align*}
We estimate the unconditional DTE bounds by integrating the estimated CDTE bounds over the empirical distribution of the covariates: Let
\[
	\widehat{\underline{\text{DTE}}}(z,c,t) = \frac{1}{N}\sum_{i=1}^N \widehat{\underline{\text{CDTE}}}(z,c,t \mid W_i)
	\qquad \text{and} \qquad
	\widehat{\overline{\text{DTE}}}(z,c,t) = \frac{1}{N}\sum_{i=1}^N \widehat{\overline{\text{CDTE}}}(z,c,t \mid W_i).
\]

We have shown in lemma \ref{lemma:conv_intpiece} that the terms $\underline{P}(c \mid w)$ and $\overline{P}(c \mid w)$ are estimated at a $\sqrt{N}$-rate by the Hadamard directional differentiability of the mapping linking empirical cdfs and these terms. We now show that the second components of the CDTE bounds are a Hadamard directionally differentiable functional as well, leading to the $\sqrt{N}$ joint convergence of the DTE bounds to a tight, random element uniformly in $c$ and $t$.

\begin{lemma}\label{lemma:conv_DTEbounds}
Fix $z \in \R$. Suppose A\ref{assn:continuity}, A\ref{assn:iid}, A\ref{assn:discreteW}, A\ref{assn:support_c_y}, and A\ref{assn:smoothness_intpiece} hold. Then 
\begin{equation}\label{eq:conv_DTEbounds}
\sqrt{N}
	\begin{pmatrix}
              \widehat{\overline{\text{DTE}}}(z,c,t) - \overline{\text{DTE}}(z,c,t) \\
              \widehat{\underline{\text{DTE}}}(z,c,t) - \underline{\text{DTE}}(z,c,t)
         \end{pmatrix}
         \rightsquigarrow \mathbf{Z}_6(c,t),
\end{equation}
a tight random element of $\ell^\infty(\mathcal{C}\times[0,1],\R^2)$ with continuous paths.
\end{lemma}

Having established the convergence in distribution of the DTE, we can now show that the breakdown frontier also converges in distribution uniformly over its arguments. Denote the estimated breakdown frontier for the conclusion that $\P(Y_1 > Y_0) \geq \underline{p}$ by
\begin{equation}\label{eq:empiricalBF}
	\widehat{\text{BF}}(c,\underline{p})
	= \min\{ \max \{ \widehat{\text{bf}}(c,\underline{p}), 0 \}, 1 \}
\end{equation}
where
\begin{equation}\label{eq:empiricalbf}
	\widehat{\text{bf}}(c,\underline{p})
	= 
	\frac{
	\widehat{\text{num}}
	}{
	\widehat{\text{denom}}
	}
\end{equation}
with
\begin{align*}
	\widehat{\text{num}} 
	&=	
	1 - \underline{p} - \frac{1}{N}\sum_{i=1}^N \widehat{\overline{P}}(c \mid W_i) \\
	\widehat{\text{denom}}
	&=
	1 + \frac{1}{N}\sum_{i=1}^N \left[\min \left\{ 
	\inf_{y \in \mathcal{Y}_0(W_i)} 
	(\overline{F}^c_{Y_1 \mid W}(y \mid W_i) - \underline{F}^c_{Y_0 \mid W}(y \mid W_i)), 0 
	\right\} 
	- \widehat{\overline{P}}(c \mid W_i)\right].
\end{align*}

By combining our previous lemmas, we can show that the estimated breakdown frontier converges in distribution.

\begin{theorem}\label{thm:BF convergence}
Suppose A\ref{assn:continuity}, A\ref{assn:iid}, A\ref{assn:discreteW}, A\ref{assn:support_c_y}, and A\ref{assn:smoothness_intpiece} hold. Let $\mathcal{P}\subset [0,1]$ be a finite grid of points. Then 
\[
	\sqrt{N}(\widehat{\text{BF}}(c,\underline{p}) - \text{BF}(c,\underline{p}))
	\rightsquigarrow \mathbf{Z}_7(c,\underline{p}),
\]
a tight random element of $\ell^{\infty}(\mathcal{C}\times\mathcal{P})$.
\end{theorem}

This result essentially follows from the convergence of the preliminary estimators established in lemma \ref{lemma:cdf_conv} in appendix \ref{sec:proofsForEstimation} and by showing that the breakdown frontier is a composition of a number of Hadamard differentiable and Hadamard directionally differentiable mappings, implying convergence in distribution of the estimated breakdown frontier. 

Breakdown frontiers for more complex conclusions can typically be constructed from breakdown frontiers for simpler conclusions. For example, consider the breakdown frontier for the joint conclusion that $\P(Y_1 > Y_0) \geq \underline{p}$ and $\text{ATE} \geq \mu$. Then the breakdown frontier for this joint conclusion is the minimum of the two individual frontier functions. Alternatively, consider the conclusion that $\Prob(Y_1 > Y_0) \geq \underline{p}$ \emph{or} $\text{ATE} \geq \mu$, or both, hold. Then the breakdown frontier for this joint conclusion is the maximum of the two individual frontier functions. Since the minimum and maximum operators are Hadamard directionally differentiable, these joint breakdown frontiers will also converge in distribution.

Since the limiting process is non-Gaussian, inference on the breakdown frontier is not based on standard errors as with Gaussian limiting theory. Our processes' distribution is characterized fully by the expressions in appendix \ref{sec:proofsForEstimation}, but obtaining analytical estimates of quantiles of functionals of these processes would be challenging. In the next subsection we give details on the bootstrap procedure we use to construct confidence bands for the breakdown frontier.

\subsection*{Bootstrap inference}

As mentioned earlier, we use a bootstrap procedure to do inference on the breakdown frontier rather than directly using its limiting process. In this subsection we discuss how to use the bootstrap to approximate this limiting process. In the next subsection we discuss its application to constructing uniform confidence bands.

First we define some general notation. Let $Z_i = (Y_i,X_i,W_i)$ and $Z^N = \{Z_1,\ldots,Z_N\}$. Let $\theta_0$ denote some parameter of interest and let $\widehat{\theta}$ be an estimator of $\theta_0$ based on the data $Z^N$. Let $\mathbf{A}_N^*$ denote $\sqrt{N}(\widehat{\theta}^* - \widehat{\theta})$ where $\widehat{\theta}^*$ is a draw from the nonparametric bootstrap distribution of $\widehat{\theta}$. Suppose $\A$ is the tight limiting process of $\sqrt{N} ( \widehat{\theta} - \theta_0)$. Denote bootstrap consistency by $\mathbf{A}_N^* \overset{P}{\rightsquigarrow} \mathbf{A}$ where $\overset{P}{\rightsquigarrow}$ denotes weak convergence in probability, conditional on the data $Z^N$. Weak convergence in probability conditional on $Z^N$ is defined as
\[
	\sup_{h\in \text{BL}_1} \left| \Exp [h(\mathbf{A}_N^*) \mid Z^N] - \Exp [h(\mathbf{A})] \right| = o_p(1)
\]
where $\text{BL}_1$ denotes the set of Lipschitz functions into $\R$ with Lipschitz constant no greater than 1.

We focus on the following specific choices of $\theta_0$ and $\widehat{\theta}$:
\[
	\theta_0 =
	\begin{pmatrix}
		F_{Y \mid X,W}(\cdot \mid \cdot,\cdot) \\
		p_{(\cdot|\cdot)}\\
		q_{(\cdot)}
	\end{pmatrix}
	\qquad \text{and} \qquad
	\widehat{\theta} =
	\begin{pmatrix}
		\widehat{F}_{Y \mid X,W}(\cdot \mid \cdot,\cdot) \\
		\widehat{p}_{(\cdot|\cdot)}\\
		\widehat{q}_{(\cdot)}
	\end{pmatrix}.	
\]
For these choices, let $\mathbf{Z}_N^* = \sqrt{N}(\widehat{\theta}^* - \widehat{\theta})$. Let $\mathbf{Z}_1$ denote the limiting distribution of $\sqrt{N}(\widehat{\theta} - \theta_0)$; see lemma \ref{lemma:cdf_conv} in appendix \ref{sec:proofsForEstimation}. Theorem 3.6.1 of \cite{VaartWellner1996} implies that $\mathbf{Z}_N^* \overset{P}{\rightsquigarrow} \mathbf{Z}_1$. Our parameters of interest are all functionals $\phi$ of $\theta_0$. For Hadamard differentiable functionals $\phi$, the nonparametric bootstrap is consistent. For example, see theorem 3.1 of \cite{FangSantos2014}. They further show that $\phi$ is Hadamard differentiable \emph{if and only if}
\[
	\sqrt{N}(\phi(\widehat{\theta}^*) - \phi(\widehat{\theta})) \overset{P}{\rightsquigarrow} \phi'_{\theta_0}(\mathbf{Z}_1)
\]
where $\phi_{\theta_0}'$ denotes the Hadamard derivative at $\theta_0$. This implies that the nonparametric bootstrap can be used to do inference on the QTE and ATE bounds since they are Hadamard differentiable functionals of $\theta_0$. A second implication is that the nonparametric bootstrap is not consistent for the DTE or for the breakdown frontier for claims about the DTE since they are Hadamard directionally differentiable mappings of $\theta_0$, but they are not ordinary Hadamard differentiable.

In such cases, \cite{FangSantos2014} show that a different bootstrap procedure is consistent. Specifically, let  $\widehat{\phi}_{\theta_0}'$ be a consistent estimator of $\phi'_{\theta_0}$. Then their results imply that
\[
	\widehat{\phi}_{\theta_0}'(\mathbf{Z}_N^*) \overset{P}{\rightsquigarrow} \phi'_{\theta_0}(\mathbf{Z}_1).
\]
Analytical consistent estimates of $\phi'_{\theta_0}$ are often difficult to obtain, so \cite{Dumbgen1993} and \cite{HongLi2015} propose using a numerical derivative estimate of $\phi'_{\theta_0}$. Their estimate of the limiting distribution of $\sqrt{N} ( \phi( \widehat{\theta}) - \phi(\theta_0) )$ is given by the distribution of
\begin{equation}\label{eq:numericalbootstrap}
	\widehat{\phi}_{\theta_0}'(\sqrt{N}(\widehat{\theta}^* - \widehat{\theta}))
	= \frac{\phi\left(\widehat{\theta} + \varepsilon_N \sqrt{N}(\widehat{\theta}^* - \widehat{\theta})\right) - \phi(\widehat{\theta})}{\varepsilon_N}
\end{equation}
across the bootstrap estimates $\widehat{\theta}^*$. Under the rate constraints $\varepsilon_N \rightarrow 0$ and $\sqrt{N} \varepsilon_N \rightarrow \infty$, and some measurability conditions stated in their appendix, \cite{HongLi2015} show
\[
	\widehat{\phi}_{\theta_0}'(\sqrt{N}(\widehat{\theta}^* - \widehat{\theta})) \overset{P}{\rightsquigarrow} \phi'_{\theta_0}(\mathbf{Z}_1).
\]
where the left hand side is defined in equation \eqref{eq:numericalbootstrap}.

This bootstrap procedure requires evaluating $\phi$ at two values, which is computationally simple. It also requires selecting the tuning parameter $\varepsilon_N$, which we discuss later. Note that the standard, or naive, bootstrap is a special case of this numerical delta method bootstrap where $\varepsilon_N = N^{-1/2}$.

\subsection*{Uniform confidence bands for the breakdown frontier}

In this subsection we combine all of our asymptotic results thus far to construct uniform confidence bands for the breakdown frontier. As in section \ref{sec:Model} we use the function $\text{BF}(\cdot,\underline{p})$ to characterize this frontier. We specifically construct one-sided \emph{lower} uniform confidence bands. That is, we will construct a lower band function $\widehat{\text{LB}}(c)$ such that
\[
	\lim_{N \rightarrow \infty} \P \left( \widehat{\text{LB}}(c) \leq \text{BF}(c,\underline{p}) \text{ for all $c \in [0,1]$} \right) = 1-\alpha.	
\]
We use a one-sided lower uniform confidence band because this gives us an \emph{inner} confidence set for the robust region. Specifically, define the set
\[
	\text{RR}_L = \{ (c,t) \in [0,1]^2 : t \leq \widehat{\text{LB}}(c) \}.
\]
Then validity of the confidence band $\widehat{\text{LB}}$ implies
\[
	\lim_{N \rightarrow \infty} \P \left( \text{RR}_L \subseteq \text{RR}(0,\underline{p}) \right) = 1-\alpha.
\]
Thus the area underneath our confidence band, $\text{RR}_L$, is interpreted as follows: Across repeated samples, approximately $100(1-\alpha)$\% of the time, every pair $(c,t) \in \text{RR}_L$ leads to a population level identified set for the parameter $\Prob(Y_1 > Y_0)$ which lies weakly above $\underline{p}$. Put differently, approximately $100(1-\alpha)$\% of the time, every pair $(c,t) \in \text{RR}_L$ still lets us draw the conclusion we want at the population level. Hence the size of this set $\text{RR}_L$ is a finite sample measure of robustness of our conclusion to failure of the point identifying assumptions. We discuss an alternative testing-based interpretation on page 5 in supplemental appendix A.

One might be interested in constructing one-sided \emph{upper} confidence bands if the goal was to do inference on the set of assumptions for which we \emph{cannot} come to the conclusion of interest. This might be useful in situations where two opposing sides are debating a conclusion. But since our focus is on trying to determine when we can come to the desired conclusion, rather than looking for when we cannot, we only describe the one-sided lower confidence band case. 

When studying inference on scalar breakdown points, \cite{KlineSantos2013} constructed one-sided lower confidence intervals. Unlike for breakdown frontiers, uniformity over different points in the assumption space is not a concern for inference on breakdown points. See supplemental appendix A for more discussion.

We consider bands of the form
\[
	\widehat{\text{LB}}(c) = \widehat{\text{BF}}(c,\underline{p}) - \widehat{k}(c)
\]
for some function $\widehat{k}(\cdot) \geq 0$. This band is an asymptotically valid lower uniform confidence band of level $1-\alpha$ if
\[
	\lim_{N \rightarrow \infty} \P \left( \widehat{\text{BF}}(c,\underline{p}) - \widehat{k}(c) \leq \text{BF}(c,\underline{p}) \text{ for all $c \in [0,1]$} \right) = 1-\alpha,	
\]
or, equivalently, if
\[
	\lim_{N \rightarrow \infty} \P \left( \sup_{c\in[0,1]} \sqrt{N} \Big( \widehat{\text{BF}}(c,\underline{p}) - \text{BF}(c,\underline{p}) - \widehat{k}(c) \Big) \leq 0\right) = 1-\alpha.	
\]

In our theoretical analysis, we consider $\widehat{k}(c) = \widehat{z}_{1-\alpha} \sigma(c)$ for a scalar $\widehat{z}_{1-\alpha}$ and a function $\sigma$. We focus on known $\sigma$ for simplicity. We start by deriving a uniform band over a grid $\mathcal{C}$, then extend it over an interval using monotonicity of the breakdown frontier. As discussed earlier, we only derive uniformity of the band over $c \in [0,\overline{C}]$ rather than over $c \in [0,1]$, but this is also for brevity and can be relaxed. The choice of $\sigma$ affects the shape of the confidence band, and there are many possible choices of the function $\sigma$ which yield valid level $1-\alpha$ uniform confidence bands. See \cite{FreybergerRai2016} for a detailed analysis. A simple choice of $\sigma$ is the constant function: $\sigma(c) = 1$, which delivers an equal width uniform band. Alternatively, as we do below, one could choose $\sigma(c)$ to construct a minimum width confidence band (equivalently, maximum area of $\text{RR}_L$).

\begin{proposition}\label{prop:bootstrap validity}
Suppose A\ref{assn:continuity}, A\ref{assn:iid}, A\ref{assn:support_c_y}, and A\ref{assn:smoothness_intpiece} hold. Define $\phi : \ell^\infty(\R \times \{0,1\},\R^2) \rightarrow \ell^\infty(\mathcal{C})$ such that 
\[
	\widehat{\text{BF}}(c,\underline{p}) = [\phi(\widehat{\theta})](c).
\]
Then $\phi$ is Hadamard directionally differentiable. Suppose that $\varepsilon_N \rightarrow 0$ and $\sqrt{N} \varepsilon_N \rightarrow\infty$. Let $\widehat{\theta}^*$ denote a draw from the nonparametric bootstrap distribution of $\widehat{\theta}$. Then
\begin{align}\label{eq:bootstrap_conv_BF}
	 \left[ \widehat{\phi}_{\theta_0}'(\sqrt{N}(\widehat{\theta}^* - \widehat{\theta})) \right]
	 \overset{P}{\rightsquigarrow}  \left[ \phi'_{\theta_0}(\mathbf{Z}_1) \right]
	 &\equiv \mathbf{Z}_7.
\end{align}
For a given function $\sigma(\cdot)$ such that $\inf_{c\in\mathcal{C}}\sigma(c)>0$, define
\begin{equation}\label{eq:empirical critical value}
	\widehat{z}_{1-\alpha}
	= 
	\inf\left\{ z \in \R : \P\left(\sup_{c\in\mathcal{C}}\frac{ \left[ \widehat{\phi}_{\theta_0}'(\sqrt{N}(\widehat{\theta}^* - \widehat{\theta})) \right] (c,\underline{p})}{\sigma(c)} \leq z \mid Z^N\right)\geq 1-\alpha\right\}.
\end{equation}	 
Finally, suppose also that the cdf of
\[
	\sup_{c\in\mathcal{C}}\frac{[\phi_{\theta_0}'(\mathbf{Z}_1) ](c,\underline{p})}{\sigma(c)} = \sup_{c\in \mathcal{C}}\frac{\mathbf{Z}_7(c,\underline{p})}{\sigma(c)}
\]
is continuous and strictly increasing at its $1-\alpha$ quantile, denoted $z_{1-\alpha}$. Then $\widehat{z}_{1-\alpha} = z_{1-\alpha} + o_p(1)$.
\end{proposition}

This proposition is a variation of corollary 3.2 in \cite{FangSantos2015workingPaper}. As a consequence of this result, the lower $1-\alpha$ band $\widehat{\text{LB}}(c) = \widehat{\text{BF}}(c,\underline{p}) - \widehat{z}_{1-\alpha}\sigma(c)$ is valid uniformly on the grid $\mathcal{C}$. To extend the uniformity to all of $[0,\overline{C}]$ we propose the following lower confidence band:
\[
	\widetilde{\text{LB}}(c)
		= 
		\begin{cases}
			\widehat{\text{LB}}(c_1) 
			&\text{ if $c\in [0,c_1]$} \\
			\quad \vdots \\
			\widehat{\text{LB}}(c_j) 
			&\text{ if $c \in (c_{j-1},c_j]$, for $j=2,\ldots,J$} \\
			\quad \vdots \\
			\quad 0
			&\text{ if $c \in (c_J,  \overline{C}]$}.
		\end{cases}
\]
This band is a step function which interpolates between grid points using the least monotone interpolation. The following result shows its validity.

\begin{corollary}\label{prop:uniform band interval}
Let the assumptions of proposition \ref{prop:bootstrap validity} hold. Then, $\widetilde{\text{LB}}(c)$ is a uniform lower $1-\alpha$ band for $\text{BF}(c,\underline{p})$ over $c\in[0,\overline{C}]$.
\end{corollary}

Corollary \ref{prop:uniform band interval} shows that, for any fixed $J \geq 1$, the interpolated lower confidence band preserves exact the $1-\alpha$ coverage on the grid points. This follows by monotonicity of the breakdown frontier; see lemma \ref{lemma:monotonicBandCoverage} in appendix \ref{sec:appendixProofs}. That said, this interpolated band might not be taut, in the sense that there may exist other lower bands with $1-\alpha$ coverage that are weakly larger than $\widetilde{\text{LB}}(c)$ for all $c$ and strictly larger at some values of $c$. See \cite{FreybergerRai2016} for further discussion of taut confidence bands.

Proposition \ref{prop:bootstrap validity} can be extended to estimated functions $\sigma$, although we leave the details for future work. We use an estimated $\sigma$ in our application, as described next. When both $z_{1-\alpha}$ and $\sigma$ are estimated, we work directly with $\widehat{k}(c) = \widehat{z}_{1-\alpha} \widehat{\sigma}(c)$. We choose $\widehat{k}(c)$ to minimize an approximation to the area between the confidence band and the estimated function; equivalently, to maximize the area of $\text{RR}_L$. Specifically, we let $\widehat{k}(c_1),\ldots,\widehat{k}(c_J)$ solve
\[
	\min_{k(c_1),\ldots,k(c_J) \geq 0} \; \sum_{j=2}^J k(c_j)(c_{j} - c_{j-1})
\]
subject to
\[
	\P\left(\sup_{c\in\{c_1,\ldots,c_J\}} \sqrt{N} \Big(\widehat{\text{BF}}(c,\underline{p}) - \text{BF}(c,\underline{p}) - k(c) \Big) \leq 0 \right) = 1-\alpha,
\]
where we approximate the left-hand side probability via the numerical delta method bootstrap. The criterion function here is just a right Riemann sum over the grid points. This optimization is not computationally costly: It is only performed once per value of $\underline{p}$ and $\varepsilon_N$. Moreover, in our empirical illustration it takes an average of 15 seconds per run on a mid-2013 MacBook Air.

\subsection*{Choosing the grid points $\mathcal{C}$}\label{sec:choosingGridPoints}

Here we suggest three approaches for choosing the number and location of the grid points $\mathcal{C} = \{ c_1,\ldots,c_J \}$. First, one can let $\mathcal{C} = \{ \overline{c}_1,\ldots, \overline{c}_K \}$ where $K$ is the number of observed covariates and, for each $k \in \{1,\ldots,K\}$, $\overline{c}_k$ is the maximal deviation between the observed propensity score and the ``leave out variable $k$'' propensity score, which we define and discuss in our empirical illustration on page \pageref{paragraph:leaveOutVariableK}. There we argue that these are natural points to consider.

Second, researchers can choose equally spaced grid points, for a fixed $J$. Third, researchers can randomly select grid points from a continuous distribution, for a fixed $J$. As mentioned on page \pageref{paragraph:randomChoiceOfGridPoints}, this random selection ensures that assumption A\ref{assn:smoothness_intpiece} holds with probability one. Both of these approaches are commonly used in the literature. Moreover, both of these approaches can be used in combination with the first approach. In this case, researchers may want to begin with $\{ \overline{c}_1,\ldots, \overline{c}_K \}$ and then add a multiple $m$ of $K$ additional points, so that the total number of points $J$ is $K + mK$ for some positive integer $m$.

Finally, we emphasize that for our asymptotic approximations to be valid, the key restriction is that the grid $\mathcal{C}$ is not too dense around the points $c$ of nondifferentiability. Picking a sufficiently sparse fixed finite grid is one way to ensure this. Thus, although we state our formal results for a fixed grid, one could instead let $J \rightarrow \infty$ sufficiently slowly as $N \rightarrow \infty$. Alternatively, one could pre-estimate the points of nondifferentiability and then let $\mathcal{C}$ equal the entire domain, minus sufficiently large intervals around these points. The length of these removed intervals then shrinks asymptotically. \cite{HorowitzLee2012,HorowitzLee2017} discuss approaches like these in different settings.

\subsection*{Bootstrap selection of $\varepsilon_N$}

While \cite{Dumbgen1993} and \cite{HongLi2015} provide rate constraints on $\varepsilon_N$, they do not recommend a procedure for picking $\varepsilon_N$ in practice. In this section, we suggest a heuristic bootstrap method for picking $\varepsilon_N$. We use this method for our empirical illustration in section \ref{sec:empiricalApplication}; we also present the full range of bands considered. Since the question of choosing $\varepsilon_N$ goes beyond the purpose of the present paper, we defer a formal analysis of this method to future research. For discussions of bootstrap selection of tuning parameters in other problems, see \cite{Taylor1989}, \cite{LegerRomano1990}, \cite{Marron1992}, and \cite{CaoCuevasManteiga1994}.

Fix a $\underline{p}$. Let $\text{CP}_N(\varepsilon ; F_{Y,X,W})$ denote the finite sample coverage probability of our confidence band as described above, for a fixed $\varepsilon$. This statistic depends on the unknown distribution of the data, $F_{Y,X,W}$. The bootstrap replaces $F_{Y,X,W}$ with an estimator $\widehat{F}_{Y,X,W}$. We pick a grid $\{ \varepsilon_1,\ldots,\varepsilon_K \}$ of $\varepsilon$'s and let $\widehat{\varepsilon}_N$ solve
\[
	\min_{k =1,\ldots, K} | \text{CP}_N(\varepsilon_k ; \widehat{F}_{Y,X,W}) - (1-\alpha) |.
\]
We compute $\text{CP}_N$ by simulation. In our empirical illustration, we take $B=500$ draws. We use the same grid of $\varepsilon$'s as in our Monte Carlo simulations in supplemental appendix C. Larger grids and larger values of $B$ can be chosen subject to computational constraints. We furthermore must choose an estimator $\widehat{F}_{Y,X,W}$. The nonparametric bootstrap uses the empirical distribution. We use the smoothed bootstrap (\citealt{DeAngelisYoung1992}, \citealt{PolanskySchucany1997}). Specifically, we estimate the distribution of $(X,W)$ by its empirical distribution. We then let $\widehat{F}_{Y \mid X,W}$ be a kernel smoothed cdf estimate of the conditional cdf of $Y|X,W$. We use the standard logistic cdf kernel and the method proposed by \cite{Hansen2004} to choose the smoothing bandwidths. We divide these bandwidths in half since this visually appears to better capture the shape of the conditional empirical cdfs, and since smaller order bandwidths are recommended for the smoothed bootstrap (section 4 of \citealt{DeAngelisYoung1992}).

Bootstrap consistency requires sufficient smoothness of the functional of interest in the underlying cdf. It may be that the lack of smoothness that requires us to use the methods of \cite{FangSantos2014} and \cite{HongLi2015} in the first place also cause the naive bootstrap to be inconsistent for approximating the distribution of $\text{CP}_N(\varepsilon; F_{Y,X,W})$. As mentioned earlier, formally investigating this issue is beyond the scope of this paper. Our goal here is merely to suggest a simple first-pass approach at choosing $\varepsilon_N$.

\section{Empirical illustration: The effects of child soldiering}\label{sec:empiricalApplication}

In this section we use our results to examine the impact of assumptions in determining the effects of child soldiering on wages. We first briefly discuss the background and then we present our analysis.

\subsection*{Background}

We use data from phase 1 of SWAY, the Survey of War Affected Youth in northern Uganda, conducted by principal researchers Jeannie Annan and Chris Blattman (see \citealt{AnnanBlattmanHorton2006}). As \cite{BlattmanAnnan2010} discuss on page 882, a primary goal of this survey was to understand the effects of a twenty year war in Uganda, where ``an unpopular rebel group has forcibly recruited tens of thousands of youth''. In that paper, they use this data to examine the impacts of abduction on educational, labor market, psychosocial, and health outcomes. In our illustration, we focus solely on the impact of abduction on wages.

Blattman and Annan note that self-selection into the military is a common problem in the literature studying the effects of military service on outcomes. They argue that forced recruitment in Uganda led to random assignment of military service in their data. They first provide qualitative evidence for this, based on interviews with former rebels who led raiding parties. After murdering and mutilating civilians, the rebels had no public support, making abduction the only means of recruitment. Youths were generally taken during nighttime raids on rural households. According to the former rebel leaders, ``targets were generally unplanned and arbitrary; they raided whatever homesteads they encountered, regardless of wealth or other traits.''

This qualitative evidence is supported by their survey data, where Blattman and Annan show that most pre-treatment covariates are balanced across the abducted and nonabducted groups (see their table 2). Only two covariates are not balanced: year of birth and prewar household size. They say this is unsurprising because
\begin{quote}
``a youth's probability of ever being abducted depended on how many years of the conflict he fell within the [rebel group's] target age range. Moreover, abduction levels varied over the course of the war, so youth of some ages were more vulnerable to abduction than others. The significance of household size, meanwhile, is driven by households greater than 25 in number. We believe that rebel raiders, who traveled in small bands, were less likely to raid large, difficult-to-control households.'' (Page 887)
\end{quote}
Hence they use a selection-on-observables identification strategy, conditioning on these two variables.

While their evidence supporting the full conditional independence assumption is compelling, this assumption is still nonrefutable. Hence they apply the methods of \cite{Imbens2003} to analyze the sensitivity to this assumption. In this analysis they only consider one outcome variable, years of education. Likewise, as in \cite{Imbens2003}, they only look at one parameter, the constant treatment effect in a fully parametric model.

We complement their results by applying the breakdown frontier methods we develop in this paper. We focus on the log-wage outcome variable. We look at both the average treatment effect and $\Prob(Y_1 > Y_0)$, which was not studied in \cite{BlattmanAnnan2010}.

\subsection*{Analysis}

The original phase 1 SWAY data has 1216 males born between 1975 and 1991. Of these, wage data is available for 504 observations. 56 of these earned zero wages; we drop these and only look at people who earned positive wages. This leaves us with our main sample of 448 observations. In addition to this outcome variable, we let our treatment variable be an indicator that the person was \emph{not} abducted. We include the two covariates discussed above, age when surveyed and household size in 1996. Additional covariates can be included, but we focus on just these two for simplicity.

Table \ref{empiricalSummaryStats} shows summary statistics for these four variables. 36\% of our sample were not abducted. Age ranges from 14 years old to 30 years old, with a median of 22 years old. Household size ranges from 2 people to 28, with a median of 8 people. Wages range from as low as 36 shillings to as high as about 83,300 shillings, with a median of 1,400 shillings.

\begin{table}
\centering\caption{Summary statistics \label{empiricalSummaryStats}} 
\setlength{\linewidth}{.1cm} 
\newcommand{\contents}{ 
\begin{tabular}{l r r r r r} \hline\hline 
Variable Name & Mean & Median & Stddev & Min & Max  \\ \hline 
 Daily wage in Uganda shillings &  2957.60 &  1400.00 &  6659.76 &    35.71 & 83333.34 \\ 
  Log wage &     7.23 &     7.24 &     1.18 &     3.58 &    11.33 \\ 
  Not abducted? &     0.36 &     0.00 &     0.48 &     0.00 &     1.00 \\ 
  Age when surveyed &    22.11 &    22.00 &     4.88 &    14.00 &    30.00 \\ 
  Household size in 1996 &     8.31 &     8.00 &     4.19 &     2.00 &    28.00 \\ 
 \hline\hline 
\multicolumn{6}{p{0.95\linewidth}}{\footnotesize Sample size is 448. 1 USD is approximately 1800 Uganda shillings (Exchange rate at time of survey, 2005-2006; source: World Bank).} \\ 
\end{tabular}
} 
\setbox0=\hbox{\contents} 
\setlength{\linewidth}{\wd0\tabcolsep0.6em} 
\contents 
\end{table}

Age has 17 support points and household size has 21 support points. Hence there are 357 total covariate cells. Including the treatment variable, this yields 714 total cells, compared to our sample size of 448 observations. Since we focus on unconditional parameters, having small or zero observations per cell is not a problem in principle. However, in the finite sample we have, to ensure that our estimates of the cdf bounds $\overline{F}_{Y_x \mid W}^c(y \mid w)$ and $\underline{F}_{Y_x \mid W}^c(y \mid w)$ are reasonably smooth in $y$, we collapse our covariates as follows. We replace age with a binary indicator of whether one is above or below the median age. Likewise, we replace household size with a binary indicator of whether one lived in a household with above or below median household size. This reduces the number of covariate cells to 4, giving 8 total cells including the treatment variable. This yields approximately 55 observations per cell. While this crude approach suffices for our illustration, in more extensive empirical analyses one may want to use more sophisticated methods. For example, we could use discrete kernel smoothing, as discussed in \cite{LiRacine2008}, who also provide additional references. We also consider alternative coarsenings in supplemental appendix E.

Table \ref{empiricalMeanComparisons} shows unconditional comparisons of means of the outcome and the original covariates across the treatment and control groups. Wages for people who were not abducted are 702 shillings larger on average. People who were not abducted are also about 1.4 years younger than those who were abducted. People who were not abducted also had a slightly larger household size than those who were abducted. Only the difference in ages is statistically significant at the usual levels, but as in tables 2 and 3 of \cite{BlattmanAnnan2010} the standard errors can be decreased by including additional controls. These extra covariates are not essential for illustrating our breakdown frontier methods, however.

\begin{table}[t] \centering\caption{Comparison of means \label{empiricalMeanComparisons}} 
\setlength{\linewidth}{.1cm} 
\newcommand{\contents}{ 
\begin{tabular}{l r r r} \hline\hline 
Variable Name & Not Abducted & Abducted & Difference  \\ \hline 
 Daily wage in Uganda shillings &  3409.12 &  2706.75 &   702.36 \, [725.49] \\ 
  Log wage &     7.33 &     7.18 &     0.15 \, [0.12] \\ 
  Age when surveyed &    21.23 &    22.60 &    -1.37 \,  [0.48] \\ 
  Household size in 1996 &     8.53 &     8.19 &     0.34 \, [0.42] \\ 
 \hline 
Observations & 160 & 288 & \\ 
\hline\hline 
\multicolumn{4}{p{0.95\linewidth}}{\footnotesize Sample size is 448. 1 USD is approximately 1800 Uganda shillings (Exchange rate at time of survey, 2005-2006; source: World Bank). Standard errors in brackets.} \\ 
\end{tabular}
} 
\setbox0=\hbox{\contents} 
\setlength{\linewidth}{\wd0\tabcolsep0.6em} 
\contents 
\end{table}

The point estimates in table \ref{empiricalMeanComparisons} are unconditional on the two covariates. Next we consider the conditional independence assumption, with age and household size in 1996 as our covariates. Under this assumption, our estimate of ATE is 890 [726.13] shillings when the outcome variable is level of wages, and is 0.21 [0.11] when the outcome variable is log wage.\footnote{Using their full set of control variables, \cite{BlattmanAnnan2010} estimate ATE to be 0.33 [0.15] when the outcome is log wage. See column 1 of their table 3.} To check the robustness of these point estimates to failure of conditional independence, we estimate the breakdown point $c^*$ for the conclusion $\text{ATE} \geq 0$, where we use log-wages as our outcome variable. We measure relaxations of conditional independence by our conditional $c$-dependence distance. The estimated breakdown point is $\widehat{c}^* = 0.041$. Based on this point estimate, for all $x \in \{0,1\}$ and $w \in \supp(W)$ we can allow the conditional propensity scores $\Prob(X=x \mid Y_x = y, W=w)$ to vary $\pm 4$ percentage points around the observed propensity scores $\Prob(X=x \mid W=w)$ without changing our conclusion.

Is this a big or small amount of variation? Well, as a baseline, the upper bound on $c$ is about 0.73. This is an estimate of
\[
	\max_{w \in \supp(W)} \max \{ \Prob(X=1 \mid W=w), \Prob(X=0 \mid W=w) \}.
\]
Any $c \geq 0.73$ would lead to the no assumptions identified set for ATE. In this sense, $0.041$ is quite small, which would suggest that our results are quite fragile. Next we examine variation in the observed propensity scores as we suggested in \cite{MastenPoirier2017}. Specifically, we consider the difference between the ``full'' propensity score and the ``leave out variable $k$'' propensity score which omits variable $k$: Define\label{paragraph:leaveOutVariableK}
\[
	\overline{c}_\texttt{age} = \sup_{s=0,1} \sup_{a=0,1} | \widehat{\Prob}(X=1 \mid \texttt{age}=a,\texttt{hhSize}=s) - \widehat{\Prob}(X=1 \mid \texttt{hhSize} = s) |
\]
and
\[
	\overline{c}_\texttt{hhSize} = \sup_{a=0,1} \sup_{s=0,1} | \widehat{\Prob}(X=1 \mid \texttt{age}=a,\texttt{hhSize}=s) - \widehat{\Prob}(X=1 \mid \texttt{age} = a) |.
\]
Using these numbers as a reference, a robust result would have a breakdown point above one or both of the $\overline{c}$'s. In the data, we obtain $\overline{c}_\texttt{age} = 0.0625$ and $\overline{c}_\texttt{hhSize} = 0.0403$. The estimated breakdown point $\widehat{c}^* = 0.041$ is below $\overline{c}_\texttt{age}$ and approximately equal to $\overline{c}_\texttt{hhSize}$. This latter result suggests that perhaps our conclusion could be considered somewhat robust. Accounting for sampling uncertainty in the breakdown point, however, shows that the true breakdown point may be less than $\overline{c}_\texttt{hhSize}$. Overall, this suggests that our conclusion that $\text{ATE} \geq 0$ is not robust to relaxations of full conditional independence.

This argument for judging the plausibility of specific values of $c$ relies on using variation in the observed propensity score to ground our beliefs about reasonable variation in the unobserved propensity scores. The general question here is how one should quantitatively distinguish `large' and `small' relaxations of an assumption. This is an old and ongoing question in the sensitivity analysis literature, and much work remains to be done. For discussions on this point for different measures of deviations or relaxations from independence in various settings, see \cite{RotnitzkyRobinsScharfstein1998}, \cite{Robins1999}, \cite{Imbens2003}, \cite{AltonjiElderTaber2005,AltonjiElderTaber2008}, and \cite{Oster2016}.

Next consider the parameter $\Prob(Y_1 > Y_0)$. Since we define treatment as \emph{not} being abducted, this parameter measures the proportion of people who earn higher wages when they are not abducted, compared to when they are abducted. For this parameter, we must make both the full conditional independence assumption and the conditional rank invariance assumption to obtain point identification. Under these assumptions, our point estimate is $0.67$ with a one-sided lower 95\% CI of $[0.48, 1]$. 

Is this point estimate robust to failures of full conditional independence and conditional rank invariance? We examine this question by estimating breakdown frontiers and corresponding confidence bands for the conclusion that $\Prob(Y_1 > Y_0) \geq \underline{p}$. We do this for $\underline{p} = 0.1$, $0.25$, $0.5$ as in our Monte Carlo simulations in supplemental appendix C. We do not consider $\underline{p} = 0.75$ or $0.9$ since these values are larger than our point estimate under the baseline assumptions; they yield empty estimated robust regions. Besides picking a grid of $\underline{p}$'s a priori, one could let $\underline{p} = \widehat{p}_{0,0} / 2$, half the value of the parameter estimated under the baseline point identifying assumptions. In our application this is $0.34$; we omit this choice for brevity. \cite{Imbens2003} suggests a similar choice of cutoff in his approach. We use the same eight ratios of $\varepsilon_N / \varepsilon_N^\text{naive}$ as in our Monte Carlo simulations in supplemental appendix C.

\begin{figure}[t]
\centering
\includegraphics[width=54mm]{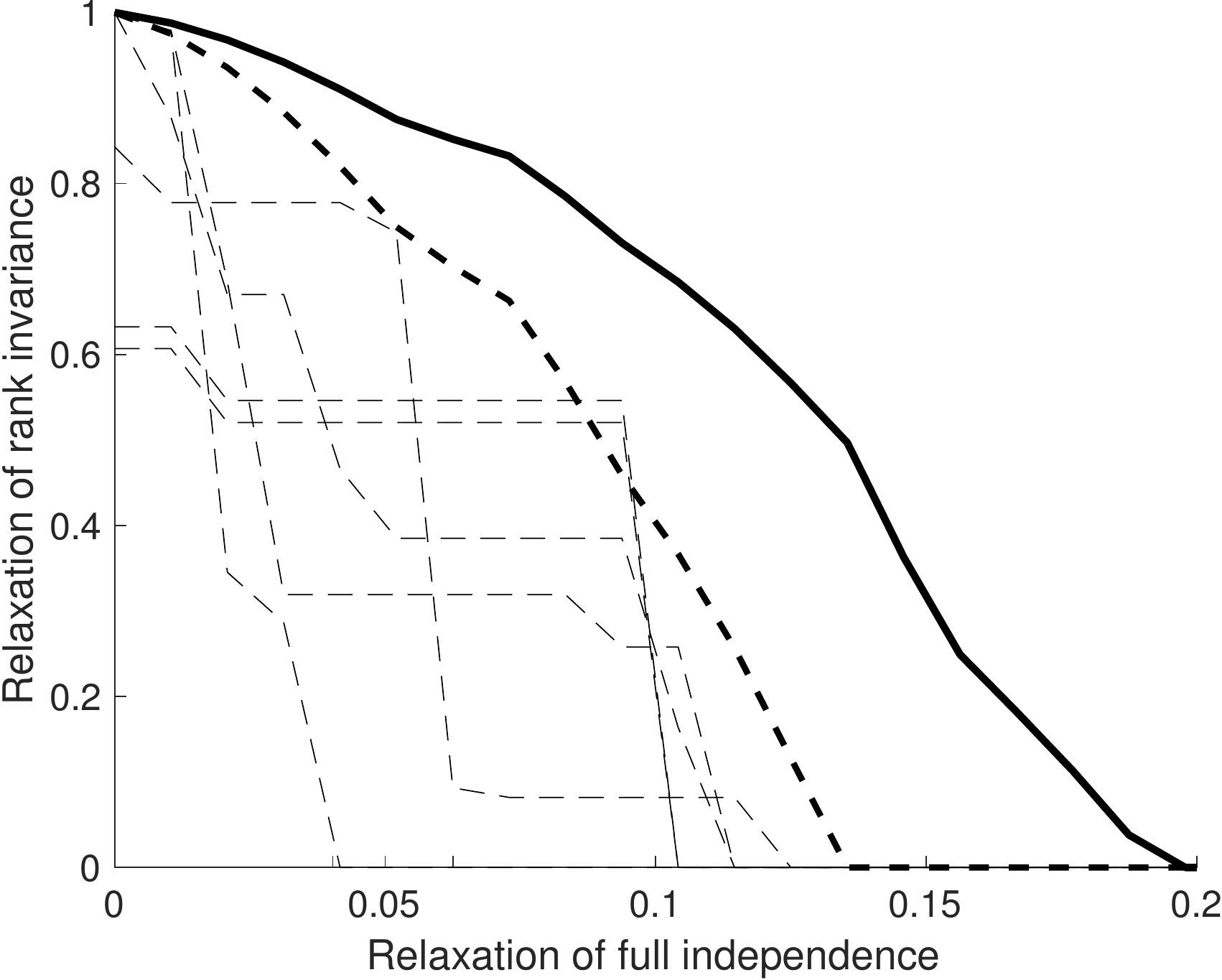}
\includegraphics[width=54mm]{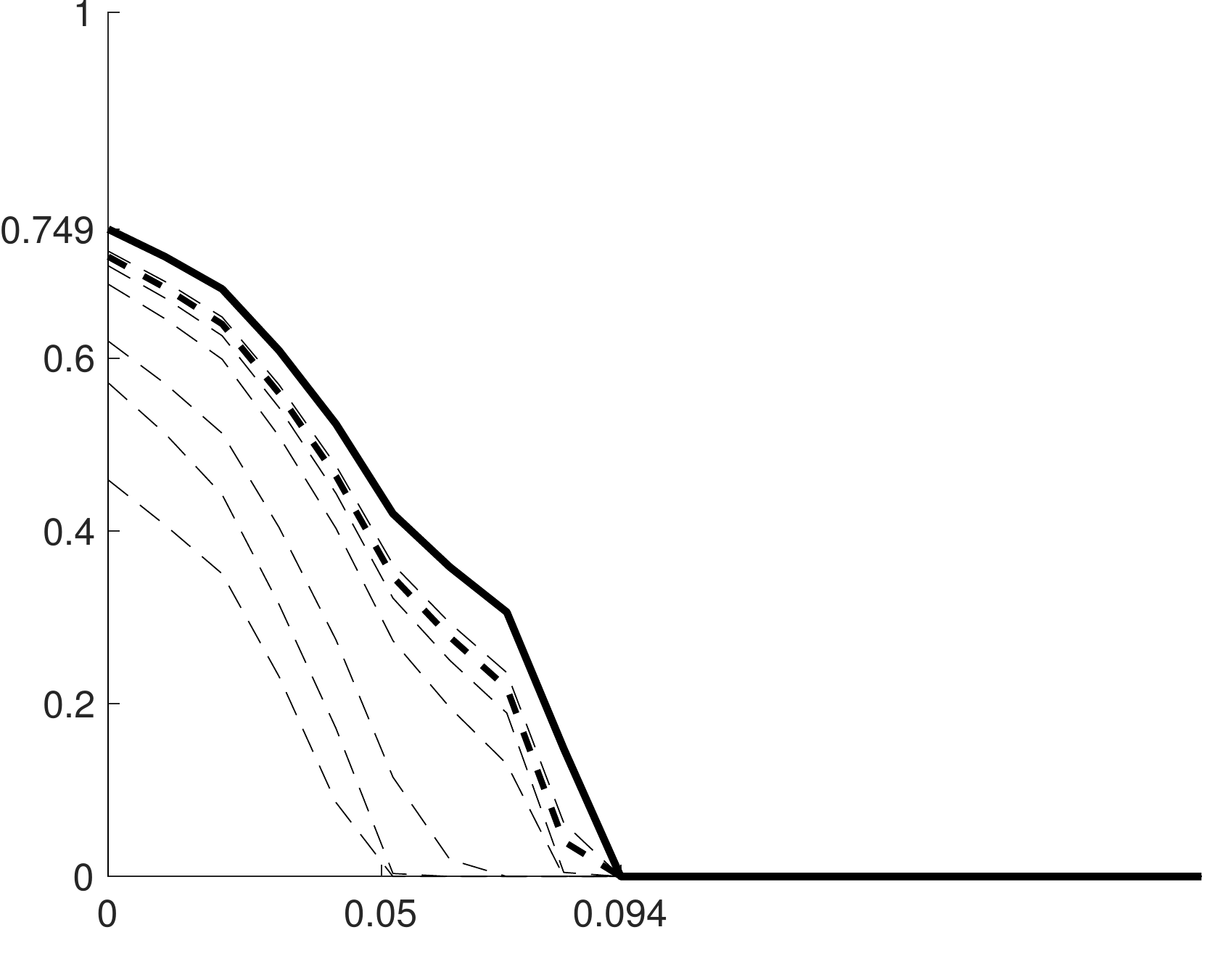}
\includegraphics[width=54mm]{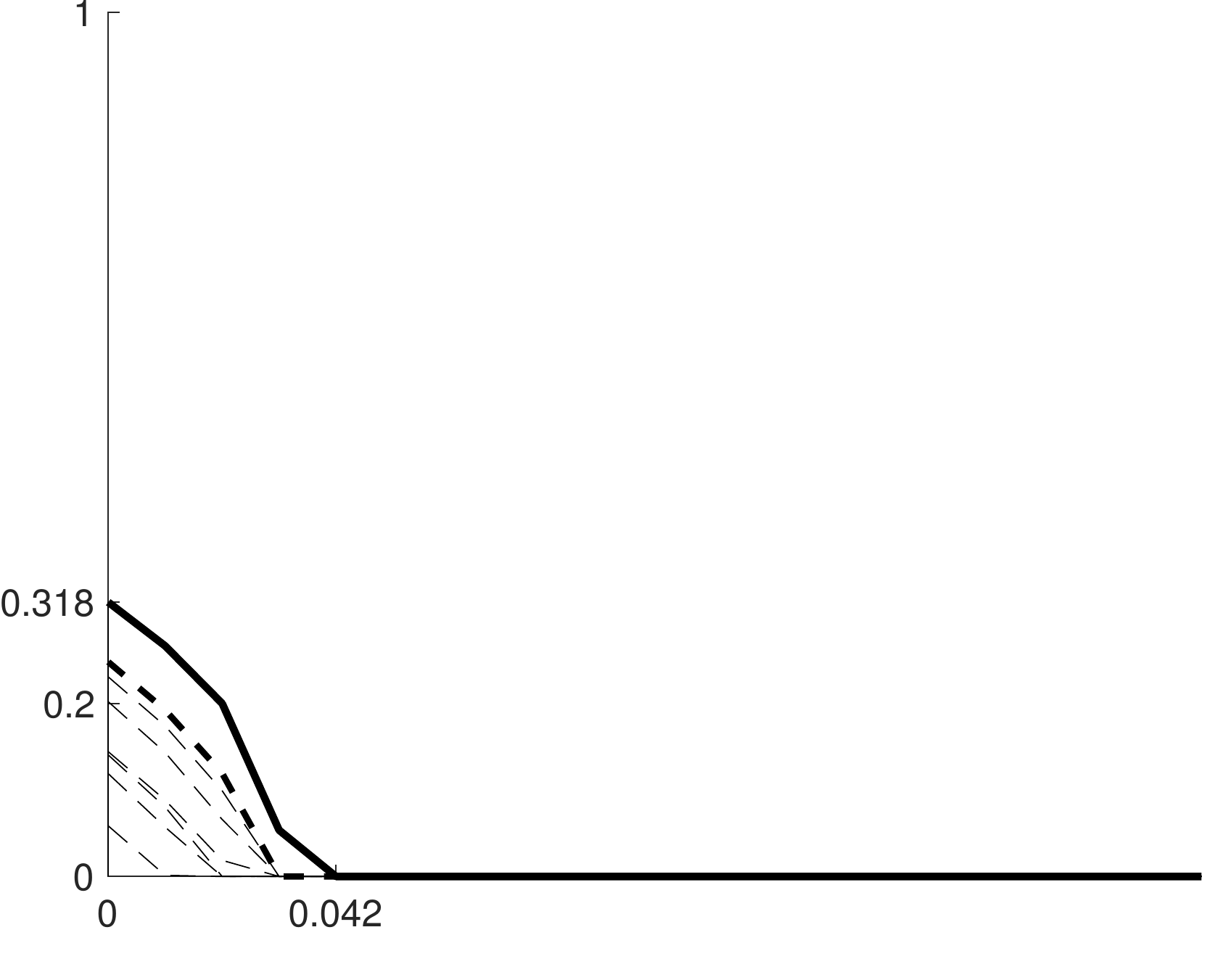}
\caption{Estimated breakdown frontiers (solid lines) and confidence bands (dashed lines) for the claim $\Prob(Y_1 > Y_0) \geq \underline{p}$. Left to right: $\underline{p} = 0.1$, $0.25$, $0.5$. Light dashed lines are confidence bands for all eight values of $\varepsilon_N$ considered. The darker dashed line is the band selected by our bootstrap procedure.}
\label{empiricalBFs}
\end{figure}

Figure \ref{empiricalBFs} shows the results. As in our earlier plots, the horizontal axis plots $c$, relaxations of full conditional independence, while the vertical axis plots $t$, relaxations of conditional rank invariance. As mentioned earlier, the natural upper bound for $c$ is about $0.73$. Since all of the breakdown frontiers intersect the horizontal axis at much smaller values, we have cut off the part of the overall assumption space with $c \geq 0.2$. Remember that, for the following analysis, it's valid to examine various $(c,t)$ combinations since we use uniform confidence bands.

First consider the left plot, $\underline{p} = 0.1$. Since this is the weakest conclusion of the five we consider, the estimated breakdown frontier and the corresponding robust region are the largest among the three plots. If we impose full conditional independence, then our estimated frontier suggests that we can completely relax conditional rank invariance and still conclude that at least 10\% of people benefit from not being forced into military service. Even accounting for sampling uncertainty, we can still draw this conclusion. Moreover, looking at all choices of $\varepsilon_N$---not just our selected one---the lowest the vertical intercept ever gets is about 61\%. Next suppose we relax full conditional independence. Recall that the maximal relaxation between the observed propensity score and the ``leave out variable $k$'' propensity scores gave $\overline{c}_\texttt{age} = 0.0625$ and $\overline{c}_\texttt{hhSize} = 0.0403$. Both of these numbers are substantially smaller than the horizontal intercept of our selected confidence band. Hence, if we impose full conditional rank invariance, our conclusion that $\Prob(Y_1 > Y_0) \geq 0.1$ is robust to relaxations of full conditional independence. Suppose instead that we think selection on unobservables is at most the largest $\overline{c}$ value, about $0.06$. Then for $c$'s in the range $[0,0.06]$, and accounting for sampling uncertainty, we can still conclude $\Prob(Y_1 > Y_0) \geq 0.1$ so long as at least 30\% of the population satisfies rank invariance. Thus we can relax full independence within this range without paying too high a cost in terms of requiring stronger rank invariance assumptions.

If we are willing to restrict selection on unobservables to be smaller than the largest $\overline{c}$ value, then we can allow for larger relaxations of conditional rank invariance. To quantify this trade-off, we can compute the difference between the values of the estimated breakdown frontier at two different points. As a starting point we recommend computing
\[
	\widehat{\text{BF}} (\overline{c}_{(K)}, \underline{p}) - \widehat{\text{BF}}(\overline{c}_{(K-1)}, \underline{p}) ,
\]
where $K$ denotes the number of observed regressors, $\overline{c}_{(K)}$ denotes the largest value of $\overline{c}$, and $\overline{c}_{(K-1)}$ denotes the second largest value. One could also divide this difference by $\overline{c}_{(K)} - \overline{c}_{(K-1)}$ to get a secant line. In our empirical application, $\overline{c}_{(K)} = \overline{c}_\texttt{age}$ and $\overline{c}_{(K-1)} = \overline{c}_\texttt{hhSize}$. Thus
\[
	\widehat{\text{BF}}(\overline{c}_\texttt{age}, 0.1) - \widehat{\text{BF}}(\overline{c}_\texttt{hhSize}, 0.1) = -6\%.
\]
The corresponding secant line has slope about $-3$. Thus if we assume selection on unobservables is at most as large as the \emph{second} largest amount of variation in ``leave out variable $k$'' propensity scores, we can allow for an additional 6\% of the population to violate conditional rank invariance. Put differently, around these values of $c$, allowing latent conditional propensity scores to vary an extra 1 percentage point requires us to impose that an additional 3\% of the population must satisfy conditional rank invariance. This rate of substitution generally increases as $c$ gets larger. Our ability to quantify this kind of trade-off between assumptions is a primary goal of our breakdown frontier analysis.

Overall, our results from this top left plot suggest that the conclusion that at least 10\% of people benefit from not being forced into military service is robust to relaxations of full conditional independence up to twice the size we see between the observed and leave out variable $k$ propensity scores, depending on how much conditional rank invariance failure we allow. For relaxations of full conditional independence up to the largest value of $\overline{c}$, we can allow up to 70\% of the population to deviate from conditional rank invariance, accounting for sampling uncertainty.

Next consider the middle plot, $\underline{p} = 0.25$. Since this is a stronger conclusion than the previous one, all the frontiers are shifted towards the origin. Consequently, by construction, this conclusion is not as robust as the other one. Our qualitative conclusions, however, as similar to those obtained for $\underline{p} = 0.1$. If we impose full conditional independence we can allow conditional rank invariance to fail for about 70\% of the population. Conversely, if we impose full conditional rank invariance, we can allow the latent conditional propensity scores to vary by about 10 percentage points---well beyond the largest observed variation $\overline{c}$. For $\underline{p} = 0.25$, we have
\[
	\widehat{\text{BF}}(\overline{c}_\texttt{age}, 0.25) - \widehat{\text{BF}}(\overline{c}_\texttt{hhSize}, 0.25) = -17\%.
\]
Hence the slope around our observed maximal $\overline{c}$'s is much larger for $\underline{p}=0.25$ as compared to $\underline{p}=0.1$. An important caveat to our conclusions for both $\underline{p} = 0.1$ and $\underline{p} = 0.25$ is that there is substantial variation in confidence bands as $\varepsilon_N$ changes. This point underscores the need for future work on the choice of $\varepsilon_N$.

Next consider the right plot, $\underline{p} = 0.5$. Here we consider the conclusion that at least half of people benefit from not being forced into military service. If we impose full conditional independence, and accounting for sampling uncertainty, then we can allow conditional rank invariance to fail for about 25\% of the population. This is quite large, but it relies on full conditional independence holding exactly. If we also relax conditional independence to $c = 0.03$ then we need conditional rank invariance to hold for everyone if we still want to conclude that at least 50\% of people benefit from not being forced into military service. $0.03$ is smaller than both $\overline{c}_\texttt{age}$ and $\overline{c}_\texttt{hhSize}$. Hence we might not be comfortable with such small values of $c$. This suggests the data do not definitively support the conclusion $\Prob(Y_1 > Y_0) \geq 0.5$, even though our point estimate under the baseline assumptions is $0.67$.

In this section we used our breakdown frontier methods to study the robustness of conclusions about ATE and $\Prob(Y_1 > Y_0)$ to failures of conditional independence and conditional rank invariance. We first considered the conclusion that the average treatment effect of not being abducted on log wages is nonnegative. Our point estimates suggest that this conclusion is robust to deviations in unobserved latent propensity scores up to the same value as $\overline{c}_\texttt{age}$, which is also about two-thirds as large as $\overline{c}_\texttt{hhSize}$; this robustness does not hold up when accounting for sampling uncertainty, however. We then considered the conclusion that at least $\underline{p}$\% of people earn higher wages when they are not abducted. This conclusion is robust to large simultaneous relaxations of conditional rank invariance and conditional independence for $\underline{p} =$ 10\%. For $\underline{p} =$ 25\%, This conclusion continues to be robust to reasonable relaxations, although after accounting for the variation in confidence bands over $\varepsilon_N$, this conclusion appears to be more sensitive to conditional independence than to conditional rank invariance. This robustness to rank invariance matches the findings of \cite{HeckmanSmithClements1997}, who imposed full independence and studied deviations from rank invariance. In their table 5B they found that, in their empirical application, one could generally conclude that $\Prob(Y_1 > Y_0)$ was at least 50\%, regardless of the assumption on rank invariance. In our empirical application our results are not quite as robust to rank invariance failures, which could be because we use a different measure of relaxation of rank invariance, and also because of differences in the empirical applications.

\section{Conclusion}

\subsection*{Summary}

In this paper we advocated the \emph{breakdown frontier} approach to sensitivity analysis. Given a set of baseline assumptions, this approach defines the population breakdown frontier as the weakest set of assumptions such that a specific conclusion of interest holds. Sample analog estimates and lower uniform confidence bands allow researchers to do inference on this frontier. The area under the confidence band is a quantitative, finite sample measure of the robustness of a conclusion to relaxations of point-identifying assumptions. To examine this robustness, empirical researchers can present these estimated breakdown frontiers and their accompanying confidence bands along with traditional point estimates and confidence intervals obtained under point identifying assumptions. We illustrated this general approach in the context of a treatment effects model, where the robustness of conclusions about ATE and $\Prob(Y_1 > Y_0)$ to relaxations of random assignment and rank invariance are examined. We applied these results in an empirical study of the effect of child soldiering on wages. We found that weak conclusions about $\Prob(Y_1 > Y_0)$ are fairly robust to failures of both rank invariance and random assignment, but stronger conclusions are more sensitive to relaxations of random assignment.

\subsection*{Breakdown frontier analysis for other models and other relaxations}

As discussed in section \ref{sec:intro}, breakdown frontier analysis can in principle be done in most models. In that section we outlined the six main steps required for any breakdown frontier analysis. In this paper we illustrated this general approach by studying a single important and widely used model: the potential outcomes model with a binary treatment. In future work it would be helpful to perform breakdown frontier analyses in other models. In particular, it may be possible to do breakdown frontier analyses in a large class of models by using the general identification analysis in \cite{ChesherRosen2015} or \cite{Torgovitsky2015}.

A key conceptual step in any breakdown frontier analysis is deciding how to define the indexed classes of assumptions such that the magnitude of the relaxation can be reasonably interpreted. This is not easy, and will generally depend on the model, the specific kind of assumption being relaxed, and the empirical context. Moreover, this choice may affect our findings: A conclusion can be robust with respect to one measure of relaxation but not another. Thus one goal of future research is to explore this space of assumption relaxations, to understand their substantive interpretations, and to chart their implications for the robustness of empirical findings. In \cite{MastenPoirier2016} we have already compared three different measures of relaxation of the random assignment assumption, including the one used here. We further studied quantile independence, a common relaxation of random assignment, in \cite{MastenPoirier2018QI}. In the present paper, we also used a general method for spanning two discrete assumptions by defining a (1$-t$)-percent relaxation, as we did with rank invariance. But much work still remains to be done.

\singlespacing
\bibliographystyle{econometrica}
\bibliography{BF_paper}

\appendix

\section{Related literature}\label{sec:relatedLit}

In this section, we review the related literature. We begin with the identification literature on breakdown points; as mentioned earlier, here we use ``breakdown'' in the same sense as Horowitz and Manski's \citeyearpar{HorowitzManski1995} identification breakdown point. This breakdown point idea goes back to the one of the earliest sensitivity analyses, performed by \cite{CornfieldEtAl1959}. They essentially asked how much correlation between a binary treatment and an unobserved binary confounder must be present to fully explain an observed correlation between treatment and a binary outcome, in the absence of any causal effects of treatment. This level of correlation between treatment and the confounder is a kind of breakdown point for the conclusion that some causal effects of treatment are nonzero. Their approach was substantially generalized by \cite{RosenbaumRubin1983sensitivity}, which is discussed in detail in chapter 22 of \cite{ImbensRubin2015}. Neither \cite{CornfieldEtAl1959} nor \cite{RosenbaumRubin1983sensitivity} formally defined breakdown points. 

\cite{HorowitzManski1995} gave the first formal definition and analysis of breakdown points. They studied a ``contaminated sampling'' model, where one observes a mixture of draws from the distribution of interest and draws from some other distribution. An upper bound $\lambda$ on the unknown mixing probability indexes identified sets for functionals of the distribution of interest. They focus on a single conclusion: That this functional is not equal to its logical bounds. They then define the breakdown point $\lambda^*$ as the largest $\lambda$ such that this conclusion holds. Put differently, $\lambda^*$ is the largest mixing probability we can allow while still obtaining a nontrivial identified set for our parameter of interest. They also relate this ``identification breakdown point'' to the earlier breakdown point concepts studied in the robust statistics literature (e.g., \citealt{HRRS1986} pages 96--98 and \citealt{HuberRonchetti2009} section 1.4 and chapter 11).

More generally, much work by Manski distinguishes between informative and noninformative bounds (which the literature also sometimes calls tight and non-tight bounds; see section 7.2 of \citealt{HoRosen2016}). The breakdown point is the boundary between the informative and noninformative cases. For example, see his analysis of bounds on quantiles with missing outcome data on page 40 of \cite{Manski2007}. There the identification breakdown point for the $\tau$th quantile occurs when $\max \{ \tau, 1 - \tau \}$ is the proportion of missing data. Similar discussions are given throughout the book.

\cite{Stoye2005,Stoye2010} generalizes the formal identification breakdown point concept by noting that breakdown points can be defined for any claim about the parameter of interest. He then studies a specific class of relaxations of the missing-at-random assumption in a model of missing data. \cite{KlineSantos2013} study a different class of relaxations of the missing-at-random assumption and also define a breakdown point based on that class.

While all of these papers study a scalar breakdown point, \cite{Imbens2003} studies a model of treatment effects where deviations from conditional random assignment are parameterized by two numbers $r = (r_1,r_2)$. His parameter of interest $\theta(r)$ is point identified given a fixed value of $r$. Imbens' figures 1--4 essentially plot estimated level sets of this function $\theta(r)$, in a transformed domain. While suggestive, these level sets do not generally have a breakdown frontier interpretation. This follows since non-monotonicities in the function $\theta(r)$ lead to level sets which do not always partition the space of sensitivity parameters into two connected sets in the same way that our breakdown frontier does.

\cite{ManskiPepper2018} also study a model where relaxations of baseline assumptions are parameterized by a vector of numbers $r$. Unlike Imbens, however, they derive identified sets indexed by $r$. These sets are weakly increasing (in the set inclusion order) in each component of $r$, and hence the non-monotonicity issue does not arise. For a two-dimensional relaxation, their table 2 presents identified sets as a function of a grid of $r = (r_1,r_2)$ values. The boundary between the italicized identified sets in that table and the non-highlighted sets is essentially a discrete approximation to the breakdown frontier in their model, for the claim that the parameter of interest is positive. Similarly, the boundary between the bold identified sets in that table and the non-highlighted sets is essentially a discrete approximation to the breakdown frontier in their model, for the claim that the parameter of interest is negative.

Neither \cite{HorowitzManski1995} nor \cite{Stoye2005,Stoye2010} discuss estimation or inference of breakdown points. \cite{Imbens2003} estimates his level sets in an empirical application, but does not discuss inference. \cite{ManskiPepper2018} also do not discuss estimation of or inference on breakdown frontiers, although inference in their setting is conceptually complicated---see their discussion on pages 234--235. \cite{KlineSantos2013}, on the other hand, is the first and only paper we're aware of that explicitly suggests doing inference on a breakdown point. We build on their work by proposing to do inference on the multi-dimensional breakdown frontier. This allows us to study the trade-off between different assumptions in drawing conclusions. They do study something they call a `breakdown curve', but this is a collection of scalar breakdown points for many different claims of interest, analogous to the collection of frontiers presented in figures \ref{BF_illustration2}, \ref{BF_illustration3}, and \ref{BF_illustration4}. Inference on a frontier rather than a point also raises additional issues they did not discuss; see our supplemental appendix A for more details. Moreover, we study a model of treatment effects while they look at a model of missing data, hence our identification analysis is different.

Building on \cite{HorowitzManski1995}, \cite{KreiderPepperGundersenJolliffe2012} combine a continuous relaxation sensitivity analysis for assumptions regarding measurement error with various discrete relaxations of assumptions regarding treatment selection. This allows them to study the interaction between these two kinds of assumptions in drawing conclusions. For inference they present confidence intervals for partially identified parameters for a variety of values of the relaxations, rather than doing inference on breakdown frontiers. See \cite{GundersenKreiderPepper2012} and \cite{KreiderPepperRoy2016} for further examples of identification analysis combining discrete and continuous relaxations.

Our breakdown frontier is a known functional of the distribution of outcomes given treatment and covariates and the observed propensity scores. This functional is not Hadamard differentiable, however, which prevents us from applying the standard functional delta method to obtain its asymptotic distribution. Instead, we show that it is Hadamard directionally differentiable, which allows us to apply the results of \cite{FangSantos2014}. We then use the numerical bootstrap of \cite{Dumbgen1993} and \cite{HongLi2015} to construct our confidence bands. For other applications of Hadamard directional differentiability, see \cite{LeeBhattacharya2016}, \cite{Kaido2016}, and \cite{Hansen2017}.

Our identification analysis builds on two strands of literature. First is the literature on relaxing statistical independence assumptions. There is a large literature on this, including important work by \cite{RosenbaumRubin1983sensitivity},  \cite{RobinsRotnitzkyScharfstein2000}, and \cite{Rosenbaum1995,Rosenbaum2002}. We apply results from our paper \cite{MastenPoirier2017}, which discusses that literature in more detail. In that paper we did not study estimation or inference. Second is the literature on identification of the distribution of treatment effects $Y_1 - Y_0$, especially without the rank invariance assumption. In their introduction, \cite{FanGuerreZhu2017} provide a comprehensive discussion of this literature; also see \cite{AbbringHeckman2007} section 2. Here we focus on the papers most related to our sensitivity analysis. \cite{HeckmanSmithClements1997} performed a sensitivity analysis to the rank invariance assumption by fixing the value of Kendall's $\tau$ for the joint distribution of potential outcomes, and then varying $\tau$ from $-1$ to $1$; see tables 5A and 5B. Their analysis is motivated by a search for breakdown points, as evident in their section 4 title, ``How far can we depart from perfect dependence and still produce plausible estimates of program impacts?'' Nonetheless, they do not formally define identified sets for parameters given their assumptions on Kendall's $\tau$, and they do not formally define a breakdown point. Moreover, they do not suggest estimating or doing inference on breakdown points. \cite{Gechter2016} performs a sensitivity analysis to the rank invariance assumption by fixing a lower bound on the value of Spearman's $\rho$. Under this assumption he derives the identified set for a certain average treatment effect. He then studies estimation of and inference on this set for a fixed value of the sensitivity parameter. \cite{FanPark2009} provide formal identification results for the joint cdf of potential outcomes and the distribution of treatment effects under the known Kendall's $\tau$ assumption. They also discuss how to extend those results to known Spearman's $\rho$ in their remark 1. They provide estimation and inference methods for their bounds, but do not study breakdown points. Finally, none of these papers studies the specific relaxation of rank invariance we consider (as defined in section \ref{sec:Model}).

In this section we have focused narrowly on the papers most closely related to ours. We situate our work more broadly in the literature on inference in sensitivity analyses in supplemental appendix A. In that section we also briefly discuss Bayesian inference, although we use frequentist inference in this paper.

\section{Estimation and inference with continuous covariates}\label{sec:contcovariates}

The estimation and inference theory in section \ref{sec:EstimationAndInference} assumes that the covariates $W$ are discretely distributed (via A\ref{assn:discreteW}). Those results are nonparametric in the sense that they do not impose any restrictions on the conditional distribution of $Y \mid X,W$ or on the propensity score $p_{x|w}$. But they rule out continuous covariates. In this section, we briefly discuss how to do estimation and inference with continuous covariates.

When some components of $W$ are continuously distributed, a simple solution is to discretize $W$ and then apply the previous estimator. Alternatively, one can smooth over different covariate values. This can be done using parametric, semiparametric, or nonparametric estimators. 

For example, especially if the dimension of $W$ is large, one could use the usual logit propensity score estimator
\[
	\widehat{p}_{1 \mid w} = \widehat{\Prob}(X=1 \mid W=w) = \Lambda(\widehat{\beta}'w),
\]
where $\Lambda(a) = \exp(a)/(1+\exp(a))$ is the standard logit cdf and $\widehat{\beta}$ are the maximum likelihood estimated index coefficients. The conditional quantile function $Q_{Y \mid X,W}(\tau \mid x,w)$ can be estimated by a linear quantile regression of $Y$ on $(1,X,W)$, so that
\[
	\widehat{Q}_{Y \mid X,W}(\tau \mid x,w) = \widehat{\gamma}(\tau)'
	\begin{pmatrix}
		1 \\
		x \\
		w
	\end{pmatrix},
\]
where $\widehat{\gamma}(\tau)$ are estimated linear quantile regression coefficients. 

Using these parametric estimators, define
\begin{equation*}
	\widehat{\overline{Q}}^c_{Y_x \mid W}(\tau \mid w)
	= \widehat{Q}_{Y \mid X,W}\left(\tau + \frac{c}{\widehat{p}_{x \mid w}}\min\{\tau,1-\tau\}\mid x,w\right)
\end{equation*}
and 
\begin{equation*}
	\widehat{\underline{Q}}^c_{Y_x \mid W}(\tau \mid w)
	=
	\widehat{Q}_{Y \mid X,W}\left(\tau - \frac{c}{\widehat{p}_{x \mid w}}\min\{\tau,1-\tau\}\mid x,w\right)
\end{equation*}
as before. Since the asymptotic properties of $\widehat{\beta}$ and $\widehat{\gamma}(\cdot)$ are well known, it should be feasible to derive the asymptotic distribution of the functionals we consider in section \ref{sec:EstimationAndInference}. Alternatively, one could use semiparametric or nonparametric estimators of the propensity score $p_{x|w}$ and the conditional quantile function $Q_{Y \mid X,W}$. Again, such first step estimators can be plug-ins to obtain estimates of the various bounds we consider in section \ref{sec:EstimationAndInference}. We leave a full analysis of the asymptotic properties of these estimators to future work.

\section{Proofs}\label{sec:appendixProofs}

\subsection*{Proofs for section \ref{sec:Model}}

\begin{proof}[Proof of theorem \ref{thm:dte_bounds}]
Let $F_1(\cdot \mid w)$ and $F_0(\cdot \mid w)$ be any strictly increasing cdfs conditional on $W = w$ for any $w\in\supp(W)$. Suppose $(Y_1,Y_0) \mid W$ have joint cdf
\[
	F_{Y_1,Y_0 \mid W}(y_1,y_0 \mid w) = C(F_1(y_1 \mid w),F_0(y_0 \mid w) \mid w).
\]
Then
\begin{align*}
	&\P(Y_1 - Y_0 \leq z \mid W=w) \\
	&= \int_{\{y_1-y_0 \leq z\}} dC(F_1(y_1 \mid w),F_0(y_0 \mid w) \mid w) \\
	&= (1-t) \int_{\{y_1-y_0 \leq z\}} dM(F_1(y_1 \mid w),F_0(y_0 \mid w))
		+ t \int_{\{y_1-y_0 \leq z\}} dH(F_1(y_1 \mid w),F_0(y_0 \mid w) \mid w).
\end{align*}
For fixed distributions $(F_1(\cdot \mid w),F_0(\cdot \mid w))$, the first integral is the probability that $\{Y_1 - Y_0 \leq z \}$ given $W=w$, where $(Y_1,Y_0) \mid W$ are random variables that satisfy conditional rank invariance. Hence for these random variables the corresponding conditional ranks are equal almost surely: Conditional on $W=w$, $U_1 = U_0$ a.s., let $U \sim \text{Unif}[0,1]$ denote this almost sure common random variable. Using A\ref{assn:continuity}.\ref{A1_1}, we can thus write
\[
	(Y_1,Y_0) \mid W \overset{d}{=} (F_1^{-1}(U \mid W), F_0^{-1}(U \mid W))
\]
and therefore
\[
	\int_{\{ y_1-y_0 \leq z \}} dM(F_1(y_1 \mid w),F_0(y_0 \mid w))
	=
	\P(F^{-1}_{1}(U \mid w) - F^{-1}_{0}(U \mid w) \leq z).
\]
\cite{Makarov1982} derived sharp bounds on
\[
	\int_{\{ y_1 - y_0 \leq z \}} dH(F_1(y_1),F_0(y_0) \mid w).
\]
Applying these bounds yields
\begin{align*}
   & \int_{\{y_1 - y_0 \leq z\}} dH(F_1(y_1 \mid w),F_0(y_0 \mid w)) \\
    &\in \left[\max\left\{\sup_{y\in\mathcal{Y}_z(w)} (F_1(y \mid w) - F_0(y-z \mid w)), 0 \right\},
    1+ \min\left\{\inf_{y\in\mathcal{Y}_z(w)}(F_1(y \mid w) - F_0(y-z \mid w)), 0 \right\}\right].
\end{align*}
Therefore, for given $w\in\supp(W)$ and given $(F_1(\cdot \mid w),F_0(\cdot \mid w))$, sharp bounds for $\P(Y_1 - Y_0 \leq z \mid W=w)$ are given by
\[
	\left[ \underline{\theta}(F_1(\cdot \mid w),F_0(\cdot \mid w)), \ \overline{\theta}(F_1(\cdot \mid w),F_0(\cdot \mid w)) \right],
\]
where
\begin{multline*}
	\underline{\theta}(F_1(\cdot \mid w),F_0(\cdot \mid w)) \\
		= (1-t)\P(F^{-1}_{1}(U \mid w) - F^{-1}_{0}(U \mid w)\leq z) + t\max\left\{\sup_{y\in\mathcal{Y}_z(w)} (F_{1}(y \mid w) - F_{0}(y-z \mid w)),0\right\}
\end{multline*}
and
\begin{align*}
	&\overline{\theta}(F_1(\cdot \mid w),F_0(\cdot \mid w)) \\
		&= (1-t) \P(F^{-1}_{1}(U \mid w) - F^{-1}_{0}(U \mid w)\leq z) + t \left(1+ \min\left\{\inf_{y\in\mathcal{Y}_z(w)}(F_{1}(y \mid w) - F_{0}(y-z \mid w)), 0 \right\} \right).
\end{align*}

Define the first order stochastic dominance ordering as follows: For two cdfs $F$ and $G$, let $F \preceq_\textsc{fsd} G$ if $F(t) \geq G(t)$ for all $t \in \R$. All of the following statements refer to this ordering. For any fixed $F_1(\cdot \mid w)$,
\[
	\tilde{F}_0(\cdot \mid w) \preceq_\textsc{fsd} F_0(\cdot \mid w)
	\qquad \text{implies} \qquad
	\underline{\theta}(F_1(\cdot \mid w),\tilde{F}_0(\cdot \mid w)) \leq \underline{\theta}(F_1(\cdot \mid w),F_0(\cdot \mid w)).
\]
That is, the lower bound function $\underline{\theta}(F_1(\cdot \mid w),F_0(\cdot \mid w))$ is weakly increasing in $F_0(\cdot \mid w)$. This can be shown in two steps. First, the expression
\[
	\P(F^{-1}_{1}(U \mid w) - F^{-1}_{0}(U \mid w)\leq z)
\]
is weakly increasing in $F_0(\cdot \mid w)$ since, for $\tilde{F}_{0}(\cdot \mid w) \preceq_\textsc{fsd}  F_0(\cdot \mid w)$, we have $\tilde{F}_0^{-1}(u \mid w) \leq F_0^{-1}(u \mid w)$ for $u\in(0,1)$, and therefore,
\[
	\P(F^{-1}_{1}(U \mid w) - \tilde{F}^{-1}_{0}(U \mid w)\leq z) \leq \P(F^{-1}_{1}(U \mid w) - F^{-1}_{0}(U \mid w)\leq z).
\]
Second, the expression
\[
	\max\left\{\sup_{y\in\mathcal{Y}_z(w)} (F_{1}(y \mid w) - F_{0}(y-z \mid w)),0\right\}
\]
is weakly increasing in $F_0(\cdot \mid w)$ since the supremum and maximum operators are weakly increasing. Thus both components of $\underline{\theta}$ are weakly increasing in $F_0(\cdot \mid w)$. Therefore their linear combination is also weakly increasing in $F_0(\cdot \mid w)$.

We can similarly show that $\underline{\theta}(F_1(\cdot \mid w),F_0(\cdot \mid w))$ is weakly decreasing in $F_1(\cdot \mid w)$. Thus substituting
\[
	(F_1(\cdot \mid w),F_0(\cdot \mid w)) = (\underline{F}_{Y_1 \mid W}^c(\cdot \mid w),\overline{F}_{Y_0 \mid W}^c(\cdot \mid w))
\]
yields the lower bound $\underline{\text{CDTE}}(z,c,t \mid w)$. The upper bound function $\overline{\theta}(F_1(\cdot \mid w),F_0(\cdot \mid w))$ is also weakly increasing in $F_0(\cdot \mid w)$ and weakly decreasing in $F_1(\cdot \mid w)$. Thus substituting
\[
	(F_1(\cdot \mid w),F_0(\cdot \mid w)) = (\overline{F}_{Y_1 \mid W}^c(\cdot \mid w),\underline{F}_{Y_0 \mid W}^c(\cdot \mid w))
\]
yields the upper bound $\overline{\text{CDTE}}(z,c,t \mid w)$. In making these substitutions we applied proposition 2 from \cite{MastenPoirier2017}. In that paper we defined functions $F_{Y_x \mid W}^c(\cdot \mid w; \epsilon,\eta)$, which we now use to sharpness of the DTE bounds.

Substitute
\[
	(F_{Y_1 \mid W}^c(\cdot \mid w;\epsilon,0),F_{Y_0 \mid W}^c(\cdot \mid w; 1-\epsilon,0))
\]
into the bound functionals and continuously vary $\epsilon$ between $[0,1]$. Note that we let $\eta = 0$ since $c < \min \{ p_{1|w},p_{0|w} \}$. By continuity of $\underline{\theta}(\cdot,\cdot)$ and $\overline{\theta}(\cdot,\cdot)$ in their arguments and continuity of $(F_{Y_1 \mid W}^c(\cdot \mid w;\epsilon,0),F_{Y_0 \mid W}^c(\cdot \mid w; 1-\epsilon,0))$ in $\epsilon$, the intermediate value theorem implies that every element between the bounds can be attained.

By integrating these CDTE bounds over the marginal distribution of $W$, we obtain the DTE bounds:
\begin{multline*}
	[ \underline{\text{DTE}}(z,c,t), \overline{\text{DTE}}(z,c,t) ] \\
	= \left[ \int_{\supp(W)} \underline{\text{CDTE}}(z,c,t \mid w)dF_W(w), \int_{\supp(W)}\overline{\text{CDTE}}(z,c,t \mid w)dF_W(w) \right].
\end{multline*}
Sharpness of these bounds results from the sharpness of the CDTE bounds for every $w\in\supp(W)$ and the joint attainability of
\[
	\{(\underline{F}_{Y_1 \mid W}^c(\cdot \mid w),\overline{F}_{Y_0 \mid W}^c(\cdot \mid w)): w \in \supp(W)\}
\]
and of
\[
	\{(\overline{F}_{Y_1 \mid W}^c(\cdot \mid w),\underline{F}_{Y_0 \mid W}^c(\cdot \mid w)): w \in \supp(W)\}.
\]
\end{proof}

\subsection*{Proofs for section \ref{sec:EstimationAndInference}}\label{sec:proofsForEstimation}

The following lemma shows that $(\widehat{F}_{Y \mid X,W}(\cdot\mid \cdot,\cdot),\widehat{p}_{(\cdot|\cdot)}, \widehat{q}_{(\cdot)})$ converges uniformly in $y$, $x$, and $w$ to a mean-zero Gaussian process. This result follows by applying the delta method.

\begin{lemma}\label{lemma:cdf_conv}
Suppose A\ref{assn:iid} and A\ref{assn:discreteW} hold. Then
    \begin{align*}
        \sqrt{N}
        \begin{pmatrix}
                    \widehat{F}_{Y \mid X,W}(y \mid x,w) - F_{Y \mid X,W}(y \mid x,w) \\
                    \widehat{p}_{x|w} - p_{x|w} \\
                    \widehat{q}_w - q_w
	\end{pmatrix}
         &\rightsquigarrow \mathbf{Z}_1(y,x,w),
    \end{align*}
    \noindent a mean-zero Gaussian process in $\ell^\infty(\R \times \{0,1\} \times\supp(W) , \R^3)$ with continuous paths and covariance kernel equal to
\begin{align*}
	&\mathbf{\Sigma}_1(y,x,w,\tilde{y},\tilde{x},\tilde{w}) \\[0.5em]
        &= \Exp[\mathbf{Z}_1(y,x)\mathbf{Z}_1(\tilde{y},\tilde{x})'] \\[0.5em]
	&= \text{diag}
	\begin{pmatrix}
	\dfrac{F_{Y \mid X,W}(\min\{y,\tilde{y}\} \mid x,w) - F_{Y \mid X,W}(y \mid x,w)F_{Y \mid X,W}(\tilde{y} \mid x,w)}{p_{x|w}q_{w}} \ind(x=\tilde{x}, w = \tilde{w}) \\[1.5em]
	\dfrac{p_{x|w}}{q_w}\ind(x=\tilde{x},w=\tilde{w}) - \dfrac{p_{x|w}p_{\tilde{x}|w}}{q_w}\ind(w = \tilde{w}) \\[1.5em]
	q_w \ind(w = \tilde{w}) - q_w q_{\tilde{w}}
	\end{pmatrix}.
\end{align*}
\end{lemma}

\begin{proof}[Proof of lemma \ref{lemma:cdf_conv}]
By a second order Taylor expansion,
\begin{align*}
	&\widehat{F}_{Y \mid X,W}(y \mid x,w) - F_{Y \mid X,W}(y \mid x,w)\\
	&= \frac{\avg \ind(Y_i\leq y)\ind(X_i=x, W_i = w)}{\avg \ind(X_i=x, W_i = w)} - \frac{\P(Y\leq y, X=x, W=w)}{\P(X=x, W=w)}\\
	&= \frac{\avg \ind(Y_i\leq y)\ind(X_i=x, W_i = w) - \P(Y\leq y,X=x, W=w)}{\P(X=x, W=w)}\\
	&\hspace{5mm} - \frac{F_{Y \mid X,W}(y \mid x,w)}{\P(X=x, W=w)}\left(\avg \ind(X_i = x, W_i = w) - \P(X=x, W=w)\right)\\
	&\hspace{5mm} + O_p \Bigg[\left(\avg \ind(Y_i \leq y)\ind(X_i = x, W_i = w) - F_{Y \mid X,W}(y \mid x,w)\P(X=x, W = w)\right)\\
	&\hspace{60mm} \cdot \left(\avg \ind(X_i=x, W_i = w) - \P(X=x, W=w)\right) \Bigg] \\
	&\hspace{5mm} + O_p\left[\left(\avg \ind(X_i=x, W_i = w) - \P(X=x, W=w)\right)^2\right].
  \end{align*}

By standard bracketing entropy results (e.g., example 19.6 on page 271 of \citealt{Vaart2000}) the function classes $\{\ind(Y\leq y)\ind(X=x)\ind(W=w): y\in\R, x\in\{0,1\}, w\in\supp(W)\}$ and $\{\ind(X=x)\ind(W = w) :x\in\{0,1\}, w\in\supp(W)\}$ are both $P$-Donsker. Hence the residual is of order $O_p(N^{-1})$ uniformly in $(y,x,w)\in\R \times \{0,1\}\times\supp(W)$. Combining this with Slutsky's theorem we get the uniform over $y$, $x$, and $w$ asymptotically linear representation
\begin{multline*}
	\widehat{F}_{Y \mid X,W}(y \mid x,w) - F_{Y \mid X,W}(y \mid x,w)
	\\
	= \avg \frac{\ind(X_i=x, W_i = w)(\ind(Y_i \leq y) - F_{Y \mid X,W}(y \mid x,w))}{\P(X=x, W=w)} + o_p(N^{-1/2}).
\end{multline*}
By the same bracketing entropy arguments, the class
\[
	\left\{\frac{\ind(X=x, W=w)(\ind(Y \leq y) - F_{Y \mid X,W}(y \mid x,w))}{\P(X=x,W =w)}: y\in\R, x\in\{0,1\}, w\in\supp(W)\right\}
\]
is $P$-Donsker and hence $\sqrt{N}(\widehat{F}_{Y \mid X,W}(\cdot \mid \cdot,\cdot) - F_{Y \mid X,W}(\cdot \mid \cdot, \cdot))$ converges in distribution to a mean-zero Gaussian process with continuous paths. 

A similar argument yields the asymptotically linear representations
\[
	\widehat{p}_{x|w} - p_{x|w} = \avg \frac{\ind(W_i = w) (\ind(X_i = x) - p_{x|w})}{q_w} +o_p(N^{-1/2})
\]
and
\[
	\widehat{q}_w - q_w = \avg (\ind(W_i = w) - q_w).
\]

The covariance kernel $\mathbf{\Sigma}_1$ can be calculated as follows:
\begin{align*}
	&[\mathbf{\Sigma}_1(y,x,w,\tilde{y},\tilde{x},\tilde{w})]_{1,1} \\
	&= \Exp\left[\frac{\ind(X_i=x,W_i = w)\ind(X_i=\tilde{x}, W_i = \tilde{w})(\ind(Y_i \leq y) - F_{Y \mid X,W}(y \mid x,w))(\ind(Y_i \leq \tilde{y}) - F_{Y \mid X,W}(\tilde{y} \mid \tilde{x},\tilde{w}))}{\P(X=x,W=w)\P(X=\tilde{x}, W = \tilde{w})}\right] \\
	&= \frac{F_{Y \mid X,W}(\min\{y,\tilde{y}\} \mid x,w) - F_{Y \mid X,W}(y \mid x,w)F_{Y \mid X,W}(\tilde{y} \mid x,w)}{p_{x|w}q_{w}} \ind(x=\tilde{x}, w = \tilde{w}).
\end{align*}
\begin{flalign*}
	[\mathbf{\Sigma}_1(y,x,w,\tilde{y},\tilde{x},\tilde{w})]_{1,2}
	= \Exp\left[\frac{\ind(X_i=x, W_i = \tilde{w})(\ind(X_i=\tilde{x})-p_{\tilde{x}|\tilde{w}})(\ind(Y_i \leq y) - F_{Y \mid X,W}(y \mid x,w))}{p_{x|w}q_w q_{\tilde{w}}} \right]
	= 0
	&&
\end{flalign*}
\begin{flalign*}
	[\mathbf{\Sigma}_1(y,x,w,\tilde{y},\tilde{x},\tilde{w})]_{1,3}
	= \Exp\left[\frac{(\ind(W_i = \tilde{w}) - q_{\tilde{w}})\ind(X_i=x, W_i = w)(\ind(Y_i \leq y) - F_{Y \mid X,W}(y \mid x,w))}{\P(X=x, W=w)} \right]
	= 0
	&&
\end{flalign*}
\begin{flalign*}
	[\mathbf{\Sigma}_1(y,x,w,\tilde{y},\tilde{x},\tilde{w})]_{2,1}
	= [\mathbf{\Sigma}_1(\tilde{y},\tilde{x},\tilde{w},y,x,w)]_{1,2}
	= 0
	&&
\end{flalign*}
\begin{flalign*}
	[\mathbf{\Sigma}_1(y,x,w,\tilde{y},\tilde{x},\tilde{w})]_{2,2}
	&= \Exp\left[\frac{\ind(W_i = w)\ind(W_i = \tilde{w})(\ind(X_i = x) - p_{x|w})(\ind(X_i = \tilde{x}) - p_{\tilde{x}|\tilde{w}})}{q_w q_{\tilde{w}} }\right] \\
	&= \frac{p_{x|w}}{q_w}\ind(x=\tilde{x},w=\tilde{w}) - \frac{p_{x|w}p_{\tilde{x}|w}}{q_w}\ind(w = \tilde{w})
	&&
\end{flalign*}
\begin{flalign*}
	[\mathbf{\Sigma}_1(y,x,w,\tilde{y},\tilde{x},\tilde{w})]_{2,3}
	= \Exp\left[\frac{(\ind(W_i = \tilde{w}) - q_{\tilde{w}})(\ind(X_i = x) - p_{x|w})}{q_w}\right]
	= 0
	&&
\end{flalign*}
\begin{flalign*}
	[\mathbf{\Sigma}_1(y,x,w,\tilde{y},\tilde{x},\tilde{w})]_{3,1}
	= [\mathbf{\Sigma}_1(\tilde{y},\tilde{x},\tilde{w},y,x,w)]_{1,3}
	= 0
	&&
\end{flalign*}
\begin{flalign*}
	[\mathbf{\Sigma}_1(y,x,w,\tilde{y},\tilde{x},\tilde{w})]_{3,2}
	= [\mathbf{\Sigma}_1(\tilde{y},\tilde{x},\tilde{w},y,x,w)]_{2,3}
	= 0
	&&
\end{flalign*}
\begin{flalign*}
	[\mathbf{\Sigma}_1(y,x,w,\tilde{y},\tilde{x},\tilde{w})]_{3,3}
	= \Exp[(\ind(W_i = w) - q_w)(\ind(W_i = \tilde{w}) - q_{\tilde{w}})]
	= q_w \ind(w = \tilde{w}) - q_w q_{\tilde{w}}.
	&&
\end{flalign*}

\end{proof}

\begin{lemma}[Chain Rule for Hadamard directionally differentiable functions] \label{lemma:hadamard_diff_chainrule}
Let $\mathbb{D}$, $\mathbb{E}$, and $\mathbb{F}$ be Banach spaces with norms $\|\cdot\|_{\mathbb{D}}$, $\|\cdot\|_{\mathbb{E}}$, and $\| \cdot \|_{\mathbb{F}}$. Let $\mathbb{D}_\phi \subseteq \mathbb{D}$ and $\mathbb{E}_\psi \subseteq \mathbb{E}$. Let $\phi:\mathbb{D}_\phi \to \mathbb{E}_\psi$ and $\psi:\mathbb{E}_\psi \to \mathbb{F}$ be functions. Let $\theta \in \mathbb{D}_\phi$ and $\phi$ be Hadamard directionally differentiable at $\theta$ tangentially to $\mathbb{D}_0 \subseteq \mathbb{D}$. Let $\psi$ be Hadamard directionally differentiable at $\phi(\theta)$ tangentially to the range $\phi_\theta'(\mathbb{D}_0) \subseteq \mathbb{E}_\psi$. Then, $\psi \circ \phi: \mathbb{D}_\phi \to \mathbb{F}$ is Hadamard directionally differentiable at $\theta$ tangentially to $\mathbb{D}_0$ with Hadamard directional derivative equal to $\psi_{\phi(\theta)}' \circ \phi_\theta'$.
\end{lemma}

This result is a version of proposition 3.6 in \cite{Shapiro1990}, who omits the proof. We give the proof here because this result is key to our paper.

\begin{proof} [Proof of lemma \ref{lemma:hadamard_diff_chainrule}]
Let $\{h_n\}_{n\geq 1}$ be in $\mathbb{D}$ and $h_n \to h\in\mathbb{D}_0$. By Hadamard directional differentiability of $\phi$ tangentially to $\mathbb{D}_0$
\[
	\left\|\frac{\phi(\theta + t_n h_n) - \phi(\theta)}{t_n} - \phi'_\theta(h)\right\|_{\mathbb{E}} = o(1)
\]
as $n\to\infty$ for any $t_n \searrow 0$. That is,
\[
	g_n \equiv \frac{\phi(\theta + t_n h_n) - \phi(\theta)}{t_n} \overset{\mathbb{E}}{\rightarrow} \phi'_\theta(h) = g
\]
where $\phi_\theta'\in \phi_\theta'(\mathbb{D}_0)$. Therefore, by Hadamard directional differentiability of $\psi$, we have
\begin{align*}
	\frac{\psi(\phi(\theta + t_n h_n)) - \psi(\phi(\theta))}{t_n} 
	&= \frac{\psi(\phi(\theta) + t_n g_n) - \psi(\phi(\theta))}{t_n} \\
	&\overset{\mathbb{F}}{\rightarrow} \psi'_{\phi(\theta)}(g) \\
	&= \psi'_{\phi(\theta)}(\phi'_\theta(h)).
\end{align*}
By Hadamard directional differentiability of $\phi$ at $\theta$ and $\psi$ at $\phi(\theta)$, $\phi'_\theta$ and $\psi'_{\phi(\theta)}$ are continuous mappings. Hence their composition $\psi'_{\phi(\theta)} \circ \phi_\theta'$ is continuous. This combined with our derivations above imply that $\psi\circ\phi$ is Hadamard directionally differentiable tangentially to $\mathbb{D}_0$ at $\theta$.
\end{proof}

\begin{proof}[Proof of lemma \ref{lemma:conv_cdf_PI}]
  Let $\theta_0 = (F_{Y \mid X,W}(\cdot \mid \cdot,\cdot),p_{(\cdot|\cdot)},q_{(\cdot)})$ and $\widehat{\theta} = (\widehat{F}_{Y \mid X,W}(\cdot \mid \cdot,\cdot),\widehat{p}_{(\cdot|\cdot)},\widehat{q}_{(\cdot)})$. For fixed $y$ and $c$ define the mapping
\[
	\phi_1:\quad \ell^\infty(\R \times\{0,1\}\times\supp(W))\times \ell^\infty(\{0,1\}\times\supp(W)) \times \ell^\infty(\supp(W)) \rightarrow \ell^\infty(\{0,1\}\times\supp(W),\R^2)
\]
by
\[
	[\phi_1(\theta)](x,w) =
	\begin{pmatrix}
                        \min\left\{ \dfrac{\theta^{(1)}(y,x,w)\theta^{(2)}(x,w)}{\theta^{(2)}(x,w)-c} , \dfrac{\theta^{(1)}(y,x,w)\theta^{(2)}(x,w)+c}{\theta^{(2)}(x,w)+c} \right\} \\
                        \\
                        \max\left\{ \dfrac{\theta^{(1)}(y,x,w)\theta^{(2)}(x,w)}{\theta^{(2)}(x,w)+c} , \dfrac{\theta^{(1)}(y,x,w)\theta^{(2)}(x,w)-c}{\theta^{(2)}(x,w)-c} \right\}
	\end{pmatrix}
\]
where $\theta^{(j)}$ is the $j$th component of $\theta$. Note that
\[
\begin{pmatrix}
              \overline{F}^c_{Y_x \mid W}(y \mid w) \\
              \underline{F}^c_{Y_x \mid W}(y \mid w)
\end{pmatrix}
          = [\phi_1(\theta_0)](x,w) 
          \qquad \text{and} \qquad
\begin{pmatrix}
              \widehat{\overline{F}}^c_{Y_x \mid W}(y \mid w) \\
              \widehat{\underline{F}}^c_{Y_x \mid W}(y \mid w)
\end{pmatrix}
          = [\phi_1(\widehat{\theta})](x,w).
\]
The maps $(a_1,a_2) \mapsto \min\{a_1,a_2\}$ and $(a_1,a_2) \mapsto \max\{a_1,a_2\}$ are Hadamard directionally differentiable with Hadamard directional derivatives at $(a_1,a_2)$ equal to
\[
   h \mapsto
   \begin{cases}
    h^{(1)} &\text{ if } a_1 < a_2\\
    \min\{h^{(1)},h^{(2)}\} &\text{ if } a_1 = a_2\\
    h^{(2)} &\text{ if } a_1 >a_2
    \end{cases}
\]
and
\[
    h \mapsto
    \begin{cases}
    h^{(2)} &\text{ if } a_1 < a_2\\
    \max\{h^{(1)},h^{(2)}\} &\text{ if } a_1 = a_2\\
    h^{(1)} &\text{ if } a_1 >a_2
  \end{cases}
\]
respectively, where $h\in\R^2$; for example, see equation (18) in \cite{FangSantos2015workingPaper}. The mapping $\phi_1$ is comprised of compositions of these $\min$ and $\max$ operators, along with four other functions. We can show that these four mappings are ordinary Hadamard differentiable. Here we compute these Hadamard derivatives with respect to $\theta$: 
\begin{align*}
	[\delta_1(\theta)](x,w) &= \frac{\theta^{(1)}(y,x,w)\theta^{(2)}(x,w)}{\theta^{(2)}(x,w)+c}  \text{ has Hadamard derivative equal to}\\
	[\delta_{1,\theta}'(h)](x,w) &= \frac{\theta^{(1)}(y,x,w)h^{(2)}(x,w) + h^{(1)}(y,x,w)\theta^{(2)}(x,w)}{\theta^{(2)}(x,w) + c} - \frac{\theta^{(1)}(y,x,w)\theta^{(2)}(x,w)h^{(2)}(x,w)}{(\theta^{(2)}(x,w)+c)^2},\\
	[\delta_2(\theta)](x,w) &=  \frac{\theta^{(1)}(y,x,w)\theta^{(2)}(x,w)-c}{\theta^{(2)}(x,w)-c}  \text{ has Hadamard derivative equal to}\\
	[\delta_{2,\theta}'(h)](x,w) &= \frac{\theta^{(1)}(y,x,w)h^{(2)}(x,w) + h^{(1)}(y,x,w)\theta^{(2)}(x,w)}{\theta^{(2)}(x,w) - c} - \frac{(\theta^{(1)}(y,x,w)\theta^{(2)}(x,w)-c)h^{(2)}(x,w)}{(\theta^{(2)}(x,w)-c)^2},\\
	[\delta_3(\theta)](x,w) &= \frac{\theta^{(1)}(y,x,w)\theta^{(2)}(x,w)}{\theta^{(2)}(x,w)-c}  \text{ has Hadamard derivative equal to}\\
	[\delta_{3,\theta}'(h)](x,w) &= \frac{\theta^{(1)}(y,x,w)h^{(2)}(x,w) + h^{(1)}(y,x,w)\theta^{(2)}(x,w)}{\theta^{(2)}(x,w) - c} - \frac{\theta^{(1)}(y,x,w)\theta^{(2)}(x,w)h^{(2)}(x,w)}{(\theta^{(2)}(x,w)-c)^2},\\
	[\delta_4(\theta)](x,w) &= \frac{\theta^{(1)}(y,x,w)\theta^{(2)}(x,w)+c}{\theta^{(2)}(x,w)+c}  \text{ has Hadamard derivative equal to}\\
	[\delta_{4,\theta}'(h)](x,w) &= \frac{\theta^{(1)}(y,x,w)h^{(2)}(x,w) + h^{(1)}(y,x,w)\theta^{(2)}(x,w)}{\theta^{(2)}(x,w) + c} - \frac{(\theta^{(1)}(y,x,w)\theta^{(2)}(x,w)+c)h^{(2)}(x,w)}{(\theta^{(2)}(x,w)+c)^2}.
\end{align*}
All these derivatives are well defined at $\theta_0$ because $\theta^{(2)}_0(x,w) = p_{x|w} > \overline{C} \geq c$. With this notation, we can write the functional $\phi_1$ as
\[
	\phi_1(\theta) =
	\begin{pmatrix}
                        \min\left\{ \delta_3(\theta),\delta_4(\theta) \right\} \\
                        \max\left\{ \delta_1(\theta),\delta_2(\theta) \right\}
         \end{pmatrix}.
\]
By the chain rule (lemma \ref{lemma:hadamard_diff_chainrule}), the map $\phi_1$ is Hadamard directionally differentiable at $\theta_0$ with Hadamard directional derivative evaluated at $\theta_0$ equal to
  \begin{align*}
    \phi'_{1,\theta_0}(h) &= \left(
                               \begin{aligned}
                                 &\ind(\delta_3(\theta_0)<\delta_4(\theta_0)) \cdot \delta_{4,\theta_0}'(h) \\
                                 + &\ind(\delta_3(\theta_0)=\delta_4(\theta_0))
                                 \cdot \min\{\delta_{3,\theta_0}'(h),\delta_{4,\theta_0}'(h)\} \\
                                  + &\ind(\delta_3(\theta_0)>\delta_4(\theta_0)) \cdot \delta_{3,\theta_0}'(h) \\
				\\                                  
                                 &\ind(\delta_1(\theta_0)<\delta_2(\theta_0)) \cdot \delta_{1,\theta_0}'(h)\\
                                 + &\ind(\delta_1(\theta_0)=\delta_2(\theta_0)) \cdot \max\{\delta_{1,\theta_0}'(h),\delta_{2,\theta_0}'(h)\} \\
                                 + &\ind(\delta_1(\theta_0)>\delta_2(\theta_0)) \cdot \delta_{2,\theta_0}'(h)
                               \end{aligned}
                             \right).
  \end{align*}
By lemma \ref{lemma:cdf_conv}, $\sqrt{N}(\widehat{\theta}(y,x,w) - \theta_0(y,x,w)) \rightsquigarrow \mathbf{Z}_1(y,x,w)$. Hence we can use the delta method for Hadamard directionally differentiable functions (see theorem 2.1 in \citealt{FangSantos2014}) to find that
  \begin{align*}
    \left[ \sqrt{N}(\phi_1(\widehat{\theta}) - \phi_1(\theta_0)) \right](x,w)
    	&\rightsquigarrow [\phi'_{1,\theta_0}(\mathbf{Z}_1) ](x,w)\\
	&\equiv \tilde{\mathbf{Z}}_2(x,w).
  \end{align*}
This result holds uniformly over any finite grid of values for $y\in\R$ and $c\in\mathcal{C}$ by considering the Hadamard directional differentiability of a vector of these mappings indexed at different values of $y$ and $c$, which yields the process $\mathbf{Z}_2(y,x,w,c)$.
\end{proof}

\begin{proof}[Proof of lemma \ref{lemma:conv_quantiles_PI}]
Let $\mathcal{S} = \{ (y,x,w) \in \R^2 : y \in [\underline{y}_x(w), \overline{y}_x(w)], x \in \{0,1 \}, w \in \supp(W) \}$. Let $\mathscr{D}(\mathcal{S}) \subset \ell^\infty(\mathcal{S})$ denote the set of functions that are c\`{a}dl\`{a}g in the first argument for each $x \in \{0,1\}$ and $w \in \supp(W)$. Define the mapping
\[
	\tilde\phi_2:\quad \mathscr{D}(\mathcal{S}) \times \ell^\infty(\{0,1\}\times\supp(W))\times\ell^\infty(\supp(W)) \rightarrow \ell^\infty((0,1) \times\{0,1\}\times\supp(W),\R^2)
\]
by
\[
	[\tilde\phi_2(\theta)](\tau,x,w) 
	=
	\begin{pmatrix}
		(\theta^{(1)})^{-1}(\tau,x,w) \\
		\theta^{(2)}(x,w)
	\end{pmatrix}.
\]
By A\ref{assn:continuity}, A\ref{assn:iid}, A\ref{assn:support_c_y}, and lemma 21.4(ii) in \cite{Vaart2000} this mapping is Hadamard differentiable at $\theta_0$ tangentially to $\mathscr{C} (\mathcal{S})\times \ell^\infty(\{0,1\}\times\supp(W))\times\ell^\infty(\supp(W))$, where $\mathscr{C} (\mathcal{S}) \subset \ell^\infty(\mathcal{S})$ is the set functions that are continuous in the first argument for each $x\in\{0,1\}$ and $w\in\supp(W)$. Its Hadamard derivative at $\theta_0 = (F_{Y \mid X,W}(\cdot \mid \cdot,\cdot), p_{(\cdot|\cdot)}, q_{(\cdot)})$ is
\[
     [\tilde\phi_{2,\theta_0}'(h)](\tau,x,w) \mapsto \left(-\frac{h^{(1)}(Q_{Y \mid X,W}(\tau \mid x,w),x,w)}{f_{Y \mid X,W}(Q_{Y \mid X,W}(\tau \mid x,w) \mid x,w)},h^{(2)}(x,w)\right).
\]
By the functional delta method and theorem 7.3.3 part (iii) of \cite{BickelDoksum2015}, 
\begin{align*}
	[\sqrt{N}(\tilde\phi_2(\widehat\theta) - \tilde\phi_2(\theta_0))](\tau,x,w) &\rightsquigarrow \tilde{\mathbf{Z}}_3(\tau,x,w),
\end{align*}
where $\tilde{\mathbf{Z}}_3$ is a mean-zero Gaussian process in $\ell^\infty((0,1)\times\{0,1\}\times\supp(W),\R^2)$ with uniformly continuous paths.

Now define the mapping
\[
	\phi_2:\quad \ell^\infty((0,1) \times\{0,1\}\times\supp(W)) \times\ell^\infty(\{0,1\}\times\supp(W)) \rightarrow \ell^\infty((0,1) \times\{0,1\}\times\supp(W)\times[0,\overline{C}],\R^2)
\]
by
\[
	[\phi_2(\psi)](\tau,x,w,c) = 
                      \begin{pmatrix}
                       \psi^{(1)}\left(\tau + \frac{c}{\psi^{(2)}(x,w)}\min\{\tau,1-\tau\},x,w\right) \\[0.5em]
                       \psi^{(1)}\left(\tau - \frac{c}{\psi^{(2)}(x,w)}\min\{\tau,1-\tau\},x,w\right)
                      \end{pmatrix}.
\]
Then
  \begin{align*}
            \begin{pmatrix}
              \overline{Q}^c_{Y_x \mid W}(\tau \mid w) \\[0.5em]
              \underline{Q}^c_{Y_x \mid W}(\tau \mid w)
            \end{pmatrix}
          = [\phi_2(\tilde\phi_2(\theta_0))](\tau,x,w,c)
          \quad \text{and} \quad 
            \begin{pmatrix}
              \widehat{\overline{Q}}^c_{Y_x \mid W}(\tau \mid w) \\[0.5em]
               \widehat{\underline{Q}}^c_{Y_x \mid W}(\tau \mid w)
            \end{pmatrix}
          = 
          [\phi_2(\tilde\phi_2(\widehat\theta))](\tau,x,w,c).
\end{align*}

We will show that $\phi_2$ is Hadamard differentiable tangentially to the space $\mathscr{C}_U( (0,1)\times\{0,1\}\times\supp(W)) \times \ell^\infty(\{0,1\}\times\supp(W))$, where $\mathscr{C}_U(\mathcal{A})$ denotes the set of uniformly continuous functions on $\mathcal{A}$. The Hadamard derivative of the first component of $\phi_2$ evaluated at $\psi_0 \equiv \tilde\phi_2(\theta_0)$ is
\begin{align*}
	[\phi_{2,\psi_0}^{(1)\prime}(h)](\tau,x,w,c) &= h^{(1)}\left(\tau + \frac{c}{\psi_0^{(2)}(x,w)}\min\{\tau,1-\tau\},x,w\right)\\
	&\qquad - \psi_0^{(1)\prime}\left(\tau + \frac{c}{\psi_0^{(2)}(x,w)}\min\{\tau,1-\tau\},x,w\right)\frac{c\min\{\tau,1-\tau\}}{(\psi_0^{(2)}(x,w))^2}h^{(2)}(x,w).
\end{align*}
To see this, a Taylor expansion gives
\begin{align*}
	&\left[\frac{\phi_2^{(1)}(\psi_0 + t_nh_n)-\phi_2^{(1)}(\psi_0)}{t_n}\right](\tau,x,w,c) \\
	&= h_n^{(1)}\left(\tau + \frac{c}{\psi_0^{(2)}(x,w) + t_nh_n^{(2)}(x,w)}\min\{\tau,1-\tau\},x,w\right)\\
	&\qquad - \psi_0^{(1)\prime}\left(\tau + \frac{c}{\psi_0^{(2)}(x,w) + a_n(x,w)}\min\{\tau,1-\tau\},x,w\right)\frac{c\min\{\tau,1-\tau\}}{(\psi_0^{(2)}(x,w) + a_n(x,w))^2}h_n^{(2)}(x,w)
\end{align*}
using the fact that $\psi_0^{(1)}(\tau,x,w) = Q_{Y \mid X}(\tau \mid x,w)$ is continuously differentiable in $\tau$ by assumption A\ref{assn:support_c_y}.\ref{A4_2}, and noting that term $a_n(x,w)$ satisfies $|a_n(x,w)| \leq |t_nh_n^{(2)}(x,w)| = O(t_n)$. Next,
\begin{align*}
	&\sup_{\tau,x,w,c}
	\Bigg|h_n^{(1)}\left(\tau + \frac{c \min\{\tau,1-\tau\}}{\psi_0^{(2)}(x,w) + t_nh_n^{(2)}(x,w)},x,w\right)
	- h^{(1)}\left(\tau + \frac{c \min\{\tau,1-\tau\}}{\psi_0^{(2)}(x,w)},x,w\right)\Bigg|\\
	 &\leq \sup_{\tau,x,w,c} \left|h_n^{(1)}\left(\tau + \frac{c \min\{\tau,1-\tau\}}{\psi_0^{(2)}(x,w) + t_nh_n^{(2)}(x,w)},x,w\right)
	 - h^{(1)}\left(\tau + \frac{c \min\{\tau,1-\tau\}}{\psi_0^{(2)}(x,w) + t_nh_n^{(2)}(x,w)},x,w\right)\right|\\
	&\hspace{15mm} + \sup_{\tau,x,w,c} \left|h^{(1)}\left(\tau + \frac{c \min\{\tau,1-\tau\}}{\psi_0^{(2)}(x,w) + t_nh_n^{(2)}(x,w)},x,w\right) - h^{(1)}\left(\tau + \frac{c \min\{\tau,1-\tau\}}{\psi_0^{(2)}(x,w)},x,w\right)\right|\\
	&\leq \| h_n^{(1)} - h^{(1)} \|_\infty + o(1)\\
	&= o(1)
\end{align*}
where all three suprema are taken over $\tau\in(0,1),x\in\{0,1\},w\in\supp(W),c\in[0,\overline{C}]$. The last inequality follows from uniform continuity of $h^{(1)}$. The last line follows from uniform convergence of $h_n$ to $h$.

Similarly, we have that
\begin{align*}
	&\sup_{\tau,x,w,c} \left|\psi_0^{(1)\prime}\left(\tau + \frac{c}{\psi_0^{(2)}(x,w) + a_n(x,w)}\min\{\tau,1-\tau\},x,w\right)\frac{c\min\{\tau,1-\tau\}}{(\psi_0^{(2)}(x,w) + a_n(x,w))^2}h_n^{(2)}(x,w)\right.\\
	&\hspace{30mm} - \left.\psi_0^{(1)\prime}\left(\tau + \frac{c}{\psi_0^{(2)}(x,w)}\min\{\tau,1-\tau\},x,w\right)\frac{c\min\{\tau,1-\tau\}}{(\psi_0^{(2)}(x,w) )^2}h^{(2)}(x,w)\right|
	= o(1)
\end{align*}
by uniform continuity of $\psi_0^{(1)\prime}$ (implied by A\ref{assn:support_c_y}.\ref{A4_2}) and by $a_n(x,w) = o(1)$. Again, the sup is over $\tau\in(0,1),x\in\{0,1\},w\in\supp(W),c\in[0,\overline{C}]$. Therefore $\phi_2^{(1)}$ is Hadamard differentiable tangentially to the space of uniformly continuous functions. A similar argument can be made for $\phi_2^{(2)}$. By composition, $\phi_2 \circ \tilde\phi_2$ is Hadamard differentiable tangentially to $\mathscr{C}(\mathcal{S})$. 

By the functional delta method and the fact that $\tilde{\mathbf{Z}}_3(y,x,w)$ has uniformly continuous paths, we have that
\begin{align*}
	[\sqrt{N}(\phi_2(\tilde\phi_2(\widehat{\theta})) - \phi_2(\tilde\phi_2(\theta_0)))](\tau,x,w,c) &\rightsquigarrow [\phi'_{2,\psi_0} \circ \tilde\phi_{2,\theta_0}'(\mathbf{Z}_1)](\tau,x,w,c)\\
	&\equiv \mathbf{Z}_3(\tau,x,w,c),
\end{align*}
a mean-zero Gaussian process with continuous paths in $\tau\in(0,1)$ and $c\in[0,\overline{C}]$.
\end{proof}

\begin{proof}[Proof of proposition \ref{prop:conv_breakdownpoint}]
Consider the lower CQTE bound of equation \eqref{eq:QTE bounds} as a function of $c$. Its first component is the lower bound of the conditional quantile of $Y_1 \mid W=w$. By assumption A\ref{assn:support_c_y}.2, the derivative of that conditional quantile with respect to $c$ equals
\begin{multline*}
	\frac{\partial}{\partial c} Q_{Y \mid X,W}\left(\tau - \frac{c}{p_{1|w}}\min\left\{\tau,1-\tau\right\} \mid 1,w\right) \\
	= \frac{-\min\{\tau,1-\tau\}}{p_{1|w} f_{Y \mid X,W}\left(Q_{Y \mid X,W}\left(\tau - \frac{c}{p_{x|w}}\min\left\{\tau,1-\tau\right\} \mid 1,w\right)\mid 1,w\right)}.
\end{multline*}
The second component of the lower CQTE bound is the upper bound of the conditional quantile of $Y_0 \mid W=w$. The derivative of that conditional quantile with respect to $c$ equals
\begin{multline*}
	\frac{\partial}{\partial c} Q_{Y \mid X,W}\left(\tau + \frac{c}{p_{0|w}}\min\left\{\tau,1-\tau\right\} \mid 0,w\right) \\
	= \frac{\min\{\tau,1-\tau\}}{p_{0|w} f_{Y \mid X,W}\left(Q_{Y \mid X,W}\left(\tau + \frac{c}{p_{0|w}}\min\left\{\tau,1-\tau\right\} \mid 0,w\right)\mid 0,w\right)}.
\end{multline*}
Moreover, these derivatives are bounded away from zero and infinity uniformly over $c \in (0,\overline{C}]$. This implies that the derivative of the CQTE is negative and uniformly bounded away from zero. 

Next recall that
\[
	\underline{\text{CATE}}(c \mid w)
	= \int_0^1 \underline{\text{CQTE}}(\tau, c \mid w) \; d\tau.
\]
Its derivative with respect to $c$ exists by the dominated convergence theorem (by A\ref{assn:continuity} and A\ref{assn:support_c_y}). Moreover, it is bounded away from zero for all $c \in (0,\overline{C}]$. By taking another expectation over the marginal distribution of $W$, $\partial \underline{\text{ATE}}(c) / \partial c$ exists (by A\ref{assn:discreteW}), is negative, and is bounded away from zero for all $c \in (0,\overline{C}]$.

$c^*$ is defined implicitly by $\underline{\text{ATE}}(c^*) = \mu$. We have shown that the function $\underline{\text{ATE}}(c)$ satisfies the assumptions of lemma 21.3 on page 306 of \cite{Vaart2000}. Thus the mapping $\underline{\text{ATE}}(\cdot) \mapsto c^*$ is Hadamard differentiable tangentially to the set of c\`adl\`ag functions on $(0,\overline{C}]$ with derivative
\[
	\frac{-h(c^*)}{\frac{\partial}{\partial c}\underline{\text{ATE}}(c^*)}.
\]
By the discussion following lemma \ref{lemma:conv_quantiles_PI}, $\sqrt{N}(\widehat{\underline{\text{ATE}}}(c) - \underline{\text{ATE}}(c))$ converges in distribution to a random element of $\ell^{\infty}([0,\overline{C}])$ with continuous paths.

Let
\[
	\tilde{c}^* = \inf\{c\in[0,\overline{C}]:\widehat{\underline{\text{ATE}}}(c) \leq \mu\}.
\]
We can then apply the functional delta method to see that $\sqrt{N}(\tilde{c}^* - c^*)$ converges in distribution to a Gaussian variable we denote by $\mathbf{Z}_4$.

Since $c^*\in(0,\overline{C}]$ and by monotonicity of $\underline{\text{ATE}}(\cdot)$, we have $\underline{\text{ATE}}(\overline{C}) \leq \mu$. By $\sqrt{N}$-convergence of the ATE bounds, 
\begin{align*}
	\P \big( \widehat{\underline{\text{ATE}}}(\overline{C}) > \mu \big)
	&= \P \left( -\sqrt{N} \big( \underline{\text{ATE}}(\overline{C})-\mu \big) < \sqrt{N} \big( \widehat{\underline{\text{ATE}}}(\overline{C})- \underline{\text{ATE}}(\overline{C}) \big) \right) \\ 
	&\rightarrow 0.
\end{align*} 
Therefore, the set $\{c\in[0,\overline{C}]:\widehat{\underline{\text{ATE}}}(c) \leq \mu\}$ is non-empty with probability approaching one. This implies that $\tilde{c}^*\in[0,\overline{C}]$ with probability approaching one, and therefore $\P(\tilde{c}^* = \widehat{c}^*)$ also approaches one as $N \rightarrow \infty$. Using these results, we obtain
\begin{align*}
	\sqrt{N}(\widehat{c}^* - c^*) 	&= \sqrt{N}(\widehat{c}^* - \tilde{c}^*) + \sqrt{N}(\tilde{c}^* - c^*)\\
								&= o_p(1) + \sqrt{N}(\tilde{c}^* - c^*)\\
								&\rightsquigarrow \mathbf{Z}_4.
\end{align*}
\end{proof}

The following result extends proposition 2(i) of \cite{ChernozhukovFernandez-ValGalichon2010} to allow for input functions which are directionally differentiable, but not fully differentiable, at one point. It can be extended to allow for multiple points of directional differentiability, but we omit this since we do not need it for our application.

\begin{lemma}\label{lemma:conv_intoperator}
Let $\theta_0(u,c,w) = (\theta_0^{(1)}(u,c,w),\theta_0^{(2)}(u,c,w))$ where for $j \in\{1,2\}$ we have that $\theta_0^{(j)}(u,c,w)$ is strictly increasing in $u\in[0,1]$, bounded above and below, and differentiable everywhere except at $u = u^*$, where it is directionally differentiable. Further, assume that the two components satisfy A\ref{assn:smoothness_intpiece}. Then, for fixed $z\in\R$, the mapping $\phi_3:\ell^\infty((0,1)\times\supp(W)\times\mathcal{C},\R^2) \rightarrow \ell^\infty(\supp(W)\times \mathcal{C},\R^2)$ defined by
\[
	[\phi_3(\theta)](w,c)
	=
	\begin{pmatrix}
		\int_0^1 \ind(\theta^{(2)}(u,c,w) \leq z) \; du \\
                 \int_0^1 \ind(\theta^{(1)}(u,c,w) \leq z) \; du
	\end{pmatrix}
\]
is Hadamard directionally differentiable tangentially to $\mathscr{C}((0,1)\times\supp(W)\times\mathcal{C},\R^2)$ with Hadamard directional derivative given by equations \eqref{eq:directional_had_der_Galichon} and \eqref{eq:directional_had_der_Galichon_secondEq} below.
\end{lemma}

\begin{proof}[Proof of lemma \ref{lemma:conv_intoperator}]
For clarity we suppress the dependence on $w$ in the expressions below. Uniformity of convergence over $w \in \supp(W)$ follows from the discreteness of $\supp(W)$ (assumption A\ref{assn:discreteW}). Our proof follows that of proposition 2(i) in \cite{ChernozhukovFernandez-ValGalichon2010}. Let
\[
	\mathcal{U}_1(c) = \{u\in(0,1): \theta_0^{(1)}(u,c) = z\}
\]
denote the set of roots to the equation $\theta_0^{(1)}(u,c) = z$ for fixed $z$ and $c$. By A\ref{assn:smoothness_intpiece}.\ref{A5_1} this set contains a finite number of elements. We denote these by
\[
	\mathcal{U}_1(c) = \{u_k^{(1)}(c), \text{ for }k=1,2,\ldots,K^{(1)}(c) < \infty\}.
\]
A\ref{assn:smoothness_intpiece}.\ref{A5_1} also implies that $\mathcal{U}_1(c)\cap \mathcal{U}^*_1(c) = \emptyset$ for any $c\in\mathcal{C}$.

We will show the first component of the Hadamard directional derivative is given by
\begin{equation}\label{eq:directional_had_der_Galichon}
	[\phi_{3,\theta_0}^{(1)\prime}(h)](c)
	= - \sum_{k=1}^{K^{(1)}(c)} h(u_k^{(1)}(c),c)\left(\frac{\ind(h(u_k^{(1)}(c),c) > 0)}{|\partial_u^- \theta_0^{(1)}(u_k^{(1)}(c),c)|} + \frac{\ind(h(u_k^{(1)}(c),c) < 0)}{|\partial_u^+ \theta_0^{(1)}(u_k^{(1)}(c),c)|}\right),
\end{equation}
where $h\in \mathscr{C}((0,1) \times \mathcal{C})$.

First suppose $u^*\notin\mathcal{U}_1(c)$ for any $c\in\mathcal{C}$. In this case we can apply proposition 2(i) of \cite{ChernozhukovFernandez-ValGalichon2010} directly to obtain
\[
	\left|\frac{ [\phi^{(1)}_3(\theta_0 + t_n h_n)](c) - [\phi^{(1)}_3(\theta_0)](c)}{t_n} - \left( - \sum_{k=1}^{K^{(1)}(c)} \frac{h(u_k^{(1)}(c),c)}{|\partial_u \theta_0^{(1)}(u_k^{(1)}(c),c)|} \right) \right| = o(1)
\]
for any $c \in \mathcal{C}$, where $t_n \searrow 0$, $h_n\in\ell^\infty((0,1)\times\mathcal{C})$, and
\[
	\sup_{(u,c)\in(0,1)\times \mathcal{C}}|h_n(u,c) - h(u,c)| = o(1)
\]
as $n\rightarrow\infty$. Hence
\[
	[\phi^{(1)\prime}_{3,\theta_0}(h)](c)
	= -\sum_{k=1}^{K^{(1)}(c)} \frac{h(u_k^{(1)}(c),c)}{|\partial_u \theta_0^{(1)}(u_k^{(1)}(c),c)|},
\]
a linear map in $h$.

Now suppose $u^* \in \mathcal{U}_1(c)$ for some $c \in \mathcal{C}$. Without loss of generality, let $u_1^{(1)}(c) = u^*$. Let $B_\epsilon(u)$ denote a ball of radius $\epsilon$ centered at $u$. By equation (A.1) in \cite{ChernozhukovFernandez-ValGalichon2010}, for any $\delta > 0$ there exists an $\epsilon > 0$ and a large enough $n$ such that
\begin{align*}
	&\frac{[\phi^{(1)}_3(\theta_0 + t_n h_n)](c) - [\phi^{(1)}_3(\theta_0)](c)}{t_n} \\
	&\leq \sum_{k=1}^{K^{(1)}(c)}\int_{B_\epsilon(u_k^{(1)}(c))} \frac{\ind(\theta_0(u,c) + t_n(h(u_k^{(1)}(c),c) - \delta)\leq z) - \ind(\theta_0(u,c)\leq z)}{t_n} \; du.
\end{align*}
Likewise, for any $\delta>0$ there exists $\epsilon>0$ and large enough $n$ such that
\begin{align*}
	&\frac{[\phi^{(1)}_3(\theta_0 + t_n h_n)](c) - [\phi^{(1)}_3(\theta_0)](c)}{t_n} \\
	&\geq \sum_{k=1}^{K^{(1)}(c)}\int_{B_\epsilon(u_k^{(1)}(c))} \frac{\ind(\theta_0(u,c) + t_n(h(u_k^{(1)}(c),c) + \delta)\leq z) - \ind(\theta_0(u,c)\leq z)}{t_n} \; du.
\end{align*}
The $k=1$ element in the first sum is
\[
	\int_{B_\epsilon(u^*)} \frac{\ind(\theta_0(u,c) + t_n(h(u^*,c) - \delta)\leq z) - \ind(\theta_0(u,c)\leq z)}{t_n} \; du.
\]
$\theta_0(u,c)$ is absolutely continuous in $u$ and, by the change of variables formula for absolutely continuous functions, the transformation $z' = \theta_0(u,c)$ implies that this $k=1$ term is
\[
	\frac{1}{t_n} \int_{J_1 \cap [z,z-t_n(h(u^*,c) - \delta)]} \frac{1}{|\partial_u \theta_0(\theta_0^{-1}(z',c),c)|} \; dz',
\]
where $J_1$ is the image of $B_{\epsilon}(u^*)$ under $\theta_0(\cdot,c)$ and the change of variables follows from the monotonicity of $\theta_0$ in $B_\epsilon(u^*)$ for small enough $\epsilon$ (this monotonicity follows from A\ref{assn:smoothness_intpiece}.\ref{A5_1}, which implies that the derivative of $\theta_0$ changes sign a finite number of times). The closed interval $[z,z-t_n(h(u^*,c) - \delta)]$ should be interpreted as $[z-t_n(h(u^*,c) - \delta),z]$ when $z-t_n(h(u^*,c) - \delta) < z$. Next consider three cases:
\begin{enumerate}
\item When $h(u^*,c)>0$, the interval $[z,z-t_n(h(u^*,c)-\delta)]$ has the form $[z - \psi_n,z]$ for an arbitrarily small $\psi_n>0$. Therefore, the denominator $|\partial_u \theta_0(\theta_0^{-1}(z',c),c)|$ converges to $|\partial^-_u \theta_0(u^*,c)|$ as $n\rightarrow \infty$, by continuous differentiability on $(0,u^*)$ and directional differentiability as $u = u^*$ and by $\theta_0^{-1}(z',c) = u^* + o(1)$. This holds by $z'\in [z-t_n(h(u^*,c)-\delta),z]$, an interval shrinking to $\{z\}$. Therefore,
\begin{align*}
	\frac{1}{t_n} \int_{J_1 \cap [z,z-t_n(h(u^*,c) - \delta)]} \frac{1}{|\partial_u \theta_0(\theta_0^{-1}(z',c),c)|} \; dz' 
	&= \frac{1}{t_n} \int_{z-t_n(h(u^*,c) - \delta)}^z \frac{1}{|\partial^-_u \theta_0(u^*,c)| + o(1)} \; dz' \\
	&= \frac{-h(u^*,c) + \delta}{|\partial^-_u \theta_0(u^*,c)|} + o(1).
\end{align*}
By a similar argument,
\[
	\int_{B_\epsilon(u^*)} \frac{\ind(\theta_0(u,c) + t_n(h(u^*,c) + \delta)\leq z) - \ind(\theta_0(u,c)\leq z)}{t_n} \; du
	=
	\frac{-h(u^*,c) - \delta}{|\partial^-_u \theta_0(u^*,c)|} + o(1).
\]
Letting $\delta>0$ be arbitrarily small and by the squeeze theorem, we obtain 
\[
	\frac{[\phi^{(1)}_3(\theta_0 + t_n h_n)](c) - [\phi^{(1)}_3(\theta_0)](c)}{t_n} 
	= -\sum_{k=1}^{K^{(1)}(c)} \frac{h(u_k^{(1)}(c),c)}{|\partial^-_u \theta_0^{(1)}(u_k^{(1)}(c),c)|} + o(1).
\]

\item When $h(u^*,c)<0$, the interval $[z,z-t_n(h(u^*,c)-\delta)]$ is of the form $[z,z+\psi_n]$ for arbitrarily small $\psi_n>0$. Using the same argument as in case 1, $|\partial_u \theta_0(\theta_0^{-1}(z',c),c)|$ converges to $|\partial^+_u \theta_0(u^*,c)|$ as $n\rightarrow \infty$. Therefore, proceeding as in the previous case, we obtain that
\begin{align*}
	\frac{[\phi^{(1)}_3(\theta_0 + t_n h_n)](c) - [\phi^{(1)}_3(\theta_0)](c)}{t_n} &= -\sum_{k=1}^{K^{(1)}(c)} \frac{h(u_k^{(1)}(c),c)}{|\partial^+_u \theta_0^{(1)}(u_k^{(1)}(c),c)|} + o(1).
\end{align*}

\item When $h(u^*,c) = 0$, this $k=1$ term converges to zero.
\end{enumerate}
Combining these three cases into a single expression we find that
\begin{multline*}
  \frac{1}{t_n}\int_{J_1\cap[z,z-t_n(h(u^*,c) - \delta)]} \frac{1}{|\partial_u \theta_0(\theta_0^{-1}(z',c),c)|} \; dz' \\
  = - h(u^*,c)\left(\frac{\ind(h(u^*,c) > 0)}{|\partial_u^- \theta_0(u^*,c)|} + \frac{\ind(h(u^*,c) < 0)}{|\partial_u^+ \theta_0(u^*,c)|}\right) + o(1).
\end{multline*}
This expression coincides with the Hadamard derivative under continuous differentiability at $u=u^*$, since that implies $\partial^-_u \theta_0(u^*,c) = \partial^+_u \theta_0(u^*,c)$. It follows from the remainder of the proof in \cite{ChernozhukovFernandez-ValGalichon2010} that
\[
  \sup_{c\in\mathcal{C}} \left| \frac{ [\phi^{(1)}_3(\theta_0 + t_n h_n)](c) - [\phi^{(1)}_3(\theta_0)](c)}{t_n} -[\phi^{(1)\prime}_{3,\theta_0}(h)](c) \right| = o(1),
\]
where $\| \cdot \|_e$ is the Euclidean norm, and where $\phi^{(1)\prime}_{3,\theta_0}$ is defined in equation \eqref{eq:directional_had_der_Galichon}. Note that $\phi^{(1)\prime}_{3,\theta_0}$ is continuous in $h$, and therefore it is a Hadamard directional derivative. 

That completes our analysis of the first component of the Hadamard directional derivative of $\phi_3$ with respect to $\theta$ at $\theta_0$. By similar arguments, the second component is
\begin{equation}\label{eq:directional_had_der_Galichon_secondEq}
	[\phi^{(2)\prime}_{3,\theta_0}(h)](c) = - \sum_{k=1}^{K^{(2)}(c)} h(u_k^{(2)}(c),c)\left(\frac{\ind(h(u_k^{(2)}(c),c) > 0)}{|\partial_u^- \theta_0^{(2)}(u_k^{(2)}(c),c)|} + \frac{\ind(h(u_k^{(2)}(c),c) < 0)}{|\partial_u^+ \theta_0^{(2)}(u_k^{(2)}(c),c)|}\right).
\end{equation}
\end{proof}

\begin{proof}[Proof of lemma \ref{lemma:conv_intpiece}]
Let
\[
	\theta_0(\tau,w,c) 
	=
	\begin{pmatrix}
                            \underline{Q}^c_{Y_1 \mid W}(\tau \mid w) - \overline{Q}^c_{Y_0 \mid W}(\tau \mid w) \\
                            \overline{Q}^c_{Y_1 \mid W}(\tau \mid w) - \underline{Q}^c_{Y_0 \mid W}(\tau \mid w)
	\end{pmatrix}   
	\quad \text{and} \quad
	\widehat{\theta}(\tau,w,c) 
	=
	\begin{pmatrix}
		\widehat{\underline{Q}}^c_{Y_1 \mid W}(\tau \mid w) - \widehat{\overline{Q}}^c_{Y_0 \mid W}(\tau \mid w) \\
                            \widehat{\overline{Q}}^c_{Y_1 \mid W}(\tau \mid w) - \widehat{\underline{Q}}^c_{Y_0 \mid W}(\tau \mid w)
	\end{pmatrix}.
\]
Therefore
\[
	\begin{pmatrix}
              \overline{P}(c \mid w) \\
              \underline{P}(c \mid w)
	\end{pmatrix}
          = [\phi_3(\theta_0)](w,c)
          \qquad \text{and} \qquad
          \begin{pmatrix}
              \widehat{\overline{P}}(c \mid w) \\
              \widehat{\underline{P}}(c \mid w)
	\end{pmatrix}
          = [\phi_3(\widehat{\theta})](w,c).
\]
By lemma \ref{lemma:conv_quantiles_PI},
\begin{multline*}
  \sqrt{N}
\begin{pmatrix}
       \widehat{\underline{Q}}^c_{Y_1 \mid W}(\tau \mid w) - \widehat{\overline{Q}}^c_{Y_0 \mid W}(\tau \mid w) - (\underline{Q}^c_{Y_1 \mid W}(\tau \mid w) - \overline{Q}^c_{Y_0 \mid W}(\tau \mid w)) \\
       \widehat{\overline{Q}}^c_{Y_1 \mid W}(\tau \mid w) - \widehat{\underline{Q}}^c_{Y_0 \mid W}(\tau \mid w) - (\overline{Q}^c_{Y_1 \mid W}(\tau \mid w) - \underline{Q}^c_{Y_0 \mid W}(\tau \mid w))
\end{pmatrix} \\
   \rightsquigarrow
	\begin{pmatrix}
   	\mathbf{Z}_3^{(2)}(\tau,1,w,c) - \mathbf{Z}_3^{(1)}(\tau,0,w,c) \\
	\mathbf{Z}_3^{(1)}(\tau,1,w,c) - \mathbf{Z}_3^{(2)}(\tau,0,w,c)
	\end{pmatrix},
\end{multline*}
a mean-zero Gaussian processes in $\ell^\infty((0,1)\times \supp(W)\times \mathcal{C},\R^2)$ with continuous paths.

By lemma \ref{lemma:conv_intoperator} with $u^* = 1/2$, the mapping $\phi_3$ is Hadamard directionally differentiable tangentially to $\mathscr{C}((0,1)\times \supp(W)\times \mathcal{C},\R^2)$. By the functional delta method for Hadamard directionally differentiable functions (e.g., theorem 2.1 in \citealt{FangSantos2014}), we obtain
\begin{align*}
  \sqrt{N}
  \begin{pmatrix}
              \widehat{\overline{P}}(c \mid w) - \overline{P}(c \mid w) \\
              \\
              \widehat{\underline{P}}(c \mid w) - \underline{P}(c \mid w)
	\end{pmatrix}
&\rightsquigarrow
\begin{pmatrix}
                                    \left[ \phi^{(2)\prime}_{3,\underline{Q}^{(\cdot)}_{Y_1 \mid W}(\cdot \mid \cdot) - \overline{Q}^{(\cdot)}_{Y_0 \mid W}(\cdot \mid \cdot)}(\mathbf{Z}^{(2)}_3(\cdot,1,\cdot,\cdot) - \mathbf{Z}^{(1)}_3(\cdot,0,\cdot,\cdot)) \right](w,c) \\
                                    \\
                                    \left[ \phi^{(1)\prime}_{3,\overline{Q}^{(\cdot)}_{Y_1 \mid W}(\cdot \mid \cdot) - \underline{Q}^{(\cdot)}_{Y_0 \mid W}(\cdot \mid \cdot)}(\mathbf{Z}^{(1)}_3(\cdot,1,\cdot,\cdot) - \mathbf{Z}^{(2)}_3(\cdot,0,\cdot,\cdot)) \right](w,c)
	\end{pmatrix} \\
                               &\equiv \mathbf{Z}_5(w,c),
\end{align*}
a tight random element of $\ell^\infty(\supp(W) \times \mathcal{C},\R^2)$.
\end{proof}

The following lemma shows that the sup operator is Hadamard directionally differentiable. It is a very minor extension of lemma B.1 in \cite{FangSantos2015workingPaper}, where we take the supremum over just one of two arguments.

\begin{lemma}\label{lemma:hadamard_diff_sup}
Let $\mathcal{A}$ and $\mathcal{C}$ be compact subsets of $\R$. Define the map $\phi: \ell^\infty(\mathcal{A} \times \mathcal{C}) \rightarrow \ell^\infty(\mathcal{C})$ by
\[
	[\phi(\theta)](c) = \sup_{a\in\mathcal{A}} \; \theta(a,c).
\]
Let
\[
	\Psi_{\mathcal{A}}(\theta,c) = \argmax_{a\in\mathcal{A}} \; \theta(a,c)
\]
be a set-valued function. Then $\phi$ is Hadamard directionally differentiable tangentially to $\mathscr{C}(\mathcal{A} \times \mathcal{C})$ at any $\theta\in \mathscr{C}(\mathcal{A} \times \mathcal{C})$, and $\phi_\theta': \mathscr{C}(\mathcal{A} \times \mathcal{C}) \rightarrow \mathscr{C}(\mathcal{C})$ is given by
\[
	[\phi_{\theta}'(h)](c) = \sup_{a\in\Psi_{\mathcal{A}}(\theta,c)} h(a,c)
\]
for any $h\in \mathscr{C}(\mathcal{A} \times \mathcal{C})$.
\end{lemma}

\begin{proof}[Proof of lemma \ref{lemma:hadamard_diff_sup}]
This proof follows that of Lemma B.1 in \cite{FangSantos2015workingPaper}. Let $t_n \searrow 0$, and $h_n\in\ell^\infty(\mathcal{A} \times \mathcal{C})$ such that
\[
	\sup_{(a,c)\in \mathcal{A} \times \mathcal{C}}|h_n(a,c) - h(a,c)| \equiv \|h_n - h\|_\infty = o(1)
\]
for $h\in \mathscr{C}(\mathcal{A} \times \mathcal{C})$. Since $\mathcal{A}$ is a closed and bounded subset of $\R$, their lemma shows that tangential Hadamard directional differentiability holds for any fixed $c\in\mathcal{C}$. We show that this holds uniformly in $c\in\mathcal{C}$ as well. First, by their equation (B.1), we note that for some $t_n \searrow 0$,
\begin{align}
	\sup_{c\in\mathcal{C}} \left|
		\sup_{a\in\mathcal{A}} \big( \theta(a,c) + t_n h_n(a,c) \big) - \sup_{a\in\mathcal{A}} \big( \theta(a,c) + t_n h(a,c) \big) 
	\right| 
	&\leq \sup_{c\in\mathcal{C}}t_n \sup_{a\in\mathcal{A}}|h_n(a,c) - h(a,c)|\notag\\
    &= t_n \| h_n - h \|_\infty \notag \\
    &= o(t_n). \label{eq:lemma_hadamard_1}
  \end{align}
  Second, by their equations leading to (B.3)
  \begin{align}
    &\sup_{c\in\mathcal{C}} \left| \sup_{a\in\mathcal{A}} \big( \theta(a,c) + t_n h(a,c) \big) - \sup_{a\in\Psi_{\mathcal{A}}(\theta,c)} \big(\theta(a,c) + t_n h(a,c) \big) \right| \notag\\
     &\leq t_n\sup_{c\in\mathcal{C}} \sup_{a_0,a_1\in\mathcal{A}:|a_0 - a_1|\leq \delta_n} |h(a_0,c) - h(a_1,c)|\notag\\
    &= o(t_n)\label{eq:lemma_hadamard_2}
  \end{align}
by uniform continuity of $h(a,c)$ in $a$ and $c$, which follows from the continuity of $h$ on its compact support $\mathcal{A} \times \mathcal{C}$. Finally, combining equations \eqref{eq:lemma_hadamard_1} and \eqref{eq:lemma_hadamard_2} as in equation (B.4) from \cite{FangSantos2014}, it follows that
\begin{align*}
    &\sup_{c\in\mathcal{C}} \left| \sup_{a\in\mathcal{A}} \big( \theta(a,c) + t_n h_n(a,c) \big) - \sup_{a\in\mathcal{A}}\theta(a,c) - t_n\sup_{a\in\Psi_{\mathcal{A}}(\theta,c)}h(a,c) \right| \\
    &\leq \sup_{c\in\mathcal{C}} \left| \sup_{a\in\Psi_{\mathcal{A}}(\theta,c)} \big( \theta(a,c) + t_n h(a,c) \big) - \sup_{a\in\Psi_{\mathcal{A}}(\theta,c)}\theta(a,c) - t_n\sup_{a\in\Psi_{\mathcal{A}}(\theta,c)}h(a,c) \right| + o(t_n) \\
    &= 0 + o(t_n),
\end{align*}
which completes the proof.
\end{proof}

\begin{proof}[Proof of lemma \ref{lemma:conv_DTEbounds}]
We begin by showing that the first component in equation \eqref{eq:conv_DTEbounds} converges to a tight random element of $\ell^\infty(\mathcal{C}\times[0,1])$. Fix $c$ and $w$ and define
\[
	\phi_4: \ell^\infty(\R) \to \R
\]
by
\[
	\phi_4(\theta) = \max \left\{ \sup_{a\in\mathcal{Y}_z(w)}\theta(a,w,c),0 \right\}.
\]
As in the proof of lemma \ref{lemma:conv_cdf_PI}, the four mappings $(\delta_1,\delta_2,\delta_3,\delta_4)$ when considered from $\ell^\infty(\R \times\{0,1\} \times \supp(W))\times\ell^\infty(\{0,1\}\times\supp(W)) \times\ell^\infty(\supp(W))$ to $\ell^\infty(\R\times\{0,1\}\times\supp(W))$ are all Hadamard differentiable when evaluated at $\theta_0$. 

We can write
\begin{align*}
	\phi_4(\theta_0)
	&= \max\left\{\sup_{a\in\mathcal{Y}_z(w)} (\underline{F}^c_{Y_1 \mid W}(a \mid w) - \overline{F}^c_{Y_0 \mid W}(a-z \mid w)),0\right\}\\ 
	&= \max \Bigg\{\sup_{a\in\mathcal{Y}_z(w)} \Big(\max\{[\delta_1(\theta_0)](a,1,w),[\delta_2(\theta_0)](a,1,w)\} \\
	&\hspace{35mm} - \min\{[\delta_3(\theta_0)](a-z,0,w),[\delta_4(\theta_0)](a-z,0,w)\} \Big),0\Bigg\}\\
	&= \max\left\{\sup_{a\in\mathcal{Y}_z(w)} \Big( [\delta_1(\theta_0)](a,1,w)-[\delta_3(\theta_0)](a-z,0,w) \Big), \right. \\
	&\hspace{15mm} \sup_{a\in\mathcal{Y}_z(w)} \Big( [\delta_1(\theta_0)](a,1,w)-[\delta_4(\theta_0)](a-z,0,w) \Big), \\
	&\hspace{15mm} \sup_{a\in\mathcal{Y}_z(w)} \Big( [\delta_2(\theta_0)](a,1,w)-[\delta_3(\theta_0)](a-z,0,w) \Big), \\[-0.6em]
	&\hspace{15mm} \sup_{a\in\mathcal{Y}_z(w)} \Big( [\delta_2(\theta_0)](a,1,w)-[\delta_4(\theta_0)](a-z,0,w) \Big), \ 0\Bigg\}.
\end{align*}

By linearity $(\delta_j - \delta_k)(\theta)$ is Hadamard differentiable at $\theta_0$ for $j=1,2$ and $k=3,4$. By the chain rule (lemma \ref{lemma:hadamard_diff_chainrule}) and lemma \ref{lemma:hadamard_diff_sup}, the mappings
\[
	\theta \mapsto \sup_{a\in\mathcal{Y}_z(w)} \Big( [\delta_j(\theta)](a,1,w)-[\delta_k(\theta)](a-z,0,w) \Big)
\]
are Hadamard directionally differentiable at $\theta_0$ for $j=1,2$ and $k=3,4$. Finally, the maximum operator over five arguments is Hadamard directionally differentiable, and by another application of the chain rule, $\phi_4$ is Hadamard directionally differentiable for fixed $c$ and $w$. Uniformity over $c\in\mathcal{C}$ and $w\in\supp(W)$ is obtained from considering the vector of Hadamard directional derivatives for all $c\in\mathcal{C}$ and $w\in\supp(W)$.

By lemma \ref{lemma:conv_intpiece}, the mapping $(F_{Y \mid X,W}(\cdot \mid \cdot,\cdot),p_{(\cdot|\cdot)}, q_{(\cdot)}) \mapsto \underline{P}(\cdot \mid \cdot)$ is Hadamard directionally differentiable. Linearity of the Hadamard directional derivative operator yields that the mapping $(F_{Y \mid X,W}(\cdot \mid \cdot,\cdot),p_{(\cdot|\cdot)}, q_{(\cdot)}) \mapsto \underline{\text{CDTE}}(z,\cdot,\cdot \mid \cdot)$ is Hadamard directionally differentiable.

Since
\[
	\inf_{a \in \mathcal{A}} \theta(a,c,w) = -\sup_{a \in \mathcal{A}} (-\theta(a,c,w)),
\]
the infimum operator is Hadamard directionally differentiable. As in the proof of lemma \ref{lemma:conv_cdf_PI}, the minimum operator is Hadamard directionally differentiable. Following lemma \ref{lemma:conv_intpiece}, the mapping $(F_{Y \mid X,W}(\cdot \mid \cdot,\cdot),p_{(\cdot|\cdot)}, q_{(\cdot)}) \mapsto \overline{P}(\cdot \mid \cdot)$ is Hadamard directionally differentiable. A similar argument as above implies the mapping $(F_{Y \mid X,W}(\cdot \mid \cdot,\cdot),p_{(\cdot|\cdot)}, q_{(\cdot)}) \mapsto \overline{\text{CDTE}}(z,\cdot,\cdot \mid \cdot)$ is Hadamard directionally differentiable.

Combining these results with lemma \ref{lemma:cdf_conv} allows us to conclude that 
\begin{align*}
\sqrt{N}\left(
            \begin{array}{c}
              \widehat{\overline{\text{CDTE}}}(z,c,t \mid w) - \overline{\text{CDTE}}(z,c,t \mid w) \\
              \widehat{\underline{\text{CDTE}}}(z,c,t \mid w) - \underline{\text{CDTE}}(z,c,t \mid w)
            \end{array}
          \right) &\rightsquigarrow \tilde{\mathbf{Z}}_6(c,t,w),
\end{align*}
a tight random element of $\ell^\infty(\mathcal{C}\times[0,1] \times \supp(W),\R^2)$ with continuous paths.

Finally, to see that equation \eqref{eq:conv_DTEbounds} holds, consider the lower bound estimator. We have
\begin{align*}
	\sqrt{N}(\widehat{\underline{\text{DTE}}}(z,c,t) - \underline{\text{DTE}}(z,c,t) ) &= 
	\sum_{k=1}^K \sqrt{N} (\widehat{\underline{\text{CDTE}}}(z,c,t \mid w_k) - \underline{\text{CDTE}}(z,c,t \mid w_k))q_{w_k}\\
	&\qquad + \sum_{k=1}^K\underline{\text{CDTE}}(z,c,t \mid w_k) \sqrt{N}(\widehat{q}_{w_k} - q_{w_k})\\
	&\rightsquigarrow \sum_{k=1}^K \tilde{\mathbf{Z}}_6(c,t,w_k) q_{w_k} + \sum_{k=1}^K \underline{\text{CDTE}}(z,c,t \mid w_k) \mathbf{Z}_1^{(3)}(0,0,w_k).
\end{align*}
A similar derivation holds for the upper bound estimator.
\end{proof}

\begin{proof}[Proof of theorem \ref{thm:BF convergence}]
By lemmas \ref{lemma:conv_intpiece} and \ref{lemma:cdf_conv}, the numerator of equation \eqref{eq:empiricalbf} converges uniformly over $c\in \mathcal{C}$. By lemmas \ref{lemma:conv_intpiece} and \ref{lemma:conv_DTEbounds}, the denominator also converges uniformly over $c\in\mathcal{C}$. By the delta method, 
\begin{align*}
	&\sqrt{N}\left(\widehat{\text{bf}}(c,\underline{p}) - \text{bf}(c,\underline{p})\right)\\
	&= \sqrt{N}\left( \frac{1 - \underline{p} - \sum_{k=1}^K\widehat{\overline{P}}(c \mid w_k)\widehat{q}_{w_k}}{
	1 + \sum_{k=1}^K \left[\min \left\{\inf_{y\in\mathcal{Y}_0(w_k)} (\widehat{\overline{F}}^c_{Y_1 \mid W}(y \mid w_k) - \widehat{\underline{F}}^c_{Y_0 \mid W}(y \mid w_k)),0 \right\} - \widehat{\overline{P}}(c \mid w_k)\right]\widehat{q}_{w_k}}\right.\\[1em]
	&\hspace{15mm}  \left.- \frac{
		1 - \underline{p} 
		- \sum_{k=1}^K \overline{P}(c \mid w_k)q_{w_k}
	}{
	1 + \sum_{k=1}^K \left[\min \left\{ 
	\inf_{y \in \mathcal{Y}_0(w_k)} 
	(\overline{F}^c_{Y_1 \mid W}(y \mid w_k) - \underline{F}^c_{Y_0 \mid W}(y \mid w_k)), 0 
	\right\} 
	- \widehat{\overline{P}}(c \mid w_k)\right]q_{w_k}}\right) \\[1em]
	&\rightsquigarrow \frac{-\sum_{k=1}^K \mathbf{Z}_5^{(1)}(w_k,c)q_{w_k} - \sum_{k=1}^K \overline{P}(c \mid w_k) \mathbf{Z}_1^{(3)}(0,0,w_k)}{1 + \sum_{k=1}^K \left[\min \left\{ 
	\inf_{y \in \mathcal{Y}_0(w_k)} 
	(\overline{F}^c_{Y_1 \mid W}(y \mid w_k) - \underline{F}^c_{Y_0 \mid W}(y \mid w_k)), 0 
	\right\} 
	- \overline{P}(c \mid w_k)\right]q_{w_k}}\\[1em]
	&\hspace{10mm}  - \frac{
		1 - \underline{p} 
		- \sum_{k=1}^K \overline{P}(c \mid w_k)q_{w_k}
	}{
	\left(1 + \sum_{k=1}^K \left[\min \left\{ 
	\inf_{y \in \mathcal{Y}_0(w_k)} 
	(\overline{F}^c_{Y_1 \mid W}(y \mid w_k) - \underline{F}^c_{Y_0 \mid W}(y \mid w_k)), 0 
	\right\} 
	- \overline{P}(c \mid w_k)\right]q_{w_k}\right)^2}\tilde{\mathbf{Z}}(c)
\end{align*}
where 
\begin{multline*}
	\sqrt{N}\left(1-\underline{p} - \sum_{k=1}^K\widehat{\overline{P}}(c \mid w_k)\widehat{q}_{w_k} -  \left[1- \underline{p} - \sum_{k=1}^K \overline{P}(c \mid w_k)q_{w_k} \right] \right) \\
	\rightsquigarrow -\sum_{k=1}^K \mathbf{Z}_5^{(1)}(w_k,c)q_{w_k} - \sum_{k=1}^K \overline{P}(c \mid w_k) \mathbf{Z}_1^{(3)}(0,0,w_k)
\end{multline*}
and
\begin{align*}
	&\sqrt{N}\left(\sum_{k=1}^K \left[\min \left\{\inf_{y\in\mathcal{Y}_0(w_k)} (\widehat{\overline{F}}^c_{Y_1 \mid W}(y \mid w_k) - \widehat{\underline{F}}^c_{Y_0 \mid W}(y \mid w_k)),0 \right\} - \widehat{\overline{P}}(c \mid w_k)\right]\widehat{q}_{w_k} \right. \\
	&\quad \qquad \left. - \sum_{k=1}^K \left[\min \left\{ 
	\inf_{y \in \mathcal{Y}_0(w_k)} 
	(\overline{F}^c_{Y_1 \mid W}(y \mid w_k) - \underline{F}^c_{Y_0 \mid W}(y \mid w_k)), 0 
	\right\} 
	- \overline{P}(c \mid w_k)\right]q_{w_k}\right) \rightsquigarrow \tilde{\mathbf{Z}}(c).
\end{align*}
Here $\tilde{\mathbf{Z}}(c)$ is a random element of $\ell^{\infty}(\mathcal{C})$ by lemmas \ref{lemma:conv_intpiece} and \ref{lemma:conv_DTEbounds}. Therefore,
\[
	\sqrt{N}\left(\widehat{\text{bf}}(c,\underline{p}) - \text{bf}(c,\underline{p})\right)
\]
converges to a random element in $\ell^{\infty}(\mathcal{C}\times\mathcal{P})$.

As discussed in the proof of lemma \ref{lemma:conv_cdf_PI}, the maximum and minimum operators in equation \eqref{eq:empiricalBF} are Hadamard directionally differentiable. By lemma \ref{lemma:hadamard_diff_chainrule}, their composition is Hadamard directionally differentiable. Therefore, by the delta method for Hadamard directionally differentiable functions, $\sqrt{N}(\widehat{\text{BF}}(c,\underline{p}) - \text{BF}(c,\underline{p}))$ converges in process as in the statement of the theorem.
\end{proof}

\begin{lemma}\label{lemma:lipschitz sup operator}
Let $h:A \rightarrow \R$ where $A\subseteq \R$. Let $F(h) = \sup_{x\in A} h(x)$. Let $\|\cdot\|_\infty$ denote the sup-norm $\|h\|_{\infty} = \sup_{x\in A}|h(x)|$. Then $F$ is Lipschitz continuous with respect to the sup-norm $\|\cdot\|_\infty$ and has Lipschitz constant equal to one.
\end{lemma}

\begin{proof}[Proof of lemma \ref{lemma:lipschitz sup operator}]
For functions $h$ and $h'$,
	\begin{align*}
		\sup_{x \in A}h(x) - \sup_{\tilde{x} \in A} h'(\tilde{x}) &= \sup_{x\in A} \left( h(x) - \sup_{\tilde{x} \in A}h'(\tilde{x}) \right)\\
												&\leq \sup_{x\in A}(h(x) - h'(x))\\
												&\leq \sup_{x\in A}|h(x) - h'(x)|.
	\end{align*}
By a symmetric argument,
\begin{align*}
	\sup_{x\in A}h'(x) - \sup_{\tilde{x} \in A} h(\tilde{x})
		&\leq \sup_{x\in A} | h'(x) - h(x) | \\
		&= \sup_{x\in A} | h(x) - h'(x) |.
\end{align*}
Therefore $|F(h) - F(h')| \leq \|h - h'\|_\infty$.
\end{proof}

\begin{proof}[Proof of proposition \ref{prop:bootstrap validity}]
Hadamard directional differentiability of $\phi$ follows from the chain rule (lemma \ref{lemma:hadamard_diff_chainrule}) and from the proof of theorem \ref{thm:BF convergence}, since the breakdown frontier is a Hadamard directionally differentiable mapping of $\overline{P}(\cdot) = \Exp[\overline{P}(\cdot \mid W)]$ and $\overline{\text{DTE}}(z,\cdot,\cdot)$, which are themselves Hadamard directionally differentiable mappings of $\theta_0$.

Lemma \ref{lemma:cdf_conv} combined with theorem 3.6.1 of \cite{VaartWellner1996} implies consistency of the nonparametric bootstrap for our underlying parameters: $\mathbf{Z}_N^* = \sqrt{N}(\widehat{\theta}^* - \widehat{\theta}) \overset{P}{\rightsquigarrow} \mathbf{Z}_1$. By this result, $\varepsilon_N \rightarrow 0$, $\sqrt{N}\varepsilon_N\rightarrow\infty$, and theorem 3.1 in \cite{HongLi2015}, equation \eqref{eq:bootstrap_conv_BF} holds.

By $1/\sigma(c)$ being uniformly bounded, we have that 
\[
	\frac{ \left[ \widehat{\phi}_{\theta_0}' ( \sqrt{N}(\widehat{\theta}^* - \widehat{\theta}) ) \right] (c,\underline{p}) }{\sigma(c)}
	\overset{P}{\rightsquigarrow} 
	\frac{\mathbf{Z}_7(c,\underline{p})}{\sigma(c)}.
\]
By lemma \ref{lemma:lipschitz sup operator}, the $\sup$ operator is Lipschitz with Lipschitz constant equal to 1. Therefore, by proposition 10.7 on page 189 of \cite{Kosorok2007}, we can apply a continuous mapping theorem to get
\[
		\sup_{c\in\mathcal{C}} \frac{ \left[ \widehat{\phi}_{\theta_0}'(\sqrt{N}(\widehat{\theta}^* - \widehat{\theta})) \right] (c,\underline{p})}{\sigma(c)}
		\overset{P}{\rightsquigarrow}
		\sup_{c\in\mathcal{C}} \frac{\mathbf{Z}_7(c,\underline{p})}{\sigma(c)}.
\]
The rest of the proof follows from corollary 3.2 of \cite{FangSantos2015workingPaper}.
\end{proof}

\begin{proof}[Proof of corollary \ref{prop:uniform band interval}]
This follows immediately from proposition \ref{prop:bootstrap validity} and lemma \ref{lemma:monotonicBandCoverage} below.
\end{proof}

\begin{lemma}\label{lemma:monotonicBandCoverage}
Let $\overline{C} > 0$ and $\mathcal{C} = \{ c_1,\ldots,c_J \} \subseteq [0,\overline{C}]$ be a finite grid of points. Let $f : [0,\overline{C}] \rightarrow \R_+$ be a nonincreasing function. Let $\widehat{\text{LB}}(\cdot)$ be an asymptotically exact uniform lower $1-\alpha$ confidence band for $f$ on the grid points: 
\[
	\lim_{N \rightarrow \infty} \Prob \left( \widehat{\text{LB}}(c_j) \leq f(c_j) \text{ for $j=1,\ldots,J$} \right) = 1-\alpha.
\]
Define $\widetilde{\text{LB}} : [0,\overline{C}] \rightarrow \R_+$ by
\[
	\widetilde{\text{LB}}(c)
		= 
		\begin{cases}
			\widehat{\text{LB}}(c_1) 
			&\text{ if $c\in [0,c_1]$} \\
			\quad \vdots \\
			\widehat{\text{LB}}(c_j) 
			&\text{ if $c \in (c_{j-1},c_j]$, for $j=2,\ldots,J$} \\
			\quad \vdots \\
			\quad 0
			&\text{ if $c \in (c_J, \overline{C}]$}.
		\end{cases}
\]
Then $\widetilde{\text{LB}}(\cdot)$ is an asymptotically exact uniform lower $1-\alpha$ confidence band on $[0,\overline{C}]$:
\[
	\lim_{N \rightarrow \infty} \Prob \left( \widetilde{\text{LB}}(c) \leq f(c) \text{ for all $c \in [0,\overline{C}]$} \right) = 1-\alpha.
\]
\end{lemma}

\begin{proof}[Proof of lemma \ref{lemma:monotonicBandCoverage}]
Define the events
\[
	A = \{ \widehat{\text{LB}}(c_j) \leq f(c) \text{ for all $c\in (c_{j-1},c_j]$, for $j=1,\ldots,J$} \}
\]
and
\[
	B = \{ \widehat{\text{LB}}(c_j) \leq f(c_j) \text{ for $j=1,\ldots,J$} \}.
\]
$A$ immediately implies $B$. Since $f$ is nonincreasing, $B$ implies $A$. Thus
\begin{align*}
	\P(\widetilde{\text{LB}}(c) \leq f(c) \text{ for all } c\in [0,\overline{C}])
	&= \P(\widehat{\text{LB}}(c_j) \leq f(c) \text{ for all } c\in (c_{j-1},c_j], \text{ for } j=1,\ldots,J)\\
	&= \P(\widehat{\text{LB}}(c_j) \leq f(c_j) \text{ for } j=1,\ldots,J).
\end{align*}
The first line follows by definition of $\widetilde{\text{LB}}$ and since $f$ is nonnegative. Taking limits as $N\rightarrow\infty$ yields
\begin{align*}
	\lim_{N\rightarrow\infty} \P(\widetilde{\text{LB}}(c) \leq f(c) \text{ for all } c\in [0,\overline{C}])
	&= \lim_{N\rightarrow\infty}\P(\widehat{\text{LB}}(c_j) \leq f(c_j)\text{ for } j=1,\ldots,J)\\
	&= 1-\alpha,
\end{align*}
where the last equality follows from the validity of the band $\widehat{\text{LB}}(\cdot)$ on $\mathcal{C}$.
\end{proof}

\end{document}


\title{\textbf{Supplemental Appendix to \\ ``Inference on Breakdown Frontiers''}}

\author{Matthew A. Masten\footnote{Department of Economics, Duke University,
        \texttt{matt.masten@duke.edu}} \qquad Alexandre Poirier\thanks{
    Department of Economics, Georgetown University,
    \texttt{alexandre.poirier@georgetown.edu}}
}

\maketitle
\begin{abstract}
This supplemental appendix provides an extended comparison of our results with the sensitivity analysis inference literature, a discussion of higher dimensional breakdown frontiers, Monte Carlo simulations for our proposed estimation and inference procedures, an alternative inference approach based on population smoothing, and additional empirical analyses. 
\end{abstract}

\onehalfspacing
\normalsize
\appendix

\section{Inference in sensitivity analyses}\label{InferenceLitDetails}

In this section we provide additional details explaining how our results compare to several approaches in the literature. We focus on the different inference methods used in sensitivity analyses. Most methods can be grouped by whether the population level sensitivity analysis is a parametric path or nonparametric neighborhood approach. In \cite{MastenPoirier2016} we compare and contrast these population level approaches in more detail. The parametric path approach has two key features: (1) a specific parametric deviation $r$ from a baseline assumption of $r = 0$ and (2) a parameter $\theta(r)$ that is point identified given that deviation. The nonparametric neighborhood approach specifies increasing nested neighborhoods around a baseline assumption of $r=0$ such that $\Theta(r)$ is the identified set for the parameter given a specific neighborhood $r$. Typically $\Theta(r) = [\Theta_L(r), \Theta_U(r)]$ for point identified lower and upper bound functions $\Theta_L$ and $\Theta_U$.

\subsection*{Parametric paths}

The most common approach for a parametric path analysis is to report the estimated function $\widehat{\theta}(r)$ along with pointwise confidence bands. For example, see figure 1 of \cite{RotnitzkyRobinsScharfstein1998}, figure 1 of \cite{Robins1999}, and figure 1 of \cite{VansteelandtEtAl2006}. Uniform confidence bands can be used instead, as in figure 3 of \cite{TodemFinePeng2010}. Those authors use their uniform confidence bands to test hypotheses about $\theta(r)$ uniformly over $r$. They also suggest projecting these bands onto their domain to obtain confidence sets for the set $\{ r : | \theta(r) | > 0 \}$, although they do not discuss this in detail (see the last few sentences of page 562). They emphasize that using uniform confidence bands is important since the functions $\theta(r)$ are often not monotonic, as we discussed earlier with respect to \cite{Imbens2003}. A similar method is proposed by \cite{RotnitzkyRobinsScharfstein1998}. They study a model with two scalar sensitivity parameters $r = (r_1,r_2)$ and two parameters $\theta_1(r)$ and $\theta_2(r)$. They construct a standard test statistic $T(r)$ for testing the null that $\theta_1(r) = \theta_2(r)$. They then plot the contours
\[
	\{ r \in \R^2 : T(r) = -1.96 \}
	\qquad \text{and} \qquad
	\{ r \in \R^2 : T(r) = 1.96 \}.
\]
See their figure 2. Unlike \cite{TodemFinePeng2010}, they do not account for multiple testing concerns. Also see figures 2--4 of \cite{RotnitzkyScharfsteinSuRobins2001}. Several papers also suggest picking a set $\mathcal{R}$ to form an identified set $\{ \theta(r) : r \in \mathcal{R} \}$ and then doing inference on this identified set. For example, see \cite{VansteelandtEtAl2006}. \cite{EscancianoZhu2013} consider the diameter of such identified sets,
\[
	d = \sup_{r,r' \in \mathcal{R} } \| \theta(r) - \theta(r') \|,
\]
and study estimation and inference on $d$.

Finally, \cite{Rosenbaum1995,Rosenbaum2002} proposes a sensitivity analysis within the context of finite sample randomization inference for testing the sharp null hypotheses of no unit level treatment effects for the units in our dataset. This is a very different approach to the approaches discussed above and what we do in the present paper.

\subsection*{Nonparametric neighborhoods}

Our population level sensitivity analysis uses nonparametric neighborhoods, not parametric paths. Thus for each $r$ we obtain an identified set $\Theta(r)$. There is a large literature on how to do inference on a single identified set; see \cite{CanayShaikh2017} for an overview. Few papers discuss inference on a continuous sequence of identified sets $\Theta(r)$, however. The simplest approach arises when the identified set is characterized by point identified upper and lower bounds: $\Theta(r) = [\Theta_L(r),\Theta_U(r)]$. In this case one can plot estimated bound functions $\widehat{\Theta}_L$ and $\widehat{\Theta}_U$ along with outer confidence bands for these functions. For example, see figure 2 of \cite{RichardsonHudgensGilbertFine2014}. They informally discuss how to use these bands to check robustness of the claim that the true parameter is nonzero, but they do not formally discuss breakdown points or inference on them.

\cite{KlineSantos2013} similarly begin by constructing pointwise confidence bands for the bound functions. They then use level sets of these bands to construct their confidence intervals for a breakdown point (see equation 41 on page 249). In their remark 4.4 on page 250 they mention the approach we take---doing inference based directly on the asymptotic distribution of breakdown point estimators. In order to compare these two approaches, we discuss the approach of projecting confidence bands for lower bound functions in more detail here.

Let the sensitivity parameter $r$ be in $[0,1]^{d_r}$ for some integer $d_r \geq 1$. Let $\Theta_L(r)$ denote the lower bound function for a scalar parameter $\theta$. By construction, $\Theta_L(\cdot)$ is weakly decreasing in its components. Suppose it is also continuous. Suppose we are interested in the conclusion that $\theta_\text{true} \geq \underline{\theta}$. Suppose for simplicity that it is known that $\Theta_L(0) \geq \underline{\theta}$. This allows us to ignore the upper bound function and its confidence band. Define the breakdown frontier for the claim that $\theta_\text{true} \geq \underline{\theta}$ by
\[
	\text{BF} = \{ r \in [0,1]^{d_r} : \Theta_L(r) = \underline{\theta} \}.
\]
Let
\[
	\text{RR} = \{ r \in [0,1]^{d_r} : \Theta_L(r) \geq \underline{\theta} \}.
\]
denote the robust region, the set of sensitivity parameters that lie on or below the breakdown frontier. The following proposition shows that, in general, projections of uniform lower confidence bands for $\Theta_L$ produce valid uniform lower confidence bands for the breakdown frontier.

\begin{proposition}\label{prop:ProjectionOfUniformBands}
Let $\text{LB}(\cdot)$ be an asymptotically exact uniform lower (1$-\alpha)$-confidence band for $\Theta_L(\cdot)$. That is,
\[
	\lim_{N \rightarrow \infty} \Prob \left( \text{LB}(r) \leq \Theta_L(r) \text{ for all $r \in [0,1]^{d_r}$} \right) = 1-\alpha.
\]
(We call $\text{LB}(\cdot)$ a `band' even though it's really a hypersurface.) Define the projections
\[
	\text{BF}_L = \{ r \in [0,1]^{d_r} : \text{LB}(r) = \underline{\theta} \}
\]
and
\[
	\text{RR}_L = \{ r \in [0,1]^{d_r} : \text{LB}(r) \geq \underline{\theta} \}.
\]
Then
\[
	\lim_{N \rightarrow \infty} \Prob ( \text{RR}_L \subseteq \text{RR} ) \geq 1-\alpha.
\]
\end{proposition}

\begin{proof}[Proof of proposition \ref{prop:ProjectionOfUniformBands}]
We have
\begin{align*}
	\Prob(\text{RR}_L \subseteq \text{RR})
		&= \Prob(\text{For all $r \in [0,1]^{d_r}$ s.t. $\text{LB}(r) \geq \underline{\theta}$, we have $\Theta_L(r) \geq \underline{\theta}$}) \\
		&\geq \Prob(\text{LB}(r) \leq \Theta_L(r) \text{ for all $r \in [0,1]^{d_r}$} ).
\end{align*}
Now take limits on both sides as $N \rightarrow \infty$. The inequality arises essentially because the functional inequality $\text{LB}(\cdot) \leq \Theta_L(\cdot)$ is a sufficient, but not necessary, condition for $\text{RR}_L \subseteq \text{RR}$.
\end{proof}

Proposition \ref{prop:ProjectionOfUniformBands} shows that projecting a uniform band always yields a confidence band for the breakdown frontier which has size at least $1-\alpha$. Notice that although we did not use monotonicity of $\Theta_L(\cdot)$ here, this monotonicity implies that we can always take $\text{LB}(\cdot)$ to be weakly decreasing without loss of generality. This follows since monotonicity of $\Theta_L(\cdot)$ allows us to convert any non-monotonic lower confidence band into a monotonic one without any loss of coverage.

There are two downsides to this projection approach, compared to our direct approach:
\begin{enumerate}
\item In general, this projection approach may be conservative.

\item Relatedly, one must choose the lower confidence band $\Theta_L(\cdot)$. There are many such choices. The standard ones, such as equal width or inversion of a sup $t$-statistic (e.g., see \citealt{FreybergerRai2016}), will likely yield conservative projection bands, since they are not chosen with the goal of doing inference on the breakdown frontier in mind.
\end{enumerate}

\cite{KlineSantos2013} do not propose projections of \emph{uniform} confidence bands. They propose projections of pointwise confidence bands. As we discuss next, projection of pointwise bands produces valid confidence intervals for breakdown points. But it does not generally produce valid confidence bands for breakdown frontiers. Hence in the multidimensional $r$ case one either must use our direct approach, or appeal to proposition \ref{prop:ProjectionOfUniformBands} above.

To see that pointwise band projections are valid in the scalar $r$ case, we expand on Kline and Santos' \citeyearpar{KlineSantos2013} analysis. Define the population breakdown point by
\[
	r^* = \inf \{ r \in [0,1] : \Theta_L(r) \leq \underline{\theta} \}.
\]
Let $c_{1-\alpha}(r)$ be the $1-\alpha$ quantile of the asymptotic distribution of
\[
	\sqrt{N} \big[ \widehat{\Theta}_L(r) - \Theta_L(r) \big].
\]
Define the pointwise one-sided lower confidence band for $\Theta_L(\cdot)$ by
\[
	\text{LB}(r) = \widehat{\Theta}_L(r) - \frac{c_{1-\alpha}(r)}{\sqrt{N}}.
\]
Let
\[
	r_L = \inf \{ r \in [0,1] : \text{LB}(r) \leq \underline{\theta} \}
\]
be the projection of this confidence band. The following result is a minor generalization of example 2.1 in \cite{KlineSantos2013}.

\begin{proposition}\label{prop:KS2013pointwiseProjection}
Assume that the cdf of the asymptotic distribution of $\sqrt{N} \big[ \widehat{\Theta}_L(r^*) - \Theta_L(r^*) \big]$ is continuous and strictly increasing at its $1-\alpha$ quantile. Then 
\[
	\lim_{N \rightarrow \infty} \Prob(r_L \leq r^*) \geq 1-\alpha.	
\]
If $\text{LB}(\cdot)$ is weakly decreasing with probability one then this inequality holds with equality.
\end{proposition}

\begin{proof}[Proof of proposition \ref{prop:KS2013pointwiseProjection}]
We have
\begin{align*}
	\Prob(r_L \leq r^*)
		&= \Prob(r^* \geq \inf \{ r \in [0,1] : \text{LB}(r) \leq \underline{\theta} \}) \\
		&\geq \Prob( \text{LB}(r^*) \leq \underline{\theta} ) \\
		&= \Prob \left( \widehat{\Theta}_L(r^*) - \frac{c_{1-\alpha}(r^*)}{\sqrt{N}} \leq \underline{\theta} \right) \\
		&= \Prob \left( \sqrt{N} \big( \widehat{\Theta}_L(r^*) - \underline{\theta} \big) \leq c_{1-\alpha}(r^*) \right) \\
		&= \Prob \left( \sqrt{N} \big( \widehat{\Theta}_L(r^*) - \Theta_L(r^*) \big) \leq c_{1-\alpha}(r^*) \right).
\end{align*}
The first line follows by definition of $r_L$. For the second line, notice that $\text{LB}(r^*) \leq \underline{\theta}$ implies that $r^* \in \{ r \in [0,1] : \text{LB}(r) \leq \underline{\theta} \}$ and hence $r^* \geq \inf \{ r \in [0,1] : \text{LB}(r) \leq \underline{\theta} \}$ by the definition of the infimum. This gives us
\[	
	\Prob( \text{LB}(r^*) \leq \underline{\theta} )
	\leq
	\Prob(r^* \geq \inf \{ r \in [0,1] : \text{LB}(r) \leq \underline{\theta} \}).	
\]
If $\text{LB}(\cdot)$ is weakly decreasing with probability one, then the reverse inequality holds, and hence we have an equality in the second line. To see this, suppose $r^* \geq r_L$ holds. Then $\text{LB}(r^*) \leq \text{LB}(r_L)$ since $\text{LB}(\cdot)$ is weakly decreasing. But now notice that $\text{LB}(r_L) \leq \underline{\theta}$ by definition of $r_L$. Hence $\text{LB}(r^*) \leq \underline{\theta}$.

The third line follows by definition of $\text{LB}$. The fifth line by definition of the population breakdown point, as the solution to $\Theta_L(r) = \underline{\theta}$. 
The result now follows by taking limits as $N \rightarrow \infty$ on both sides, and by definition of $c_{1-\alpha}(r^*)$ and the invertibility of the limiting cdf at its $1-\alpha$ quantile.
\end{proof}

Proposition \ref{prop:KS2013pointwiseProjection} shows that, for doing inference on scalar breakdown points, projections of monotonic lower pointwise confidence bands for the lower bound function yields a one-sided confidence interval $[r_L,1]$ for the breakdown point $r^*$ which has asymptotically exact size. If the lower band function is not always monotonic, however, this projection can be conservative. Moreover, since we're constructing one-sided pointwise confidence bands, we do not have any flexibility to choose the shape of this confidence band. Hence whether it is monotonic or not will be determined by the distribution of the data. Furthermore, it does not appear that this proof strategy extends to multidimensional $r$. Hence projections of pointwise bands are unlikely to yield uniform confidence bands for the breakdown frontier.

Overall, our analysis above shows that the projection of confidence bands approach to doing inference on breakdown points and frontiers will likely yield conservative inference. This is not surprising since, unlike our approach, these bands are not designed specifically for doing inference on the breakdown frontier. Finally, we note that if one nonetheless wants to use a projection approach, our asymptotic results in section 3 can be used to do so.

\subsection*{A testing interpretation of lower confidence bands for breakdown frontiers}\label{sec:testingInterpretation}

Consider the scalar $r$ case, as above. Suppose we want to test
\[
	H_0 : \Theta_L(r) \leq \underline{\theta}
	\qquad \text{versus} \qquad
	H_1 : \Theta_L(r) > \underline{\theta}
\]
for a fixed $r \in [0,1]$. By definition of the breakdown point, $H_0$ is true if and only if $r \geq r^*$. Let $[r_L,1]$ denote a one-sided lower confidence interval for the breakdown point $r^*$; that is, $\Prob([r_L,1] \ni r^*) = 1-\alpha$. Define the test
\[
	\phi
	=
	\begin{cases}
		\text{Choose $H_0$} &\text{if $r_L < r$} \\
		\text{Choose $H_1$} &\text{if $r \leq r_L$}.
	\end{cases}
\]
Then
\begin{align*}
	\Prob(\text{Choose $H_1$} \mid \text{$H_0$ true})
		&= \Prob(r \leq r_L) \\
		&\leq \Prob(r^* \leq r_L) \\
		&= \alpha.
\end{align*}
The second line follows since $r \geq r^*$. The last line follows by construction of $r_L$. Hence $\phi$ has size at most $\alpha$. This result holds for any $r \in [0,1]$. Thus we can interpret the robust region inner confidence set $[0,r_L]$ as the set of sensitivity parameters $r$ such that we reject the null that the true parameter might be below $\underline{\theta}$. That is, for $r \in [0,r_L]$, our test concludes that $\theta > \underline{\theta}$. For $r$ outside the robust region inner confidence set, we do not reject the null that $\theta$ might be at or below $\underline{\theta}$. 

Here we considered the scalar $r$ case for simplicity, but this argument extends to the general case of interpreting lower confidence bands for an arbitrary dimensional breakdown frontier.

\subsection*{Local analyses}

Since \cite{Pitman1949}, local asymptotics are sometimes used to study the behavior of a given estimator under small deviations from model assumptions. Several papers use this approach to study deviations from exogeneity-type assumptions. In a missing data model, \cite{CopasEguchi2001} consider local-to-full-independence asymptotic distributions of MLEs. \cite{ConleyHansenRossi2012} derive the asymptotic bias of the 2SLS estimator in an IV model along sequences where violations of the exclusion restriction converge to zero. The asymptotic bias depends on a local parameter. By placing a prior on this local parameter, they do Bayesian inference on the coefficient of interest. \cite{AndrewsGentzkowShapiro2017} generalize this local-to-zero asymptotic result to the GMM estimator for a given system of moment equalities. Unlike this literature, our approach is global: Breakdown frontier analysis focuses on the largest relaxations of an assumption under which one's conclusions still hold.

\subsection*{Bayesian inference and breakdown frontiers}

Although we focus on frequentist inference, here we briefly discuss Bayesian approaches. In section 11 of \cite{RobinsRotnitzkyScharfstein2000}, Robins studied Bayesian inference in a parametric path approach to sensitivity analysis. Let $r$ denote the sensitivity parameter and $\theta(r)$ the parameter of interest, which is point identified given $r$. Holding $r$ fixed, one can do standard Bayesian inference on $\theta(r)$. Thus Robins simply suggests placing a prior on $r$ and averaging posteriors conditional on $r$ over this prior.
Indeed, this approach is essentially just Bayesian model averaging, where $r$ indexes the class of models under consideration. See \cite{HoetingEtAl1999} for a survey of Bayesian model averaging, and \cite{Leamer1978} for important early work. Among other approaches, \cite{ConleyHansenRossi2012} apply these ideas to do a sensitivity analysis in an IV model. See \cite{DitragliaGarciaJimeno2016} for a generalization and a detailed analysis of priors in that IV setting.

Next consider the nonparametric neighborhood approach. Here the parameter of interest is only partially identified for a fixed $r$, and thus even holding $r$ fixed leads to non-standard Bayesian analysis. \cite{GiacominiKitagawaVopicella2016} study Bayesian model averaging where one of the models is partially identified. They study averaging of a finite number of models. If their results can be extended to a continuum of models, then this method could be applied to the model and assumptions we consider in this paper.

A subtlety arises in both \cite{RobinsRotnitzkyScharfstein2000} and \cite{GiacominiKitagawaVopicella2016}: Depending on how one specifies the joint prior for the sensitivity parameters and the remaining parameters, it may be possible to obtain some updating of the prior for the sensitivity parameters (a point mentioned more generally by \citealt{Lindley1972} in his footnote 34 on page 46; also see \citealt{KoopPoirier1997}). As \cite{GiacominiKitagawaVopicella2016} discuss, however, the model posterior will not converge to the truth unless the model is refutable. None of the assumptions $(c,t)$ in the model we study are refutable. Hence the prior over $(c,t)$ generally matters even asymptotically. That said, the breakdown frontier determines exactly how much the model priors matter for a specific claim. For instance, suppose the model prior places all of its mass below the breakdown frontier for a specific claim. Then we conjecture that the Bayesian model averaged posterior probability that the claim is true will converge to one as $N \rightarrow \infty$, regardless of the specific choice of prior. \cite{KlineTamer2016} provide results like this in the single model case. More generally, we conjecture that the proportion of model prior mass that falls below the breakdown frontier partially determines the tightness of the corresponding asymptotic posterior probability of the conclusion being true: The more mass outside the breakdown frontier, the more the model priors matter. Consequently, a sample analog estimate and perhaps even frequentist inference on the breakdown frontier can be useful even in a Bayesian analysis, to help determine the importance of one's model priors. This is similar to Moon and Schorfheide's \citeyearpar{MoonSchorfheide2012} recommendation that one report estimated identified sets along with Bayesian posteriors. Here we have just sketched the relationship between Bayesian analysis and breakdown frontiers. We leave a complete analysis of these issues to future work.

\section{Higher dimensional breakdown frontiers}

In this paper we focus on breakdown frontiers where the assumption space---the domain of the sensitivity parameters that index relaxations---is two dimensional. In two dimensions, the breakdown frontier can be viewed as a function on a one dimensional domain. Hence presenting estimated breakdown frontiers and corresponding confidence bands is conceptually straightforward. As in other settings, however, summarizing nonparametric functions of two or more parameters is difficult. In this section, we briefly discuss four ways to summarize higher dimensional breakdown frontiers.
\begin{enumerate}
\item \textit{Compute the size of the robust region.} We call the area under the breakdown frontier the robust region. The area of this region provides a quantitative measure of the robustness of the conclusion of interest. This area is a scalar statistic that can be computed regardless of the dimension of the space of assumptions. Although this statistic does not capture the trade-off between assumptions' identifying power, it can be used to compare the overall robustness of conclusions across different studies.

\item \textit{Compute directional breakdown points.} Let the sensitivity parameter $r$ be in $[0,1]^{\text{dim}(r)}$ for some integer $\text{dim}(r) \geq 2$. One can focus on values $r = m \cdot d$ where $m \in \R_+$ is a scalar and $d \in \R_+^{\text{dim}(r)}$ is a known vector. Given the direction $d$, the largest possible value of $m$ such that $m \cdot d \in [0,1]^{\text{dim}(r)}$ is
\[
	\overline{m} = \sup \{ m \in \R_+ : m \cdot d_\ell \leq 1 \text{ for all $\ell=1,\ldots,\text{dim}(r)$ } \}.
\]
Suppose we are interested in the conclusion that $\theta \in \mathcal{C}$ for some pre-specified set $\mathcal{C} \subseteq \Theta$. Then the breakdown point in the direction $d$ is
\[
	m^* = \sup \{ m \in [0,\overline{m}] : \Theta_I(m \cdot d) \subseteq \mathcal{C} \}.
\]
This is the largest we can relax the assumptions in the direction $d$ while still being able to conclude that $\theta \in \mathcal{C}$. The point $r^* = m^* \cdot d$ is on the breakdown frontier for this conclusion.

\item \textit{Use parametric shapes to construct confidence sets.} Characterizing the breakdown frontier is equivalent to characterizing the robust region. In this paper, we study a setting where we construct inner confidence sets for the robust region. We presented these inner confidence sets graphically. It is well known, however, that summarizing higher dimensional confidence sets is difficult. A common solution is to use rectangular confidence regions. These regions can then be summarized by a set of intervals. Here we could similarly construct rectangular inner confidence sets for the robust region. One downside of using rectangles is that they may not well approximate any curvature in the shape of the breakdown frontier. To capture such curvature, while retaining the ability to summarize a higher dimensional set, we could instead construct ellipsoidal inner confidence sets.

\item \textit{Compute average derivatives.} Suppose the spacecraft of assumptions is three dimensional. Let $r = (r_1,r_2,r_3)$ denote the sensitivity parameters. Suppose we are interested in the trade off between $r_1$ and $r_2$. Fix the value of $r_3$ and compute the breakdown frontier for the conclusion that $\theta \in \mathcal{C}$ as
\[
	\text{BF}(r_2 \mid r_3) = \sup \{ r_1 \in [0,1] : \Theta_I(r_1,r_2,r_3) \subseteq \mathcal{C} \}.
\]
Holding $r_3$ fixed, this is a function from $[0,1]$ to $[0,1]$. Suppose that for each $r_3 \in [0,1]$, it is differentiable almost everywhere with respect to the Lebesgue measure. Then we can compute its average derivative
\[
	\int_0^1 \text{BF}'(r_2 \mid r_3) \omega_2(r_2) \; dr_2
\]
where $\omega_2(\cdot)$ is some fixed weight function. We could present this average derivative as a function of $r_3$, or we could also average over $r_3$. This approach allows us to summarize the average rate of substitution between assumptions 1 and 2. If the function $\text{BF}(\cdot \mid r_3)$ is not differentiable, we could instead use secant functions to summarize the trade-offs.

\end{enumerate}
Here we have just briefly discussed each method. We leave a full analysis and implementation of these to future work.

\section{Monte Carlo simulations}\label{sec:MonteCarlo}

In this section we study the finite sample performance of our estimation and inference procedures proposed in section 3. We consider the following dgp. For $x =0,1$, $Y \mid X = x$ has a truncated normal distribution, with density
\[
	f_{Y \mid X}(y \mid x) = \frac{1}{\gamma x + 1} \phi_{[-4,4]} \left( \frac{y - \pi x}{\gamma x + 1} \right),
\]
where $\phi_{[-4,4]}$ is the truncated standard normal density. We let $\gamma = 0.1$ and $\pi = 1$. We set $\Prob(X=1) = 0.5$. This dgp implies a joint distribution of $(Y,X)$, which we draw independently from. Figure 2a in the introduction shows population breakdown frontiers for this dgp.

We consider two sample sizes, $N=500$ and $N=2000$. For each sample size we generate $S = 500$ simulated datasets. In each dataset we compute the estimated breakdown frontier and a 95\% lower bootstrap uniform confidence band, as discussed in section 3. We use $B=1000$ bootstrap draws. We consider the same five values of $\underline{p}$ used in the introduction: $0.1$, $0.25$, $0.5$, $0.75$, and $0.9$.

\begin{figure}[t]
\centering
\includegraphics[width=75mm]{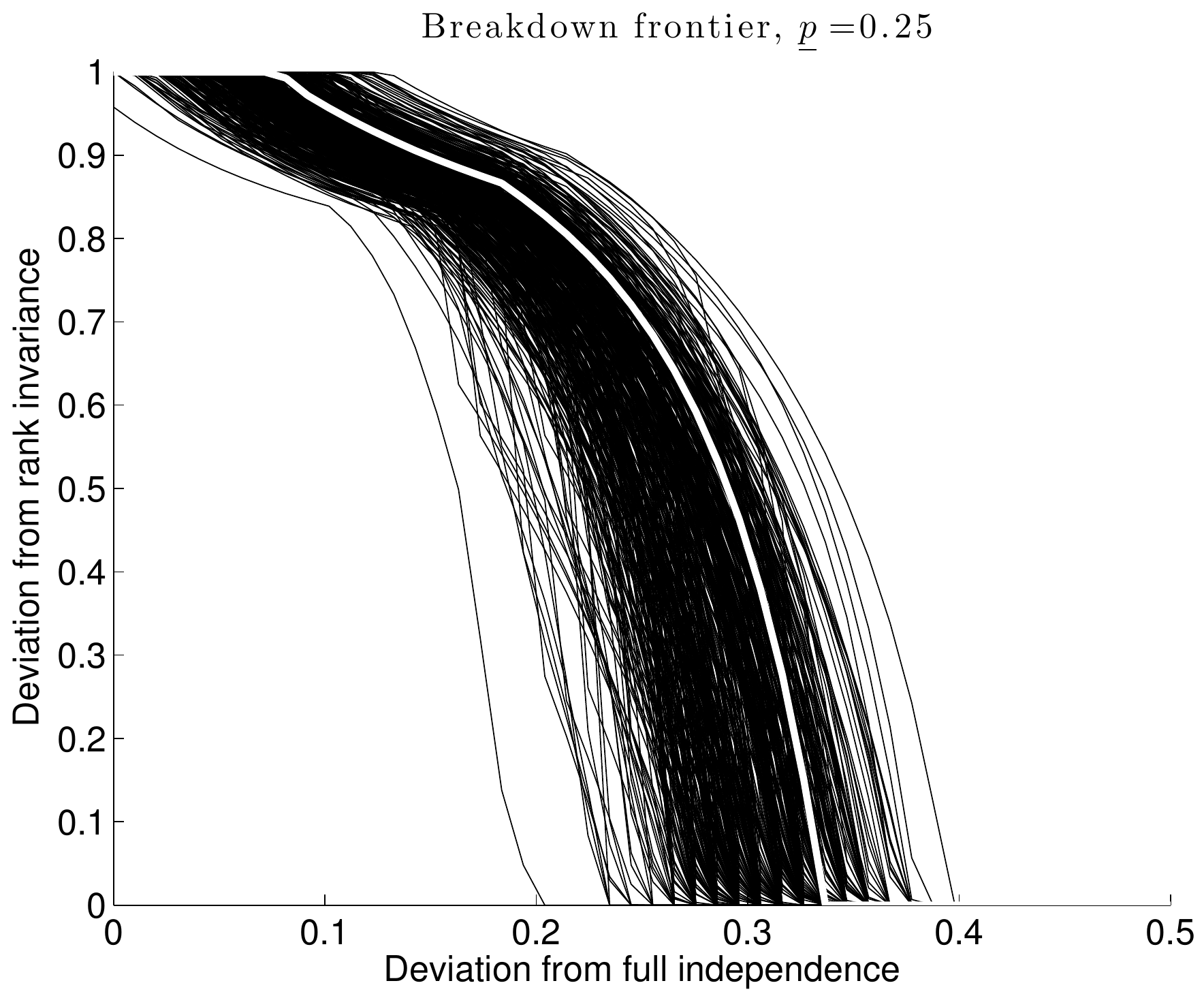}
\hspace{2mm}
\includegraphics[width=75mm]{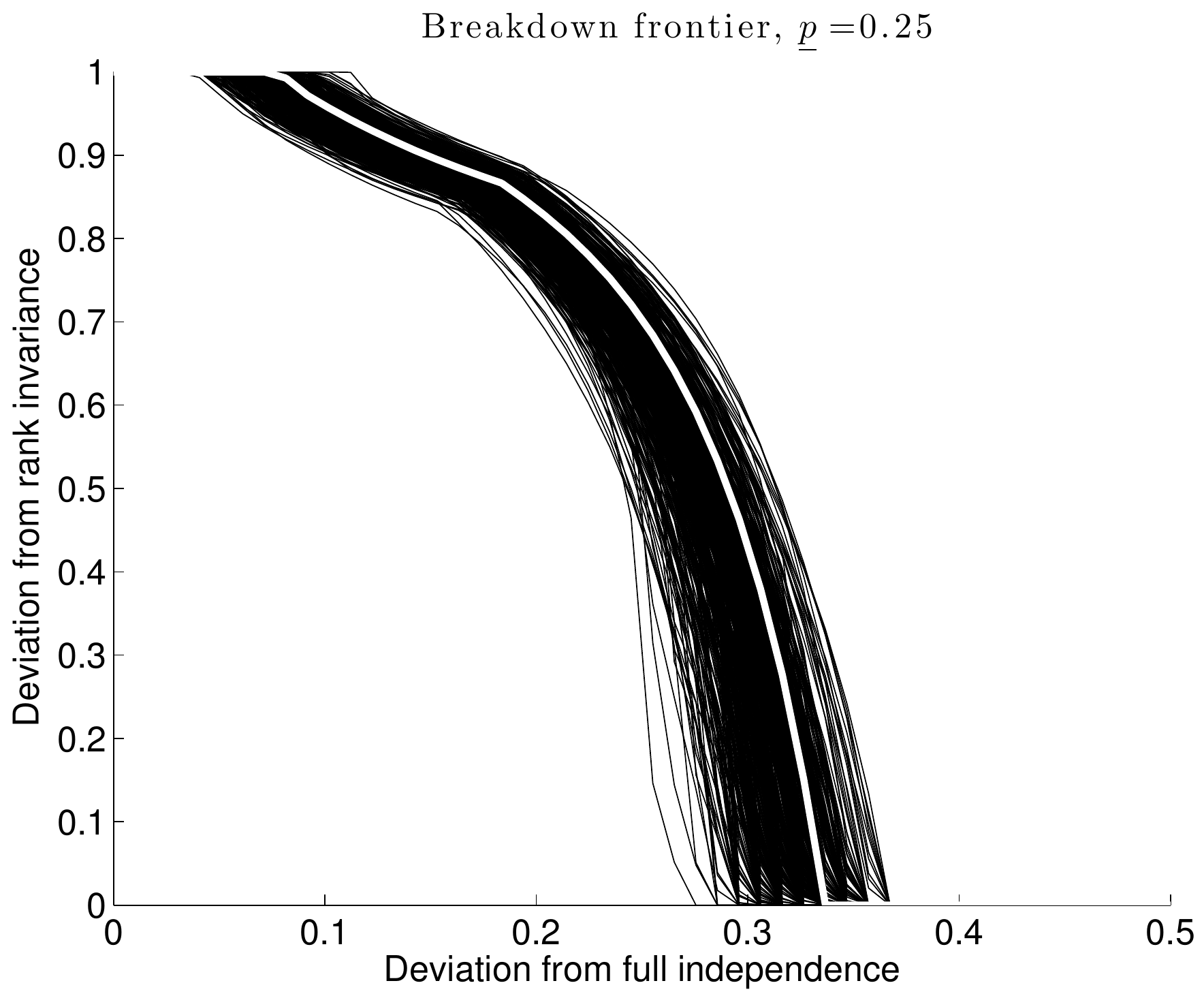}
\caption{Left: $N=500$. Right: $N=2000$. These plots show the sampling distribution of our breakdown frontier estimator by gathering the point estimates of the breakdown frontier across all Monte Carlo simulations into one plot. The true breakdown frontier is shown on top in white.}
\label{MC_BFallEstimates_plot}
\end{figure}

\begin{figure}[H]
\centering
\includegraphics[width=30mm]{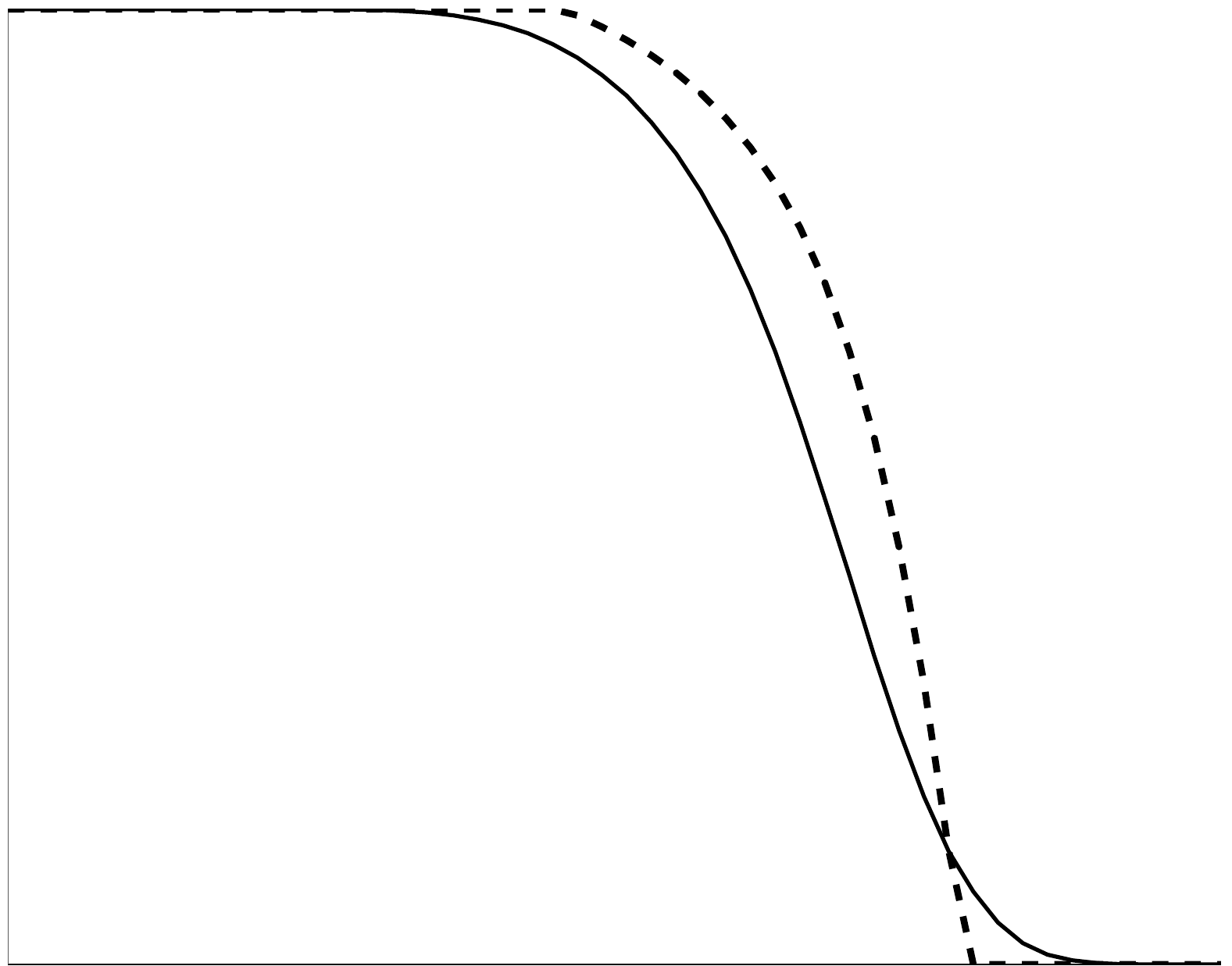}
\includegraphics[width=30mm]{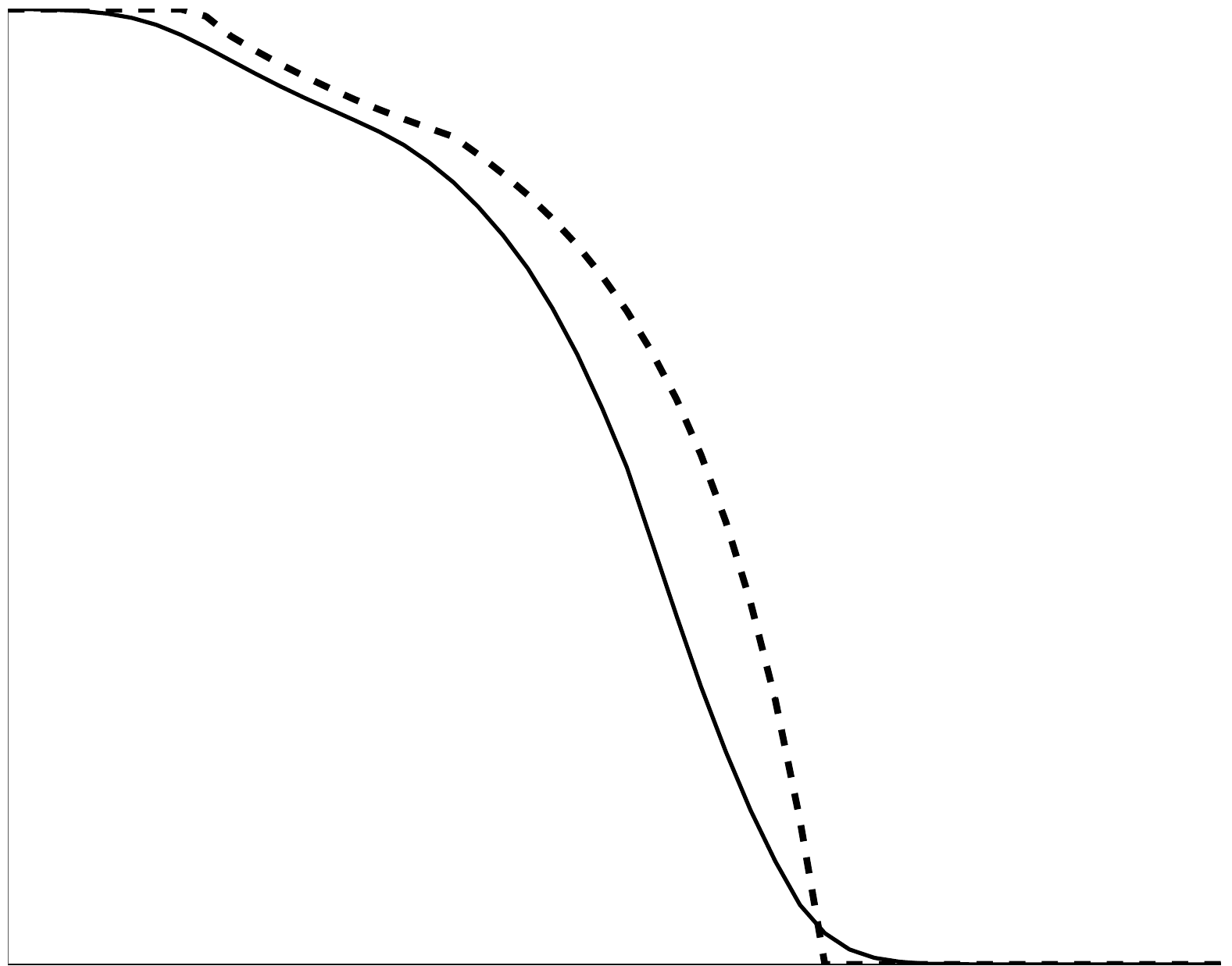}
\includegraphics[width=30mm]{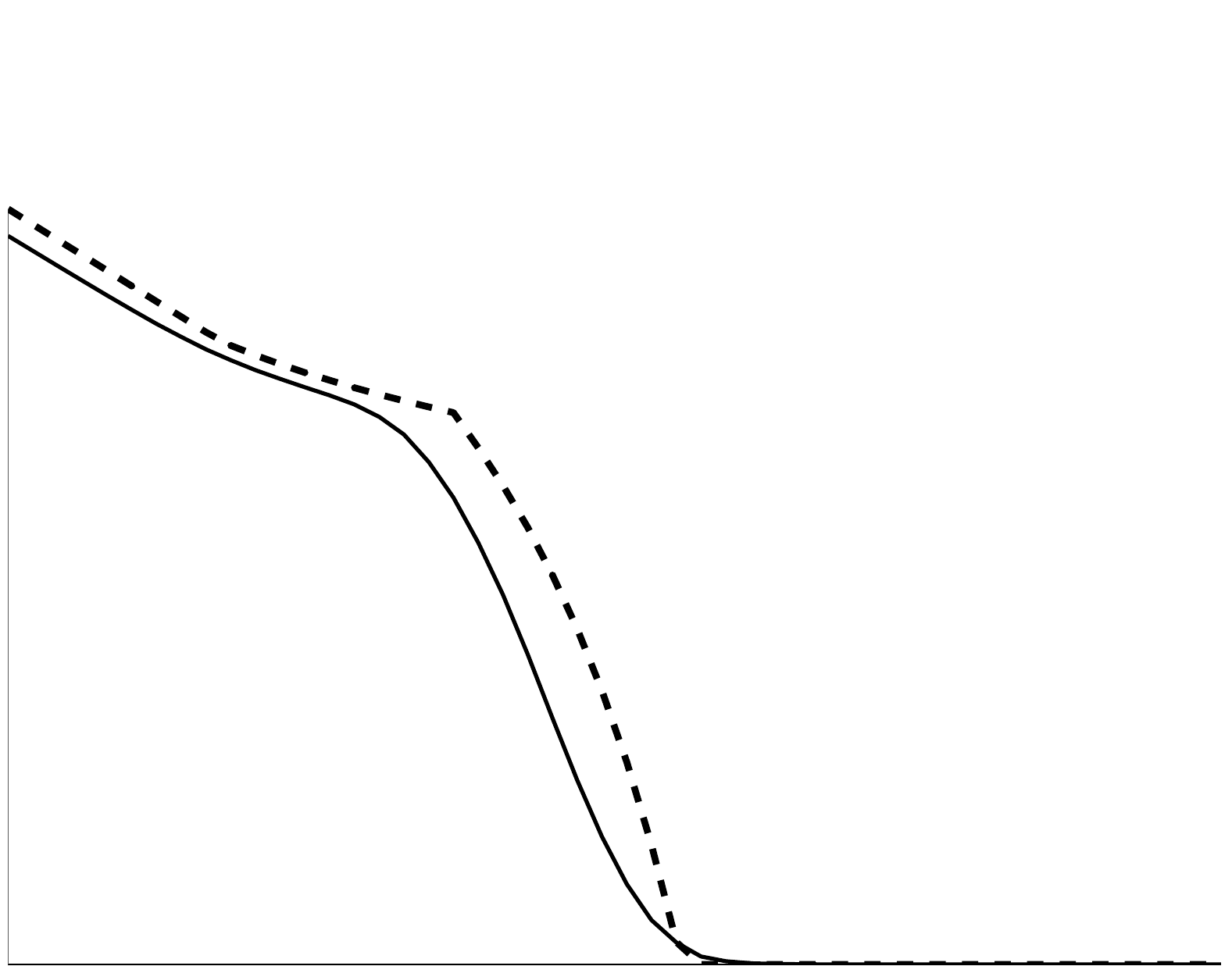}
\includegraphics[width=30mm]{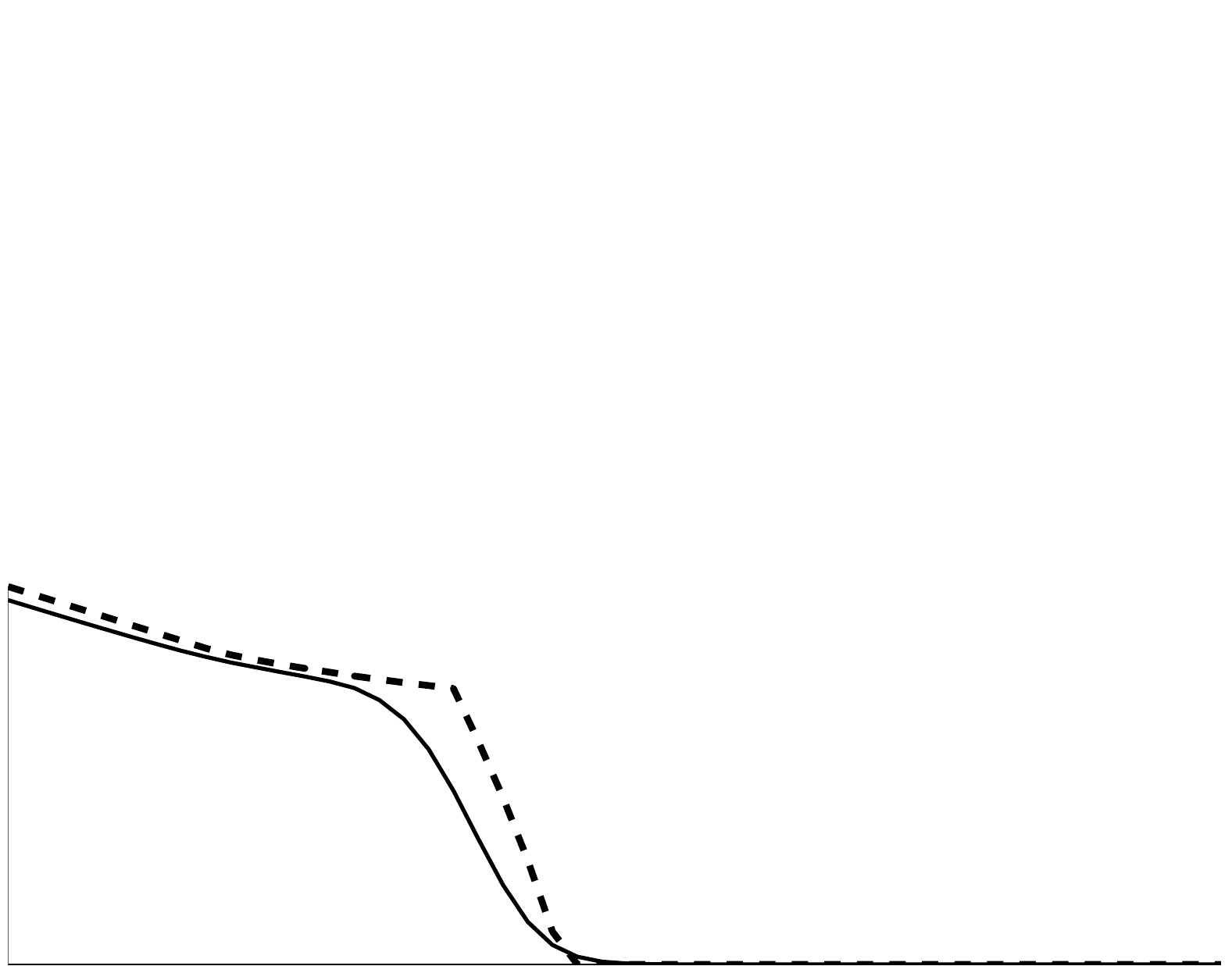}
\includegraphics[width=30mm]{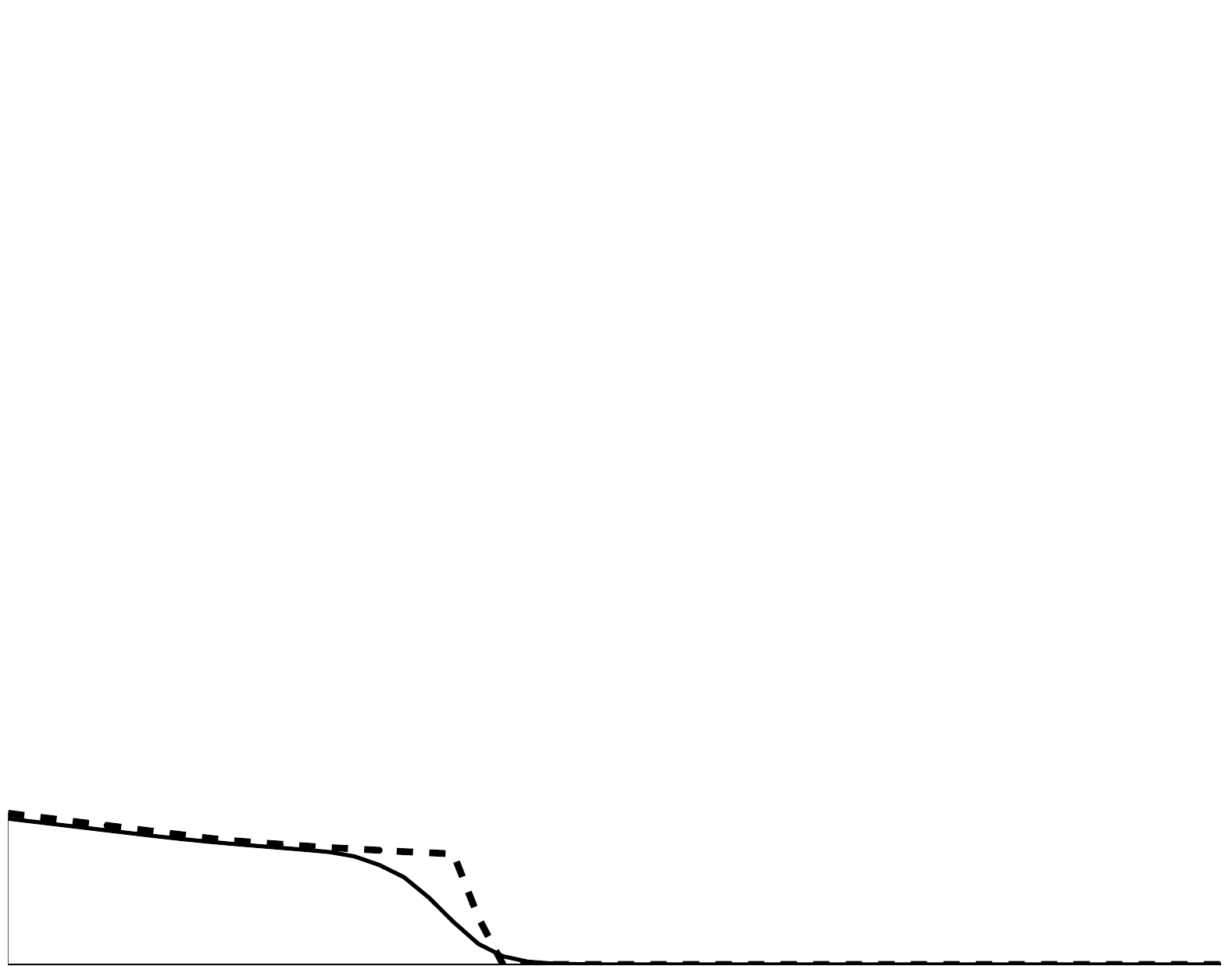} \\
\vspace{2mm}
\includegraphics[width=30mm]{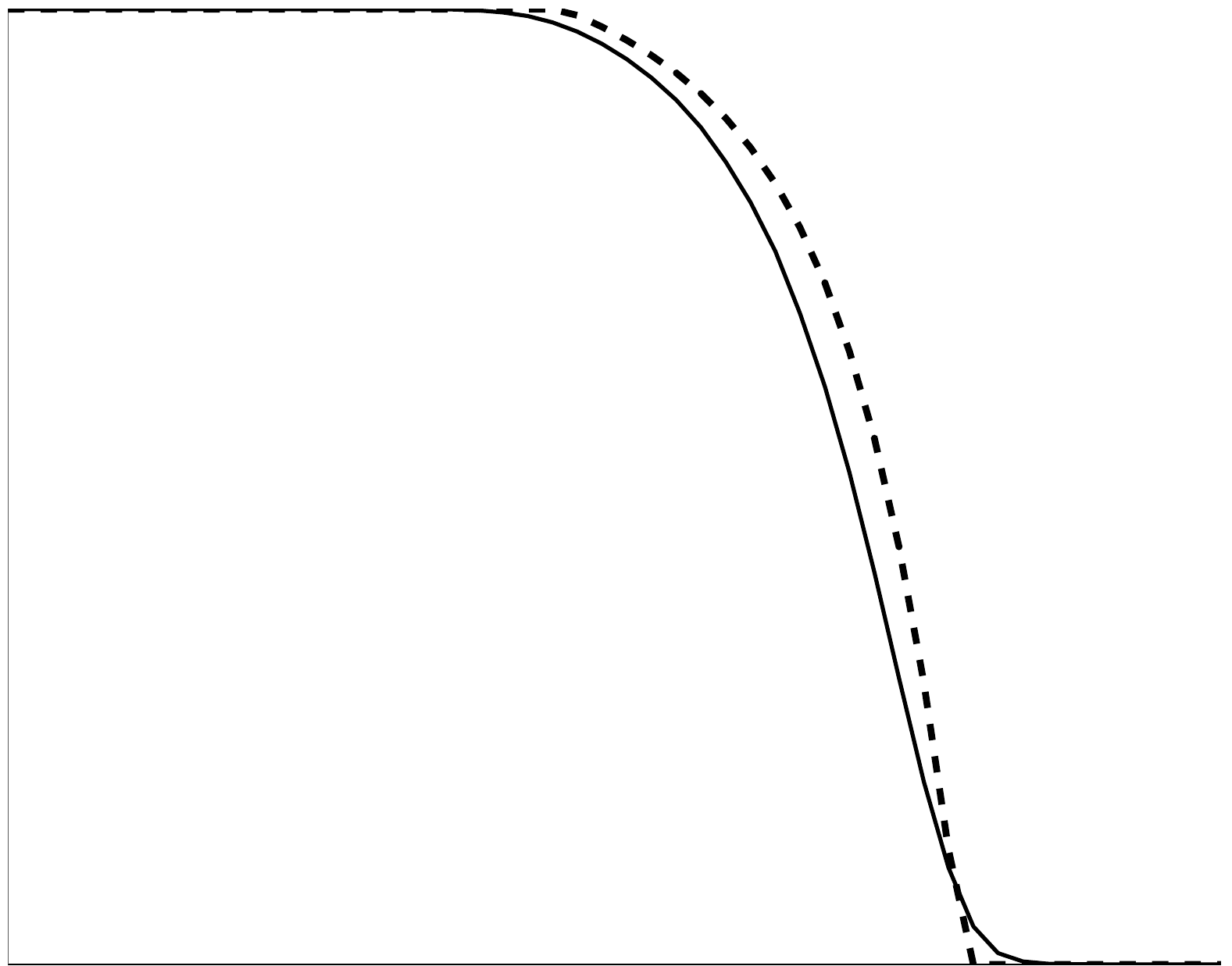}
\includegraphics[width=30mm]{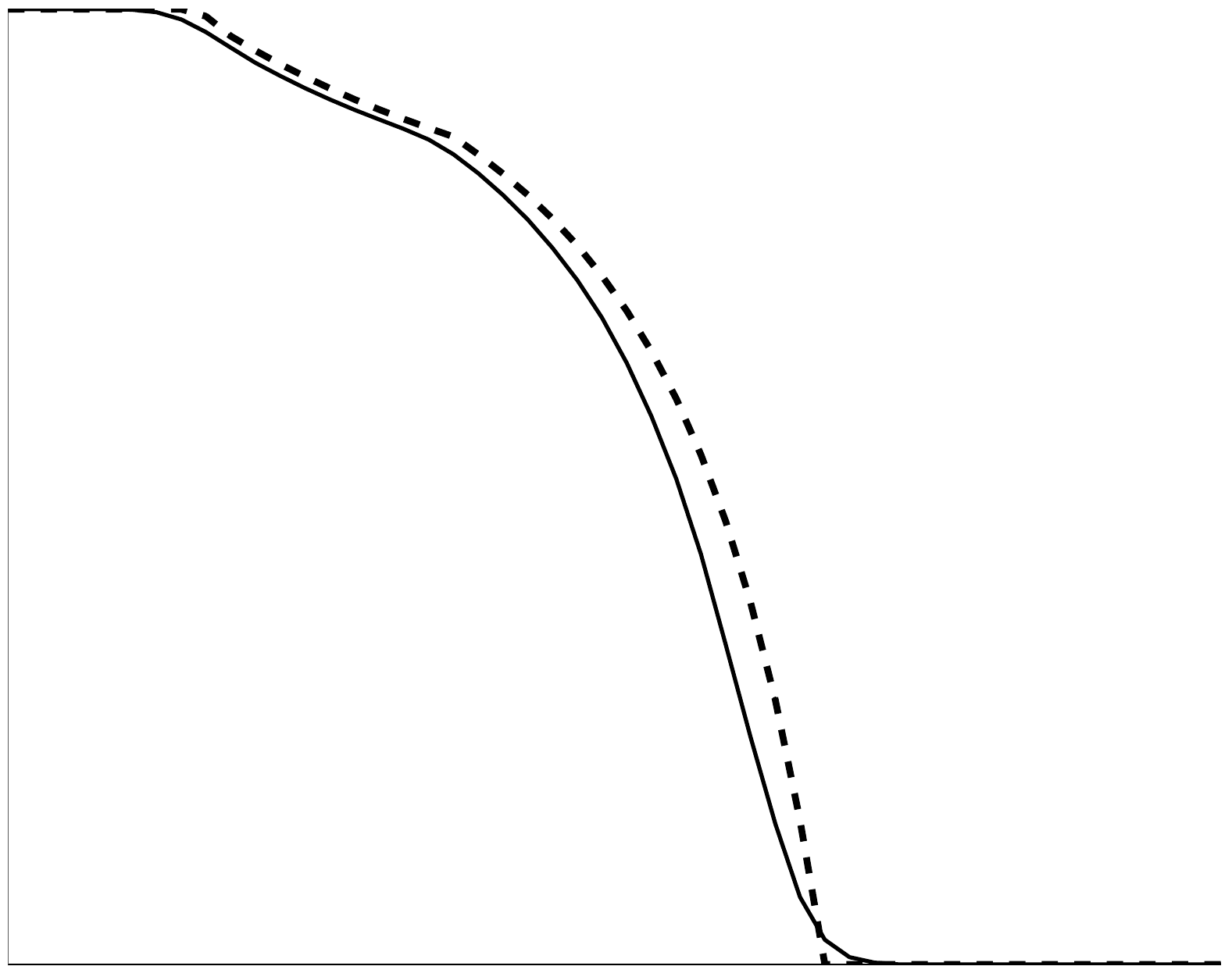}
\includegraphics[width=30mm]{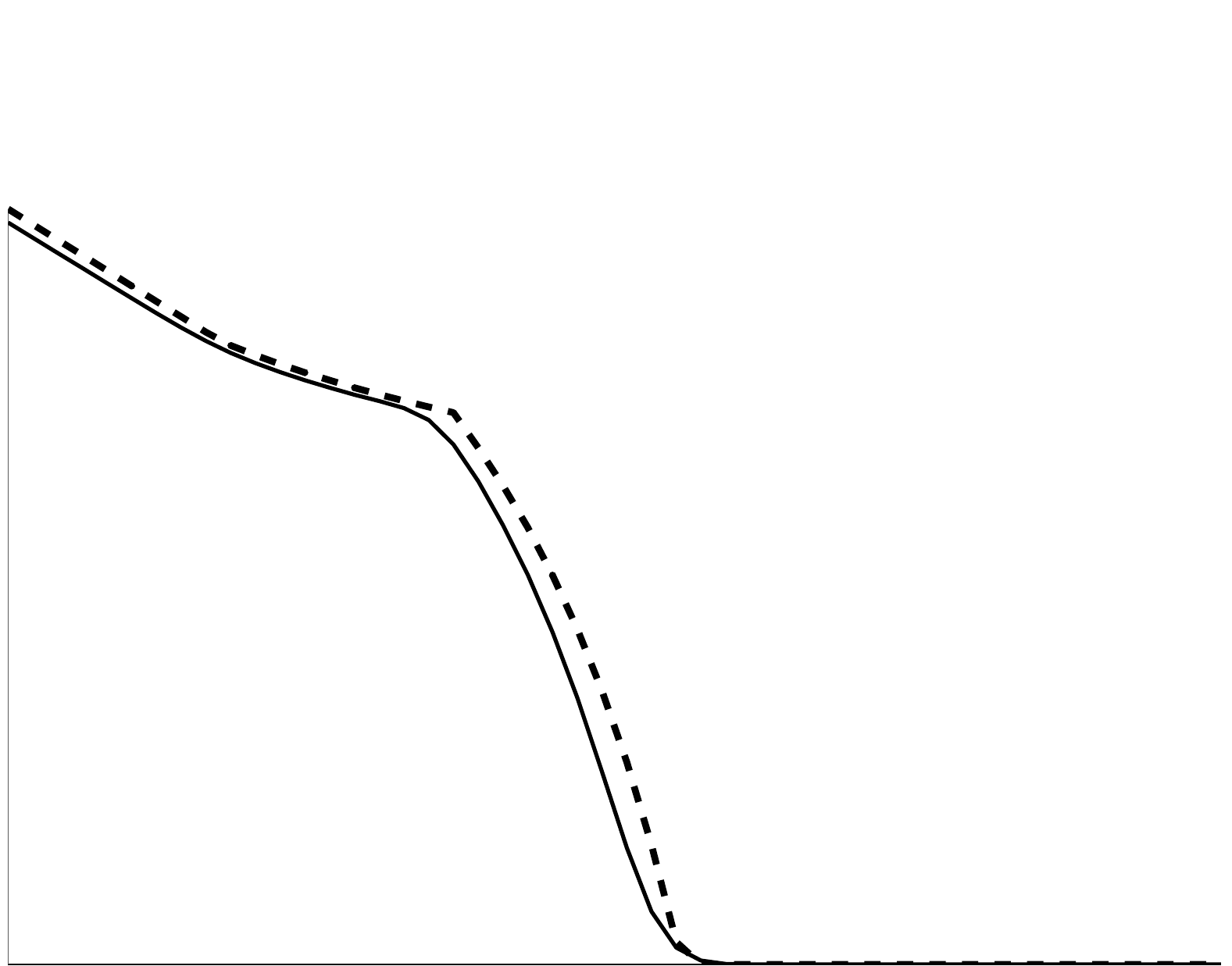}
\includegraphics[width=30mm]{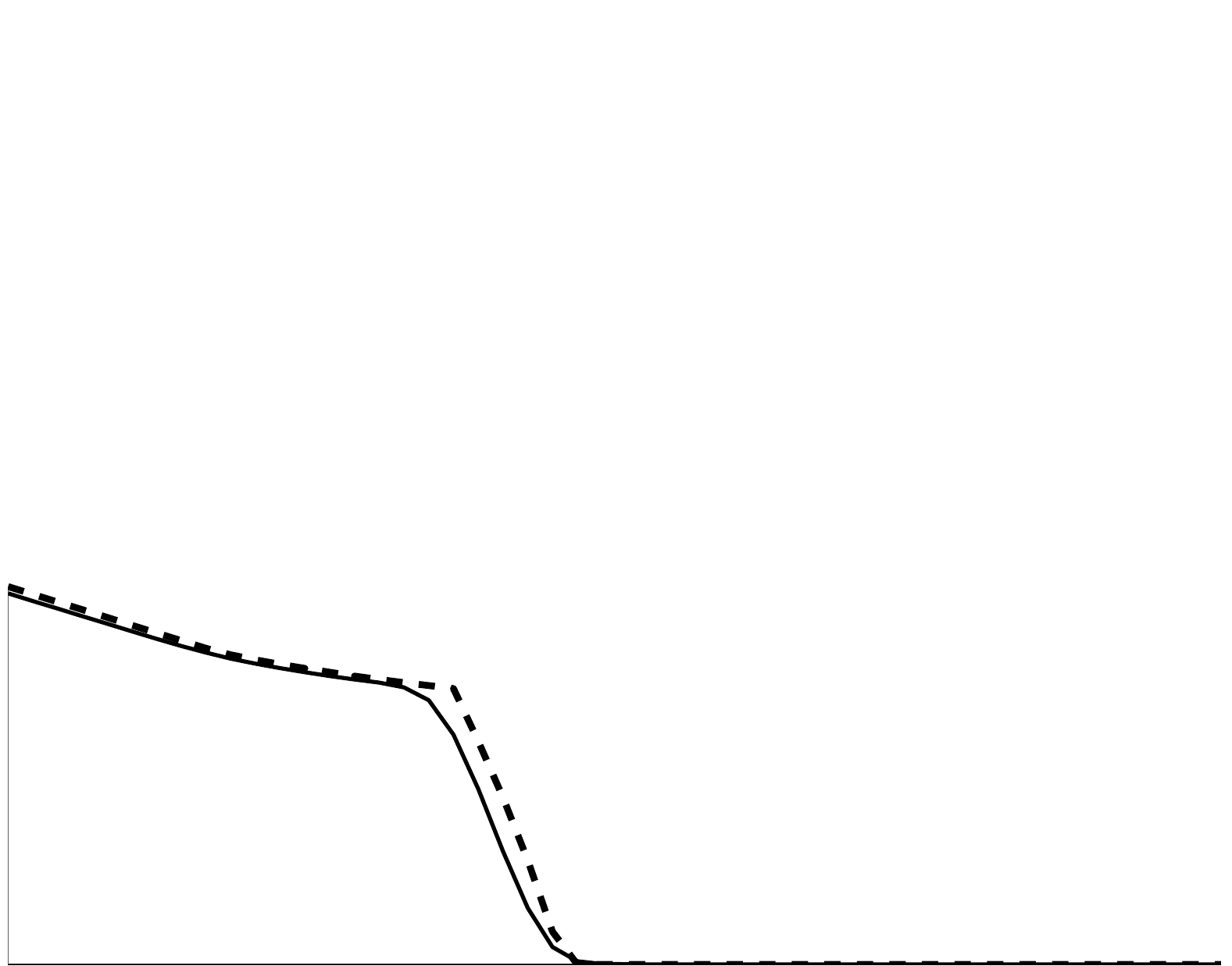}
\includegraphics[width=30mm]{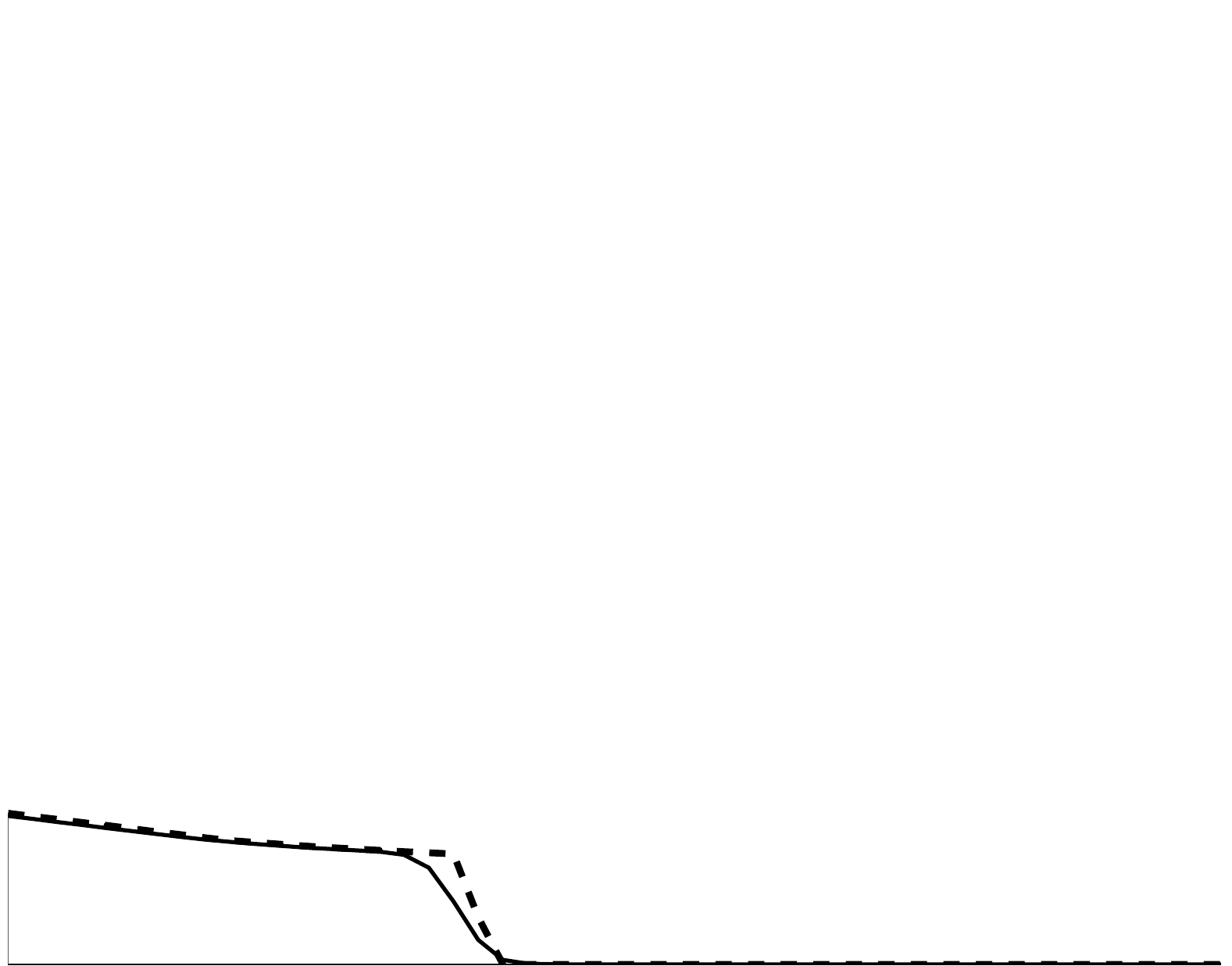} \\
\caption{Rows are sample sizes (top is $N=500$, bottom is $N=2000$). Columns are five values of $\underline{p}$ (from left to right: $\underline{p} = 0.1$, $0.25$, $0.5$, $0.75$, and $0.9$). Dotted lines are the true breakdown frontiers. The solid lines are the Monte Carlo estimates of $\Exp[ \widehat{\text{BF}}(c) ]$. This plot shows the finite sample bias of our breakdown frontier estimator.}
\label{MC_bias_plot}
\end{figure}

First we consider the performance of our point estimator of the breakdown frontier. Figure \ref{MC_BFallEstimates_plot} shows the sampling distribution of our breakdown frontier estimator. We show only $\underline{p} = 0.25$, but the other values of $\underline{p}$ yield similar figures. For this $\underline{p}$, we gather all point estimates of the breakdown frontier in the same plot. These plots show several features. First, as predicted by consistency, the sampling distribution becomes tighter around the truth as sample size increases. Second, the sampling distribution is not symmetric around the true frontier---it generally appears biased downwards. This is confirmed in figure \ref{MC_bias_plot} which plots the estimated finite sample mean function, $\widehat{\Exp}[ \widehat{\text{BF}}(c) ]$. This mean is estimated as the sample mean across all of our Monte Carlo datasets; that is, across all estimates shown in figure \ref{MC_BFallEstimates_plot}. The figure also shows the true breakdown frontier as a dotted line. In general the truth lies above the mean function. Again by consistency, this finite sample bias converges to zero as sample size increases, which we see when comparing the top row to the bottom row.

\begin{figure}[t]
\centering
\includegraphics[width=75mm]{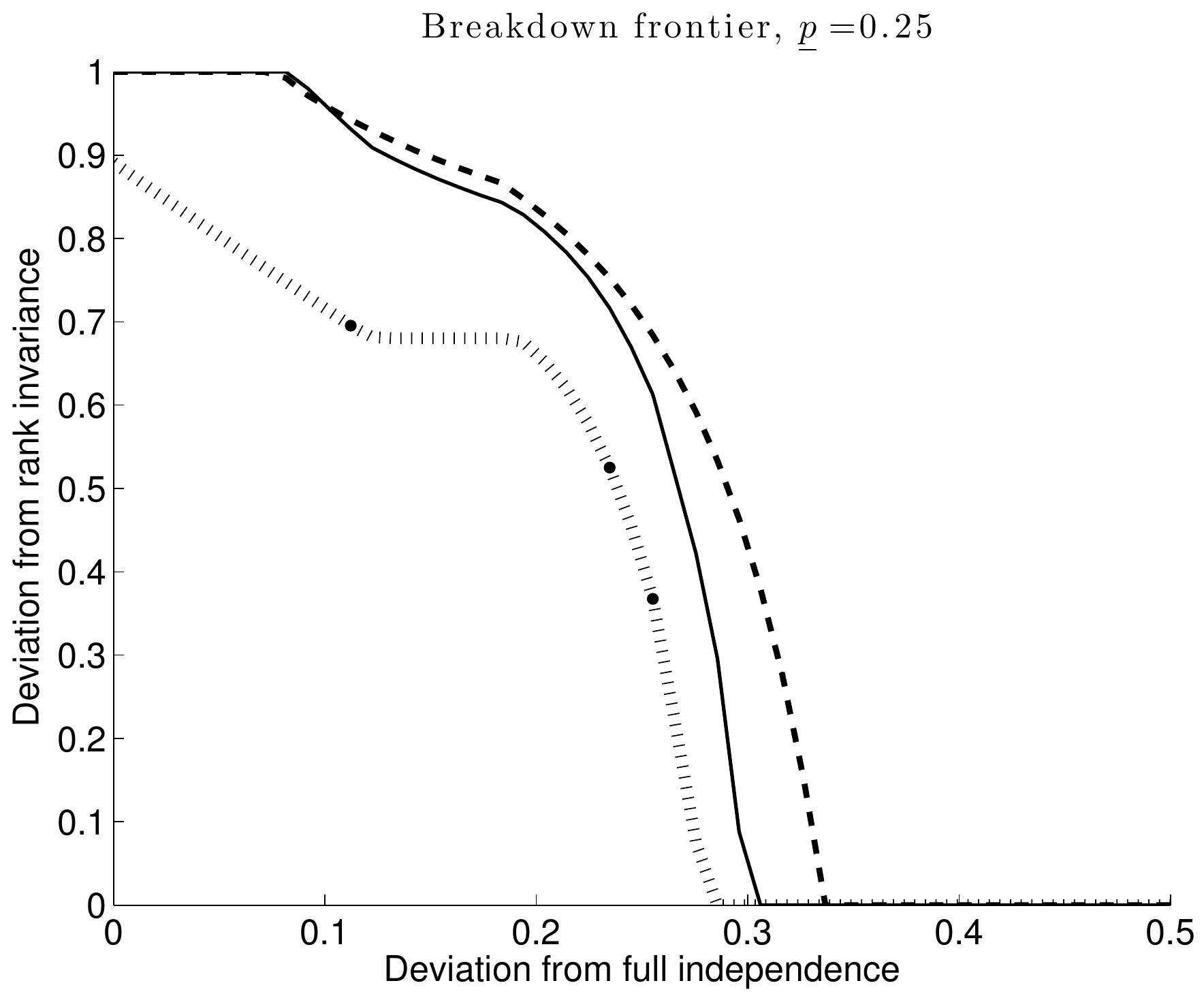}
\caption{$N=500$. Example 95\% lower uniform confidence band (dotted line), estimated breakdown frontier (solid line), true breakdown frontier (dashed line).}
\label{MCsection_exampleConfidenceBand}
\end{figure}

\begin{table}[htb]
\centering\caption{Coverage Probabilities \label{MCresultsTable_coverageProbs_dgp1}} 
\setlength{\linewidth}{.1cm} 
\newcommand{\contents}{ 
\begin{tabular}{c c c | c c c c c} \hline\hline 
& & & \multicolumn{5}{c}{$\underline{p}$} \\ 
$N$ & $\varepsilon_N$ & $\varepsilon_N / \varepsilon_N^{\text{naive}}$ & $    0.10$ & $    0.25$ & $    0.50$ & $    0.75$ & $    0.90$ \\ \hline 
$500$ &   0.0224 &     0.50 &    1.000 &    1.000 &    0.998 &    0.966 &    0.898 \\ 
  &   0.0447 &     1.00 &    0.986 &    0.992 &    0.990 &    0.928 &    0.892 \\ 
  &   0.0671 &     1.50 &    0.970 &    0.990 &    0.988 &    0.922 &    0.884 \\ 
  &   0.0894 &     2.00 &    0.956 &    0.990 &    0.990 &    0.936 &    0.884 \\ 
  &   0.1789 &     4.00 &    0.974 &    0.994 &    0.994 &    0.980 &    0.956 \\ 
  &   0.2683 &     6.00 &    0.998 &    1.000 &    1.000 &    1.000 &    1.000 \\ 
  &   0.3578 &     8.00 &    1.000 &    1.000 &    1.000 &    1.000 &    1.000 \\ 
  &   0.4472 &    10.00 &    1.000 &    1.000 &    1.000 &    1.000 &    1.000 \\ 
 \\ \hline
$2000$ &   0.0112 &     0.50 &    0.994 &    1.000 &    0.992 &    0.934 &    0.934 \\ 
  &   0.0224 &     1.00 &    0.986 &    0.992 &    0.990 &    0.934 &    0.918 \\ 
  &   0.0335 &     1.50 &    0.980 &    0.988 &    0.986 &    0.932 &    0.900 \\ 
  &   0.0447 &     2.00 &    0.980 &    0.976 &    0.982 &    0.930 &    0.882 \\ 
  &   0.0894 &     4.00 &    0.952 &    0.970 &    0.984 &    0.926 &    0.870 \\ 
  &   0.1342 &     6.00 &    0.960 &    0.982 &    0.986 &    0.942 &    0.906 \\ 
  &   0.1789 &     8.00 &    0.980 &    0.996 &    1.000 &    0.990 &    0.978 \\ 
  &   0.2236 &    10.00 &    0.994 &    1.000 &    1.000 &    1.000 &    1.000 \\ 
 \hline\hline 
\multicolumn{8}{p{0.95\linewidth}}{\footnotesize Nominal coverage is $1-\alpha =     0.95$. As discussed in the body text, the choice $\varepsilon_N^{\text{naive}} = 1/\sqrt{N}$ yields the naive bootstrap. Cell values show uniform-over-$c$ coverage probabilities of one-sided lower confidence bands, computed to maximize total area under the band.} \\ 
\end{tabular}
} 
\setbox0=\hbox{\contents} 
\setlength{\linewidth}{\wd0\tabcolsep0.6em} 
\contents 
\end{table}

Next we consider the performance of our confidence bands. Figure \ref{MCsection_exampleConfidenceBand} shows an example band along with the estimated frontier and the true frontier. To evaluate the performance of bands like this, we compute uniform coverage probabilities. We use 50 grid points for computing and evaluating uniform coverage of the confidence band. Table \ref{MCresultsTable_coverageProbs_dgp1} shows the results. Here we present a range of choices for $\varepsilon_N$. Since $\varepsilon_N^\text{naive} = 1/ \sqrt{N}$ yields the naive bootstrap, we use this choice as our baseline. We then consider seven other choices by rescaling the naive $\varepsilon_N$. Specifically, we consider $\varepsilon_N = K \varepsilon_N^\text{naive}$ for $K \in \{ 0.5, 1.5, 2, 4, 6, 8, 10 \}$. Recall that \cite{HongLi2015} impose the rate constraints that $\varepsilon_N \rightarrow 0$ and $\sqrt{N} \varepsilon_N \rightarrow \infty$. Hence asymptotically the ratio $\varepsilon_N / \varepsilon_N^\text{naive}$ must diverge.

First consider $N=500$. For $\underline{p} = 0.1$, $0.25$, and $0.75$, the choice of $\varepsilon_N$ which yields coverage probabilities closest to the nominal coverage of $0.95$ is twice the naive choice. This is also approximately true for $\underline{p} = 0.5$. For $\underline{p} = 0.9$, the next largest $\varepsilon_N$ has the coverage probability closest to the nominal coverage. Focusing on these choices of $\varepsilon_N$, the coverage probabilities are relatively close to the nominal for the `outside' columns $\underline{p} = 0.1$, $0.25$, and $0.9$. For the `inside' columns $\underline{p} = 0.25$ and $\underline{p} = 0.5$, we have substantial over-coverage. Indeed, for $\underline{p} = 0.1$, $0.25$, and $0.5$, all choices of $\varepsilon_N$'s considered lead to over-coverage. For the two larger values of $\underline{p}$, some values of $\varepsilon_N$ lead to under-coverage. Finally, with $\varepsilon_N$'s large enough, we obtain 100\% coverage for all $\underline{p}$'s.

Next consider $N=2000$. Here we obtain similar results. For $\underline{p} = 0.1$ and $0.25$, the choice of $\varepsilon_N$ which yields coverage probabilities closest to the nominal coverage of $0.95$ is four times the naive choice. This is also approximately true for $\underline{p} = 0.5$. For $\underline{p} = 0.75$, the next largest $\varepsilon_N$ is the best (six times the naive choice). For $\underline{p} = 0.9$, an even larger $\varepsilon_N$ is the best (eight times the naive, with the optimal scaling probably around seven). And for $\varepsilon_N$'s large enough, we obtain essentially 100\% coverage for all $\underline{p}$'s.

\begin{table}[htb]
\centering\caption{Proportional area under the confidence bands \label{MCresultsTable_areaRatios_dgp1}} 
\setlength{\linewidth}{.1cm} 
\newcommand{\contents}{ 
\begin{tabular}{c c c | c c c c c} \hline\hline 
& & & \multicolumn{5}{c}{$\underline{p}$} \\ 
$N$ & $\varepsilon_N$ & $\varepsilon_N / \varepsilon_N^{\text{naive}}$ & $    0.10$ & $    0.25$ & $    0.50$ & $    0.75$ & $    0.90$ \\ \hline 
$500$ &   0.0224 &     0.50 &    0.644 &    0.643 &    0.637 &    0.672 &    0.734 \\ 
  &   0.0447 &     1.00 &    0.759 &    0.716 &    0.705 &    0.740 &    0.774 \\ 
  &   0.0671 &     1.50 &    0.780 &    0.734 &    0.722 &    0.751 &    0.776 \\ 
  &   0.0894 &     2.00 &    0.779 &    0.730 &    0.722 &    0.746 &    0.763 \\ 
  &   0.1789 &     4.00 &    0.604 &    0.552 &    0.541 &    0.529 &    0.468 \\ 
  &   0.2683 &     6.00 &    0.252 &    0.174 &    0.117 &    0.069 &    0.024 \\ 
  &   0.3578 &     8.00 &    0.022 &    0.007 &    0.001 &    0.000 &    0.000 \\ 
  &   0.4472 &    10.00 &    0.000 &    0.000 &    0.000 &    0.000 &    0.000 \\ 
 \\ \hline
$2000$ &   0.0112 &     0.50 &    0.869 &    0.832 &    0.808 &    0.834 &    0.884 \\ 
  &   0.0224 &     1.00 &    0.894 &    0.865 &    0.841 &    0.862 &    0.896 \\ 
  &   0.0335 &     1.50 &    0.901 &    0.876 &    0.853 &    0.873 &    0.901 \\ 
  &   0.0447 &     2.00 &    0.904 &    0.882 &    0.859 &    0.879 &    0.902 \\ 
  &   0.0894 &     4.00 &    0.906 &    0.879 &    0.862 &    0.877 &    0.890 \\ 
  &   0.1342 &     6.00 &    0.875 &    0.840 &    0.829 &    0.837 &    0.833 \\ 
  &   0.1789 &     8.00 &    0.814 &    0.755 &    0.732 &    0.717 &    0.665 \\ 
  &   0.2236 &    10.00 &    0.704 &    0.615 &    0.563 &    0.499 &    0.387 \\ 
 \hline\hline 
\multicolumn{8}{p{0.95\linewidth}}{\footnotesize Nominal coverage is $1-\alpha =     0.95$. As discussed in the body text, the choice $\varepsilon_N^{\text{naive}} = 1/\sqrt{N}$ yields the naive bootstrap. Cell values show the average (across simulations) ratio of the area under the confidence band to the area under the estimated breakdown frontier.} \\ 
\end{tabular}
} 
\setbox0=\hbox{\contents} 
\setlength{\linewidth}{\wd0\tabcolsep0.6em} 
\contents 
\end{table}

Before we interpret these results, we discuss one more table, table \ref{MCresultsTable_areaRatios_dgp1}. While table \ref{MCresultsTable_coverageProbs_dgp1} showed coverage probabilities, table \ref{MCresultsTable_areaRatios_dgp1} gives us an idea of the power of our confidence bands. For each simulation, we compute the ratio of the area under the confidence band to the area under the estimated breakdown frontier. By definition our confidence bands are all below the estimated breakdown frontier and hence this ratio can never be larger than one. Although we do not perform a formal analysis of power, this ratio gives us an idea of the main trade-off in obtaining our confidence bands: We want them to be as large as possible subject to the constraint that they have correct coverage. This is how we defined our band in section 3, for a fixed $\varepsilon_N$. Here we compare the properties of these bands across different $\varepsilon_N$'s. First consider $N=500$ and $\underline{p} = 0.1$. From table \ref{MCresultsTable_coverageProbs_dgp1}, twice the naive choice of $\varepsilon_N$ yields the closest to nominal coverage. All other choices gave over-coverage. We see this in table \ref{MCresultsTable_areaRatios_dgp1} since twice the naive choice gives essentially the largest area---all but one other choice have smaller area. Similarly, for $\underline{p} = 0.9$, the best choice based on table \ref{MCresultsTable_coverageProbs_dgp1} is four times the naive choice, which gives an area under the confidence band of 47\% that of the area under the estimated breakdown frontier. Smaller $\varepsilon_N$'s give larger areas, but under-cover. Larger $\varepsilon_N$'s give smaller areas, but over-cover. For large enough $\varepsilon_N$, the confidence bands get close to zero everywhere, and hence have very small area and 100\% coverage. The results for $N=2000$ are similar.

In table \ref{MCresultsTable_coverageProbs_dgp1} we saw that most combinations of $\underline{p}$ and $\varepsilon_N$ led to over-coverage. This is caused by a downward bias in our estimated breakdown frontiers, as shown in figure \ref{MC_bias_plot}. Since we are constructing lower confidence bands, this downward bias causes our confidence bands to over-cover. Although this finite-sample bias disappears asymptotically, one may wish to do a finite-sample bias correction to obtain higher-order refinements. \cite{FanPark2009} previously studied this specific bias problem, in the case with random assignment (our $c=0$) and no assumptions on rank invariance (our $t=1$). They propose analytical and bootstrap bias corrected estimators of the bounds. \cite{ChernozhukovLeeRosen2013} study a related problem. 
Constructing such bias corrected estimators of nondifferentiable functionals, however, is delicate due to the results of \citet[theorem 1]{DossSethuraman1989} and \citet[theorem 2(a)]{HiranoPorter2012}. They show that achieving asymptotic unbiasedness for estimators of nondifferentiable functionals generally requires the variance to diverge. This result motivates consideration of alternative criteria, like the half-median unbiasedness property used by \cite{ChernozhukovLeeRosen2013}. We leave a full analysis of such corrections to future work.

\section{Inference via population smoothing}\label{sec:SmoothingAppendix}

In this section we develop an alternative approach to constructing lower uniform confidence bands for the breakdown frontier. As discussed in section 3, the population breakdown frontier $\text{BF}(\cdot,\underline{p})$ evaluated on a finite grid of $c$'s is a Hadamard directionally differentiable functional of the underlying parameters $\theta_0 = (F_{Y \mid X}(\cdot \mid \cdot),p_{(\cdot)})$, but it is not necessarily ordinary Hadamard differentiable. We therefore applied the work of \cite{Dumbgen1993}, \cite{FangSantos2014}, and \cite{HongLi2015} to do inference. In this section, we instead replace $\text{BF}(\cdot,\underline{p})$ by a smoother lower envelope function. We then construct uniform lower confidence bands for this smoothed breakdown frontier, which are asymptotically valid---but potentially conservative---for the original breakdown frontier. We compare and contrast these two approaches to inference at the end of this section. For simplicity, we omit covariates throughout this section.

Specifically, recall from section 2 that 
\[
	\text{BF}(c,\underline{p})
	= 
	\min\{ \max \{ \text{bf}(c,\underline{p}), 0 \}, 1 \}
\]
where
\[
	\text{bf}(c,\underline{p})
	= 
	\frac{
		1 - \underline{p} 
		- \Prob(\underline{Q}^c_{Y_1}(U) - \overline{Q}^c_{Y_0}(U) \leq 0)
	}{
	1 + \min \left\{ 
	\inf_{y \in \mathcal{Y}_0} 
	(\overline{F}^c_{Y_1}(y) - \underline{F}^c_{Y_0}(y-0)), 0 
	\right\} 
	- \Prob(\underline{Q}^c_{Y_1}(U) - \overline{Q}^c_{Y_0}(U) \leq 0)
	}.
	\tag{(9)}
\]
For simplicity, we fix $\underline{p} \in [0,1]$ throughout this section. We use $\kappa$ throughout to denote a scalar or vector of smoothing parameters. We replace $\text{BF}(\cdot,\underline{p})$ by a \emph{smooth lower approximation} $\text{SBF}_\kappa(\cdot,\underline{p})$, defined as follows.

\begin{definition}
Let $(\Theta, \| \cdot \|_\Theta)$ and $(\mathscr{G},\| \cdot \|_\mathscr{G})$ be Banach spaces. Let $\leq$ be a partial order on $\mathscr{G}$. Let $f : \Theta \rightarrow \mathscr{G}$ be a function. Consider a function $f_\kappa : \Theta \rightarrow \mathscr{G}$, where $\kappa \in \R^{\dim(\kappa)}_+$ is a vector of bandwidths. We say $f_\kappa$ is a \emph{smooth lower approximation} of $f$ if it satisfies the following:
\begin{enumerate}
\item (Lower envelope) $f_\kappa(\theta) \leq f(\theta)$ for all $\theta \in \Theta$ and $\kappa \in \R_+^{\dim(\kappa)}$.

\item (Approximation) For each $\theta \in \Theta$, $f_\kappa(\theta) \rightarrow f(\theta)$ as all components of $\kappa$ converge to infinity. Here we take $\rightarrow$ to mean pointwise convergence.

\item (Smoothness) $f_\kappa$ is Hadamard differentiable, possibly only tangentially to a specified set.
\end{enumerate}
Define smooth \emph{upper} approximations analogously.
\end{definition}

Throughout this section we let $\leq$ denote the component-wise order when applied to functions with Euclidean codomain. Recall our notation $\theta_0 = (F_{Y \mid X}(\cdot \mid \cdot),p_{(\cdot)})$, $\widehat{\theta} = (\widehat{F}_{Y \mid X}(\cdot \mid \cdot),\widehat{p}_{(\cdot)})$, and $\Z_1$ as the limiting distribution of $\sqrt{N}(\widehat{\theta} - \theta_0)$. Below we show how to construct a functional $\psi : \ell^\infty(\R \times \{0,1\}) \times \ell^\infty(\{0,1\}) \rightarrow \ell^\infty([0,\overline{C}])$ which maps $\theta_0$ into $\text{SBF}_\kappa(\cdot,\underline{p})$ such that this functional is a smooth lower approximation of the functional mapping $\theta_0$ to $\text{BF}(\cdot,\underline{p})$. Given such a functional, we estimate the smoothed breakdown frontier by sample analog:
\[
	\widehat{\text{SBF}}_\kappa(c,\underline{p}) = [ \psi(\widehat{\theta}) ](c).
\]
We then construct uniform confidence bands for the breakdown frontier as follows. As in section 3, consider bands of the form
\[
	\widehat{\text{LB}}(c) = \widehat{\text{SBF}}_\kappa(c,\underline{p}) - \widehat{k}(c)
\]
for some function $\widehat{k}(\cdot) \geq 0$. We specifically focus on $\widehat{k}(c) = \widehat{z}_{1-\alpha} \sigma(c)$ for a scalar $\widehat{z}_{1-\alpha}$ and a known function $\sigma$, for simplicity. We now immediately obtain the following result.

\begin{proposition}\label{prop:smoothingInference}
Suppose A1, A3, and A5 hold. Let $\psi$ denote the functional described above, a smooth lower approximation to the breakdown frontier functional. Let $\widehat{\theta}^*$ denote a draw from the nonparametric bootstrap distribution of $\widehat{\theta}$. Then
\begin{align}\label{eq:bootstrap_conv_SBF}
	\sqrt{N} ( \psi(\widehat{\theta}^*) - \psi(\widehat{\theta}) ) \overset{P}{\rightsquigarrow} \psi_{\theta_0}'(\Z_1)
\end{align}
where $\psi_{\theta_0}'$ denotes the Hadamard derivative of $\psi$ at $\theta_0$. For a given function $\sigma(\cdot)$ such that $\inf_{c\in[0,\overline{C}]}\sigma(c)>0$, define
\begin{equation}\label{eq:empirical critical value_SBF}
	\widehat{z}_{1-\alpha}
	= 
	\inf\left\{ z \in \R : \Prob\left(\sup_{c\in[0,\overline{C}]}\frac{ 
	\sqrt{N}( [\psi(\widehat{\theta}^*)](c) - [\psi(\widehat{\theta})](c) )
	}{\sigma(c)} \leq z \mid Z^N\right)\geq 1-\alpha\right\}.
\end{equation}	 
Finally, suppose also that the cdf of
\[
	\sup_{c\in[0,\overline{C}]}\frac{[\psi_{\theta_0}'(\mathbf{Z}_1) ](c)}{\sigma(c)}
\]
is continuous and strictly increasing at its $1-\alpha$ quantile, denoted $z_{1-\alpha}$. Then $\widehat{z}_{1-\alpha} = z_{1-\alpha} + o_p(1)$.
\end{proposition}

\begin{corollary}\label{corr:SBFtoBF}
Suppose the assumptions of proposition \ref{prop:smoothingInference} hold. Let $\widehat{k}(c) = \widehat{z}_{1-\alpha} \sigma(c)$, where $\widehat{z}_{1-\alpha}$ is defined in equation \eqref{eq:empirical critical value_SBF}. Then
\[
	\lim_{N \rightarrow \infty} \Prob \left( \widehat{\text{SBF}}_\kappa(c,\underline{p}) - \widehat{k}(c) \leq \text{BF}(c,\underline{p}) \text{ for all $c \in [0,\overline{C}]$} \right) \geq 1-\alpha.
\]
\end{corollary}

Importantly, the level of smoothing $\kappa$ is \emph{fixed} asymptotically. This is analogous to the requirement that the grid of $c$'s must be fixed asymptotically in the approach discussed in section 3. It is also similar to a proposal by \cite{ChernozhukovFernandez-ValGalichon2010} in their corollary 4. They suggested replacing the non-smooth function with a smoothed version and doing inference on the smoothed version. Their approach delivers valid inference on the smoothed function, but not the original function. This follows since their smoothed function does not satisfy an envelope property. Our modification of their suggestion, however, delivers valid inference on both the smoothed and original functions.

All that remains is to construct such a function $\text{SBF}_\kappa(c,\underline{p})$. We consider each piece composing the function $\text{BF}(c,\underline{p})$ in turn. First consider $\overline{F}_{Y_1}^c(y)$. This bound is a minimum of two terms (see equation 7). In general, consider the minimum of a finite number of terms $x_1,\ldots,x_n$. There are many smooth approximations of this function. Here we just consider one:
\[
	\text{sm}_\kappa \{ x_1,\ldots,x_n \} = \sum_{i=1}^n x_i \frac{\exp(\kappa x_i) }{\sum_{j=1}^n \exp(\kappa x_j)}
\]
for $\kappa < 0$. This same function approximates $\max \{ x_1,\ldots,x_n \}$ for $\kappa > 0$. Let $\mathcal{D}$ be a subset of a Euclidean space. Let $\mathbb{D}_1$ denote the set of functions in $\ell^\infty(\mathcal{D})^n$ with range contained in some compact set $\mathcal{Y} \subseteq \R^n$. In our application, we are interested using the functional $\psi_{1,\kappa}: \mathbb{D}_1 \rightarrow \ell^\infty(\mathcal{D})$ defined by
\[
	[\psi_{1,\kappa}(f_1,\ldots,f_n)](y) = \text{sm}_\kappa \{ f_1(y),\ldots,f_n(y) \}
\]
to approximate the functionals $\psi_{1,\text{max}} : \mathbb{D}_1 \rightarrow \ell^\infty(\mathcal{D})$ defined by
\[
	[\psi_{1,\text{max}}(f_1,\ldots,f_n)](y) = \max \{ f_1(y),\ldots,f_n(y) \}
\]
and $\psi_{1,\text{min}} : \mathbb{D}_1 \rightarrow \ell^\infty(\mathcal{D})$ defined by
\[
	[\psi_{1,\text{min}}(f_1,\ldots,f_n)](y) = \min \{ f_1(y),\ldots,f_n(y) \}.
\]

\begin{lemma}\label{lemma:smoothMax}
Let $\kappa \in \R_+$.
\begin{enumerate}
\item $\psi_{1,\kappa}$ is a smooth lower approximation of $\psi_{1,\text{max}}$.

\item $\psi_{1,-\kappa}$ is a smooth upper approximation of $\psi_{1,\text{min}}$.
\end{enumerate}
\end{lemma}

Since $\overline{F}_{Y_1}^c(y)$ enters the denominator of equation (9), and since $\text{sm}_\kappa$ for $\kappa < 0$ is an upper envelope for the minimum, replacing the minimum in the definition of $\overline{F}_{Y_1}^c(y)$ with $\text{sm}_\kappa$ for $\kappa < 0$ decreases the value of equation (9). Similarly, replacing the maximum in the definition of $\underline{F}_{Y_0}^c(y)$ by $\text{sm}_\kappa$ for some $\kappa > 0$ decreases the value of equation (9).

Next consider the $\Prob( \underline{Q}_{Y_1}^c(U) - \overline{Q}_{Y_0}^c(U) \leq 0)$ term. As discussed in section 3, this term is the pre-rearrangement operator $\text{pr} : \ell^\infty((0,1) \times \mathcal{C}) \rightarrow \ell^\infty(\mathcal{C})$ defined by
\[
	[\text{pr}(f)](c) = \int_0^1 \indicator[ f(u,c) \leq 0 ] \; du
\]
evaluated at the difference of the quantile bounds, where $\mathcal{C} \subseteq [0,1]$. Define the smoothed pre-rearrangement operator $\text{spr}_\kappa : \ell^\infty((0,1) \times \mathcal{C}) \rightarrow \ell^\infty(\mathcal{C})$ by
\[
	[\text{spr}_\kappa(f)](c) = 1 - \int_0^1 \text{ss}_\kappa( f(u,c) ) \; du
\]
where $\text{ss}_\kappa$ is a smooth (upper or lower) approximation to the step function $\indicator(x \geq 0)$.

\begin{lemma}\label{lemma:smoothPrerearrangement}
Let $\text{ss}_\kappa : \R \rightarrow \R$ be a smooth upper (lower) approximation to the step function. Suppose further that $\text{ss}_\kappa$ approximates the step function in the $L_1$-norm and $\text{ss}_\kappa$'s derivative is uniformly continuous on its domain. Then $\text{spr}_\kappa$ is a smooth lower (upper) approximation to $\text{pr}$.
\end{lemma}

As with the maximum and minimum, there are many ways to construct smooth approximations to the step function $\indicator(x \geq 0)$. Here we mention just one:
\begin{equation}\label{eq:smoothStep}
	\text{ss}_\kappa^+(x) = S_1(\kappa x - 1)
	\qquad \text{and} \qquad
	\text{ss}_\kappa^-(x) = S_1(\kappa x)
\end{equation}
where
\[
	S_1(x) =
	\begin{cases}
		0 &\text{if $x \leq 0$} \\
		3x^2 - 2 x^3 &\text{if $x \in (0,1)$} \\
		1 &\text{if $x \geq 1$}.
	\end{cases}
\]

\begin{lemma}\label{lemma:smoothStep}
Let $\kappa \in \R_+$. Consider $\text{ss}_\kappa^+$ and $\text{ss}_\kappa^-$ defined in equation \eqref{eq:smoothStep}.
\begin{enumerate}
\item $\text{ss}_\kappa^+ : \R \rightarrow \R$ is a smooth upper approximation of $\indicator(x \geq 0)$. 

\item $\text{ss}_\kappa^- : \R \rightarrow \R$ is a smooth lower approximation of $\indicator(x \geq 0)$.
\end{enumerate}
Moreover, both $\text{ss}_\kappa^-$ and $\text{ss}_\kappa^+$ approximate the step function in the $L_1$-norm and have uniformly continuous derivative on $\R$.
\end{lemma}

We can now replace the pre-rearrangement operator in the numerator of equation (9) by $\text{spr}_\kappa(f)$ where the step function is approximated by $\text{ss}_\kappa^-$. Likewise we replace the pre-rearrangement operator in the denominator by $\text{spr}_\kappa(f)$ where the step function is approximated by $\text{ss}_\kappa^+$. Both of these changes decrease the value of equation (9).

Next consider the infimum piece in the denominator of equation (9). First notice that
\[
	\inf_{y \in \mathcal{Y}_0} 
	(\overline{F}^c_{Y_1}(y) - \underline{F}^c_{Y_0}(y-0))
	=
	1 + \inf_{y \in \mathcal{Y}_0} 
	(-1 + \overline{F}^c_{Y_1}(y) - \underline{F}^c_{Y_0}(y-0)).
\]
This ensures the argument of the infimum is nonpositive, a property we use below. As \cite{FangSantos2015workingPaper} note in their example 2.3 on page 10, if this infimum always has a unique optimizer, then the infimum operator is actually ordinary Hadamard differentiable. To avoid assuming that such a unique optimizer always exists, however, we will also replace the infimum by a smooth approximation. Specifically, let
\[
	\| f \|_p = \left( \int_{\mathcal{Y}_0} | f(y) |^p \; dy \right)^{1/p}
\]
denote the $L_p(\mathcal{Y}_0)$-norm. As $p \rightarrow \infty$, the $L_p$-norm converges to the sup-norm. We use this result to construct our smooth approximation to the infimum. Let $\underline{y} = \inf \mathcal{Y}_0$ and $\overline{y} = \sup \mathcal{Y}_0$, which are both finite since $\mathcal{Y}_0$ is compact. Let $\mathbb{D}_2$ denote the set of all nonpositive functions in $\ell^\infty(\R \times \mathcal{C})$ with $L_p$-norm bounded away from zero and range contained in $[-2,0]$. Let $p \geq 1$. Define $\psi_{2,p} : \mathbb{D}_2 \rightarrow \ell^\infty(\mathcal{C})$ by
\[
	[\psi_{2,p}(f)](c) = - \frac{1}{(\overline{y} - \underline{y})^{1/p}}\| - f(\cdot,c) \|_p.
\]
We scale the $L_p$-norm to ensure that we obtain a lower approximation to the supremum. The two minus signs then switch this to an upper approximation to the infimum.

\begin{lemma}\label{lemma:smoothInfimum}
$\psi_{2,p}$ is a smooth upper approximation to the infimum function, as $p \rightarrow \infty$.
\end{lemma}

Using this result, we replace
\[
	1 + \inf_{y \in \mathcal{Y}_0} 
	(-1 + \overline{F}^c_{Y_1}(y) - \underline{F}^c_{Y_0}(y-0))
\]
by its smooth upper approximation
\[
	1 + \left[\psi_{2,p} \left( -1 + \overline{F}^{(\cdot)}_{Y_1}(\cdot) - \underline{F}^{(\cdot)}_{Y_0}(\cdot-0) \right) \right](c)
\]
in equation (9), which decreases the value of the breakdown frontier. In this step we require the argument of $\psi_{2,p}$ to have nonzero $L_p$-norm. This assumption rules out extreme cases, such as when both $0 \in \mathcal{C}$ and $F_{Y \mid X}(\cdot \mid 0) = F_{Y \mid X}(\cdot \mid 1)$.

Finally, the maximum in the definition of $\text{BF}(c,\underline{p})$ can be replaced with $\text{sm}_\kappa$. For the minimum in this definition we want a smooth \emph{lower} approximation. Since $\min \{ 0, x \} = x [ 1 - \indicator(x \geq 0) ]$, one such approximation is $x [ 1 - \text{ss}_\kappa^+(x) ]$.

In section 3 we showed that the population breakdown frontier is a composition of Hadamard directionally differentiable functionals. In this section, we showed how to replace each functional in this composition by an ordinary Hadamard differentiable functional in such a way that the overall function is weakly smaller than the original breakdown frontier. Moreover, the difference between the original and smoothed frontiers can be made arbitrarily small by choosing appropriate values of the tuning parameters. Corollary \ref{corr:SBFtoBF} above shows how to use this construction to do valid inference on the original breakdown frontier.

We conclude by comparing our two approaches to inference: The first based on Hadamard directional differentiability and the second based on smoothing the population frontier. For any fixed finite grid of $c$ values, the first approach provides asymptotically exact inference while the second approach will always be possibly conservative. Visually, the first approach uses step functions to obtain a uniform band while the second approach produces a smoother appearing frontier. The first approach required assumption A6 while the second approach does not. The first approach requires choosing the $\varepsilon_N$ tuning parameter for the numerical delta method bootstrap, while the second approach does not since the ordinary bootstrap is valid. The second approach, however, requires choosing a large number of smoothing functions and bandwidths, unlike the first approach. Too much smoothing will lead to conservative inference, while too little smoothing will likely lead to poor finite sample performance. Overall, neither approach appears to strictly dominate.

\subsection*{Proofs for appendix \ref{sec:SmoothingAppendix}}

\begin{proof}[Proof of proposition \ref{prop:smoothingInference}]
Equation \eqref{eq:bootstrap_conv_SBF} follows by the functional delta method (e.g., theorem 3.1 of \citealt{FangSantos2014}), since $\psi$ is Hadamard differentiable. The rest of the proof follows as in the proof of proposition 2.
\end{proof}

\begin{proof}[Proof of corollary \ref{corr:SBFtoBF}]
This result follows immediately by the lower envelope property of the smoothed breakdown frontier.
\end{proof}

\begin{proof}[Proof of lemma \ref{lemma:smoothMax}]
We give the proof for part 1. Part 2 is analogous.
\begin{enumerate}

\item Let $x_1,\ldots,x_n \in \R$. $\text{sm}_\kappa \{ x_1,\ldots,x_n \}$ is a weighted average of $x_1,\ldots,x_n$ where the weights are in $(0,1)$. Hence it must always be weakly smaller than the maximum of $x_1,\ldots,x_n$. Thus $\text{sm}_\kappa \{ f_1(y), \ldots, f_n(y) \} \leq \max \{ f_1(y),\ldots, f_n(y) \}$ for any functions $f_1,\ldots,f_n$ and any $y \in \mathcal{D}$.

\item Let $x_1,\ldots,x_n \in \R$. Suppose $x_k = \max \{ x_1,\ldots,x_n \}$. Without loss of generality, suppose this maximum is unique. Multiplying and dividing by $\exp(-\kappa x_k)$ yields
\[
	\text{sm}_\kappa \{ x_1,\ldots,x_n \}
	= \sum_{i=1}^n x_i \frac{\exp(\kappa [ x_i - x_k])}{\sum_{j=1}^n \exp(\kappa [ x_j - x_k ])}.
\]
For all $i \neq k$, $x_i - x_k < 0$ hence $\exp(\kappa [ x_i  - x_k ]) \rightarrow 0$ as $\kappa \rightarrow \infty$. Thus the weights on all $i \neq k$ converge to zero while the weight on $x_k$ converges to one. Hence for any fixed $f_1,\ldots,f_n$ and $y \in \mathcal{D}$, $[\psi_{1,\kappa}(f_1,\ldots,f_n)](y)$ converges to $\psi_{1,\text{max}}(f_1,\ldots,f_n)](y)$.

\item For any fixed $\kappa$, the derivatives of the weights with respect to each $x_i$ are uniformly bounded. This follows by the functional form of the weights and compactness of $\mathcal{Y}$. Therefore $\psi_{1,\kappa}$ is Fr\'echet differentiable by lemma \ref{lemma:pluginDerivative} below. Finally, note that Fr\'echet differentiability implies Hadamard differentiability.

\end{enumerate}
\end{proof}

\begin{lemma}\label{lemma:pluginDerivative}
Let $g: \R^n \rightarrow \R$ be an everywhere differentiable function with uniformly continuous derivative on $\mathcal{Y} \subseteq \R^n$. Let $\mathcal{D}$ be a subset of a Euclidean space. Define the functional $\phi : \ell^\infty(\mathcal{D})^n \rightarrow \ell^\infty(\mathcal{D})$ by
\[
	[\phi(f_1,\ldots,f_n)](y) = g(f_1(y),\ldots,f_n(y)).
\]
Let $\mathbb{D}$ denote the set of functions in $\ell^\infty(\mathcal{D})^n$ with range contained in $\mathcal{Y}$. Let $f \in \mathbb{D}$. Then $\phi$ is Fr\'echet differentiable at $f$ with derivative
\[
	[\phi_{f}'(h)](y) = \sum_{i=1}^n [\nabla_i g](f_1(y),\ldots,f_n(y)) \cdot h_i(y)
\]
where $\nabla_i g$ denotes the $i$th partial derivative of $g$.
\end{lemma}

\begin{proof}[Proof of lemma \ref{lemma:pluginDerivative}]
We show the $n=1$ case, where
\[
	[\phi_{f}'(h)](y) = g'(f(y)) h(y).
\]
The $n \geq 2$ case is similar. We also suppose $\mathcal{Y} = \R$ for simplicity. We have
\begin{align*}
	\left| [\phi(f+h)](y) - \big( [ \phi(f)](y) + [\phi_{f}'(h)](y) \big) \right|
	&= \left| \big( g(f(y) + h(y)) - g(f(y)) \big) - g'(f(y)) h(y) \right| \\
	&= | g'(\bar{f}(y)) h(y) - g'(f(y)) h(y) | \\
	&= | g'(\bar{f}(y)) - g'(f(y)) | \cdot | h(y) |. 
\end{align*}
The second line follows by the mean value theorem, which says that there exists a $\bar{f}(y)$ such that
\[
	g(f(y) + h(y)) - g(f(y)) = g'(\bar{f}(y)) h(y)
\]
and $| \bar{f}(y) - f(y) | \leq | h(y) |$. We apply this argument for each $y \in \mathcal{D}$.

Next, fix $\varepsilon > 0$. By uniform continuity of $g'(\cdot)$, there is a $\delta > 0$ such that for all $\tilde{h} \in \R$ with $| \tilde{h} | < \delta$ and all $\tilde{f} \in \R$,
\[
	| g' (\tilde{f}+\tilde{h}) - g' (\tilde{f}) | < \varepsilon.
\]
Therefore,
\[
	| g'(\bar{f}(y)) - g'(f(y)) | \cdot | h(y) |
	\leq \varepsilon | h(y) |
\]
for all $\| h \|_\infty < \delta$. Hence
\[
	\sup_{y \in \mathcal{D}} \left| [\phi(f+h)](y) - \big( [ \phi(f)](y) + [\phi_{f}'(h)](y) \big) \right|
	\leq \varepsilon \sup_{y \in \mathcal{D}} | h(y) |
\]
for all $\| h \|_\infty < \delta$. That is,
\[
	\frac{\| \phi(f+h) - \big( \phi(f) + \phi_f'(h) \big) \|_\infty}{\| h \|_\infty} \leq \varepsilon
\]
for $\| h \|_\infty < \delta$. Since $\varepsilon$ was arbitrary, this shows that the left hand side is $o(1)$.
\end{proof}

\begin{proof}[Proof of lemma \ref{lemma:smoothPrerearrangement}]
We show that $\text{spr}_\kappa$ is a smooth lower approximation to $\text{pr}$ when $\text{ss}_\kappa$ is a smooth upper approximation to the step function. The second part is analogous. 
\begin{enumerate}

\item This follows immediately since $\text{ss}_\kappa$ is an upper approximation to the step function, which is then multiplied by a negative sign in the definition of $\text{spr}_\kappa$.

\item This follows by our assumption that $\text{ss}_\kappa$ approximates the step function in the $L_1$-norm:
\[
	\int_{\R} | \text{ss}_\kappa(u) - \indicator( u \geq 0) | \; du \rightarrow 0
\]
as $\kappa \rightarrow \infty$.

\item This follows immediately from lemma \ref{lemma:derivativeOfIntegralOperator} below.
\end{enumerate}
\end{proof}

\begin{lemma}\label{lemma:derivativeOfIntegralOperator}
Let $\Lambda : \R \rightarrow \R$ be an everywhere differentiable function with uniformly continuous derivative on its domain. Let $\mathcal{C} \subseteq \R$. Define the functional $\phi : \ell^\infty((0,1) \times \mathcal{C}) \rightarrow \ell^\infty(\mathcal{C})$ by
\[
	[\phi(f)](c) = \int_0^1 \Lambda[f(u,c)] \; du.
\]
Then $\phi$ is Fr\'echet differentiable, where
\[
	[\phi_f'(h)](c) = \int_0^1 \Lambda'[ f(u,c)] h(u,c) \; du
\]
is the Fr\'echet derivative of $\phi$ at $f$.
\end{lemma}

\begin{proof}[Proof of lemma \ref{lemma:derivativeOfIntegralOperator}]
This result is a modification of example 5 on page 174 of \cite{Luenberger1969}. We have
\begin{multline*}
	\left| [\phi(f+h)](c) - \big( [\phi(f)](c) + [\phi_f'(h)](c) \big) \right|
	\\
	= \left| \int_0^1 \left( \Lambda[ f(u,c) + h(u,c)] - \Lambda[f(u,c)] - \Lambda'[f(u,c)] h(u,c) \right) du \right|
\end{multline*}
By the usual mean value theorem,
\[
	\Lambda[ f(u,c) + h(u,c) ] - \Lambda[ f(u,c) ] = \Lambda' [ \bar{f}(u,c) ] h(u,c)
\]
where $| f(u,c) - \bar{f}(u,c) | \leq | h(u,c) |$. We apply this argument for each $u \in (0,1)$ and $c \in \mathcal{C}$.

Next, fix $\varepsilon > 0$. $\Lambda'(\cdot)$ is uniformly continuous by assumption. Hence there is a $\delta > 0$ such that for all $\tilde{h} \in \R$ with $| \tilde{h} | < \delta$ and all $\tilde{f} \in \R$,
\[
	| \Lambda' (\tilde{f}+\tilde{h}) - \Lambda' (\tilde{f}) | < \varepsilon.
\]
Therefore,
\begin{align*}
	&\sup_{c \in \mathcal{C}} \left| \int_0^1 \left( \Lambda[ f(u,c) + h(u,c)] - \Lambda[f(u,c)] + \Lambda'[f(u,c)] h(u,c) \right) du \right| \\
	= \; &\sup_{c \in \mathcal{C}} \left| \int_0^1 \Big( \Lambda'[ \bar{f}(u,c) ] - \Lambda'[f(u,c)] \Big) h(u,c) \; du \right| \\
	\leq \; &\sup_{c \in \mathcal{C}} \left| \int_0^1 \varepsilon h(u,c) \; du \right| \\
	\leq \; &\varepsilon \sup_{c \in \mathcal{C}} \sup_{u \in (0,1)} | h(u,c) | \\
	= \; &\varepsilon \| h \|_\infty.
\end{align*}
The first inequality holds for $\| h \|_\infty < \delta$. Thus
\[
	\frac{\| \phi(f+h) - \big( \phi(f) + \phi_f'(h) \big) \|_\infty}{\| h \|_\infty} \leq \varepsilon
\]
for $\| h \|_\infty < \delta$. Since $\varepsilon$ was arbitrary, this shows that the left hand side is $o(1)$.
\end{proof}

\begin{proof}[Proof of lemma \ref{lemma:smoothStep}]
We give the proof for part 1. Part 2 is analogous.
\begin{enumerate}

\item By construction, $0 \leq S_1(x) \leq 1$ on $(0,1)$.

\item By construction, $\text{ss}_\kappa^+(x)$ equals $\indicator(x \geq 0)$ everywhere except on $(0,1/\kappa)$. Thus we immediately obtain the desired pointwise convergence. Furthermore, note that
\[
	\int_\R | \text{ss}_\kappa^+(x) - \indicator(x \geq 0) | \; dx \leq \frac{1}{\kappa}
\]
which converges to zero as $\kappa \rightarrow \infty$. We use this property in lemma \ref{lemma:smoothPrerearrangement}.

\item This follows since $S_1 : \R \rightarrow \R$ is differentiable.
\end{enumerate}
$S_1'(x) = 6 x(1-x)$ is bounded by 3 on $x \in [0,1]$. For $x \notin [0,1]$, $S_1'(x) = 0$. Hence $S_1$ is Lipschitz. Therefore it is uniformly continuous. Thus both $\text{ss}_\kappa^+$ and $\text{ss}_\kappa^-$ are also uniformly continuous.
\end{proof}

\begin{proof}[Proof of lemma \ref{lemma:smoothInfimum}]
\hfill
\begin{enumerate}

\item This follows from
\begin{align*}
	\left( \int_{\mathcal{Y}_0} f(y,c)^p \; dx \right)^{1/p}
		&\leq \sup_{y \in \mathcal{Y}_0} f(y,c) \left( \int_{\mathcal{Y}_0} 1 \; dx \right)^{1/p} \\
		&\leq \| f(\cdot,c) \|_\infty (\overline{y} - \underline{y})^{1/p}.
\end{align*}

\item This follows, for example, from proposition 2.2 on page 8 of \cite{Stein2011}.

\item Define $\phi : \mathbb{D}_2 \rightarrow \ell^\infty(\mathcal{C})$ by $[\phi(f)](c) = \| -f(\cdot,c) \|_p$. $\psi_{2,p}$ is just a scaled version of this functional. Define $\psi : \mathbb{D}_2 \rightarrow \ell^\infty(\mathcal{C})$ by
\[
	[\psi(f)](c) = [\phi(f)](c)^p = \| -f(\cdot,c) \|_p^p = \int_0^1 [-f(u,c)]^p \; du.
\]
The last equality follows since $-f$ is nonnegative. By lemma \ref{lemma:derivativeOfIntegralOperator}, $\psi$ is Fr\'echet differentiable with Fr\'echet derivative
\[
	[\psi_f'(h)](c) 
	= - \int_0^1 p [-f(u,c)]^{p-1} h(u,c) \; du.
\]
Here we use uniform continuity of $x^p$ on the compact set $[-2,0]$. 

Let $\mathbb{D}_3$ denote the set of nonnegative functions in $\ell^\infty(\mathcal{C})$ with $L_p$-norm bounded away from zero. Note that the range of $\psi$ is contained in $\mathbb{D}_3$. Consider the functional $\theta : \mathbb{D}_3 \rightarrow \ell^\infty(\mathcal{C})$ defined by $[\theta(g)](c) = g(c)^{1/p}$. By arguments as in lemma \ref{lemma:pluginDerivative}, the Fr\'echet derivative of $\theta$ is
\[
	[\theta_g'(h)](c) = \frac{1}{p} g(c)^{1/p-1} h(c).
\]
Here we will use both the bounded range of our input functions $f$ and their $L_p$-norm bounded away from zero to ensure uniform continuity of $(1/p) x^{1/p - 1}$. Note that $[\phi(f)](c) = [\theta(\psi(f))](c)$. The result now follows by the chain rule, which further states that $[\phi_f'(h)](c) = [\theta_{\psi(f)}'( \psi_f'(h) )](c)$. Hence
\begin{align*}
	[\phi_f'(h)](c)
		&= \frac{1}{p} \left( \int_0^1 [-f(u,c)]^p \; du \right)^{1/p -1} (-1) \left( \int_0^1 p [-f(u,c)]^{p-1} h(u,c) \; du \right) \\
		&= -\frac{\| -f(\cdot,c) \|_p}{\int_0^1 [-f(u,c)]^p \; du} \int_0^1 [-f(u,c)]^{p-1} h(u,c) \; du.
\end{align*}
This derivative is not defined when $f$ has zero $L_p$-norm.
\end{enumerate}
\end{proof}

\begin{figure}[H]
\centering
\subfigure[$\underline{p}=0.1$. Binary coarsenings.]{\includegraphics[width=65mm]{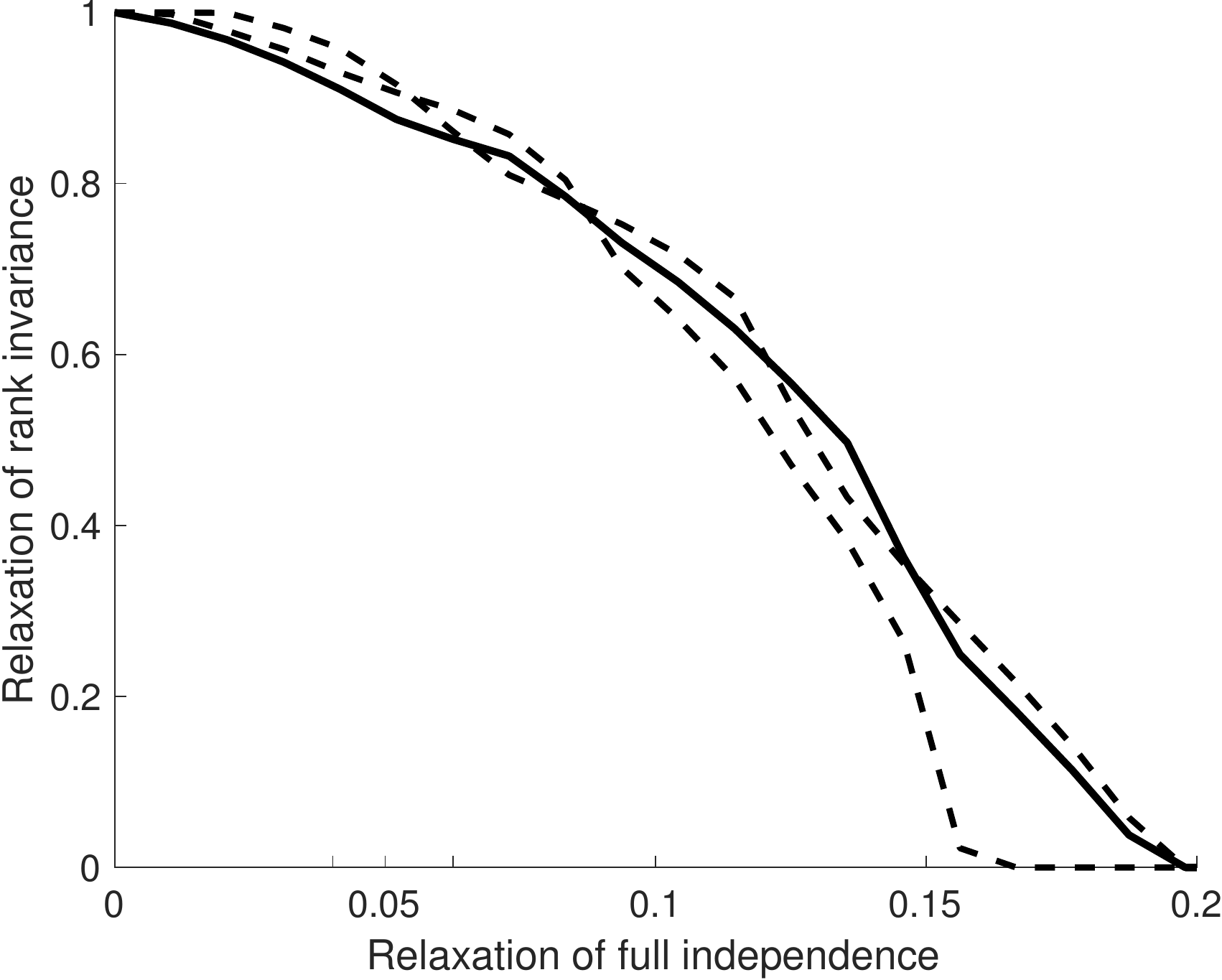}}
\hspace{.3in}
\subfigure[$\underline{p}=0.1$. Trinary coarsenings.]{ \includegraphics[width=65mm]{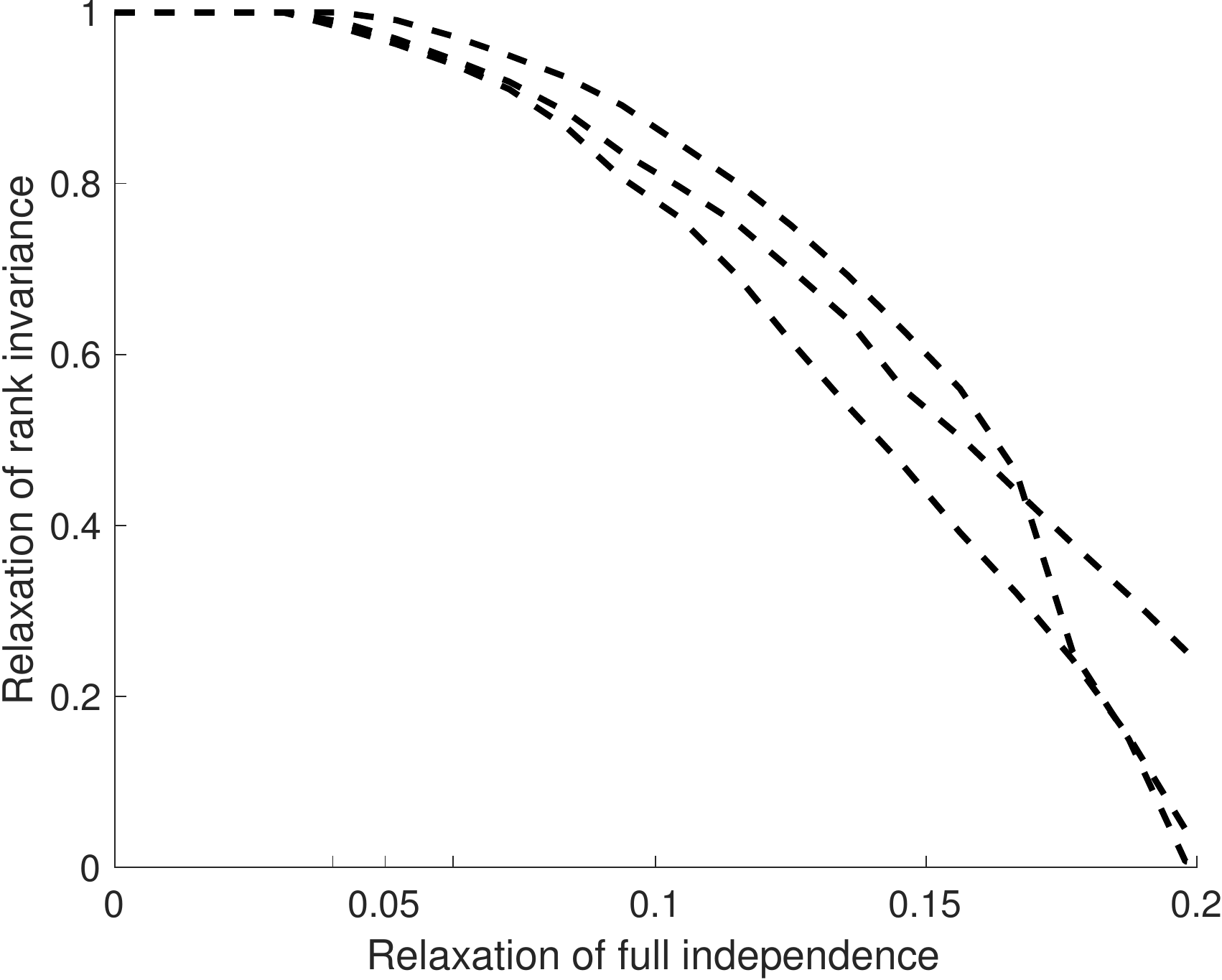}} \\ 
\vspace{.025in}
\subfigure[$\underline{p}=0.25$. Binary coarsenings.]{\includegraphics[width=65mm]{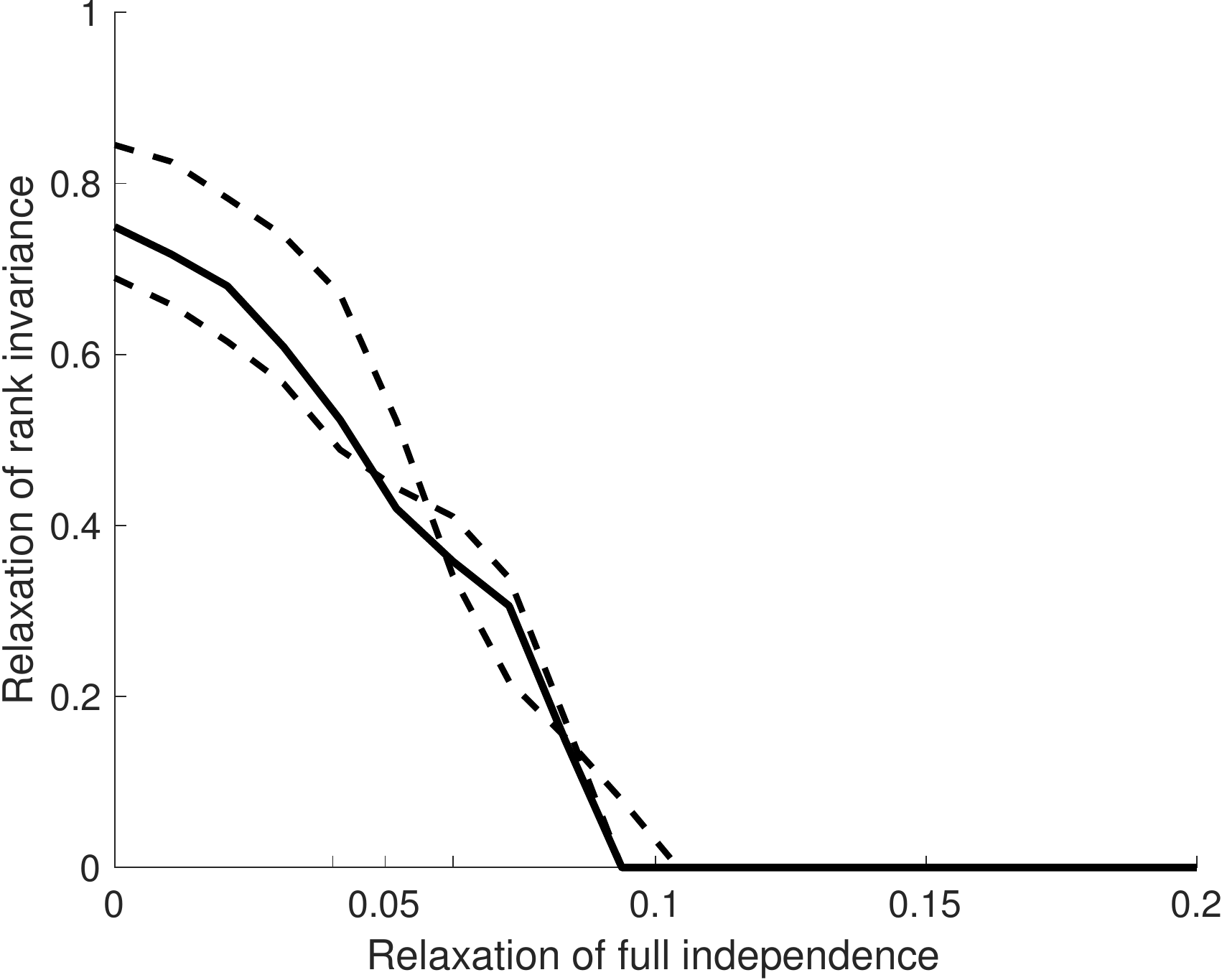}}
\hspace{.3in}
\subfigure[$\underline{p}=0.25$. Trinary coarsenings.]{ \includegraphics[width=65mm]{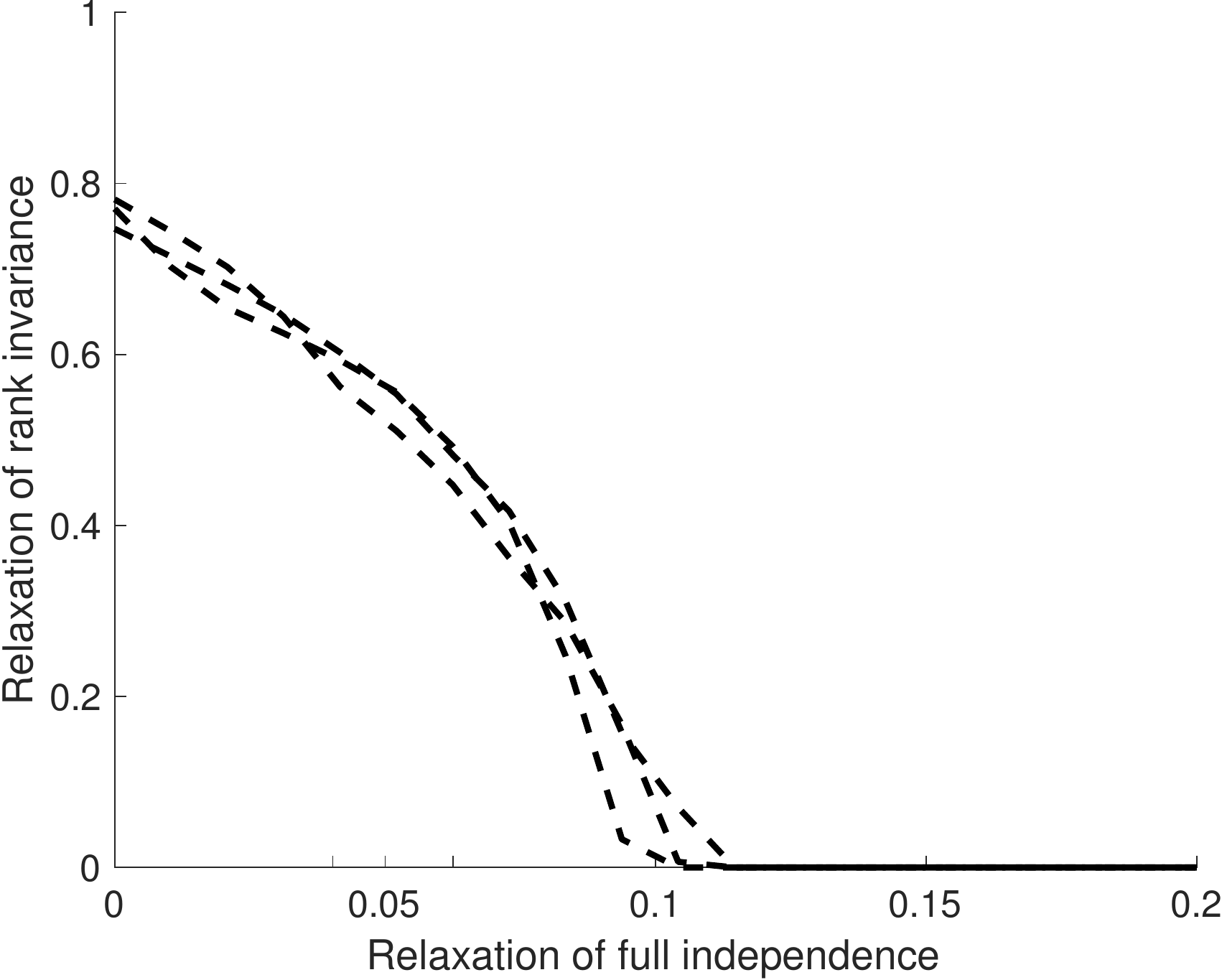}} \\ 
\vspace{.025in}
\subfigure[$\underline{p}=0.5$. Binary coarsenings.]{\includegraphics[width=65mm]{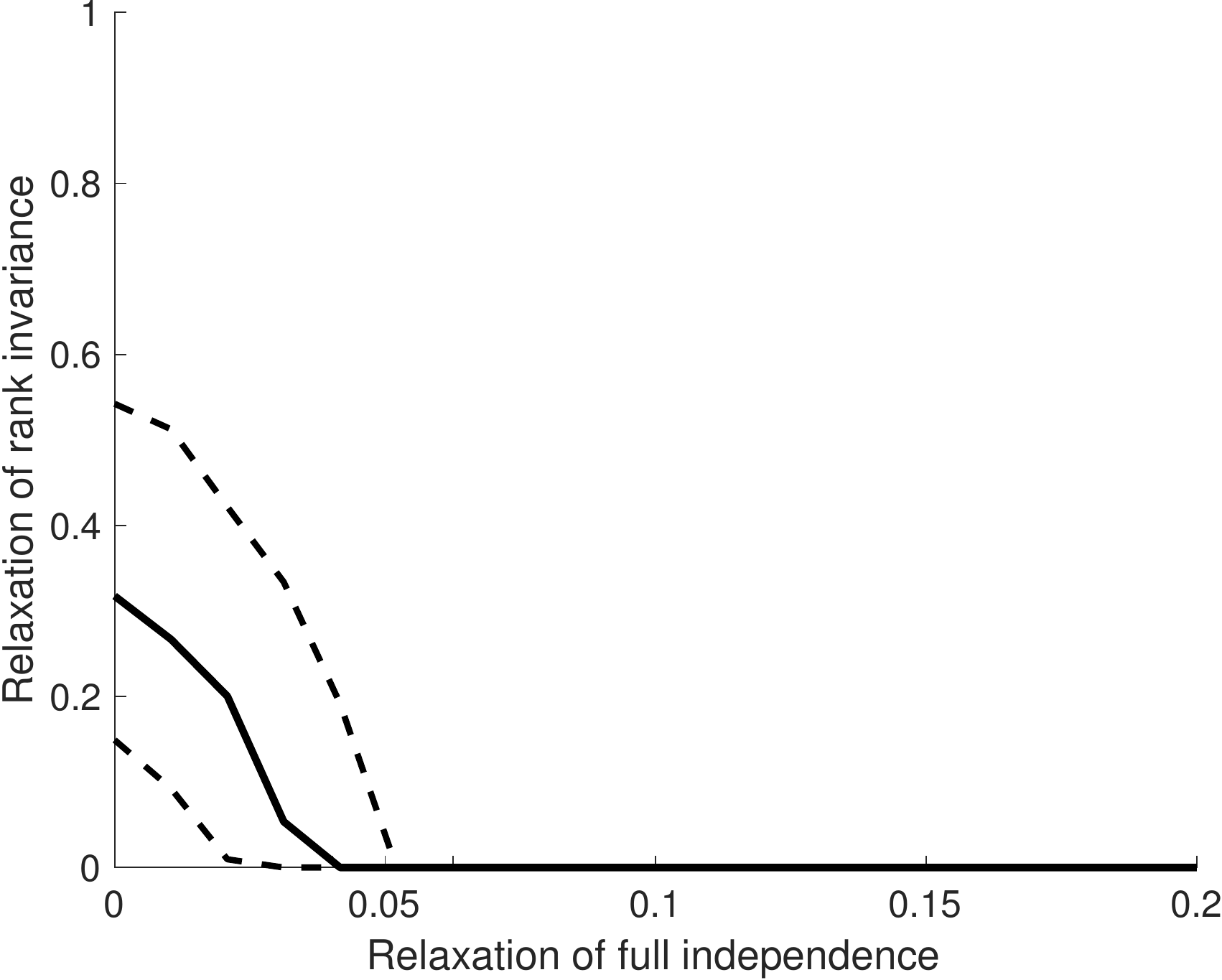}}
\hspace{.3in}
\subfigure[$\underline{p}=0.5$. Trinary coarsenings.]{ \includegraphics[width=65mm]{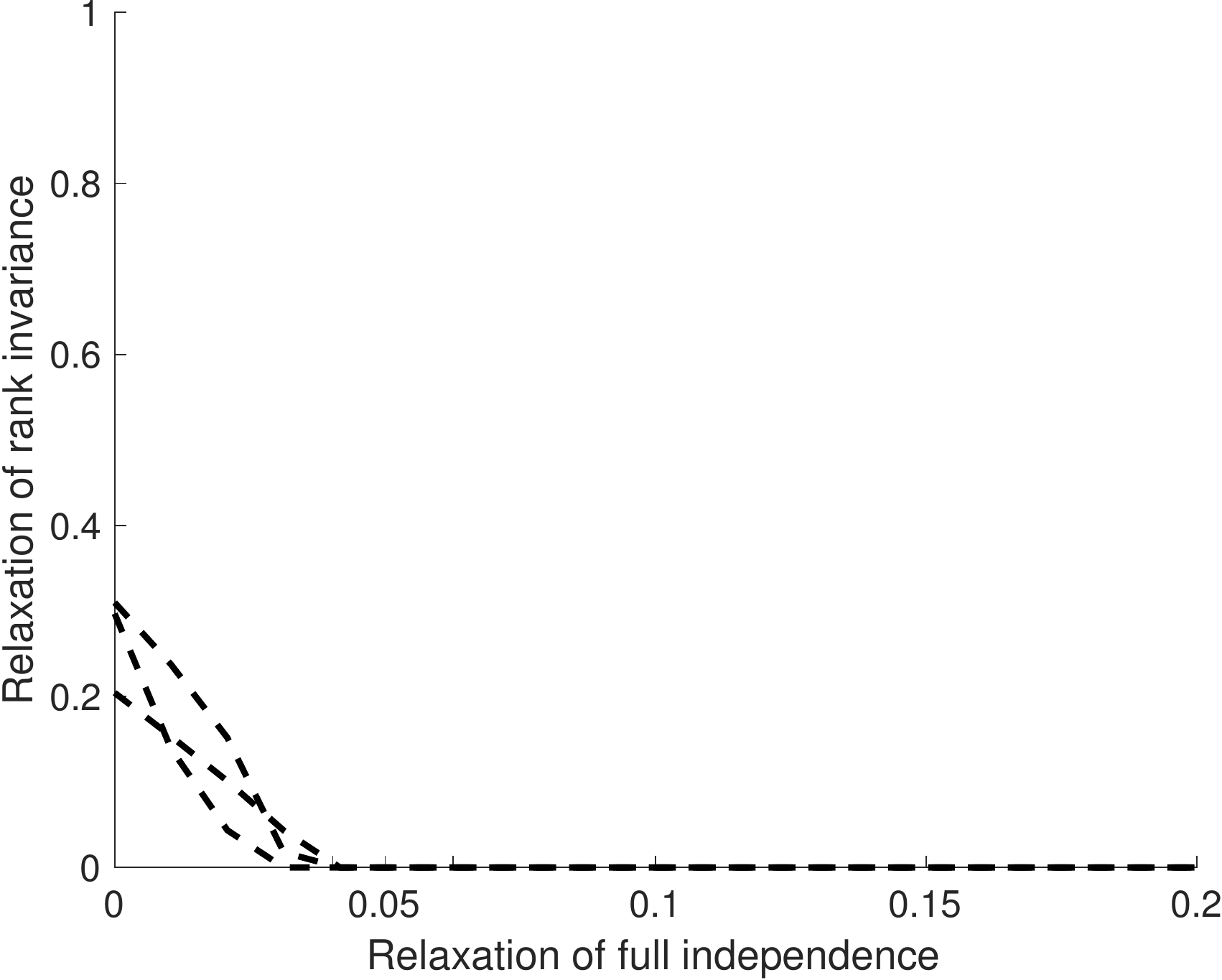}} \\ 
\caption{Estimated breakdown frontiers using various coarsenings of the conditioning variables. The solid bold line is our baseline estimate. The dotted lines represent the five alternative coarsenings we consider. \label{BF_coarsenings}}     
\end{figure}

\section{Additional empirical analyses}\label{sec:suppAppendixEmpirical}

In our empirical analysis of section 4 we collapse our two discrete covariates into two binary indicators of whether one is above or below the median value of the covariate. In this section we explore the impact of alternative coarsenings of these covariates on our empirical results. Specifically, we consider 6 different coarsenings total: 3 binary coarsenings and 3 trinary coarsenings. 

For the binary coarsenings, we collapse each covariate $W_k$ into a binary variable based on whether the value of the covariate is above or below the $\tau$th quantile $Q_{W_k}(\tau)$, for $\tau \in \{ 0.25, 0.5, 0.75 \}$. This includes our baseline case, $\tau = 0.5$. For the trinary coarsenings, we collapse each covariate $W_k$ into three bins: below $Q_{W_k}(\tau_1)$, between $Q_{W_k}(\tau_1)$ and $Q_{W_k}(\tau_2)$, and above $Q_{W_k}(\tau_2)$, where $Q_{W_k}(\tau)$ is the $\tau$th quantile of the covariate $W_k$. We use three choices of $(\tau_1,\tau_2)$: $(0.35, 0.65)$, $(0.30, 0.70)$, and $(0.25, 0.75)$. As discussed in section 4, given our overall sample size of 448 observations, using finer coarsenings (that is, more cells) or more asymmetric choices of $\tau$ yields cells with too few observations.

Figure \ref{BF_coarsenings} shows the estimated breakdown frontiers for each of the 6 different coarsenings, and for $\underline{p} \in \{ 0.1, 0.25, 0.5 \}$. The solid line is our baseline estimate---the binary coarsening split at the median. The dotted lines show our alternative coarsenings. Comparing the dotted lines with the solid line, we see that the estimated breakdown frontier generally does not change much as we vary how we coarsen the covariates. This is especially true for $\underline{p} = 0.1$ and $\underline{p} = 0.25$. Thus our overall conclusions regarding the robustness of claims about $\Prob(Y_1 > Y_0)$ do not appear to depend strongly on the specific choice of coarsening.

\singlespacing
\bibliographystyle{econometrica}
\bibliography{BF_paper}